\documentclass[aps,prd,10pt,superscriptaddress,nofootinbib,notitlepage]{revtex4-2}
\usepackage[utf8]{inputenc}
\usepackage{textcomp}
\usepackage[dvipsnames]{xcolor}
\usepackage{amssymb,amsfonts,amsmath}
\usepackage[sort&compress]{natbib}
\usepackage[english]{babel}
\usepackage{braket}
\usepackage{graphicx}
\usepackage{bm}
\usepackage{bbm}
\usepackage{braket}
\usepackage{gensymb}
\usepackage{dcolumn}
\usepackage{mathtools}
\usepackage{IEEEtrantools}
\usepackage{pstricks,pst-coil,pst-node,pst-circ,pst-plot,pst-3dplot,
pst-solides3d,pst-sigsys,pstricks-add,pst-eucl,pst-grad}
\usepackage[normalem]{ulem}
\allowdisplaybreaks[1]
\def\biggg#1{{\hbox{$\left#1\vbox to30.0pt{}\right.$}}}

\def\Biggg#1{{\hbox{$\left#1\vbox to45.0pt{}\right.$}}}

\DeclareMathOperator{\tr}{tr}
\usepackage{dcolumn}
\usepackage{lmodern}
\usepackage{microtype}
\usepackage{hyperref}
\hypersetup{
	breaklinks=true,
    colorlinks=true,
    linkcolor=Blue,      
    urlcolor=Blue,
    citecolor=Blue
}
\usepackage{uri}

\begin{document}

\title{\textit{S}-wave pion-pion scattering lengths from nucleon-meson fluctuations}

\author{J\"urgen Eser}
\email[]{juergen.eser@ipht.fr}
\affiliation{Universit\'e Paris-Saclay, CNRS, CEA, Institut de physique th\'eorique, 
91191, Gif-sur-Yvette, France}

\author{Jean-Paul Blaizot}
\email[]{jean-paul.blaizot@ipht.fr}
\affiliation{Universit\'e Paris-Saclay, CNRS, CEA, Institut de physique th\'eorique,
91191, Gif-sur-Yvette, France}

\date{\today}

\begin{abstract}
We present calculations of the $S$-wave isospin-zero and isospin-two pion-pion scattering 
lengths within a nucleon-meson model with parity doubling.\ Both scattering lengths are computed in 
various approximations, ranging from a mean-field (MF) calculation towards the inclusion of loop 
corrections by means of the functional renormalization group (FRG). The bosonic part of the 
investigated nucleon-meson model is formulated in terms of stereographic projections as a 
``natural'' set of coordinates on the respective vacuum manifold.\ We thereby elucidate 
subtleties concerning the truncation of the effective action w.r.t.\ higher-derivative pion 
interactions and the ``successful'' computation of the scattering lengths.\ As the main result, 
we find simultaneous agreement for the isospin-zero and isospin-two scattering lengths with 
experimental data within the $\mathrm{LPA}^{\prime}$-truncation of the FRG, together with 
chiral symmetry breaking (roughly) occurring at the characteristic scale of $4\pi f_{\pi}$. 
The isoscalar $\sigma$-mass is dynamically generated by the FRG integration of momentum modes,
and is a prediction of the model.\ It ends being of the order of $500\ \mathrm{MeV}$, i.e.,
much lower than the value ($> 1\ \mathrm{GeV}$) found in MF or one-loop treatment of this
or related models.\ Finally, the convergence of the corresponding low-energy expansion 
of the quantum effective action in terms of pion momenta is discussed.
\end{abstract}

\maketitle

\section{Introduction}

With only the two lightest (and idealized as mass-degenerate) quark flavors (``up'' and ``down''), 
the chiral $\mathsf{SU}(2) \times \mathsf{SU}(2)$-symmetry of quantum chromodynamics (QCD) 
is broken to an $\mathsf{SU}(2)$ flavor subgroup.\ This breaking occurs explicitly through 
the nonvanishing quark masses and spontaneously through the formation of a quark condensate.\ As 
the ultimate consequence, the color-singlet hadrons (approximately) appear as multiplets of the 
flavor subgroup, named isospin.\ Historically, the concept of isospin was first established 
from the observation of almost mass-degenerate protons and neutrons \cite{Heisenberg:1932dw, 
Wigner:1936dx, ParticleDataGroup:2020ssz}, whose quark contents are combinations of the up and 
down flavors.\ The proton and the neutron form the nucleon doublet, which is the fundamental 
representation of the $\mathsf{SU}(2)$ isospin group.\ They carry opposite projections of the 
isospin quantum number $I = 1/2$ (proton:\ $+1/2$; neutron:\ $-1/2$).

The pivotal principles of chiral and isospin symmetries governing the hadronic regime of the 
strong interaction give rise to effective models of QCD. These basically come in two distinct 
classes: linear and nonlinear models---referring to linear or nonlinear realizations of chiral
symmetry. Their respective stereotypes are given by the linear sigma model (LSM) 
\cite{Schwinger:1957em, Gell-Mann:1960mvl, Weinberg:1966fm} and the nonlinear sigma model 
(NLSM) \cite{Gursey:1959yy, Gell-Mann:1960mvl, Schwinger:1967tc, Weinberg:1968de, 
Zinn-Justin:1996khx}. The LSM incorporates all chiral partners (fields with the same 
quantum numbers except for parity), which appear in the hadronic multiplets, equally
as dynamical degrees of freedom (d.o.f.). The NLSM reduces the field content to the
(pseudo-)Nambu-Goldstone bosons of chiral symmetry breaking. It is the fundamental building block
of chiral perturbation theory (ChPT) \cite{Gasser:1983yg, Gasser:1984gg, Leutwyler:1993iq, 
Scherer:2002tk, Bijnens:2006zp, Leutwyler:2012sco}, which elevates itself as the effective field 
theory of low-energy QCD. In general, phenomenological Lagrangians obtained from these principles 
were seminally discussed in Refs.\ \cite{Weinberg:1968de, Coleman:1969sm, Callan:1969sn, 
Meetz:1969as, Isham:1969ci, Peczkis:1977xw}.

The research objective of the present work is the computation of the $S$-wave isospin-zero 
and isospin-two pion-pion scattering lengths within an effective nucleon-meson model. To
this end, we formulate the model in the framework of the quantum effective-action formalism,
with special emphasis on the integration of quantum fluctuations using the FRG technique
\cite{Wetterich:1992yh, Ellwanger:1993mw, Morris:1993qb, Pawlowski:2005xe}.\ We are interested
in the capability of this approach to reproduce experimental data on the scattering 
lengths, hence testing and exemplifying the infrared (IR) dynamics of the class of LSMs.
Our results are predictions of a non-fundamental hadronic model (with a linear realization 
of chiral symmetry). Nevertheless, they represent loop-corrected low-energy observables computed 
with a nonperturbative method.\ We systematically and conclusively compare the scattering lengths 
to (common) MF and one-loop calculations as well as to similar effective models \cite{Parganlija:2010fz, 
Parganlija:2012fy, Divotgey:2016pst, Lakaschus:2018rki}. 

Another focus lies on the relation of the nucleon-meson model to the corresponding NLSM, 
which readily emerges when restricting the dynamics of the system to the three-sphere 
$\mathsf{S}^{3}$ (that corresponds to the vacuum manifold) \cite{Bessis:1972sn, 
Jhung:1974fd, Appelquist:1980ae, Bentz:1997ry}---the regularization and perturbative renormalization 
of the NLSM with the help of the LSM (in this particular limit) was indeed proposed in the past by Refs.\ 
\cite{Bessis:1972sn, Brezin:1975sq, Brezin:1976ap, Brezin:1976qa}.\ To account for these geometric aspects 
along with the derivation of the scattering lengths, we express the bosonic part of the quantum effective 
action through stereographic projections as a ``natural'' set of coordinates on $\mathsf{S}^{3}$.\
We elucidate subtleties regarding higher-derivative interactions and their influence 
on the analytic and numeric results.\ The usage of stereographic coordinates for this 
objective originates from Refs.\ \cite{Eser:2018jqo, Divotgey:2019xea, Eser:2019pvd, Eser:2020phd, 
Divotgey:2020phd, Cichutek:2020bli}, which treat the dynamic generation of (stereographic) 
higher-derivative couplings in the $\mathsf{SO}(4)$- and $\mathsf{SO}(6)$-symmetric 
quark-meson(-diquark) models; this has the practical virtue of adopting the mathematical 
formulas and geometric knowledge developed in these publications.\ In the formulation of
stereographic coordinates, well-known low-energy relations from the NLSM are evident---yet 
other sets of coordinates are equally appropriate for the calculation of the 
coordinate-independent on-shell pion-pion scattering amplitude.

The FRG is a modern implementation of the Wilsonian renormalization group \cite{Wilson:1971bg, 
*Wilson:1971dh}. It has found diverse applications in high-energy and statistical 
physics (and other areas such as quantum gravity), see Ref.\ \cite{Dupuis:2020fhh} 
for a recent review. The method is well-suited to analyze the long-distance low-energy 
properties of theories or models of various kinds, interpolating from the ultraviolet (UV) 
classical action towards the quantum effective action in the IR. For instance, it offers
a straightforward description of QCD bound states via dynamical hadronization 
\cite{Ellwanger:1994wy, *Gies:2001nw, *Gies:2002hq, *Floerchinger:2009uf, *Mitter:2014wpa, 
*Braun:2014ata, *Cyrol:2017ewj, *Alkofer:2018guy, *Fu:2019hdw}.
Furthermore, the mapping of LSMs onto nonlinear realizations of chiral symmetry (within 
the FRG) was pioneered by Refs.\ \cite{Jungnickel:1997yu, Jendges:2006yk, Codello:2015oqa,
Eser:2018jqo, Divotgey:2019xea, Eser:2019pvd, Eser:2020phd, Divotgey:2020phd, Cichutek:2020bli}, 
where linearly realized quark-meson(-diquark) models were transformed into their nonlinear 
low-energy limits.\ The fixed-point structure and the geometry of the NLSM were thoroughly worked 
out by Refs.\ \cite{Codello:2008qq, Fabbrichesi:2010xy, Bazzocchi:2011vr, Flore:2012ma, Percacci:2013jpa, 
Eser:2020phd}, also underlining its relevance for gravitational physics. The consistent embedding 
of effective hadronic models into the FRG formalism and related developments were addressed 
and summarized in Ref.\ \cite{Braun:2009si, *Braun:2011pp, *Rennecke:2015eba, *Braun:2017srn, 
*Braun:2018bik, *Braun:2019aow, *Braun:2018svj, *Jung:2019nnr}; LSMs are oftentimes employed for
questions related to the QCD phase diagram \cite{Jungnickel:1995fp, *Berges:1997eu, 
*Berges:1998sd, *Schaefer:2004en, *Schaefer:2006sr, *Herbst:2010rf, *Herbst:2013ufa, 
*Mitter:2013fxa, *Pawlowski:2014zaa, *Eser:2015pka, *Jung:2016yxl, *Rennecke:2016tkm, 
*Fu:2016tey, *Tripolt:2017zgc, *Resch:2017vjs, *Otto:2019zjy, *Otto:2020hoz, *Jung:2021ipc}.

Effective nucleon-meson models, introduced during the ``pre-QCD era,'' were originally motivated 
by the ambition to describe the strong nuclear force between protons and neutrons.\ Within such 
models, the nuclear force is carried by light mesons, which are exchanged among nucleons, a 
picture that still holds in the framework of QCD for small momenta.\ The nucleon-meson model to 
be considered here takes up this idea of effective dynamics.\ It further features the concept of 
parity doubling \cite{Detar:1988kn, Hatsuda:1988mv, Jido:1998av, Jido:2001nt}, i.e., it comprises 
the explicit treatment of the chiral partners of the nucleons, where we assume that this applies
to the $N(1535)$-resonance \cite{ParticleDataGroup:2020ssz}.\ The parity doubling of nucleons, 
which is accompanied by the mirror assignment of Dirac spinors w.r.t.\ chiral symmetry, allows for 
a chiral-invariant ``bulk mass'' $m_{0}$ in the fermionic sector of the model. On the one hand, this 
$m_{0}$-parameter reflects the ``heterogeneous'' nature of the nucleon mass, since only a minor 
fraction comes from chiral symmetry breaking [besides the trace anomaly as well as intrinsic quark 
and gluon field energies, see e.g.\ Refs.\ \cite{Shifman:1978zn, Ji:1994av, Yang:2015uis, Yang:2018nqn}]. 
On the other side, it leads to nonzero nucleon masses even in the chiral-restored phase, where chiral 
partners become degenerate in mass.

Despite the fact that the parity-doublet model incorporates the nucleons in the ``nonstandard'' 
mirror assignment---and this property does not (necessarily) stand for example in the tradition of the 
early Walecka and Lee-Wick models for nuclear matter \cite{Walecka:1974qa, Lee:1974ma}---, it serves 
as an interesting starting point for the analysis of the pion-pion scattering amplitude (and represents 
the class of LSMs).\ The parity doubling promises (presumably) heavier fermion masses as compared 
to quark-meson or naively-assigned nucleon-meson models (in which the masses of quarks/nucleons 
do not ``survive'' chiral symmetry restoration).\ The suppression and massive decoupling of the 
fermionic fluctuations along decreasing energy-momentum scales is an important ingredient 
for the reproduction of hadronic low-energy observables \cite{Divotgey:2019xea, Cichutek:2020bli, 
Ellwanger:1994wy, *Gies:2001nw, *Gies:2002hq, *Floerchinger:2009uf, *Mitter:2014wpa, *Braun:2014ata, 
*Cyrol:2017ewj, *Alkofer:2018guy, *Fu:2019hdw}.\ This lets us (in the first instance) recede 
from analogous investigations in quark-meson models, whose low-energy couplings have proven 
to be dominated by the fermionic loop contributions even far in the IR \cite{Divotgey:2019xea}, 
and from similarly constructed nucleon-meson models without chiral-invariant nucleon mass.

The thermodynamics of the parity-doublet nucleon-meson model was extensively studied within 
the FRG approach in Refs.\ \cite{Weyrich:2015hha, Tripolt:2021jtp}. The corresponding phase
diagram shows a robust chiral phase transition inside dense nuclear matter, in addition to 
the nuclear liquid-gas transition. The mass parameter $m_{0}$ was chosen as $800\ \mathrm{MeV}$. 
This value principally matches the rather broad range of $200\ \mathrm{MeV} \lesssim m_{0} \lesssim 
900\ \mathrm{MeV}$ assumed in numerous phenomenological investigations, see e.g.\ Refs.\ \cite{Detar:1988kn, 
Jido:2001nt, Gallas:2009qp, Nemoto:1998um, *Bramon:2003xq, *Zschiesche:2006zj, *Wilms:2007uc, 
*Dexheimer:2007tn, *Dexheimer:2008cv, *Hayano:2008vn, *Sasaki:2010bp, *Sasaki:2011ff, *Giacosa:2011qd, 
*Gallas:2011qp, *Steinheimer:2011ea, *Paeng:2011hy, *Dexheimer:2012eu, *Gallas:2013ipa, *Heinz:2013hza, 
*Paeng:2013xya, *Benic:2015pia, *Motohiro:2015taa, *Olbrich:2015gln, *Mukherjee:2016nhb, *Mukherjee:2017jzi, 
*Suenaga:2017wbb, *Takeda:2017mrm, *Paeng:2017qvp, *Marczenko:2017huu, *Sasaki:2017glk,
*Marczenko:2018jui, *Takeda:2018ldi, *Yamazaki:2018stk, *Yamazaki:2019tuo, *Marczenko:2019trv, 
*Suenaga:2019urn, *Marczenko:2020jma, *Minamikawa:2020jfj, *Marczenko:2020omo, *Minamikawa:2021fln}.
Most often, slightly more precise values of $500\ \mathrm{MeV}$ to $800\ \mathrm{MeV}$ are 
discussed. Although the value of the chiral-invariant mass $m_{0}$ is lively debated in the 
current literature and directly affects the equation of state of dense nuclear matter, it is of 
subordinate importance for our purposes---albeit the computation of the scattering lengths as 
IR observables of the parity-doublet model lets us comment on it.\ Indications for parity 
doubling and nonzero nucleon masses beyond the chiral phase transition were also found in recent 
lattice-QCD calculations \cite{Glozman:2012fj, *Aarts:2015mma, *Aarts:2017rrl}.\ Generally, 
FRG applications to nuclear matter under extreme conditions of density (apart from the 
discussion of parity doubling) were carried out in Ref.\ \cite{Berges:1998ha, *Floerchinger:2012xd, 
*Drews:2013hha, *Drews:2014wba, *Drews:2014spa, *Drews:2016wpi, *Fejos:2017kpq, *Fejos:2018dyy, 
*Weise:2018ukn, *Leonhardt:2019fua, *Friman:2019ncm, *Brandes:2021pti}.

The theoretical reference of our investigations---concerning algebraic expressions and numeric
precision---is ChPT, which provides a systematic and model-independent analysis of hadronic 
correlation functions. Effective Lagrangians formulated in the context of ChPT are 
based on a nonlinear realization of chiral symmetry, leading at lowest order to exclusive 
pion self-interactions. The pions are the (pseudo-)Nambu-Goldstone bosons of spontaneous 
chiral symmetry breaking \cite{Nambu:1960xd, Goldstone:1961eq, Goldstone:1962es}. They dominate 
the pole structure of the hadronic correlation functions, which precisely makes them the most 
relevant d.o.f.\ in the low-energy regime of the strong interaction. The couplings of the 
pion self-interactions within effective Lagrangians are determined by the low-energy constants 
of QCD \cite{Gasser:1983yg, Gasser:1984gg, Bijnens:2014lea}. 

In the spectrum of physical particles, the neutral pion $\pi^{0}$ and the two charged pions 
$\pi^{\pm}$ form an isospin triplet, i.e., $I = 1$. The chiral partner of the three pions is 
the isoscalar $\sigma$-meson, which is eliminated in the nonlinear realization in favor of 
pionic interactions (whereas its presence is eponymous in the LSM). The light pion 
mass of $M_{\pi} \simeq 138\ \mathrm{MeV}$ \cite{ParticleDataGroup:2020ssz} (averaged
over $M_{\pi^{0}}$ and $M_{\pi^{\pm}}$) substantiates the special role of the pions as 
low-energy mediators of the strong interaction. 

The nonlinear effective field theory of pions is the result of the so-called chiral expansion, 
i.e., the simultaneous low-energy expansion of the generating functional of QCD in terms of pion 
momenta and quark masses. The generating functional contains mesonic currents as perturbations
to the massless QCD Lagrangian. In the sense of the external field method, this supplementation 
yields direct access to (mesonic) correlation functions. Moreover, these functions obey symmetry 
relations, which are known as the chiral Ward identities and which are equivalent to the invariance 
property of the generating functional under (local) chiral transformations \cite{Gasser:1983yg, 
Gasser:1984gg, Leutwyler:1993iq, Scherer:2002tk, Leutwyler:2012sco}.\footnote{Strictly speaking,
quantum anomalies are excluded from this statement.}

The chiral expansion relies on the small ratio $|p^2|/M^{2} < 1$, where $p$ denotes the pion 
momentum and $M$ indicates the mass scale such that (low-energy) propagators of the form 
$(p^2 - M^2)^{-1} \simeq -M^{-2} (1 + p^2/M^{2} + \cdots)$ converge \cite{Scherer:2002tk}. 
The parameter $M$ is identified with typical hadronic mass scales like the proton mass, the 
$\rho$-meson mass $M_{\rho} \simeq 775.26\ \mathrm{MeV}$ \cite{ParticleDataGroup:2020ssz}, or 
the characteristic chiral-symmetry breaking scale $4\pi f_{\pi}$ \cite{Manohar:1983md, 
Scherer:2002tk}. The latter follows from a ``naive'' dimensional analysis of loop diagrams 
contributing to the pion-pion scattering process. It involves the (weak) pion decay constant 
$f_{\pi} = (130.2 \pm 1.2)/\sqrt{2}$ \cite{ParticleDataGroup:2020ssz},\footnote{The additional 
factor of $1/\sqrt{2}$ is the convention we use in this work.} which itself is a measure for 
the strength of chiral symmetry breaking. The chiral expansion and the dynamic generation of 
low-energy couplings from renormalization-group flow equations were explored by Refs.\ 
\cite{Jungnickel:1997yu, Jendges:2006yk, Eser:2018jqo, Divotgey:2019xea, Eser:2019pvd, 
Cichutek:2020bli}; some of these ``preparatory'' works also promote the usage of 
stereographic coordinates.

An immediate application of ChPT is the computation of the $S$-wave pion-pion scattering lengths
\cite{Weinberg:1966kf, Bijnens:1995yn, Bijnens:1997vq, Ananthanarayan:2000ht, Colangelo:2000jc, 
Colangelo:2001df}. Regarding the $S$-wave (with angular momentum $l = 0$) in the partial-wave 
decomposition, the leading-order scattering amplitude is completely determined by the two scattering 
lengths $a_{0}^{0}$ and $a_{0}^{2}$ (with isospin $I = 0,2$; the subscript is given by the angular 
momentum $l$ and the superscript denotes the isospin $I$). Both scattering lengths are experimentally 
measured and theoretically predicted (beyond leading order) to high accuracy \cite{Leutwyler:2012sco},
merging into the values
\begin{equation}
	a_{0}^{0} = 0.2198 \pm 0.0126, \qquad
	a_{0}^{2} = -0.0445 \pm 0.0023 .
	\label{eq:scatteringlengths}
\end{equation}
At tree level, the scattering lengths reduce to the Weinberg limits (WLs) \cite{Weinberg:1966kf}
\begin{equation}
	a_{0}^{0} \stackrel{\mathrm{WL}}{=} 
	\left.\frac{2s - M_{\pi}^2}{32 \pi f_{\pi}^{2}}
	\right|_{s\, =\, 4M_{\pi}^{2}}
	= \frac{7 M_{\pi}^{2}}{32 \pi f_{\pi}^{2}}
	\simeq 0.1565, \qquad
	a_{0}^{2} \stackrel{\mathrm{WL}}{=} 
	- \left.\frac{s - 2 M_{\pi}^2}{32 \pi f_{\pi}^{2}}
	\right|_{s\, =\, 4M_{\pi}^{2}}
	= - \frac{M_{\pi}^{2}}{16 \pi f_{\pi}^{2}}
	\simeq -0.0447,
	\label{eq:Weinberg}
\end{equation}
which solely depend on the first Mandelstam variable $s$ (evaluated at threshold, $s = 4M_{\pi}^{2}$) 
as well as the two (geometric) constants $f_{\pi}$ and $M_{\pi}$. Thus the limit captures the 
isospin-two scattering length already fairly well, cf.\ Eq.\ (\ref{eq:scatteringlengths}).

As measure for chiral symmetry breaking, the pion decay constant $f_{\pi}$ parametrizes 
the vacuum-to-hadron transition amplitude 
\begin{equation}
	\bra{0}\mathcal{J}_{A}^{\mu,i}(x)\ket{\pi^{j}(p)} = - i\delta^{ij} f_{\pi}
	p^{\mu} e^{-i p \cdot x}, \label{eq:axialcurrent}
\end{equation}
which is encountered in the weak decay of the positively charged pion to an antimuon and a
muon-neutrino, $\pi^{+} \rightarrow \mu^{+} \nu_{\mu}$ \cite{Scherer:2002tk}. By virtue of
Eq.\ (\ref{eq:axialcurrent}), the axial-vector current $\smash{\mathcal{J}_{A}^{\mu,i}}$ creates 
on-shell pions out of the vacuum state [the indices $i,j$ are adjoint isospin indices; the
pion triplet $\pi^{i}$, $i = 1,2,3$, is the adjoint representation of the isospin symmetry 
and it is related to the physical (neutral and charged) pions $\pi^{0},\pi^{\pm}$ via complex 
linear combinations \cite{Greiner:1989eu}]. It is the algebraic complement to the (conserved) 
isospin current. For nonzero quark masses, we have $\smash{\partial_{\mu} \mathcal{J}^{\mu,i}_{A} 
\neq 0}$ \cite{Gell-Mann:1960mvl}.

In the LSM, the axial-vector current $\smash{\mathcal{J}^{\mu,i}_{A}}$ (as a function of the 
spacetime variable $x$) normally takes the form \cite{Gell-Mann:1960mvl, Koch:1997ei}
\begin{equation}
	\left.\mathcal{J}_{A}^{\mu,i}(x)\right|_{\mathrm{pions}}
	= \sigma_{0} \partial^{\mu} \pi^{i}(x)
	+ \mathcal{O}\!\left(\pi^{3}\right) . \label{eq:axialcurrent2}
\end{equation}
Using the matrix element (\ref{eq:axialcurrent}), this equation identifies (at leading order) 
the condensate $\sigma_{0}$ of the isoscalar $\sigma$-field with the pion decay constant, 
$\sigma_{0} \equiv f_{\pi}$, which we exploit in the following. The condensate, in turn, is 
the manifestation of chiral symmetry breaking within the model.

The computation of the scattering lengths $a_{0}^{0}$ and $a_{0}^{2}$ within LSMs was frequently 
discussed \cite{Sannino:1995ik, *Harada:1995dc, *Scadron:1999qd, *Lucio:1999ha, *Black:2000qq, 
*Scadron:2006mq, *Fariborz:2007km}, but in many cases ended rather inconclusive. Especially, the 
simultaneous correct prediction of the $S$-wave scattering lengths often fails \cite{Kramer:1969gw, 
Basdevant:1970nu, Geddes:1975cf, Geddes:1976qf, Aouissat:1994sx, Parganlija:2010fz, Soto:2011ap, 
Parganlija:2012fy, Divotgey:2016pst, Lakaschus:2018rki}, where some of the considered models are extended 
with (axial-)vector mesons or ``exotic'' states like tetraquarks and glueballs (as well as include or 
neglect baryonic d.o.f.). This shortcoming can be attributed to the large isoscalar mass 
$M_{\sigma}$ favored e.g.\ in Refs.\ \cite{Parganlija:2010fz, Parganlija:2012fy, Lakaschus:2018rki} 
($M_{\sigma} \sim 1370\ \mathrm{MeV}$, in contrast to $M_{\sigma} \sim 500\ \mathrm{MeV}$ 
typically needed for a successful reproduction of the isoscalar channel within LSMs).\ This 
disagreement coming with such a heavy scalar resonance is corroborated by the introduction of vector 
fields, as shown in Refs.\ \cite{Fariborz:2009wf, Divotgey:2016pst}. The vector fields shift 
the $\sigma$-mass required for reproducing the isoscalar scattering length to even smaller 
values.\ Some of the cited publications then proceed to the inverse and infer the isoscalar mass 
from a fitting procedure of model parameters to the $S$-wave scattering lengths 
(and further data, for instance, mesonic decay constants and decay widths) \cite{Fariborz:2009wf, 
Parganlija:2010fz}. Others adjust (or choose) the mass ``by hand,'' exclusively for the purpose 
of obtaining reasonable results for the pion-pion scattering lengths \cite{vanBeveren:2002mc, 
Black:2009bi}. Equal sensitivity to the isoscalar mass was also observed for pion-nucleon scattering, 
see e.g.\ Ref.\ \cite{Gallas:2009qp}.

Of great importance for the following Sections is the analysis of the pion-pion scattering process 
in Ref.\ \cite{Divotgey:2016pst}, which (partly) stimulated our work; it is a study of the low-energy 
limit of an extended LSM (with vector and axial-vector mesons). In this limit, the corresponding low-energy 
couplings and the pion-pion scattering amplitude were computed by a tree-level integration of the scalar 
and vector resonances based on the classical action.\ Due to the favored large $\sigma$-mass of the 
model and the preferable identification of the $\sigma$-meson with the $f_{0}(1370)$-resonance, the 
obtained $S$-wave scattering lengths $a_{0}^{0}$ and $a_{0}^{2}$ were not compatible with experimental 
data---inviting the consideration of loop corrections.

In contrast to the aforementioned investigations, we will demonstrate throughout this paper 
that the present approach simultaneously meets (within errorbars) both scattering 
lengths---which is verifiably not achieved in numerous other linear model calculations (as stated 
in the previous two paragraphs).\ The isoscalar $\sigma$-mass, which is a model prediction (together
with the scattering lengths), will end up comparably light when including loop corrections (from values
of $M_{\sigma} > 1\ \mathrm{GeV}$ within the MF and one-loop approximations towards $M_{\sigma} \sim 
500\ \mathrm{MeV}$ within the FRG computations).\ This finding suggests an identification of the
$\sigma$-meson with the $f_{0}(500)$-resonance. The $\sigma$-mass is derived from the FRG integration 
of quantum fluctuations within the nucleon-meson model; its value as well as those of the pion-pion 
scattering lengths are therefore dynamically generated and intrinsically predicted by the initialization 
of the integration procedure, which we tune such that chiral symmetry breaking sets in at $\Lambda_{\chi} 
= 1.2\ \mathrm{GeV} \simeq 4 \pi f_{\pi}$. Hence, the layout of argumentation of this work will have the 
structure as listed below:
\begin{equation*}
	\begin{aligned}
	\text{Model predictions:} \quad & \text{$a_{0}^{0}$, $a_{0}^{2}$, and $M_{\sigma}$} \\[0.2cm]
	\text{Model parameters (input):} \quad & \text{$f_{\pi}$, $M_{\pi}$, 
	$M_{\mathrm{fermions}}$, and chiral symmetry breaking at $\Lambda_{\chi} 
	\simeq 4\pi f_{\pi}$}\\[0.2cm]
	\text{Findings:} \quad & 
	\begin{cases}
	\text{FRG: $a_{0}^{0}$ and $a_{0}^{2}$ successfully simultaneously reproduced (in the
	$\mathrm{LPA}^{\prime}$-truncation);}\\ 
	\text{$M_{\sigma} \sim 500\ \mathrm{MeV}$, $\sigma$-meson identified as the 
	$f_{0}(500)$-resonance.}\\[0.2cm]
	\text{MF and one-loop: $a_{0}^{0}$ and $a_{0}^{2}$ not simultaneously reproduced;}\\
	\text{$M_{\sigma} > 1\ \mathrm{GeV}$, $\sigma$-meson rather associated with isoscalar
	resonances above $1\ \mathrm{GeV}$}.
	\end{cases}
	\end{aligned}
\end{equation*}

The remainder of this manuscript is organized as follows: Sec.\ \ref{sec:model} is dedicated to 
the (parity-doublet) nucleon-meson model and its formulation in terms of the effective 
average action. The subsequent Sec.\ \ref{sec:scattering} deals with the theory of pion-pion 
scattering. The numeric results are presented in Sec.\ \ref{sec:results} and a conclusion is 
given in Sec.\ \ref{sec:summary}. Lastly, some technical details are summarized in Appendices 
\ref{sec:modelconstruction} to \ref{sec:LECs}. We use natural units, $\hbar = c = 1$; Lorentz
indices in Euclidean spacetime (denoted as $\mathbb{E}^{4}$) appear as ``lower'' indices only.

\section{Nucleon-meson model}
\label{sec:model}

The central object of the FRG is the effective average action $\Gamma_{k}$.\footnote{In
the following, we refer to the ``effective average action'' simply as the ``effective action,''
and similarly for the ``quantum effective action.''} This quantity
interpolates between the classical action $S$ in the UV (with cutoff $\Lambda$) and the 
quantum effective action $\Gamma$ in the IR ($\Gamma_{k\, =\, \Lambda} = S$ and 
$\Gamma_{k\, =\, 0} = \Gamma$). The index $k$ symbolizes a sliding IR cutoff, separating 
low-energy from high-energy fluctuations in the Wilsonian manner. The $k$-dependence of 
$\Gamma_{k}$ is dictated by the Wetterich equation \cite{Wetterich:1992yh}, \vspace{-0.15cm}
\begin{equation}
	\partial_{k} \Gamma_{k} = \frac{1}{2} \tr \left[
	\left(\Gamma_{k}^{(2)} + R_{k}\right)^{-1}\partial_{k} R_{k} \right] 
	= \frac{1}{2} \! \!
	\vcenter{\hbox{
	\begin{pspicture}[showgrid=false](1.5,1.8)
		\psarc[linewidth=0.025](0.75,0.9){0.6}{115}{65}
		\pscircle[linewidth=0.03](0.75,1.5){0.25}
		\psline[linewidth=0.03](0.75,1.5)(0.92,1.67)
		\psline[linewidth=0.03](0.75,1.5)(0.58,1.67)
		\psline[linewidth=0.03](0.75,1.5)(0.58,1.33)
		\psline[linewidth=0.03](0.75,1.5)(0.92,1.33)
	\end{pspicture}
	}} \! \! ,
	\label{eq:Wetterich}
\end{equation}
where the superscript ``(2)'' indicates the second functional derivative w.r.t.\ the field 
d.o.f. The regulator function $R_{k}$ introduces the cutoff scale $k$. It typically operates 
as an additional mass contribution for the low-energy modes. From Eq.\ (\ref{eq:Wetterich}) 
arises an infinite set of differential equations, which is truncated by choosing a specific
ansatz for $\Gamma_{k}$.\ More technical aspects (like the regulator functions)
are deferred to Appendix \ref{sec:FRG}.

In Eq.\ (\ref{eq:Wetterich}), we made use of the diagrammatic Feynman representation \vspace{-0.2cm}
\begin{equation}
	\vcenter{\hbox{
	\begin{pspicture}(-0.6,-0.05)(0.6,0.05)
		\psline[linewidth=0.025](-0.6,0)(0.6,0)
	\end{pspicture}
	}} \!
	= \left(\Gamma_{k}^{(2)} + R_{k}\right)^{-1}, \qquad
	\vcenter{\hbox{
	\begin{pspicture}(-0.4,-0.3)(0.4,0.3)
		\psline[linewidth=0.025](-0.4,0)(0.4,0)
		\pscircle[linewidth=0.03,fillstyle=solid,
		fillcolor=white](0,0){0.25}
		\psline[linewidth=0.03](0,0)(0.17,0.17)
		\psline[linewidth=0.03](0,0)(-0.17,0.17)
		\psline[linewidth=0.03](0,0)(-0.17,-0.17)
		\psline[linewidth=0.03](0,0)(0.17,-0.17)
	\end{pspicture}
	}} \!
	= \partial_{k} R_{k}, \qquad
	\vcenter{\hbox{
	\begin{pspicture}[showgrid=false](-0.6,-0.6)(0.6,0.6)
		\psline[linewidth=0.03](0,0)(-0.35,0)
		\psline[linewidth=0.03](0,0)(-0.25,0.25)
		\psline[linewidth=0.03](0,0)(0.35,0)
		\pscircle[linewidth=0.03,fillstyle=solid,
		fillcolor=lightgray!25](0,0){0.20}
		\rput{*0}(0.005,-0.005){\fontsize{7pt}{0pt}{$n$}\selectfont}
		\psdots[dotsize=1.2pt](0,0.3)(0.102606,0.281908)(0.192836,0.229813)
		\rput(-0.5,0.015){\fontsize{7pt}{0pt}{$1$}\selectfont}
		\rput(-0.38,0.38){\fontsize{7pt}{0pt}{$2$}\selectfont}
		\rput(0.5,-0.005){\fontsize{7pt}{0pt}{$n$}\selectfont}
	\end{pspicture}
	}} \! \! 
	= \Gamma_{k}^{(n)} \vspace{-0.3cm}
\end{equation}
of the propagator, the regulator insertion, and the dressed $n$-point vertex functions 
(relevant for later equations). The trace operator leads to the loop structure of the
Wetterich equation (\ref{eq:Wetterich}).

\subsection{Effective action}

The effective action $\Gamma_{k}$ of the parity-doublet nucleon-meson 
model reads 
\begin{IEEEeqnarray}{rCl}
	\Gamma_{k}\left[\varphi,\bar{\psi}_{1},\psi_{1},
	\bar{\psi}_{2},\psi_{2}\right] & = & \int_{x} \bigg\lbrace 
	\frac{1}{2} Z_{k} \left(\partial_{\mu}\varphi\right) \cdot \partial_{\mu}\varphi
	+ V_{k}\!\left(\varphi^{2}\right) - h \sigma
	+ \bar{\psi}_{1} \left[Z_{k}^{\psi}\gamma_{\mu}\partial_{\mu} 
	+ y_{1,k} \left(\sigma + i \gamma_{5} \vec{\pi} \cdot \vec{\tau} \right)\right] \psi_{1}
	\nonumber\\[0.1cm] & & \qquad
	+\, \bar{\psi}_{2} \left[Z_{k}^{\psi}\gamma_{\mu}\partial_{\mu} 
	- y_{2,k} \left(\sigma - i \gamma_{5} \vec{\pi} \cdot \vec{\tau} \right)\right] \psi_{2}
	+ m_{0,k} \left(\bar{\psi}_{1}\psi_{2} + \bar{\psi}_{2}\psi_{1}\right)
	\bigg\rbrace ,
	\label{eq:truncation}
\end{IEEEeqnarray}
with the short-hand notation $\smash{\int_{x} = \int \mathrm{d}^{4}x}$ for the (Euclidean)
spacetime integration. The two Dirac fermions $\smash{ (\bar{\psi}_{i},\psi_{i})}$, 
$i = 1,2$, represent a pair of ``mirror-assigned'' flavor doublets of the chiral 
$\mathsf{SU}(2) \times \mathsf{SU}(2)$-symmetry [and its diagonal $\mathsf{SU}(2)$ isospin 
subgroup], cf.\ Appendix \ref{sec:modelconstruction}. Both are defined to be parity-even. The 
bosonic d.o.f.\ are collected in the irreducible (linear) representation $(2,2)$ of chiral symmetry, 
$\varphi = (\vec{\pi}, \sigma) = (\pi^{1}, \pi^{2}, \pi^{3}, \sigma)$. The three Pauli matrices 
are denoted as $\vec{\tau}$.

Equation (\ref{eq:truncation}) constitutes an $\mathrm{LPA}^{\prime}$-ansatz, that includes a 
scale-dependent potential $V_{k}$, Yukawa couplings $y_{1,k}$ and $y_{2,k}$, the 
chiral-invariant mass $m_{0,k}$, as well as bosonic and fermionic (field-independent)
wave-function renormalization factors $Z_{k}$ and $\smash{Z_{k}^{\psi}}$, respectively.\ 
The parameter $m_{0,k} \neq 0$ generates a nonvanishing fermion mass also in the chiral-restored 
phase.\ Moreover, the nucleon-meson model exhibits explicit chiral symmetry breaking through the
term proportional to $h \neq 0$, which accounts for nonzero quark masses (and a physical pion mass). 
Setting $\smash{Z_{k} = Z_{k}^{\psi} = 1}$ and neglecting the scale dependence of $m_{0,k}$, $y_{1,k}$, 
and $y_{2,k}$, we call truncation (\ref{eq:truncation}) the LPA (local potential approximation). 
For the effective potential $V_{k}\!\left(\varphi^{2}\right)$, we use a Taylor polynomial 
of order $N_{\alpha} = 6$ (in the invariant $\smash{\varphi^{2}} = \smash{\vec{\pi}^{\;\! 2} 
+ \sigma^{2}}$),
\begin{equation}
	V_{k}\!\left(\varphi^{2}\right) = \sum_{n\, =\, 1}^{N_{\alpha}} 
	\frac{\alpha_{n,k}}{n!} \left(\varphi^{2} - \varphi_{0}^{2}\right)^{n}, \qquad
	\alpha_{n,k} = V_{k}^{(n)}\!\left(\varphi_{0}^{2}\right),
	\label{eq:potential}
\end{equation}
with the coefficients $\alpha_{n,k}$, $n = 1, \ldots, N_{\alpha}$, and the scale-independent
expansion point $\varphi_{0}^{2}$ (to be chosen slightly larger than the IR minimum, i.e., 
$\varphi_{0}^{2} \gtrsim f_{\pi}^{2}$). Pursuant to Refs.\ \cite{Divotgey:2019xea, 
Cichutek:2020bli, Pawlowski:2014zaa}, this truncation goes beyond ordinary ``tree-level'' 
potentials with quartic field interactions ($N_{\alpha} = 2$).\ The (relative) numeric 
robustness against variations in $N_{\alpha}$ is discussed in Appendix \ref{sec:FRG}.

Upon condensation of the isoscalar $\sigma$-field ($\sigma \rightarrow \sigma_{0} + \sigma$) 
and the assumption that $m_{0} \neq 0$,\footnote{Henceforth, we frequently leave out the subscript 
``$k$'' for a more compact notation; the $k$-dependence is explicitly given in Eqs.\ (\ref{eq:truncation})
and (\ref{eq:potential}). It should be clear from the context, if we refer to the respective quantity 
in the IR limit, $k \rightarrow 0$, or not.} the fermion fields---as defined above---are entangled 
through the nondiagonal mass matrix
\begin{equation}
	\mathfrak{M}_{\psi} = \begin{pmatrix}
	y_{1} \sigma_{0} & m_{0} \\
	m_{0} & -y_{2} \sigma_{0} 
	\end{pmatrix} .
	\label{eq:massunphys}
\end{equation}
Hence, they are connected through a nondiagonal propagator structure. The fermions are
disentangled to the physical parity-opposite basis, $(\psi_{1},\psi_{2}) \rightarrow 
(N_{+},N_{-})$, by applying a parity transformation in addition to an orthogonal rotation, 
such that
\begin{equation}
	\mathfrak{M}_{N} = \begin{pmatrix}
	m^{+} & 0 \\
	0 & m^{-}
	\end{pmatrix} , \qquad
	m^{\pm} = \frac{1}{2} \left[\pm \sigma_{0} \left(y_{1} - y_{2}\right)
	+ \sqrt{\sigma_{0}^{2} (y_{1} + y_{2})^{2} +4 m_{0}^{2}}\right].
	\label{eq:massphys}
\end{equation}
For the details of this transformation, we refer again to Appendix 
\ref{sec:modelconstruction}. In the physical basis, the fermionic part of 
Eq.\ (\ref{eq:truncation}) becomes
\begin{IEEEeqnarray}{rCl}
	\left. \Gamma\!\left[\varphi,\bar{\psi}_{1},\psi_{1},
	\bar{\psi}_{2},\psi_{2}\right]\right|_{\mathrm{fermions}}
	& \longrightarrow & \left. \Gamma\!\left[\varphi,\bar{N}_{+},N_{+},
	\bar{N}_{-},N_{-}\right]\right|_{\mathrm{fermions}} \nonumber\\[0.2cm]
	& & \equiv \int_{x} \bigg\lbrace
	\bar{N}_{+} \left(Z^{\psi}\gamma_{\mu}\partial_{\mu} + m^{+} \right) N_{+}
	+ \bar{N}_{-} \left(Z^{\psi}\gamma_{\mu}\partial_{\mu} + m^{-} \right) N_{-}
	\nonumber\\
	& & \qquad\quad +\, \bar{N}_{+} \left(y_{\sigma}^{++} \sigma 
	+ i y_{\pi}^{++} \gamma_{5} \vec{\pi} \cdot \vec{\tau} \right) N_{+}
	+ \bar{N}_{-} \left(y_{\sigma}^{--} \sigma 
	+ i y_{\pi}^{--} \gamma_{5} \vec{\pi} \cdot \vec{\tau} \right) N_{-}
	\nonumber\\
	& & \qquad\quad -\, \bar{N}_{+} \left(y_{\sigma}^{+-} \sigma 
	+ i y_{\pi}^{+-} \gamma_{5} \vec{\pi} \cdot \vec{\tau} \right) N_{-}
	+ \bar{N}_{-} \left(y_{\sigma}^{-+} \sigma 
	+ i y_{\pi}^{-+} \gamma_{5} \vec{\pi} \cdot \vec{\tau} \right) N_{+}
	\bigg\rbrace . \qquad \label{eq:fermions}
\end{IEEEeqnarray}
The fermion fields $N_{+}$ and $N_{-}$ represent the parity-even isospin doublet of protons 
and neutrons and the parity-odd chiral partners of the nucleons, respectively. Together, they
form the parity doublet $N = (N_{+}, N_{-})$. In Eq.\ (\ref{eq:fermions}), the Yukawa interactions 
with the bosonic d.o.f.\ evidently split,
\begin{IEEEeqnarray}{rCl}
	y_{\sigma}^{++/--} & = & \frac{1}{2}\left[\pm (y_{1} - y_{2})
	+ \frac{\sigma_{0} (y_{1} + y_{2})^{2}}
	{\sqrt{\sigma_{0}^{2} (y_{1} + y_{2})^{2} +4 m_{0\vphantom{,k}}^{2}}}\right], 
	\label{eq:yukawa1} \\[0.3cm]
	y_{\pi}^{++/--} & = & \frac{1}{2}\left[\pm (y_{1} + y_{2})
	+ \frac{\sigma_{0} \bigl(y_{1}^{2} - y_{2}^{2}\bigr)}
	{\sqrt{\sigma_{0}^{2} (y_{1} + y_{2})^{2} +4 m_{0\vphantom{,k}}^{2}}}\right], 
	\label{eq:yukawa2} \\[0.3cm]
	y_{\sigma/\pi}^{+-} \equiv y_{\sigma/\pi}^{-+} & = & \frac{m_{0} (y_{1} \pm y_{2})}
	{\sqrt{\sigma_{0}^{2} (y_{1} + y_{2})^{2} +4 m_{0\vphantom{,k}}^{2}}}.
	\label{eq:yukawa3}
\end{IEEEeqnarray}
In the limit $m_{0} \rightarrow 0$, the fermion masses $m^{\pm}$ consistently reduce to $m^{+} = 
y_{1}\sigma_{0}$ and $m^{-} = y_{2}\sigma_{0}$, i.e., they are solely generated by spontaneous chiral 
symmetry breaking; the Yukawa couplings of Eqs.\ (\ref{eq:yukawa1}), (\ref{eq:yukawa2}), 
and (\ref{eq:yukawa3}) simplify to $\smash{y_{\sigma}^{++} = y_{\pi}^{++} = y_{1}}$, 
$\smash{y_{\sigma}^{--} = - y_{\pi}^{--} = y_{2}}$, and $\smash{y_{\sigma}^{+-} = 
y_{\pi}^{+-} = 0}$. Thus in summary, the fermions decouple and become massless in the 
chiral-restored phase, $\sigma_{0} \rightarrow 0$.\ In contrast, for $\sigma_{0} \rightarrow 0$ 
but nonzero chiral-invariant mass $m_{0} \neq 0$, one finds that $m^{\pm} = m_{0}$, $y_{\sigma}^{++} 
= y_{\pi}^{+-} = \smash{- y_{\sigma}^{--} = (y_{1} - y_{2})/2}$, and $\smash{y_{\pi}^{++} = 
y_{\sigma}^{+-} = - y_{\pi}^{--} = (y_{1} + y_{2})/2}$.

The physical observables, which are computed from the effective actions (\ref{eq:truncation}) 
and (\ref{eq:fermions}) in the IR limit, $\Gamma \equiv \Gamma_{k\, =\, 0}$, are directly 
related to the renormalization-group-invariant vertex functions of the generic form
\begin{equation}
	\left.
	\frac{\delta^{n}\Gamma_{k}\!\left[\phi\right]}{\delta\phi_{1} \cdots \delta\phi_{n}} 
	\prod_{i\, =\, 1}^{n} \left(Z_{k}^{\phi_{i}}\right)^{- 1/2}
	\right|_{k\, =\, 0} 
	\eqqcolon \left.
	\frac{\delta^{n}\tilde{\Gamma}_{k}\!\left[\tilde{\phi}\right]}
	{\rule[2ex]{0pt}{0.5ex} \delta\tilde{\phi}_{1} \cdots \delta\tilde{\phi}_{n}}
	\right|_{k\, =\, 0} , \qquad
	n \in \mathbb{N}. \label{eq:npointrenorm}
\end{equation}
The object $\phi_{i}$ symbolizes any field component (with all its different vector-space 
indices collected in the index $i$) of the model at hand and $\smash{Z_{k}^{\phi_{i}}}$ 
is the corresponding wave-function renormalization.\ This rescaling of vertex functions 
through the $\smash{Z_{k}^{\phi_{i}}}$-factors is absorbed into the renormalized fields 
$\smash{\tilde{\phi}_{i} \coloneqq (Z_{k}^{\phi_{i}})^{1/2}\;\! \phi_{i}}$, as shown on the 
right-hand side of the definition (\ref{eq:npointrenorm}).\ Furthermore, the effective
action $\tilde{\Gamma}_{k}$ is obtained by replacing the field $\phi$ accordingly.\ The physical 
squared boson and fermion masses, which stem from the two-point functions, are then given by 
the following expressions:
\begin{equation}
	M_{\pi}^{2} \equiv \frac{m_{\pi}^{2}}{Z} = \frac{2 V'}{Z}, \qquad
	M_{\sigma}^{2} \equiv \frac{m_{\sigma}^{2}}{Z}
	= \frac{2 V' + 4 \sigma_{0}^{2} V''}{Z}, \qquad
	\left(M^{\pm}\right)^{2} \equiv \frac{\bigl(m^{\pm}\bigr)^{2}}
	{(Z^{\psi})^{2}} . \label{eq:renormmasses}
\end{equation}
Likewise, the renormalized mass parameter $M_{0}$ and the renormalized Yukawa couplings 
(exemplified by the coupling $\smash{\tilde{y}_{\sigma}^{++/--}}$) are deduced as
\begin{equation}
	M_{0} \equiv \frac{m_{0}}{Z^{\psi}}, \qquad
	\tilde{y}_{\sigma}^{++/--} \equiv \frac{y_{\sigma}^{++/--}}{Z^{\psi}Z^{1/2}} 
	\equiv \frac{1}{2} \left[\pm (\tilde{y}_{1} - \tilde{y}_{2})
	+ \frac{\tilde{\sigma}_{0} (\tilde{y}_{1} + \tilde{y}_{2})^{2}}
	{\sqrt{\tilde{\sigma}_{0}^{2} (\tilde{y}_{1} + \tilde{y}_{2})^{2} 
	+ 4 M_{0\vphantom{,k}}^{2}}}\right].
\end{equation}
While renormalized couplings generally carry a tilde (``$\,\tilde{\boldsymbol{\cdot}}\,$''), we 
simply use capital letters for the renormalized masses $M_{\pi}$, $M_{\sigma}$, $M_{0}$, and $M^{\pm}$.

\subsection{Vacuum manifold, stereographic projections, and the NLSM}

With the local isomorphisms to special orthogonal groups, two-flavor chiral symmetry breaking 
is characterized by the breaking scheme $\mathsf{SO}(4) \rightarrow \mathsf{SO}(3)$.\ The vacuum 
manifold of the $\mathsf{SO}(4)$-invariant bosonic part of the effective action, i.e., 
the set of energetically degenerate vacuum states of the system, is given by the three-sphere 
$\mathsf{S}^{3}$.\ It is then convenient to consider the bosonic interactions in terms of stereographic 
projections, which is a common set of coordinates on $\mathsf{S}^{3}$.\ This rewriting foremost yields 
an intuitive understanding of the underlying geometric concepts: The radius $\theta$ of the 
three-sphere $\mathsf{S}^{3}$ is the ``massive'' mode in this picture and it physically 
corresponds to the original $\sigma$-field (which manifests itself in the fact that $M_{\theta} 
\equiv M_{\sigma}$).\ Its fluctuations are thus radial excitations of the sphere around its 
physical ground state with $\theta = f_{\pi}$. Additionally, if the dynamics of the radius is 
effectively frozen by fixing it to the pion decay constant, one readily recovers the NLSM, which is 
the fundamental building block of ChPT. The NLSM exclusively ``lives'' on $\mathsf{S}^{3}$.

The pions geometrically describe an $\mathsf{SO}(4)$-rotation, traveling on the sphere $\mathsf{S}^{3}$
with constant radius $\theta$. The (pionic) coset generators of $\mathsf{SO}(4)/\mathsf{SO}(3)$, which 
correspond to the axial charge operators derived from the axial current $\mathcal{J}_{A}^{\mu}$, rotate one 
vacuum state into another, all of which are physically equivalent. The remaining generators of the
residual $\mathsf{SO}(3)$-symmetry ``annihilate'' the vacuum, such that $\mathsf{SO}(3)$-rotations
around the axis of the position vector pointing onto $\mathsf{S}^{3}$ leave the actual vacuum state 
invariant. The pionic $\mathsf{SO}(4)$-matrix and the decomposition of the $\mathfrak{so}(4)$-algebra 
into the coset generators and the $\mathfrak{so}(3)$-subalgebra are presented in Appendix \ref{sec:algebra}.

Concretely, we use (properly rescaled) stereographic projections from the south pole
$\tilde{\varphi} = (0,0,0,-\theta)$,
\begin{equation}
	\Pi^{a} = 2 f_{\pi} \frac{\tilde{\pi}^{a}}{\theta + \tilde{\sigma}},
	\qquad a = 1,2,3, \qquad
	\theta = |\tilde{\varphi}| = \sqrt{\tilde{\pi}^{2} + \tilde{\sigma}^{2}} .
	\label{eq:stereo}
\end{equation}
Substituting the fields $\tilde{\varphi}$ with the coordinates defined in Eq.\ (\ref{eq:stereo})
and using the metric tensor $g_{\bar{a}\bar{b}}(\Pi)$ on $\mathsf{S}^{3}$ (with the ``curved''
coset indices $\bar{a},\bar{b} = 1,2,3$),
\begin{equation}
	g_{\bar{a}\bar{b}}(\Pi) = \frac{16 f_{\pi}^{2}}{(4 f_{\pi}^{2} + \Pi^{2})^{2}}
	\;\!\delta_{\bar{a}\bar{b}} = \delta_{\bar{a}\bar{b}}
	\left[\frac{1}{f_{\pi}^{2}} - \frac{\Pi^{2}}{2 f_{\pi}^{4}}
	+ \frac{3 \Pi^{4}}{16 f_{\pi}^{6}}
	+ \mathcal{O}\!\left(\Pi^{6}\right) \right] , \label{eq:metric}
\end{equation}
the bosonic part of the (renormalized) nucleon-meson model (\ref{eq:truncation}) is converted into
$\tilde{\Gamma}_{\text{stereographic}}\left[\Pi,\theta\right]$, with
\begin{equation}
	\tilde{\Gamma}_{\text{stereographic}}\left[\Pi,\theta\right] 
	\coloneqq \int_{x} \left[\frac{\theta^{2}}{2} \;\! g_{\bar{a}\bar{b}}\left(\Pi\right)
	\left(\partial_{\mu} \Pi^{\bar{a}}\right) 
	\partial_{\mu} \Pi^{\bar{b}} 
	+ \frac{1}{2}\left(\partial_{\mu}\theta\right)\partial_{\mu}\theta
	+ \tilde{V}_{k}\!\left(\theta^{2}\right)
	- \tilde{h}\theta \;\! \frac{4 f_{\pi}^{2} - \Pi^{2}}
	{4 f_{\pi}^{2} + \Pi^{2}}\right] .
	\label{eq:gammastereo}
\end{equation}
Note that the pions $\Pi$ carry the indices $\lbrace\bar{a}, \bar{b}, \ldots \rbrace$ 
in their role as mappings between (Euclidean) spacetime and the manifold $\mathsf{S}^{3}$ 
[to be distinguished from the Lie-algebra indices $\lbrace a, b, \ldots \rbrace$]. The
notation $\Pi^{2} \equiv \Pi \cdot \Pi$ means $\smash{\delta_{\bar{a}\bar{b}} 
\Pi^{\bar{a}} \Pi^{\bar{b}}}$; other technical details are found in Appendix
\ref{sec:algebra}.

From this stage one immediately recovers the effective action of the NLSM by forcing the
radius $\theta$ of the three-sphere to be equal to the pion decay constant,
\begin{equation}
	\theta = f_{\pi} = \mathrm{const.},
\end{equation}
which restricts the dynamics to the vacuum manifold $\mathsf{S}^{3}$. Up to irrelevant constants 
and setting $\tilde{h} = 0$, the effective action then reads
\begin{equation} 
	\tilde{\Gamma}_{\mathrm{NLSM}} \left[\Pi\right]
	= \int_{x} \frac{f_{\pi}^{2}}{2}\;\! g_{\bar{a}\bar{b}}\left(\Pi\right)
	\left(\partial_{\mu} \Pi^{\bar{a}}\right) 
	\partial_{\mu} \Pi^{\bar{b}}
	= \int_{x} \left[
	\frac{1}{2} \left(\partial_{\mu} \Pi \right) \cdot \partial_{\mu} \Pi
	- \frac{1}{4 f_{\pi}^{2}} {\Pi}^{2} \left(\partial_{\mu} \Pi \right) 
	\cdot \partial_{\mu} \Pi
	+ \mathcal{O}\!\left(\Pi^{6}\right) \right] .
	\label{eq:nlsm}
\end{equation}
The last term in Eq.\ (\ref{eq:gammastereo}) proportional to the parameter 
$\tilde{h}$ adds momentum-independent pion self-interactions 
to the (purely momentum-dependent) action (\ref{eq:nlsm}) of the NLSM, i.e.,
\begin{equation}
	\tilde{h} \left[f_{\pi} - \frac{\Pi^{2}}{2 f_{\pi}}
	+ \frac{\Pi^{4}}{8 f_{\pi}^{3}} + \mathcal{O}\!\left(\Pi^{6}\right)\right] .
	\label{eq:hesb}
\end{equation}
The squared pion mass is thus given by\footnote{We stick to the small Greek letter 
``$\pi$'' in the subscript of the physical (coordinate-independent) quantities $f_{\pi}$ 
and $M_{\pi} \equiv M_{\Pi}$ throughout the entire paper. The pion-pion scattering 
amplitude $\mathcal{M}_{\pi\pi}$ is analogously denoted.}
\begin{equation}
	M_{\pi}^{2} = \frac{\tilde{h}}{f_{\pi}}
	\equiv \frac{h}{Z^{1/2} f_{\pi}},
	\label{eq:mpion}
\end{equation}
whereas the squared mass of the $\theta$-field in Eq.\ (\ref{eq:gammastereo}) is calculated 
from the (renormalized) effective potential $\tilde{V}$,\footnote{We use the masses $M_{\theta}$ 
and $M_{\sigma}$ synonymously in the later Sections.}
\begin{equation}
	M_{\theta}^{2} = 2 \tilde{V}' + 4 f_{\pi}^{2} \tilde{V}''
	\equiv M_{\sigma}^{2} .
\end{equation}
The physical vacuum configuration of the nucleon-meson model is attained at the north pole 
$\tilde{\varphi} = (0,0,0,\theta)$,
\begin{equation}
	\left. \Pi^{a} \right|_{\tilde{\sigma}\, =\, \tilde{\sigma}_{0},\,
	\tilde{\pi}^{a}\, =\, 0} = 0 \quad \forall\, a, \qquad
	\left. \theta \right|_{\tilde{\sigma}\, =\, \tilde{\sigma}_{0},\,
	\tilde{\pi}^{a}\, =\, 0} = \tilde{\sigma}_{0} = f_{\pi},
	\label{eq:physconfig}
\end{equation}
where the last equation is the identification of the (renormalized) isoscalar condensate 
$\tilde{\sigma}_{0}$ with the pion decay constant [according to Eqs.\ (\ref{eq:axialcurrent}) 
and (\ref{eq:axialcurrent2})]. Of course, this configuration was already anticipated in the 
discussion of this Section. It a posteriori justifies the Taylor expansion for small $\Pi^{2}$
($\Pi^{2} \ll 1$; see also Fig.\ \ref{fig:stereo} in Appendix \ref{sec:algebra}), that 
we implicitly employed in Eqs.\ (\ref{eq:metric}), (\ref{eq:hesb}), and (\ref{eq:mpion}).

The reformulation of the (parity-doublet) nucleon-meson model through the stereographic coordinates
allows one to compute the pion-pion scattering amplitude in a ``chiral expansion'' similar to the 
NLSM and ChPT. It pointedly elaborates on common geometric features of the linear and nonlinear 
realizations of chiral symmetry. In the following, we make use of these concepts and perform an 
expansion of the effective action in terms of ``small'' (squared) pion fields and pion momenta 
(derivatives acting on the pion fields), after the integration of quantum nucleon-meson fluctuations
based on Eq.\ (\ref{eq:Wetterich}). Although other choices of coordinates, including that in the
original formulation (\ref{eq:truncation}), are still valid and do not affect the physical outcome 
of the calculation, this particular choice allows us to explicitly demonstrate the convergence 
of the above formalism and work out subtleties with higher-derivative pion self-interactions and 
their influence on the analytic and numeric accuracy.
\begin{figure}[t]
	\centering
	\includegraphics[scale=1.0]{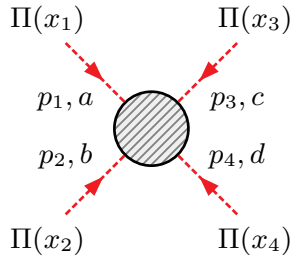}
	\caption{Pion-pion scattering process with incoming pion momenta $p_{n}$, $n = 1, 
	\ldots , 4$, and isospin indices $\lbrace a, b, c, d\rbrace$.}
	\label{fig:amplitude}
\end{figure}

\section{Pion-pion scattering}
\label{sec:scattering}

The amplitude $\mathcal{M}_{\pi\pi}$ of pion-pion scattering $\left(\Pi^{a}\ \Pi^{b} 
\longrightarrow \Pi^{c}\ \Pi^{d}\right)$, graphically illustrated in Fig.\ \ref{fig:amplitude}, 
is constrained by Lorentz and crossing symmetries to
\begin{equation}
	\mathcal{M}_{\pi\pi}^{abcd}(s,t,u) = 
	i\mathfrak{A}(s,t,u)\;\! \delta^{ab}\delta^{cd}
	+ i\mathfrak{A}(t,u,s)\;\! \delta^{ac}\delta^{bd}
	+ i\mathfrak{A}(u,s,t)\;\! \delta^{ad}\delta^{bc},
	\label{eq:amplitude}
\end{equation}
with the adjoint isospin indices $\lbrace a,b,c,d \rbrace$ and the Mandelstam variables
\begin{equation}
	\begin{aligned}
	s & = \left(p_{1} + p_{2}\right)^{2} = \left(p_{3} + p_{4}\right)^{2}, \\[0.1cm]
	t & = \left(p_{1} + p_{3}\right)^{2} = \left(p_{2} + p_{4}\right)^{2}, \\[0.1cm]
	u & = \left(p_{1} + p_{4}\right)^{2} = \left(p_{2} + p_{3}\right)^{2} 
	\end{aligned}
\end{equation}
with incoming pion momenta $p_{n}$, $n = 1,\ldots ,4$ \cite{Leutwyler:2012sco}.

Working with the effective action $\Gamma$, it is adequate to compute the scattering 
amplitude $\mathcal{M}_{\pi\pi}$ at tree level.\footnote{The quantum effective action $\Gamma$ 
corresponds to the renormalized action functional $\tilde{\Gamma}$ from Sec.\ \ref{sec:model}. 
We omit the tilde from now on.} The effects of loop corrections are entirely taken into account 
by using the ``dressed'' propagators and vertices in place of the classical ones (derived from 
the classical action $S$) \cite{Weinberg:1996kr}. The notion ``tree level'' refers to the topological 
form of the Feynman diagrams---generally speaking, irrespective of the underlying action functional 
and not restricted to the ``standard'' perturbative sense. Our strategy is therefore to integrate 
the FRG equation (\ref{eq:Wetterich}) in order to obtain the effective action $\Gamma$ and 
subsequently focus on the bosonic part of the truncation (\ref{eq:truncation}) in its 
stereographic picture (\ref{eq:gammastereo}) in order to compute the amplitude $\mathcal{M}_{\pi\pi}$
in an expansion in pion fields and pion momenta. The fermions (nucleons and their chiral partners) do 
not enter the tree-level scattering diagrams; their loop effects are captured by the dressed bosonic 
vertices. In consequence, this approach goes much beyond the common MF and one-loop levels.

At leading order of the chiral expansion (equivalent to tree level) \cite{Bijnens:1995yn, 
Bijnens:1997vq, Ananthanarayan:2000ht, Colangelo:2000jc, Colangelo:2001df, Leutwyler:2012sco}, 
the amplitude $\mathfrak{A}$ on the right-hand side of Eq.\ (\ref{eq:amplitude}) solely depends 
on the first Mandelstam variable $s$, i.e., $\mathfrak{A}(s,t,u) \equiv \mathfrak{A}(s)$. In 
the partial-wave analysis of $\mathfrak{A}$, this implies that only the $S$- and $P$-waves (with 
angular-momentum quantum numbers $l = 0$ and $l = 1$, respectively) are nonzero \cite{Colangelo:2000jc, 
Leutwyler:2012sco}. At threshold, $s = 4 M_{\pi}^{2}$, the tree-level amplitude even further
reduces to the isospin-zero ($I = 0$) and isospin-two ($I = 2$) $S$-wave contributions. More 
precisely, the two relevant scattering lengths $a_{0}^{0}$ and $a_{0}^{2}$ are related to the 
partial-wave amplitudes $t_{l}^{I}$ as follows \cite{Divotgey:2016pst}:\footnote{The scattering 
length $a_{0}^{1} \sim \mathfrak{A}(t,u,s) - \mathfrak{A}(u,s,t)$ vanishes at threshold, 
$a_{0}^{1} = 0$. The term ``scattering length'' originates from its canonical definition 
$a_{l}^{I} \sim t_{l}^{I}/M_{\pi}$, which has the dimension of length (inverse energy).
In the chiral limit, $M_{\pi} \rightarrow 0$, $a_{0}^{0}$ and $a_{0}^{2}$ are zero.}
\begin{IEEEeqnarray}{rCl}
	a_{0}^{0} & = & \left. t_{0}^{0} \right|_{s\, =\, 4 M_{\pi}^{2},\, t\, =\, u\, =\, 0}
	= \left. \frac{1}{32\pi} \left[3 \mathfrak{A}(s,t,u) + \mathfrak{A}(t,u,s) 
	+ \mathfrak{A}(u,s,t)\right] \right|_{s\, =\, 4 M_{\pi}^{2},\, t\, =\, u\, =\, 0}
	\nonumber\\[0.1cm]
	& = & \frac{1}{32\pi} \left[3 \mathfrak{A}\!\left(4 M_{\pi}^{2}\right) + 
	2 \mathfrak{A}(0)\right], \qquad \\[0.2cm]
	a_{0}^{2} & = & \left. t_{0}^{2} \right|_{s\, =\, 4 M_{\pi}^{2},\, t\, =\, u\, =\, 0}
	= \left. \frac{1}{32\pi} \left[\mathfrak{A}(t,u,s) 
	+ \mathfrak{A}(u,s,t)\right] \right|_{s\, =\, 4 M_{\pi}^{2},\, t\, =\, u\, =\, 0}
	= \frac{1}{16\pi}\;\! \mathfrak{A}(0) .
\end{IEEEeqnarray}

The tree-level pion-pion scattering amplitude (in Minkowski spacetime $\mathbb{M}^{4}$) within 
the effective action (\ref{eq:gammastereo}) is determined by the Feynman diagrams
\begin{IEEEeqnarray}{rCl}
	\mathcal{M}_{\pi\pi}^{abcd} \sim 
	\vcenter{\hbox{
	\begin{pspicture}[showgrid=false](2.0,2.0)
		\psline[linewidth=0.03,linestyle=dashed,
		dash=2pt 1pt,linecolor=Red](0.45,1.55)(1.0,1.0)
		\psline[linewidth=0.03,linestyle=dashed,
		dash=2pt 1pt,linecolor=Red](1.55,1.55)(1.0,1.0)
		\psline[linewidth=0.03,linestyle=dashed,
		dash=2pt 1pt,linecolor=Red](0.45,0.45)(1.0,1.0)
		\psline[linewidth=0.03,linestyle=dashed,
		dash=2pt 1pt,linecolor=Red](1.55,0.45)(1.0,1.0)
		\rput[b]{*0}(0.25,1.65){\small{$\Pi$}}
		\rput[b]{*0}(1.75,1.65){\small{$\Pi$}}
		\rput[b]{*0}(0.25,0.15){\small{$\Pi$}}
		\rput[b]{*0}(1.75,0.15){\small{$\Pi$}}
		\pscircle[linewidth=0.03,fillstyle=hlines*,hatchwidth=0.02,
		fillcolor=lightgray!25,hatchsep=0.05,hatchcolor=gray](1.0,1.0){0.4}
	\end{pspicture}
	}} & = &
	\vcenter{\hbox{
	\begin{pspicture}[showgrid=false](2.0,2.0)
		\psline[linewidth=0.03,linestyle=dashed,
		dash=2pt 1pt,linecolor=Red](0.45,1.55)(1.0,1.0)
		\psline[linewidth=0.03,linestyle=dashed,
		dash=2pt 1pt,linecolor=Red](1.55,1.55)(1.0,1.0)
		\psline[linewidth=0.03,linestyle=dashed,
		dash=2pt 1pt,linecolor=Red](0.45,0.45)(1.0,1.0)
		\psline[linewidth=0.03,linestyle=dashed,
		dash=2pt 1pt,linecolor=Red](1.55,0.45)(1.0,1.0)
		\rput[b]{*0}(0.25,1.65){\small{$\Pi$}}
		\rput[b]{*0}(1.75,1.65){\small{$\Pi$}}
		\rput[b]{*0}(0.25,0.15){\small{$\Pi$}}
		\rput[b]{*0}(1.75,0.15){\small{$\Pi$}}
		\pscircle[linewidth=0.03,fillstyle=solid,
		fillcolor=lightgray!25](1.0,1.0){0.20}
		\rput[b]{*0}(0.99,0.92){\fontsize{7pt}{0pt}{$4$}\selectfont}
	\end{pspicture}
	}} +
	\underbrace{
	\vcenter{\hbox{
	\begin{pspicture}[showgrid=false](3.0,2.0)
		\psline[linewidth=0.03,linestyle=dashed,
		dash=2pt 1pt,linecolor=Red](0.45,1.55)(1.0,1.0)
		\psline[linewidth=0.03,linestyle=dashed,
		dash=2pt 1pt,linecolor=Red](2.55,1.55)(2.0,1.0)
		\psline[linewidth=0.03,linestyle=dashed,
		dash=2pt 1pt,linecolor=Red](0.45,0.45)(1.0,1.0)
		\psline[linewidth=0.03,linestyle=dashed,
		dash=2pt 1pt,linecolor=Red](2.55,0.45)(2.0,1.0)
		\psline[linewidth=0.03,linestyle=dashed,
		dash=2pt 1pt,linecolor=Blue](1.0,1.0)(2.0,1.0)
		\rput[b]{*0}(0.25,1.65){\small{$\Pi$}}
		\rput[b]{*0}(2.75,1.65){\small{$\Pi$}}
		\rput[b]{*0}(0.25,0.15){\small{$\Pi$}}
		\rput[b]{*0}(2.75,0.15){\small{$\Pi$}}
		\rput[b]{*0}(1.5,0.65){\small{$\theta$}}
		\pscircle[linewidth=0.03,fillstyle=solid,
		fillcolor=lightgray!25](1.0,1.0){0.20}
		\rput[b]{*0}(1.0,0.92){\fontsize{7pt}{0pt}{$3$}\selectfont}
		\pscircle[linewidth=0.03,fillstyle=solid,
		fillcolor=lightgray!25](2.0,1.0){0.20}
		\rput[b]{*0}(2.0,0.92){\fontsize{7pt}{0pt}{$3$}\selectfont}
	\end{pspicture}
	}}
	}_{\text{$s$-, $t$-, and $u$-channel}} \nonumber\\[0.2cm]
	& = & i\Gamma^{(4)\;\! abcd}_{\Pi\Pi\Pi\Pi}(x_{1},x_{2},x_{3},x_{4})
	+ \underbrace{\int_{z_{1}} \int_{z_{2}} 
	i\Gamma^{(3)\;\! ab}_{\Pi\Pi\theta}(x_{1},x_{2},z_{1})\;
	G_{\theta\theta}(z_{1},z_2)\; 
	i\Gamma^{(3)\;\! cd}_{\Pi\Pi\theta}(x_{3},x_{4},z_{2})}_{
	\text{$s$-channel}} \nonumber\\[0.2cm]
	& & \hspace{3.6cm}
	+\ \text{$t$-channel} + \text{$u$-channel}, 
	\label{eq:treeamplitude}
\end{IEEEeqnarray}
where we left out the pion-leg specifications (cf.\ Fig.\ \ref{fig:amplitude}) in the first line 
to enhance readability. In the second line, we reintroduced the adjoint isospin indices $\lbrace 
a,b,c,d \rbrace$ as well as the spacetime variables $x_{n}$, $n = 1, \ldots , 4$. The 
propagator of the $\theta$-field, which is given by the inverse two-point function, is denoted 
by $G_{\theta\theta}$ and the imaginary unit $i$ attached to the vertex functions is a typical
convention in $\mathbb{M}^{4}$. The Minkowskian version of the (Euclidean) effective action 
$\Gamma$, sitting on the vertices in the above diagrams, is produced by a naive inverse Wick 
rotation, $\Gamma \rightarrow -i \Gamma$, which is tacitly carried out but not labeled separately
in the course of the current Section ($\mathbb{E}^{4} \rightarrow \mathbb{M}^{4}$). 

The summation of all relevant tree-level diagrams for the four-pion amplitude, i.e., with 
at most a single $\theta$-exchange, is equivalent to eliminating the $\theta$-field 
by evaluating the effective action $\Gamma$ at its stationary point (w.r.t.\ $\theta$),
\begin{equation}
	\left.\frac{\delta\Gamma}{\delta\theta}\right|_{\theta_{\mathrm{sol}}} = 0
	\qquad \Rightarrow \qquad \Gamma_{\mathrm{sol}}[\Pi] \coloneqq
	\left.\Gamma[\Pi,\theta]\right|_{\theta\, =\, \theta_{\mathrm{sol}}[\Pi]} 
	\qquad \Rightarrow \qquad
	\vcenter{\hbox{
	\begin{pspicture}[showgrid=false](2.0,2.0)
		\psline[linewidth=0.03,linestyle=dashed,
		dash=2pt 1pt,linecolor=Red](0.45,1.55)(1.0,1.0)
		\psline[linewidth=0.03,linestyle=dashed,
		dash=2pt 1pt,linecolor=Red](1.55,1.55)(1.0,1.0)
		\psline[linewidth=0.03,linestyle=dashed,
		dash=2pt 1pt,linecolor=Red](0.45,0.45)(1.0,1.0)
		\psline[linewidth=0.03,linestyle=dashed,
		dash=2pt 1pt,linecolor=Red](1.55,0.45)(1.0,1.0)
		\rput[b]{*0}(0.25,1.65){\small{$\Pi$}}
		\rput[b]{*0}(1.75,1.65){\small{$\Pi$}}
		\rput[b]{*0}(0.25,0.15){\small{$\Pi$}}
		\rput[b]{*0}(1.75,0.15){\small{$\Pi$}}
		\pscircle[linewidth=0.03,fillstyle=hlines*,hatchwidth=0.02,
		fillcolor=lightgray!25,hatchsep=0.05,hatchcolor=gray](1.0,1.0){0.4}
	\end{pspicture}
	}} \equiv
	i\Gamma^{(4)\;\! abcd}_{\mathrm{sol}\;\Pi\Pi\Pi\Pi}(x_{1},x_{2},x_{3},x_{4}).
	\label{eq:eom}
\end{equation}
Thus the summation collapses to a single dressed vertex, including bosonic and fermionic
quantum fluctuations. The equation of motion (EOM) in Eq.\ (\ref{eq:eom}) has the form
($\Box = \partial_{\mu}\partial^{\mu}$)
\begin{equation}
	\frac{\delta\Gamma}{\delta\theta} = 0 \qquad \Rightarrow \qquad
	\xi + \epsilon \left[\frac{\Box\xi}{M_{\pi}^{2}} 
	- \frac{\xi + f_{\pi}}{M_{\pi}^{2}}
	\!\; g_{\bar{a}\bar{b}} 
	\left(\partial_{\mu}\Pi^{\bar{a}}\right) 
	\partial^{\mu} \Pi^{\bar{b}}
	+ 2 f_{\pi} \frac{\Pi^2}{4 f_{\pi}^{2} - \Pi^{2}} 
	+ \frac{\mathcal{O}\!\left(\xi^{2}\right)}{M_{\pi}^{2}} 
	\right] = 0 \label{eq:eomxi}
\end{equation}
after shifting the $\theta$-field by its physical expansion point, $\xi \coloneqq
\theta - f_{\pi}$. We furthermore renamed the dimensionless ratio $M_{\pi}^{2}/M_{\sigma}^{2}$ 
as $\epsilon$, for which we take a small value as granted, $\epsilon \ll 1$. This holds at 
least for the masses quoted in Ref.\ \cite{ParticleDataGroup:2020ssz},
\begin{equation}
	\epsilon \coloneqq \frac{M_{\pi}^{2}}{M_{\sigma}^{2}}, \qquad
	\epsilon_{\mathrm{phys}} \sim \frac{(138\ \mathrm{MeV})^{2}}
	{(475\ \mathrm{MeV})^2} \simeq 0.0844 .
	\label{eq:epsilon}
\end{equation}
Following Ref.\ \cite{Appelquist:1980ae}, the solution to the EOM is a power series in 
the small parameter $\epsilon$,
\begin{equation}
	\theta_{\mathrm{sol}} = f_{\pi} + \xi_\mathrm{sol}, \qquad
	\xi_{\mathrm{sol}} = \epsilon \xi_{1} + \epsilon^{2} \xi_{2} 
	+ \epsilon^{3} \xi_{3} + \epsilon^{4} \xi_{4} + \epsilon^{5} \xi_{5} + \cdots ,
\end{equation}
which itself is a functional of the $\Pi$-fields, $\xi_{\mathrm{sol}}[\Pi]$. The expansion 
coefficients $\xi_{n}$, $n \in \mathbb{N}$, are obtained in an iterative manner, namely,
\begin{IEEEeqnarray}{rCl}
	\xi_{1} & = & \frac{f_{\pi}}{M_{\pi}^{2}} \!\; g_{\bar{a}\bar{b}} 
	\left(\partial_{\mu}\Pi^{\bar{a}}\right) 
	\partial^{\mu} \Pi^{\bar{b}}
	- 2 f_{\pi} \frac{\Pi^2}{4 f_{\pi}^{2} - \Pi^{2}}, \\[0.1cm]
	\xi_{2} & = & - \frac{1}{M_{\pi}^{2}} \left[\Box
	- g_{\bar{a}\bar{b}} 
	\left(\partial_{\mu}\Pi^{\bar{a}}\right) 
	\partial^{\mu} \Pi^{\bar{b}} \right] \xi_{1}, \\[0.1cm]
	\xi_{3} & = & - \frac{1}{M_{\pi}^{2}} \left\lbrace \left[\Box
	- g_{\bar{a}\bar{b}} 
	\left(\partial_{\mu}\Pi^{\bar{a}}\right) 
	\partial^{\mu} \Pi^{\bar{b}} \right] \xi_{2}
	+ \text{terms $\propto \xi_{1}^{2}$} \right\rbrace , \\[0.1cm]
	\xi_{4} & = & - \frac{1}{M_{\pi}^{2}} \left\lbrace \left[\Box
	- g_{\bar{a}\bar{b}} 
	\left(\partial_{\mu}\Pi^{\bar{a}}\right) 
	\partial^{\mu} \Pi^{\bar{b}} \right] \xi_{3}
	+ \text{terms $\propto \xi_{1}^{3}$} 
	+ \text{terms $\propto \xi_{1}\xi_{2}$} \right\rbrace , \\[0.1cm]
	\xi_{5} & = & - \frac{1}{M_{\pi}^{2}} \left\lbrace \left[\Box
	- g_{\bar{a}\bar{b}} 
	\left(\partial_{\mu}\Pi^{\bar{a}}\right) 
	\partial^{\mu} \Pi^{\bar{b}} \right] \xi_{4}
	+ \text{terms $\propto \xi_{1}^{4}$} 
	+ \text{terms $\propto \xi_{1}^2\xi_{2}$}
	+ \text{terms $\propto \xi_{1} \xi_{3}$}
	+ \text{terms $\propto \xi_{2}^2$} \right\rbrace ,
\end{IEEEeqnarray}
and so forth. At order $n > 1$, we deduce the general (iterative) solution
\begin{equation}
	\xi_{n} = - \frac{1}{M_{\pi}^{2}} \left\lbrace \left[\Box
	- g_{\bar{a}\bar{b}} 
	\left(\partial_{\mu}\Pi^{\bar{a}}\right) 
	\partial^{\mu} \Pi^{\bar{b}} \right] \xi_{n-1}
	+ \underbrace{\text{terms $\propto \xi_{1}^{n-1}$} 
	+ \cdots}_{\text{$\mathfrak{P}(n-1)-1$ possibilites}} \right\rbrace ,
\end{equation}
with the (number-theoretical) partition function $\mathfrak{P}(n)$ of the integer $n$. 
The solution of $\mathcal{O}(\epsilon^{n})$ manifestly involves pion derivatives up to 
$\smash{\mathcal{O}\!\left(\partial^{2(n+1)}\right)}$. This subtle point already leads to the 
expectation that we need to expand the effective action $\Gamma_{\mathrm{sol}}$ to that 
order in pion derivatives, if we want to achieve analytic accuracy at $\smash{\mathcal{O}(\epsilon^{n})}$ 
for the scattering lengths. Indeed, solving Eq.\ (\ref{eq:eomxi}) for $\xi$ along the above iterative 
scheme, expanding the resulting pionic effective action $\Gamma_{\mathrm{sol}}$ up to $\mathcal{O}
\!\left(\Pi^{4}, \partial^{12}\right)$, and extracting the four-pion amplitude $\mathfrak{A}(s)$
out of the fourth functional derivative of $\Gamma_{\mathrm{sol}}$, we find for the $S$-wave 
scattering lengths
\begin{IEEEeqnarray}{rCl}
	\mathcal{O}\!\left(p^{0}\right)\!\colon\quad 
	a_{0}^{0} & = & \frac{5 M_{\pi}^{2}}{32 \pi f_{\pi}^{2}}
	\left(1 + \epsilon\right), \label{eq:eomexpansionstart} \\[0.1cm]
	a_{0}^{2} & = & \frac{M_{\pi}^{2}}{16 \pi f_{\pi}^{2}}
	\left(1 + \epsilon\right); \\[0.2cm]
	\mathcal{O}\!\left(p^{2}\right)\!\colon\quad 
	a_{0}^{0} & = & \frac{M_{\pi}^{2}}{32 \pi f_{\pi}^{2}} 
	\left(7 + 9 \epsilon + 12 \epsilon^{2}\right), \\[0.1cm]
	a_{0}^{2} & = & - \frac{M_{\pi}^{2}}{16 \pi f_{\pi}^{2}}
	\left(1 + 3 \epsilon\right); \\[0.2cm]
	\mathcal{O}\!\left(p^{4}\right)\!\colon\quad 
	a_{0}^{0} & = & \frac{M_{\pi}^{2}}{32 \pi f_{\pi}^{2}} 
	\left(7 + 29 \epsilon + 60 \epsilon^{2} + 48 \epsilon^{3}\right), 
	\label{eq:eomexpansionp40} \\[0.1cm]
	a_{0}^{2} & = & - \frac{M_{\pi}^{2}}{16 \pi f_{\pi}^{2}}
	\left(1 - \epsilon\right); \label{eq:eomexpansionp42} \\[0.2cm]
	\mathcal{O}\!\left(p^{6}\right)\!\colon\quad 
	a_{0}^{0} & = & \frac{M_{\pi}^{2}}{32 \pi f_{\pi}^{2}} 
	\left(7 + 29 \epsilon + 108 \epsilon^{2} + 240 \epsilon^{3}
	+ 192 \epsilon^{4}\right); \\[0.2cm]
	\mathcal{O}\!\left(p^{8}\right)\!\colon\quad 
	a_{0}^{0} & = & \frac{M_{\pi}^{2}}{32 \pi f_{\pi}^{2}} 
	\left(7 + 29 \epsilon + 108 \epsilon^{2} + 432 \epsilon^{3}
	+ 960 \epsilon^{4} + 768 \epsilon^{5} \right); \\[0.2cm]
	\mathcal{O}\!\left(p^{10}\right)\!\colon\quad 
	a_{0}^{0} & = & \frac{M_{\pi}^{2}}{32 \pi f_{\pi}^{2}} 
	\left(7 + 29 \epsilon + 108 \epsilon^{2} + 432 \epsilon^{3}
	+ 1728 \epsilon^{4} + 3840 \epsilon^{5} + 3072 \epsilon^{6}\right) ;\\[0.2cm] 
	\mathcal{O}\!\left(p^{12}\right)\!\colon\quad 
	a_{0}^{0} & = & \frac{M_{\pi}^{2}}{32 \pi f_{\pi}^{2}} 
	\left(7 + 29 \epsilon + 108 \epsilon^{2} + 432 \epsilon^{3}
	+ 1728 \epsilon^{4} + 6912 \epsilon^{5} + 15360 \epsilon^{6}
	+ 12288 \epsilon^{7} \right) ;
	\label{eq:eomexpansionend} \\
	& \vdots & \nonumber
\end{IEEEeqnarray}
The indication $\mathcal{O}(p^{m})$, $m$ even, stands for the expansion order in the small pion 
momentum $p$ (via the correspondence between derivatives acting on the $\Pi$-fields and the 
pion momentum, $\partial \leftrightarrow p$). The outcome of this procedure in Eqs.\ 
(\ref{eq:eomexpansionstart}) to (\ref{eq:eomexpansionend}) can be systematically compared 
to the exact (coordinate-independent) tree-level scattering lengths of the LSM 
\cite{Bessis:1972sn, Bentz:1997ry, Black:2009bi} [cf.\ Eq.\ (\ref{eq:universal}) in
Appendix \ref{sec:fourpoint}],
\begin{IEEEeqnarray}{rCl}
	a_{0}^{0} & = & \frac{M_{\pi}^2}{32\pi f_{\pi}^{2}} 
	\left(\epsilon - 1\right) \left(\frac{9}{4\epsilon - 1} + 2\right)
	\stackrel{\epsilon\, <\, 1/4}{\equiv} \frac{M_{\pi}^2}{32\pi f_{\pi}^{2}} 
	\left(1 - \epsilon\right) \left[7 + 9 \sum_{n\, =\, 1}^{\infty} 
	\left(4\epsilon\right)^{n} \right] \label{eq:a00exact} \nonumber\\[0.1cm]
	& = & \frac{M_{\pi}^{2}}{32 \pi f_{\pi}^{2}} 
	\left(7 + 29 \epsilon + 108 \epsilon^{2} + 432 \epsilon^{3}
	+ 1728 \epsilon^{4} + 6912 \epsilon^{5} + 27648 \epsilon^{6}
	+ 110592 \epsilon^{7} \right)
	+ \mathcal{O}\!\left(\epsilon^{8}\right), \\[0.1cm]
	a_{0}^{2} & = & - \frac{M_{\pi}^{2}}{16\pi f_{\pi}^{2}}
	\left(1 - \epsilon\right) . \label{eq:a02exact}
\end{IEEEeqnarray}
In the first line of Eq.\ (\ref{eq:a00exact}), we have expressed the scattering length
$a_{0}^{0}$ through a geometric series, which is convergent for $\epsilon < 1/4$. The
second line confirms our statement about the analytic accuracy; for instance, at 
$\smash{\mathcal{O}\!\left(p^{2}\right)}$ and higher, the WLs of Eq.\ (\ref{eq:Weinberg})
are reproduced for $\epsilon \rightarrow 0$, whereas this is not the case at $\smash{\mathcal{O}
\!\left(p^{0}\right)}$.\ From $\smash{\mathcal{O}\!\left(p^{4}\right)}$ on, the isospin-two
scattering length $a_{0}^{2}$ matches the exact value (\ref{eq:a02exact}), as expected (as being 
a polynomial of order $\epsilon$).\ This also explains why we have cut out $a_{0}^{2}$
at higher orders, since it does not change anymore. In total, we have to take pion momenta of 
$\smash{\mathcal{O}\!\left(p^{12}\right)}$ into account---as predicted---in order to obtain 
the correct numeric prefactors up to $\epsilon^{5}$ during the expansion of the (nonlinear) 
pion action $\Gamma_{\mathrm{sol}}$, cf.\ Eqs.\ (\ref{eq:eomexpansionend}) and (\ref{eq:a00exact}).
The (exact) tree-level scattering lengths (\ref{eq:a00exact}) and (\ref{eq:a02exact}) are
calculated in terms of stereographic coordinates in Appendix \ref{sec:fourpoint}, where
furthermore the proof of coordinate independence of the (on-shell) four-point function is
given---thereby showing the universal character of the tree-level relations.

The deviations among the numeric prefactors at $\mathcal{O}\!\left(p^{m}\right)$ define
the ``analytic'' and ``numeric'' errors 
\begin{IEEEeqnarray}{rCl}
	\Delta_{\text{analytic}}^{I,m} & = & 
	\mathrm{abs}\left[\left. a_{0}^{I}\right|_{\text{exact; expanded 
	up to $\epsilon^{m/2+1}$}} 
	- \left. a_{0}^{I}\right|_{\mathcal{O}\left(p^{m}\right)}\right] , 
	\label{eq:error1} \\[0.2cm]
	\Delta_{\text{numeric}}^{I,m} & = & 
	\mathrm{abs}\left[\left. a_{0}^{I}\right|_{\text{exact}} 
	- \left. a_{0}^{I}\right|_{\mathcal{O}\left(p^{m}\right)}\right] .
	\label{eq:error2}
\end{IEEEeqnarray}
Per definition, we then have the following identities:
\begin{equation}
	\begin{aligned}
	\Delta_{\mathrm{numeric}}^{0,m} & \equiv \Delta_{\mathrm{analytic}}^{0,m}
	+ \mathcal{O}\!\left(\epsilon^{m/2+2}\right) , \\[0.1cm]
	\Delta_{\mathrm{numeric}}^{2,m} & \equiv \Delta_{\mathrm{analytic}}^{2,m},
	\end{aligned}
	\label{eq:errordiff}
\end{equation}
where the first implies that $\Delta_{\mathrm{numeric}}^{0,m}\rightarrow\Delta_{\mathrm{analytic}}^{0,m}$
for large $m$. The second identity in Eq.\ (\ref{eq:errordiff}) immediately holds, since the isospin-two 
scattering length is a polynomial of degree one. The analytic errors are collected in Table \ref{tab:errors}.
\begin{table*}
	\caption{\label{tab:errors}Analytic errors $\Delta^{I,m}_{\mathrm{analytic}}$ in units of
	$M_{\pi}^{2}/(32\pi f_{\pi}^{2})$ and $M_{\pi}^{2}/(16\pi f_{\pi}^{2})$ for $I = 0$ 
	and $I = 2$, respectively.}
	\begin{ruledtabular}
		\begin{tabular}{lccccccc}
		Isospin $I$ & $\mathcal{O}\!\left(p^{0}\right)$ &
		$\mathcal{O}\!\left(p^{2}\right)$ & $\mathcal{O}\!\left(p^{4}\right)$ & 
		$\mathcal{O}\!\left(p^{6}\right)$ & $\mathcal{O}\!\left(p^{8}\right)$ &
		$\mathcal{O}\!\left(p^{10}\right)$ & $\mathcal{O}\!\left(p^{12}\right)$ \\[0.1cm]
		\colrule\\[-0.25cm]
		$I = 0$ & 
		$2 + 24\epsilon$ & 
		$20 \epsilon + 96 \epsilon^{2}$ &
		$48 \epsilon^{2} + 384 \epsilon^{3}$ & 
		$192 \epsilon^{3} + 1536 \epsilon^{4}$ & 
		$768 \epsilon^{4} + 6144 \epsilon^{5}$ & 
		$3072\epsilon^{5} + 24576 \epsilon^{6}$ &
		$12288 \epsilon^{6} + 98304 \epsilon^{7}$ \\
		$I = 2$ &
		$2$ & 
		$4\epsilon$ & zero & zero & zero & zero & zero 
		\end{tabular}
	\end{ruledtabular}
\end{table*}

The elimination of the $\theta$-field with the EOM (\ref{eq:eom}) in favor of pionic interactions
generates various higher-derivative couplings in the effective action $\Gamma_{\mathrm{sol}}$, 
in addition to the ones from the metric (\ref{eq:metric}) and Eq.\ (\ref{eq:hesb}). E.g., there
principally is a single four-pion interaction at $\smash{\mathcal{O}\!\left(p^{0}\right)}$ and 
three different term structures at $\smash{\mathcal{O}\!\left(p^{2}\right)}$, respectively,
\begin{equation}
	\mathcal{O}\!\left(p^{0}\right)\!\colon \quad \Pi^{4}; \qquad
	\mathcal{O}\!\left(p^{2}\right)\!\colon \quad
	\Pi^{2} \Pi \cdot \Box \Pi, \quad
	\Pi^{2} \left(\partial_{\mu}\Pi\right) \cdot \partial^{\mu}\Pi, \quad
	\left(\Pi \cdot \partial_{\mu}\Pi\right)^{2}. 
	\label{eq:termstructures}
\end{equation}
For higher momentum orders, the amount of (possible) four-pion interaction terms ``rapidly'' increases, 
see Appendix \ref{sec:LECs}, and this might lead to intricacies when analytically computing the 
effective action $\Gamma_{\mathrm{sol}}$. In fact, this is the reason why we stopped the expansion 
at $\smash{\mathcal{O}\!\left(p^{12}\right)}$, as the number of pion terms already went up to 
$3335$.\footnote{Though the number of actually generated term structures depends on the chosen 
coordinates and turns out to be smaller.} The dimensionless expansion parameter $\epsilon$ lets
us numerically determine the ``radius of convergence'' w.r.t.\ the running energy-momentum scale $k$ 
in Sec.\ \ref{sec:results}, since the dynamic ratio $\smash{M_{\pi}^{2}/M_{\sigma}^{2}}$ asymptotically
tends towards one, $\epsilon \rightarrow 1$, in the chiral-restored phase. It thereby crosses the 
critical value of $\epsilon = 1/4$, which signals the transition from convergence ($\epsilon < 1/4$) 
to divergence ($\epsilon \ge 1/4$) of the geometric series---hence enclosing the predictability of 
the scattering lengths within the convergent energy region of the pionic effective action,
\begin{equation}
	\epsilon = \frac{M_{\pi}^{2}}{M_{\sigma}^{2}} < 1/4.
\end{equation}

Taking the limit $M_{\sigma} \rightarrow \infty$ ($\epsilon \rightarrow 0$ for nonzero and 
finite $M_{\pi}$), such that
\begin{equation}
	\theta = f_{\pi} \equiv \lim_{M_{\sigma}\, \rightarrow\, \infty} \theta_{\mathrm{sol}},
\end{equation}
we consistently arrive again at the WLs of Eq.\ (\ref{eq:Weinberg}),
\begin{equation}
	\lim_{M_{\sigma}\, \rightarrow\, \infty} a_{0}^{0} 
	= \frac{7 M_{\pi}^{2}}{32 \pi f_{\pi}^{2}}, \qquad 
	\lim_{M_{\sigma}\, \rightarrow\, \infty} a_{0}^{2} 
	= - \frac{M_{\pi}^{2}}{16 \pi f_{\pi}^{2}}. 
	\label{eq:Weinberg2}
\end{equation}
These correspond to the tree-level scattering lengths within the NLSM (with pionic mass
contribution), i.e.,
\begin{equation}
	\vcenter{\hbox{
	\begin{pspicture}[showgrid=false](2.0,2.0)
		\psline[linewidth=0.03,linestyle=dashed,
		dash=2pt 1pt,linecolor=Red](0.45,1.55)(1.0,1.0)
		\psline[linewidth=0.03,linestyle=dashed,
		dash=2pt 1pt,linecolor=Red](1.55,1.55)(1.0,1.0)
		\psline[linewidth=0.03,linestyle=dashed,
		dash=2pt 1pt,linecolor=Red](0.45,0.45)(1.0,1.0)
		\psline[linewidth=0.03,linestyle=dashed,
		dash=2pt 1pt,linecolor=Red](1.55,0.45)(1.0,1.0)
		\rput[b]{*0}(0.25,1.65){\small{$\Pi$}}
		\rput[b]{*0}(1.75,1.65){\small{$\Pi$}}
		\rput[b]{*0}(0.25,0.15){\small{$\Pi$}}
		\rput[b]{*0}(1.75,0.15){\small{$\Pi$}}
		\pscircle[linewidth=0.03,fillstyle=hlines*,hatchwidth=0.02,
		fillcolor=lightgray!25,hatchsep=0.05,hatchcolor=gray](1.0,1.0){0.4}
	\end{pspicture}
	}} = 
	\underbrace{
	\vcenter{\hbox{
	\begin{pspicture}[showgrid=false](2.0,2.0)
		\psline[linewidth=0.03,linestyle=dashed,
		dash=2pt 1pt,linecolor=Red](0.45,1.55)(1.0,1.0)
		\psline[linewidth=0.03,linestyle=dashed,
		dash=2pt 1pt,linecolor=Red](1.55,1.55)(1.0,1.0)
		\psline[linewidth=0.03,linestyle=dashed,
		dash=2pt 1pt,linecolor=Red](0.45,0.45)(1.0,1.0)
		\psline[linewidth=0.03,linestyle=dashed,
		dash=2pt 1pt,linecolor=Red](1.55,0.45)(1.0,1.0)
		\rput[b]{*0}(0.25,1.65){\small{$\Pi$}}
		\rput[b]{*0}(1.75,1.65){\small{$\Pi$}}
		\rput[b]{*0}(0.25,0.15){\small{$\Pi$}}
		\rput[b]{*0}(1.75,0.15){\small{$\Pi$}}
		\pscircle[linewidth=0.03,fillstyle=solid,
		fillcolor=lightgray!25](1.0,1.0){0.20}
		\rput[b]{*0}(0.99,0.92){\fontsize{7pt}{0pt}{$4$}\selectfont}
	\end{pspicture}
	}} +
	\vcenter{\hbox{
	\begin{pspicture}[showgrid=false](3.0,2.0)
		\psline[linewidth=0.03,linestyle=dashed,
		dash=2pt 1pt,linecolor=Red](0.45,1.55)(1.0,1.0)
		\psline[linewidth=0.03,linestyle=dashed,
		dash=2pt 1pt,linecolor=Red](2.55,1.55)(2.0,1.0)
		\psline[linewidth=0.03,linestyle=dashed,
		dash=2pt 1pt,linecolor=Red](0.45,0.45)(1.0,1.0)
		\psline[linewidth=0.03,linestyle=dashed,
		dash=2pt 1pt,linecolor=Red](2.55,0.45)(2.0,1.0)
		\psline[linewidth=0.03,linestyle=dashed,
		dash=2pt 1pt,linecolor=Blue](1.0,1.0)(2.0,1.0)
		\rput[b]{*0}(0.25,1.65){\small{$\Pi$}}
		\rput[b]{*0}(2.75,1.65){\small{$\Pi$}}
		\rput[b]{*0}(0.25,0.15){\small{$\Pi$}}
		\rput[b]{*0}(2.75,0.15){\small{$\Pi$}}
		\rput[b]{*0}(1.5,0.65){\small{$\theta$}}
		\pscircle[linewidth=0.03,fillstyle=solid,
		fillcolor=lightgray!25](1.0,1.0){0.20}
		\rput[b]{*0}(1.0,0.92){\fontsize{7pt}{0pt}{$3$}\selectfont}
		\pscircle[linewidth=0.03,fillstyle=solid,
		fillcolor=lightgray!25](2.0,1.0){0.20}
		\rput[b]{*0}(2.0,0.92){\fontsize{7pt}{0pt}{$3$}\selectfont}
	\end{pspicture}
	}}
	}_{\text{LSM}}
	\stackrel{M_{\sigma}\, \rightarrow\, \infty}{\longrightarrow}
	\underbrace{
	\vcenter{\hbox{
	\begin{pspicture}[showgrid=false](2.0,2.0)
		\psline[linewidth=0.03,linestyle=dashed,
		dash=2pt 1pt,linecolor=Red](0.45,1.55)(1.0,1.0)
		\psline[linewidth=0.03,linestyle=dashed,
		dash=2pt 1pt,linecolor=Red](1.55,1.55)(1.0,1.0)
		\psline[linewidth=0.03,linestyle=dashed,
		dash=2pt 1pt,linecolor=Red](0.45,0.45)(1.0,1.0)
		\psline[linewidth=0.03,linestyle=dashed,
		dash=2pt 1pt,linecolor=Red](1.55,0.45)(1.0,1.0)
		\rput[b]{*0}(0.25,1.65){\small{$\Pi$}}
		\rput[b]{*0}(1.75,1.65){\small{$\Pi$}}
		\rput[b]{*0}(0.25,0.15){\small{$\Pi$}}
		\rput[b]{*0}(1.75,0.15){\small{$\Pi$}}
		\pscircle[linewidth=0.03,fillstyle=solid,
		fillcolor=lightgray!25](1.0,1.0){0.20}
		\rput[b]{*0}(0.99,0.92){\fontsize{7pt}{0pt}{$4$}\selectfont}
	\end{pspicture}
	}}
	}_{\text{NLSM}},
\end{equation}
where the $\theta$-exchange diagram represents the $s$-, $t$-, and $u$-channel. Finding 
corrections to Eq.\ (\ref{eq:Weinberg2}) (at order $p^{4}$ and higher) requires 
the extension of the (leading-order) NLSM to ChPT, involving the low-energy constants of QCD 
\cite{Bijnens:1995yn, Bijnens:1997vq, Ananthanarayan:2000ht, Colangelo:2000jc, Colangelo:2001df}. 
In the linear realization, one would have to explicitly compute the flow of higher-derivative 
couplings, which translate into covariant derivatives of the (pseudo-)Nambu-Goldstone fields 
\cite{Flore:2012ma, Divotgey:2019xea, Eser:2020phd}.

\section{Numeric results}
\label{sec:results}

As announced in the Introduction, we compute and compare the scattering lengths $a_{0}^{0}$ 
and $a_{0}^{2}$ in different approximations (or truncations). Following the strategy from 
Sec.\ \ref{sec:scattering}, we distinguish four cases (the first two of which itemized together):
\begin{itemize}
	\item FRG (LPA, $\mathrm{LPA}^{\prime}$): The flow equations for the Taylor coefficients 
	$\alpha_{n}$, $n = 1,\ldots , N_{\alpha}$, of the potential (\ref{eq:potential}) are 
	solved in the LPA and $\mathrm{LPA}^{\prime}$ truncations. The latter further includes the flow 
	of the wave-function renormalization factors, the Yukawa couplings, and the chiral-invariant 
	mass parameter. After the integration procedure, we translate the resulting effective action 
	$\Gamma$ into the stereographic form (\ref{eq:gammastereo}) and deduce the scattering lengths 
	(as described above). The scaling of the effective action $\Gamma_{k}$ even permits the evaluation 
	of the scattering lengths on all $k$-scales.
	\item One-loop approximation: By replacing the second functional derivative on the
	right-hand side of Eq.\ (\ref{eq:Wetterich}) with the classical two-point function, we 
	can analytically integrate the equation, once the initial couplings are fixed,
	\begin{equation}
		\Gamma_{k}^{(2)} \longrightarrow S^{(2)} \qquad \Rightarrow \qquad
		\Gamma_{\text{one-loop}} - \Gamma_{\Lambda,\text{one-loop}} 
		= \frac{1}{2} \tr \left\lbrace \ln S^{(2)} 
		- \ln\left[S^{(2)} + R_{\Lambda}\right] \right\rbrace .
		\label{eq:oneloop}
	\end{equation}
	In Eq.\ (\ref{eq:oneloop}), we plugged in the regulator property $R_{k\, =\, 0} = 0$ 
	(see Appendix \ref{sec:FRG}) and recognize the usual one-loop ``trace-log'' formula
	(plus a regulator term) \cite{Litim:2001ky, *Litim:2002xm}; see furthermore Appendix 
	\ref{sec:analytic} for the analytic integration of the flow equation for the effective 
	potential. We then proceed with the scattering lengths as in the first item.
	\item MF: We ignore the bosonic fluctuations on the right side of the flow equation
	(\ref{eq:Wetterich}), i.e., only the fermionic loop is integrated out \cite{Strodthoff:2011tz, 
	Weyrich:2015hha, *Stoll:2021ori}.\footnote{The different approximations/truncations are taken 
	here as definitions. It is clear that these may differ from other contexts/calculations.} 
	The outcome accounts for the physics of fermions in a fixed external bosonic potential.
	The integration procedure follows again Eq.\ (\ref{eq:oneloop}).\ Once more, we eventually 
	proceed as delineated in the first item.
\end{itemize}

We adjust the integration of momentum modes such that spontaneous chiral symmetry breaking 
roughly occurs at $\Lambda_{\chi} = 1.2\ \mathrm{GeV}$ in all approximations ($\mathrm{LPA}^{\prime}$, 
LPA, one-loop, and MF). This guarantees qualitative comparability among them by setting a particular 
reference scale. With this major premise, the initial parameters are independently tuned such 
that physical values for the particle masses (pions, fermions) and the pion decay constant 
$f_{\pi}$ are achieved in the IR limit, matching published data from the Particle Data Group (PDG) 
\cite{ParticleDataGroup:2020ssz}---the $\sigma$-mass, in turn, is a prediction of the approximation/truncation,
which is intrinsically constrained by the scale $\Lambda_{\chi}$ and the aforementioned observables.
To this end, we identify the fields in truncations (\ref{eq:truncation}) and (\ref{eq:fermions}) 
with the lightest resonances in the PDG data featuring the respective quantum numbers. Note
that, in order to find an unambiguous and compatible prediction for $M_{\sigma}$, we need to fix 
the chiral-invariant mass $m_{0}$ to a common value among all investigated approximations; see also 
the discussion in Appendix \ref{sec:analytic}. A reasonable choice for $m_{0}$ in the $\mathrm{LPA}$, 
one-loop, and MF cases is given by the renormalized IR mass $M_{0} = m_{0}/Z^{\psi}$ of the 
$\mathrm{LPA}^{\prime}$-truncation, which leads to exactly the same chiral-invariant mass and 
Yukawa couplings in the IR. Further details about the numeric implementation are presented in 
Appendices \ref{sec:FRG} and \ref{sec:analytic}.
\begin{figure*}
	\centering
	\includegraphics[scale=1.0]{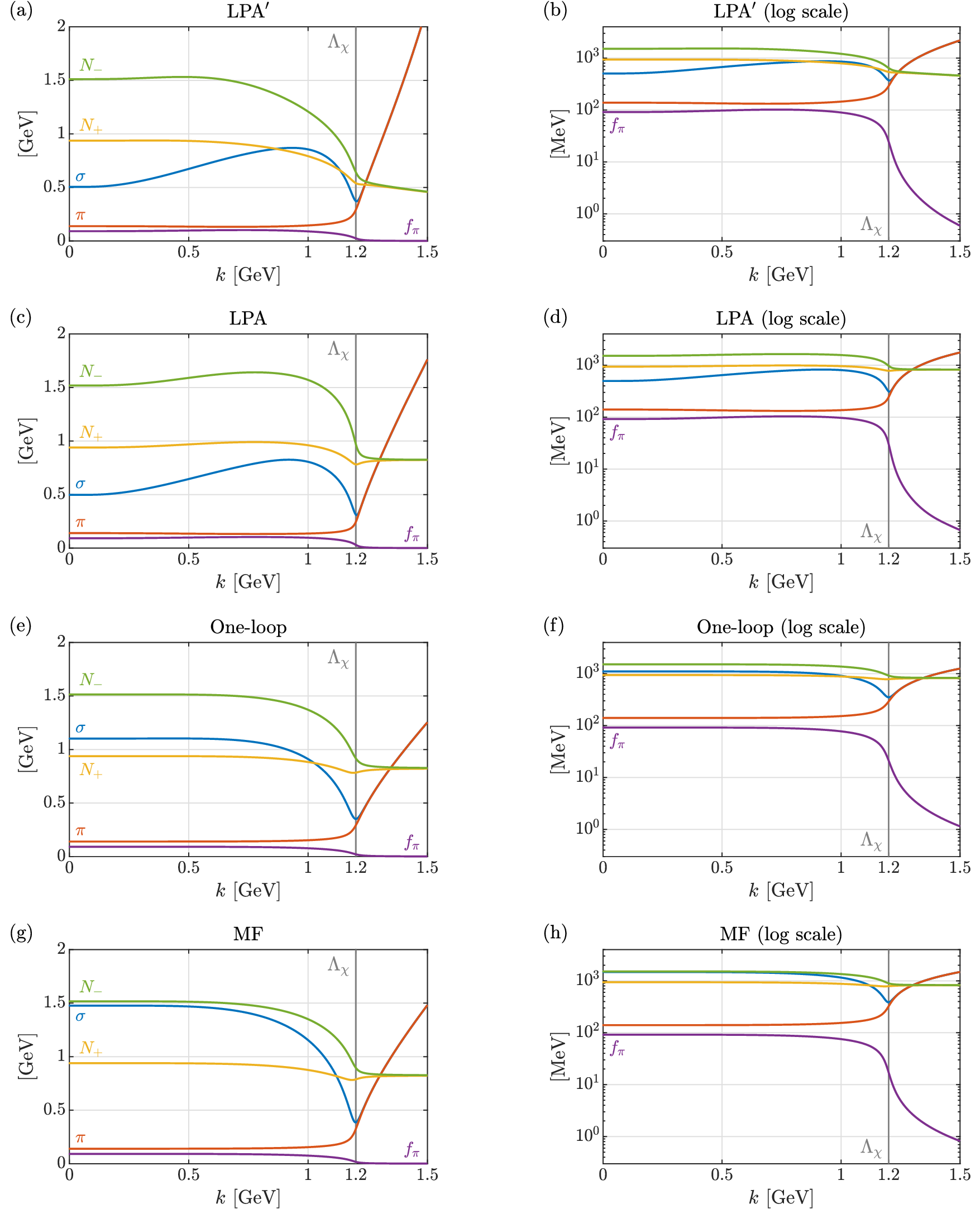}
	\caption{Scale dependence of the (renormalized) masses and the pion decay 
	constant $f_{\pi}$. The vertical gray line indicates the characteristic 
	chiral-symmetry breaking scale $\Lambda_{\chi}$. It is kept constant in 
	all approximations to achieve qualitative comparability. In each row,
	the plot on the right-hand side is the log-scale equivalent of the
	plot to its left.}
	\label{fig:masses}
\end{figure*}

Figure \ref{fig:masses} shows as a function of $k$, the integrated flow for all four cases, from 
the upper limit of $1.5\ \mathrm{GeV}$, which is the largest mass scale in the model [$M^{-} \equiv M_{N(1535)} 
\simeq 1.5\ \mathrm{GeV}$ \cite{ParticleDataGroup:2020ssz}], down to the IR limit, $k \rightarrow
0$. The IR ``observables'' produced by these flows are listed in Table \ref{tab:IR}; they are
compared to PDG data and to Ref.\ \cite{Divotgey:2016pst}.\ Intriguingly, if we require that
the scale of chiral symmetry breaking roughly lies at $\Lambda_{\chi}$ (where the mass splitting 
between the bosons becomes evident) and that the pion mass $M_{\pi}$, the pion decay constant 
$f_{\pi}$, as well as the fermion masses (mostly) account for physical values, we obtain isoscalar
$\sigma$-masses of around $500\ \mathrm{MeV}$ in the $\mathrm{LPA}$ and $\mathrm{LPA}^{\prime}$ 
(cf.\ columns five and six in Table \ref{tab:IR}). In contrast, the obtained $\sigma$-mass is relatively 
large in the MF and one-loop approximations (columns three and four; $M_{\sigma} > 1\ \mathrm{GeV}$), 
similar to the tree-level fitting result of Ref.\ \cite{Divotgey:2016pst} (second column).\ 
Taking the $f_{0}(500)$-resonance as reference (last column), the FRG-masses (LPA and 
$\mathrm{LPA}^{\prime}$) meet (within errorbars) the broad experimental value, while this 
not true for the MF and one-loop cases; those are rather associated with isoscalar $f_{0}$-resonances
above $1\ \mathrm{GeV}$.\ The same holds for the tree-level $\sigma$-mass of Ref.\ \cite{Divotgey:2016pst}, 
where the authors (and related publications therein) claim that it is rather the $f_{0}(1370)$-resonance 
that should be interpreted as the chiral partner of the pions.\ Regarding the other IR quantities,
we are able to generate reasonable values for the pion decay constant, the pion mass, and the fermion 
masses in all four approximations.\footnote{For the pion and nucleon masses, we take the arithmetic 
means of the PDG values (cf.\ the last column of Table \ref{tab:IR}). This is motivated by the exact 
$\mathsf{SU}(2)$ isospin symmetry of the nucleon-meson model (\ref{eq:truncation}).}
\begin{table}
	\caption{\label{tab:IR}IR observables obtained from the integration 
	of momentum modes as compared to the PDG \cite{ParticleDataGroup:2020ssz} 
	and to the tree-level fitting data used in Ref.\ \cite{Divotgey:2016pst}.}
	\begin{ruledtabular}
		\begin{tabular}{lcccccc}
		Observable & \multicolumn{6}{c}{} \\
		$[\mathrm{MeV}]$ & Ref.\ \cite{Divotgey:2016pst} & MF & One-loop & 
		FRG (LPA) & FRG ($\mathrm{LPA}^{\prime}$) & PDG \\
		\colrule\\[-0.3cm]
		$f_{\pi}$ & $96.3 \pm 0.7$ & $91.79$ & $91.56$ & $91.92$ & 
		$91.25$ & $(130.2 \pm 1.2)/\sqrt{2} \simeq 92.07 \pm 0.85$ \\[0.025cm]
		$M_{\pi}$ & $141.0 \pm 5.8$ & $140.0$ & $140.2$ & $139.9$ & 
		$138.2$ & $ (2 M_{\pi^{\pm}} + M_{\pi^{0}})/3 \simeq 138.1$ \\[0.025cm]
		$M_{\sigma}$ & $1363$ & $1477$ & $1102$ & $497$ & 
		$505$ & $M_{f_{0}(500)} = 475 \pm 75$ \\[0.025cm]
		$M^{+}$ & not included & $938.9$ & $938.0$ & $939.5$ & 
		$936.7$ & $(M_{\mathrm{proton}} + M_{\mathrm{neutron}})/2 \simeq 938.9$ \\[0.025cm]
		$M^{-}$ & not included & $1517$ & $1514$ & $1518$ & 
		$1511$ & $M_{N(1535)} = 1510 \pm 10$ \\[0.025cm]
		\end{tabular}
	\end{ruledtabular}
\end{table}

At the crossover transition from the chiral-broken phase in the IR ($\sigma_{0} \gg 0$) 
to the restored phase ($\sigma_{0} \rightarrow 0$), the chiral partners become degenerate 
in mass as one moves to the UV ($\pi \leftrightarrow \sigma$ and $N_{+} \leftrightarrow N_{-}$), 
according to the discussion in Sec.\ \ref{sec:model}. In the $\mathrm{LPA}^{\prime}$-truncation
[cf.\ Figs.\ \ref{fig:masses}(a) and \ref{fig:masses}(b)], the fermionic masses $M^{+}$ 
and $M^{-}$ obviously ``converge'' to a scale-dependent nonzero mass, which is precisely 
given by the (renormalized) mass parameter $450\ \mathrm{MeV} \lesssim M_{0} \lesssim 825\
\mathrm{MeV}$, as revealed in Appendix \ref{sec:FRG}. In the LPA [cf.\ Figs.\ \ref{fig:masses}(c) 
and \ref{fig:masses}(d)], they likewise tend to the constant value $m_{0} = 824.5\ \mathrm{MeV}$,
which coincides with the renormalized IR parameter $M_{0}$ of the $\mathrm{LPA}^{\prime}$-truncation.\ 
Comparing the LPA and $\mathrm{LPA}^{\prime}$ to the one-loop and MF calculations, we further 
observe that the evolution of the $\sigma$-mass w.r.t.\ the energy-momentum scale $k$ below 
$\Lambda_{\chi}$ is non-monotonous for the FRG truncations, but monotonous in Figs.\ \ref{fig:masses}(e) 
to \ref{fig:masses}(h). Apparently, the successive inclusion of quantum fluctuations (from MF and 
one-loop towards the FRG flows) bends the curve down to smaller isoscalar meson masses in the IR. 
This ``down bending'' emerges with the self-consistent treatment of the Taylor coefficients in
the integration of the flow equations; see the analysis of this feature in Fig.\ \ref{fig:LPAprime}(c)
in Appendix \ref{sec:FRG}. Furthermore, in Appendix \ref{sec:analytic}, we show that the down bending
can also occur for some parameter sets in the one-loop approximation, which however do not meet
the requirement of the chiral-symmetry breaking scale $\Lambda_{\chi}$. The bending is associated with 
the similar (non-)monotonicity in the flowing pion decay constant $f_{\pi} \sim \sigma_{0}$, which 
is however hardly visible in Figs.\ \ref{fig:masses}(a) to \ref{fig:masses}(d).\ As for the $\mathrm{LPA}$, 
the fermion masses in the MF and one-loop approximations approach the constant value of $m_{0} = 824.5\ 
\mathrm{MeV}$ in the UV (which equals the IR value of $m_{0}$ in these cases).
\begin{figure*}
	\centering
	\includegraphics[scale=1.0]{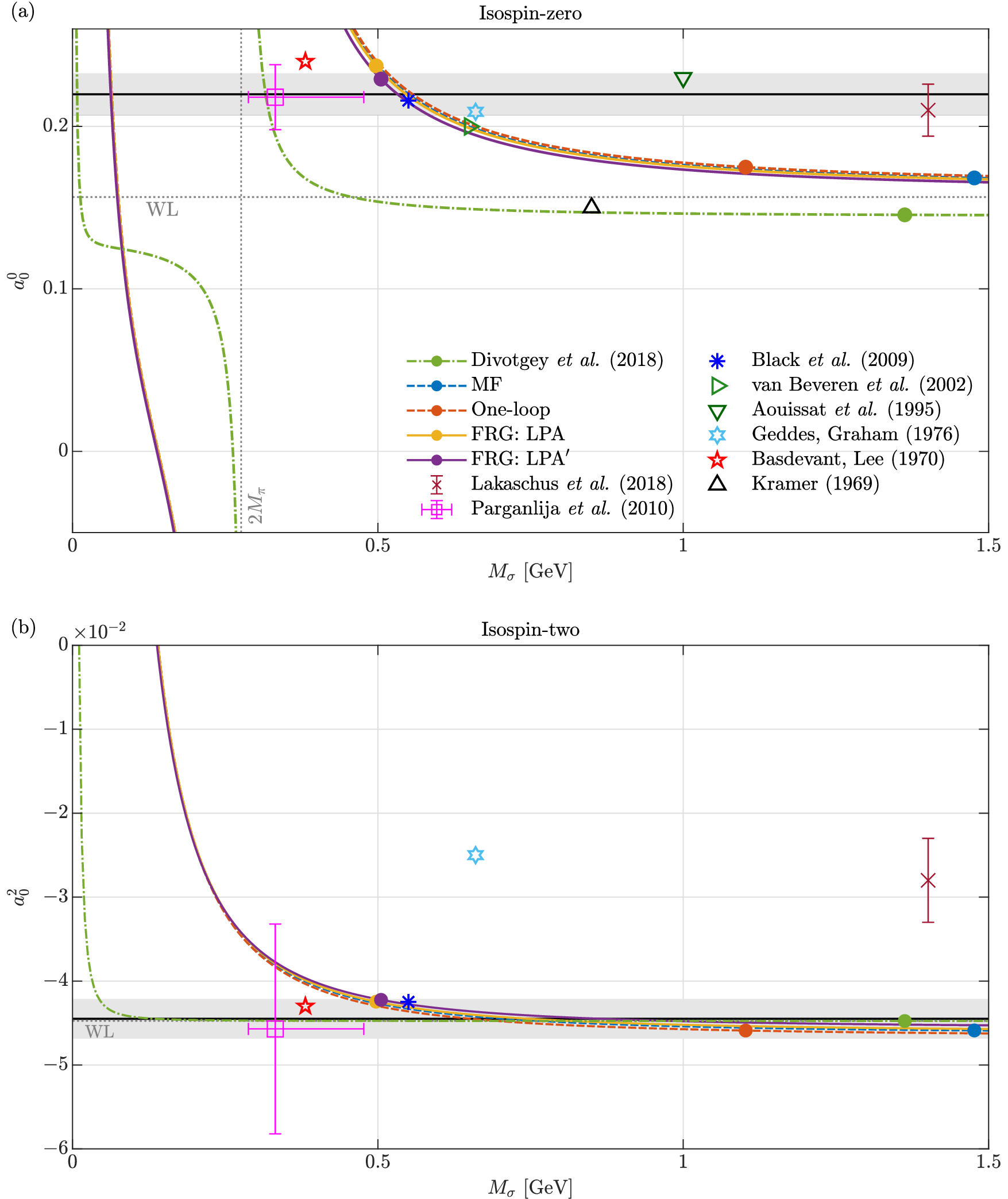}
	\caption{$S$-wave pion-pion scattering lengths as functions of the (renormalized)
	isoscalar mass $M_{\sigma}$. The legend in subfigure (a) applies to both panels;
	the markers on the solid, dashed, or dash-dotted lines, corresponding to the first
	five entries, indicate the computed or fitted pairs $(M_{\sigma},a_{0}^{I})$, 
	$I \in \lbrace 0,2 \rbrace$, of the present study and the closely related 
	Ref.\ \cite{Divotgey:2016pst} (cf.\ also Table \ref{tab:scattering}). The horizontal 
	black solid lines surrounded by the gray-shaded areas depict the respective 
	experimental values and their uncertainty. See the main text for detailed 
	explanations about the various (external) data points.}
	\label{fig:scattering}
\end{figure*}

In the linear realization of chiral symmetry, the $S$-wave scattering lengths $a_{0}^{0}$ and 
$a_{0}^{2}$ are functions of $M_{\sigma}$ via the scale-dependent ratio $\epsilon$, as 
elaborated in Sec.\ \ref{sec:scattering}. Keeping the pion mass $M_{\pi}$ and the pion decay 
constant $f_{\pi}$ fixed, we plot the scattering lengths as functions of $M_{\sigma}$ in Fig.\ 
\ref{fig:scattering}; the markers on the drawn lines represent the masses of Table \ref{tab:IR} 
and the corresponding pairs $(M_{\sigma},a_{0}^{I})$, $I \in \lbrace 0,2 \rbrace$, are 
summarized in Table \ref{tab:scattering}. As the main result, the low-energy scattering lengths 
from the $\mathrm{LPA}^{\prime}$-flow of Fig.\ \ref{fig:masses}(a) [computed with Eqs.\ (\ref{eq:a00exact}) 
and (\ref{eq:a02exact})] simultaneously agree (within errorbars) with the experimental/theoretical 
ChPT values of Refs.\ \cite{Gasser:2009zz, Leutwyler:2012sco}, see columns six to eight.\footnote{The 
reference values in columns seven and eight are obtained by an interplay of experimental data and 
theoretical ChPT-input, leading to high-accuracy scattering lengths; for details see Refs.\ 
\cite{Bijnens:1995yn, Bijnens:1997vq, Ananthanarayan:2000ht, Colangelo:2000jc, Colangelo:2001df,
Gasser:2009zz, Leutwyler:2012sco}. We quoted two different references for a thorough analysis.} 
The $\mathrm{LPA}$-flow of Fig.\ \ref{fig:masses}(c) produces an isospin-zero scattering length, 
which is close to but slightly outside of the error band, and only the isospin-two scattering 
length matches the reference value.

In the third and fourth column of Table \ref{tab:scattering}, in contrast, the values of $a_{0}^{0}$ 
significantly differ, which we ascribe to the comparatively large mass $M_{\sigma}$ in Figs.\ 
\ref{fig:masses}(e) to \ref{fig:masses}(h) (MF and one-loop).\ In general, $a_{0}^{0}$ strongly 
depends on the precise value of $M_{\sigma}$, as we infer from the curves in Fig.\ \ref{fig:scattering}(a) 
becoming steep below $M_{\sigma} \sim 500\ \mathrm{MeV}$.\ Comparing all approximations, 
the value of $a_{0}^{0}$ is successively improved by the advancement of the truncation 
and the loop corrections. On the other hand, for $I = 2$ [cf.\ Fig.\ \ref{fig:scattering}(b)], the 
curves overlap with the experimental band over a wide range of $500\ \mathrm{MeV}$ to the UV cutoff of 
$1.5\ \mathrm{GeV}$. This renders the isospin-two scattering length $a_{0}^{2}$ almost insensitive to 
$M_{\sigma}$ (for $M_{\sigma} \gtrsim 500\ \mathrm{MeV}$) and explains the good values found in 
all of the approximations (third line of Table \ref{tab:scattering}).

Note finally that the minor differences between the curves (i.e., the $M_{\sigma}$-dependence is
basically robust) in Figs.\ \ref{fig:scattering}(a) and \ref{fig:scattering}(b) come from the 
slight variation of the ratio $M_{\pi}^{2}/(4\pi f_{\pi})^{2}$ in the four cases,
\begin{IEEEeqnarray}{rCl}
	& \varpi \coloneqq \left(\frac{M_{\pi}}{4 \pi f_{\pi}}\right)^{2}
	\sim \left(\frac{M_{\pi}}{\Lambda_{\chi}}\right)^{2} , & 
	\label{eq:varpi} \\[0.1cm]
	& \varpi \stackrel{\mathrm{LPA}^{\prime}}{\simeq} 0.0145, \qquad
	\varpi \stackrel{\mathrm{LPA}}{\simeq} 0.0147, \qquad
	\varpi \stackrel{\text{One-loop}}{\simeq} 0.0148, \qquad
	\varpi \stackrel{\mathrm{MF}}{\simeq} 0.0147. &
	\label{eq:varpi2}
\end{IEEEeqnarray}
The low-energy parameter $\varpi$ is a measure for the pion mass $M_{\pi}$ in units of the characteristic
chiral-symmetry breaking scale $4\pi f_{\pi} \simeq \Lambda_{\chi}$ \cite{Colangelo:2000jc}; it is
part of the typical low-energy expansion of the Gell-Mann-Oakes-Renner relation, putting the pion
mass in direct connection to the light-flavor quark masses \cite{Gell-Mann:1968hlm, Colangelo:2000jc}. 
For the physical values from the PDG \cite{ParticleDataGroup:2020ssz}, one has $\varpi_{\mathrm{phys}} 
= 0.0142 \ll 1$.
\begin{table*}
	\caption{\label{tab:scattering}$S$-wave pion-pion scattering lengths 
	$a_{0}^{I}$, $I \in \lbrace 0,2 \rbrace$, within the (parity-doublet) nucleon-meson 
	model (for the approximations presented in Figs.\ \ref{fig:masses} and 
	\ref{fig:scattering} as well as in comparison with experimental data and ChPT).}
	\begin{ruledtabular}
		\begin{tabular}{lccccccc}
		Isospin $I$ & Ref.\ \cite{Divotgey:2016pst} & MF & One-loop & FRG (LPA) & 
		FRG ($\mathrm{LPA}^{\prime}$) &
		Experiment/ChPT \cite{Leutwyler:2012sco} & Experiment/ChPT \cite{Gasser:2009zz}\\
		\colrule\\[-0.3cm]
		$I = 0$ & $\phantom{-}0.1456$ & $\phantom{-}0.1683$ & $\phantom{-}0.1749$ & 
		$\phantom{-}0.2373$ & $\phantom{-}0.2291$ & 
		$\phantom{-}0.2198 \pm 0.0126$ & $\phantom{-}0.2196 \pm 0.0096$ \\
		$I = 2$ & $-0.0448$ & $-0.0459$ & $-0.0459$ & $-0.0425$ & $-0.0422$ & 
		$-0.0445 \pm 0.0023$ & $-0.0444 \pm 0.0024$ 
		\end{tabular}
	\end{ruledtabular}
\end{table*}

In the limit $M_{\sigma} \rightarrow 0$, which is equivalent to $\epsilon \rightarrow \infty$
(for nonzero and finite $M_{\pi}$), the scattering lengths diverge,
\begin{equation}
	\lim_{\epsilon\, \rightarrow\, \infty} a_{0}^{0} \equiv
	\lim_{\epsilon\, \rightarrow\, \infty} a_{0}^{2} = \infty. 
\end{equation}
This is easily verified, analytically and numerically, with the help of Eqs.\ (\ref{eq:a00exact})
and (\ref{eq:a02exact}) as well as Fig.\ \ref{fig:scattering}.\ Furthermore, at the other
extreme ($M_{\sigma} \rightarrow \infty$, $\epsilon \rightarrow 0$), the WLs of Eq.\ 
(\ref{eq:Weinberg2}) are shown as horizontal dotted lines in Figs.\ \ref{fig:scattering}(a)
and \ref{fig:scattering}(b); minor differences for large $\sigma$-masses between the WLs and 
the curves are inherited from the nonvanishing deviations in the pion mass $M_{\pi}$ and the
pion decay constant $f_{\pi}$ w.r.t.\ PDG data (which is employed to calculate the numeric 
WLs)---or, in other words, $\varpi \neq \varpi_{\mathrm{phys}}$. With regard to the scattering 
length $a_{0}^{0}$ in Fig.\ \ref{fig:scattering}(a), we also run into a pole at $M_{\sigma} = 
2 M_{\pi}$ (vertical dotted line; corresponding to resonant $\sigma$-exchange), i.e.,
\begin{equation}
	\begin{aligned}
	\lim_{\epsilon\, \rightarrow\, 1/4 + 0} a_{0}^{0} & = - \infty, \\[0.1cm]
	\lim_{\epsilon\, \rightarrow\, 1/4 - 0} a_{0}^{0} & = \infty ,
	\end{aligned}
\end{equation}
where the limit $\epsilon \rightarrow 1/4 \pm 0$ translates into the limit $M_{\sigma} 
\rightarrow 2M_{\pi}$ ``from below'' ($+$) and ``from above'' ($-$).\ Hence, one hits the 
pole when crossing the value of $\epsilon = 1/4$ during the integration process, accompanied 
by the onset of divergence in the geometric-series expansion (\ref{eq:a00exact}) of the pionic 
effective action.

Aside from the calculated scattering lengths, Figs.\ \ref{fig:scattering}(a) and \ref{fig:scattering}(b)
include external data from similar phenomenological models. The dash-dotted curve ``Divotgey \textit{et al.}\ 
(2018)'' corresponds to the (extended) LSMs of Refs.\ \cite{Parganlija:2012fy, Divotgey:2016pst}, 
where the (three-flavor) model parameters were fitted to mesonic masses, decay constants, and decay 
widths. As it involves the effect of vector and axial-vector mesons, the line in Fig.\ 
\ref{fig:scattering}(a) runs fairly below our results in the domain of $M_{\sigma} > 2 M_{\pi}$. 
It crosses the experimental band for values around $M_{\sigma} \sim 300\ \mathrm{MeV}$---a 
substantial shift to lower masses, as it was also detected in Ref.\ \cite{Fariborz:2009wf}.\ 
The isoscalar mass of the model, $M_{\sigma} = 1363\ \mathrm{MeV}$ (cf.\ once again column 
two of Table \ref{tab:IR}), hampers the simultaneous reproduction of the experimental scattering 
lengths; see the marker on the dash-dotted curve. For $a_{0}^{2}$, the outcome matches the
experimental band and the spin-one vector resonances remarkably augment its general agreement with
the theoretical calculation at lower $\sigma$-masses, cf.\ Fig.\ \ref{fig:scattering}(b). As repeatedly
mentioned, the analysis of Ref.\ \cite{Divotgey:2016pst} (and its Fig.\ 1) (partly) stimulated our 
work---despite the fact that the nucleons and their chiral partners were actually neglected there.

The various other data points in Figs.\ \ref{fig:scattering}(a) and \ref{fig:scattering}(b) 
embed our results into a historical and broader research context, with the linear realization
of chiral symmetry as the common footing. As stated in the Introduction, many approaches
fail to simultaneously predict both $a_{0}^{0}$ and $a_{0}^{2}$. The time-honored model calculations
of Refs.\ \cite{Kramer:1969gw, Basdevant:1970nu, Geddes:1976qf} do not meet both experimental/ChPT
values; using chiral (nucleon-)meson Lagrangians with a dynamic isoscalar $\sigma$-field, the data 
points ``Basdevant,\ Lee (1970)'' and ``Geddes,\ Graham (1976)'' match the isospin-two or isospin-zero 
scattering length, respectively, but they miss the other one.\ Together with the data point ``Kramer 
(1969),'' which is not accurate in either cases---in Fig.\ \ref{fig:scattering}(b), it even lies outside
of the plot range---, these publications document interest in the topic of pion-pion scattering in
the framework of LSMs back in the 60's and 70's; they shed light on different chiral-symmetry breaking 
schemes and investigated many parameter settings in their chiral models. In more modern studies on
LSMs, the scattering process remains a standard low-energy consistency check, as e.g.\ given by
the data points ``Aouissat \textit{et al.}\ (1995),'' ``van Beveren \textit{et al.}\ (2002),''
and ``Black \textit{et al.}\ (2009)'' \cite{Aouissat:1994sx, vanBeveren:2002mc, Black:2009bi}. 
While the former two determined the isospin-zero scattering length $a_{0}^{0}$ only (and the first 
of which solely agrees with the experimental uncertainty band by adding phenomenological form factors), 
the authors of the latter chose the mass $M_{\sigma}$ ``by hand'' as $550\ \mathrm{MeV}$ (without any 
dedicated fitting procedure); with that choice, they were able to match both scattering lengths at 
the same time and underlined the crucial dependence of the result on the isoscalar $\sigma$-mass---again, 
this is contrary to the dynamic generation of model parameters of our approach.

As an example for addressing the inverse question, i.e., determining the $\sigma$-mass from pion-pion
scattering data, the data point ``Parganlija \textit{et al.}\ (2010)'' represents the outcome of
the model fit of Ref.\ \cite{Parganlija:2010fz}. The point carries horizontal and vertical errorbars,
where its actual value and the errorbars stand for the experimental input, the resulting mass $M_{\sigma}$, 
and the employed experimental uncertainty, respectively. The $\sigma$-mass has a rather broad range 
of $M_{\sigma} \sim 300\ \mathrm{MeV}$ to $500\ \mathrm{MeV}$, caused by the large vertical errorbars 
(among other numeric errors). The latest external data point, which is named ``Lakaschus \textit{et al.}\ 
(2018),'' is an extension of Refs.\ \cite{Parganlija:2010fz, Parganlija:2012fy, Divotgey:2016pst} by
light scalar tetraquark and glueball states. In combination with the heavy $\sigma$-resonance, however,
the prediction of the scattering lengths (with vertical errorbars) is still not (simultaneously) compatible 
with the gray-shaded areas in Figs.\ \ref{fig:scattering}(a) and \ref{fig:scattering}(b).

\begin{figure*}
	\centering
	\includegraphics[scale=1.0]{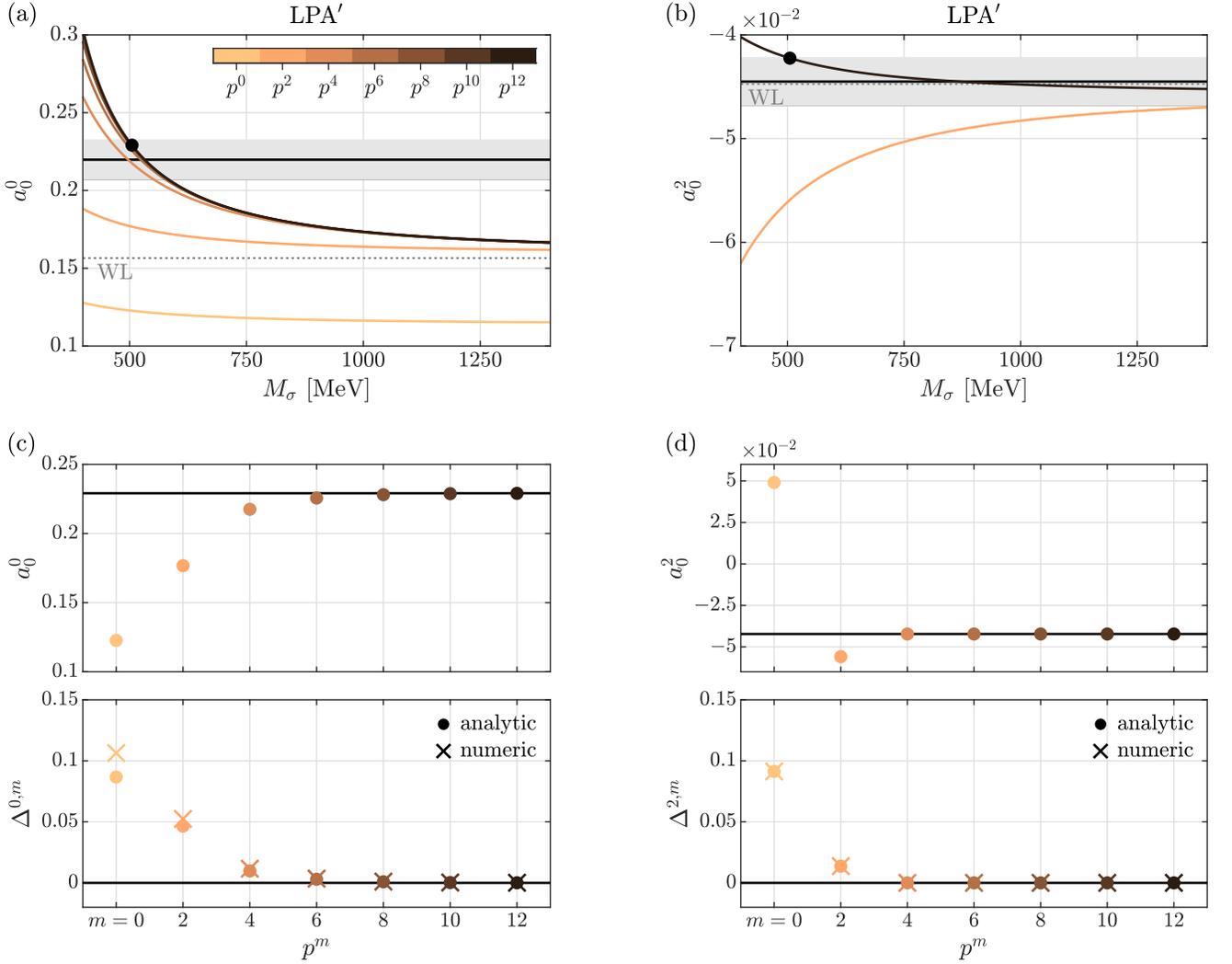}
	\caption{$S$-wave pion-pion scattering lengths within the $\mathrm{LPA}^{\prime}$ 
	and for different momentum orders in the low-energy expansion. The color bar in 
	subfigure (a) indicates the approximation order w.r.t.\ the pion momentum $p$, in 
	accordance with Eqs.\ (\ref{eq:eomexpansionstart}) to (\ref{eq:eomexpansionend}).
	The black dots in subplots (a) and (b) as well as the black horizontal lines in panels
	(c) and (d) (upper plots) represent the exact values of Table \ref{tab:scattering}.
	(c) and (d) Convergence of the computed pairs $(M_{\sigma},a_{0}^{I})$, $I \in \lbrace 
	0,2 \rbrace$, with successively advancing the $p$-order. The lower plots show the
	analytic and numeric errors $\smash{\Delta^{I,m}_{\mathrm{analytic/numeric}}}$, 
	cf.\ Eqs.\ (\ref{eq:error1}) through (\ref{eq:errordiff}) and Table \ref{tab:errors}.
	The analytic error is numerically evaluated for $\epsilon \simeq 0.0749$.}
	\label{fig:expansion}
\end{figure*}

Figure \ref{fig:expansion} demonstrates the (numeric) convergence of the expansion in Eqs.\
(\ref{eq:eomexpansionstart}) to (\ref{eq:eomexpansionend}) within the $\mathrm{LPA}^{\prime}$,
which we conceive as the most relevant approximation, as it incorporates the effective action 
$\Gamma_{k}$ in its full-fledged ansatz (\ref{eq:truncation}).\ The curves for the scattering length 
$a_{0}^{0}$ in Fig.\ \ref{fig:expansion}(a) overlap for $M_{\sigma} > 1\ \mathrm{GeV}$ already for 
momentum orders of $m \ge 4$. They expose at the same time the necessity for higher orders ($m \gtrsim 8$) 
in the ``sensitive'' range around $M_{\sigma} \sim 500\ \mathrm{MeV}$. While the lowest order ($m = 0$) 
does not even match the WL (\ref{eq:Weinberg2}) for a heavy $\sigma$-meson, the high-momentum orders 
($m \ge 6$) converge fast to the exact value of Eq.\ (\ref{eq:a00exact}) and the analytic and numeric 
errors of Eqs.\ (\ref{eq:error1}) and (\ref{eq:error2}) shrink [cf.\ Fig.\ \ref{fig:expansion}(c)].\ For
$a_{0}^{2}$ the curves and the numeric values become identical as soon as $m \ge 4$ and 
the expansion errors completely vanish [cf.\ Figs.\ \ref{fig:expansion}(b) and 
\ref{fig:expansion}(d)], as stated in Sec.\ \ref{sec:scattering}. Like for $a_{0}^{0}$, the
first-order curve ($m = 0$) is incompatible with the WL (\ref{eq:Weinberg}). Its values are positive, 
thus lying outside of the actual axis borders of Fig.\ \ref{fig:expansion}(b). The numeric evaluation
of the low-energy expansion and the errors for the $\mathrm{LPA}^{\prime}$ are also given in 
Table \ref{tab:numerrors}. To restrain the errors to orders of $10^{-2}$ and $10^{-3}$ for
the isospin-zero and isospin-two scattering lengths, respectively, one should consider the
low-energy expansion up to $\smash{\mathcal{O}\!\left(p^{4}\right)}$. This roughly yields
equal uncertainties as for the experimental/ChPT values \cite{Leutwyler:2012sco, Gasser:2009zz}
(for $I = 2$, the analytic and numeric errors are zero). At $\smash{\mathcal{O}\!\left(p^{12}\right)}$, 
the errors {\small $\smash{\Delta_{\mathrm{analytic/numeric}}^{0,12}}$} are smaller than $10^{-4}$, 
hence below the significant digits of the values in Table \ref{tab:scattering}. In leading order and
next-to leading order of the low-energy expansion [$\smash{\mathcal{O}\!\left(p^{0}\right)}$
and $\smash{\mathcal{O}\!\left(p^{2}\right)}$], the results are unreliable, since the errors 
are of the same order as the $S$-wave scattering lengths themselves.\footnote{These statements are 
concerned with truncation errors w.r.t.\ the generated higher-derivative pion self-interactions 
from the elimination of the isoscalar field; other error sources are still present.}
\begin{table*}
	\caption{\label{tab:numerrors}Numeric evaluation of the errors $\Delta^{I,m}_{
	\mathrm{analytic/numeric}}$ (upper part: analytic; lower part: numeric) for 
	the $\mathrm{LPA}^{\prime}$ ($\epsilon \simeq 0.0749$).}
	\begin{ruledtabular}
		\begin{tabular}{lccccccc}
		Isospin $I$ & $\mathcal{O}\!\left(p^{0}\right)$ &
		$\mathcal{O}\!\left(p^{2}\right)$ & $\mathcal{O}\!\left(p^{4}\right)$ & 
		$\mathcal{O}\!\left(p^{6}\right)$ & $\mathcal{O}\!\left(p^{8}\right)$ &
		$\mathcal{O}\!\left(p^{10}\right)$ & $\mathcal{O}\!\left(p^{12}\right)$ \\[0.1cm]
		\colrule\\[-0.25cm]
		$I = 0$ & 
		$0.0867$ & 
		$0.0465$ &
		$9.8 \times 10^{-3}$ & 
		$2.9 \times 10^{-3}$ & 
		$8.8 \times 10^{-3}$ & 
		$2.6 \times 10^{-4}$ &
		$7.9 \times 10^{-5}$ \\
		$I = 2$ &
		$0.0913$ & 
		$0.0137$ & zero & zero & zero & zero & zero \\
		\colrule\\[-0.25cm]
		$I = 0$ & 
		$0.1064$ & 
		$0.0524$ &
		$0.0116$ & 
		$3.5 \times 10^{-3}$ & 
		$1.1 \times 10^{-3}$ & 
		$3.1 \times 10^{-4}$ &
		$9.3 \times 10^{-5}$ \\
		$I = 2$ &
		$0.0913$ & 
		$0.0137$ & zero & zero & zero & zero & zero
		\end{tabular}
	\end{ruledtabular}
\end{table*}

\begin{figure*}
	\centering
	\includegraphics[scale=1.0]{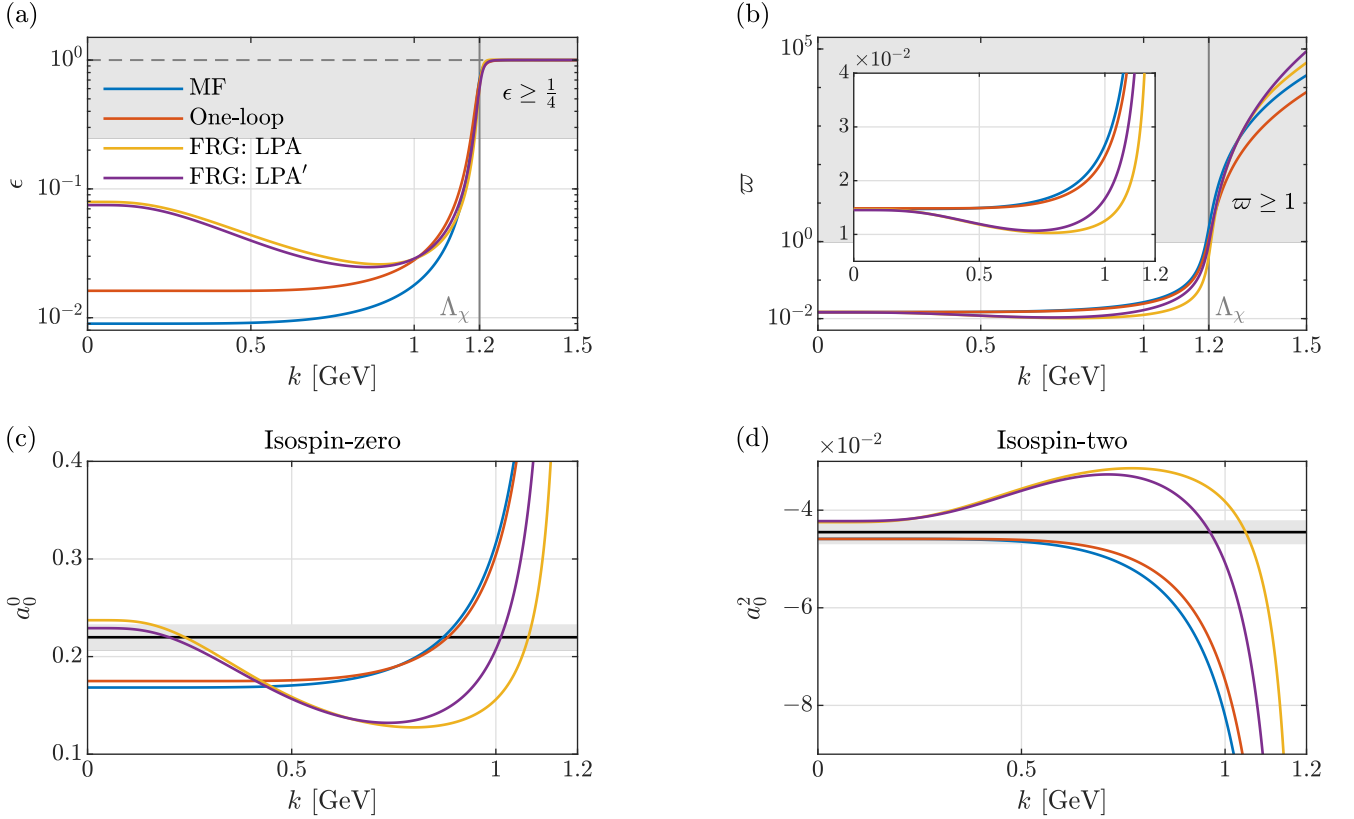}
	\caption{Flow of the low-energy parameters $\epsilon$ and $\varpi$
	and the $S$-wave scattering lengths as functions of the energy-momentum scale $k$.
	The legend in subfigure (a) is valid for all panels. (a) The parameter $\epsilon$ 
	approaches the value of one, $\epsilon \rightarrow 1$ (horizontal dashed line), 
	for large $k$. In the gray-shaded area at $\epsilon \ge 1/4$, the low-energy
	expansion (\ref{eq:a00exact}) necessarily fails. (b) The parameter $\varpi$ 
	drastically grows in the vicinity of the chiral-symmetry breaking scale $\Lambda_{\chi}$. 
	It further signals for $\varpi \ge 1$ the breakdown of the low-energy description 
	(purely) in terms of pion fields. (c) and (d) Isospin-zero and isospin-two scattering 
	lengths $a_{0}^{0}$ and $a_{0}^{2}$ and their dependence on the energy-momentum scale 
	$k < \Lambda_{\chi} = 1.2\ \mathrm{GeV}$. As before, the experimental/ChPT results
	are given as black horizontal lines on a gray background (errors).}
	\label{fig:scattering_evolution}
\end{figure*}

The integration of momentum modes gives direct access to the $\epsilon$-parameter and to the 
$\varpi$-ratio over the entire energy-momentum range, $0\ \mathrm{GeV} \le k \le 1.5\ \mathrm{GeV}$.\
This is shown in Figs.\ \ref{fig:scattering_evolution}(a) and \ref{fig:scattering_evolution}(b); in those 
panels, the gray-shaded areas depict the regimes, where the low-energy parameters become large,
i.e., $\epsilon \ge 1/4$ and $\varpi \ge 1$, and the low-energy $\epsilon$-expansion (\ref{eq:a00exact}) 
as well as analogous series in $\varpi$ break down.\ This happens close to the crossover 
scale $\Lambda_{\chi}$, where the bosons are (almost) degenerate in mass ($\epsilon 
\rightarrow 1$ for $k > \Lambda_{\chi}$) and the pion mass surpasses the characteristic 
scale of $4\pi f_{\pi}$ ($\varpi > 1$ for $k \gtrsim \Lambda_{\chi}$), cf.\ Eq.\ (\ref{eq:varpi}). 
In consequence, the scale evolution of the two parameters sets the ``radius of convergence'' 
of the low-energy expansion approximately to $\Lambda_{\chi}$. Additionally, 
Fig.\ \ref{fig:scattering_evolution}(a) and the inset in Fig.\ \ref{fig:scattering_evolution}(b) 
confirm the (non-)monotonicity in the scaling of the $\sigma$-mass, as originally observed 
in Fig.\ \ref{fig:masses}. Most notably, this pushes the FRG curves ``upwards'' to the 
experimental band for $a_{0}^{0}$ and corrects them ``downwards'' for $a_{0}^{2}$, see Fig.\ 
\ref{fig:scattering_evolution}(c) and Fig.\ \ref{fig:scattering_evolution}(d), respectively. The 
scale evolution is plotted for $0 \le k < \Lambda_{\chi}$, the region beyond which the scattering 
lengths take on ``unrealistically'' large or small values [as also seen in the heat-kernel approach 
of Ref.\ \cite{Schaefer:1997nd} at finite temperature]. In the IR, the $\epsilon$-parameters for 
the four different approximations are read off Fig.\ \ref{fig:scattering_evolution}(a) as
\begin{equation}
	\epsilon \stackrel{\mathrm{LPA}^{\prime}}{\simeq} 0.0749, \qquad
	\epsilon \stackrel{\mathrm{LPA}}{\simeq} 0.0792, \qquad
	\epsilon \stackrel{\text{One-loop}}{\simeq} 0.0162, \qquad
	\epsilon \stackrel{\mathrm{MF}}{\simeq} 0.0090.
\end{equation}
In all cases, they are smaller than $\epsilon_{\mathrm{phys}}$ of Eq.\ (\ref{eq:epsilon}), 
$\epsilon < \epsilon_{\mathrm{phys}} \ll 1$, which justifies the expansion in powers of 
$\epsilon$.

\section{Summary and conclusion}
\label{sec:summary}

We have investigated the dynamic generation of the $S$-wave pion-pion scattering lengths 
within a nucleon-meson model featuring parity doubling. To this end, the powerful framework of 
the FRG has been combined with a systematic expansion of the bosonic part of the quantum 
effective action in terms of stereographic coordinates and pion momenta, underlining
similarities to the NLSM and the chiral expansion of ChPT. The FRG technique allowed us to 
produce fluctuation-enhanced results, which were compared to MF and one-loop calculations.
In particular, this ansatz enabled us to
\begin{itemize}
	\item bring together significant low-energy scales and parameters in the dynamic
	context of the renormalization-group idea; we deliberately chose as the initialization 
	scale for the integration of momentum modes the characteristic scale of chiral symmetry 
	breaking, $\Lambda_{\chi} \simeq 4\pi f_{\pi}$.
	\item get access to the effective action of the nucleon-meson model and compute 
	the pion-pion scattering amplitude via tree diagrams consisting of dressed 
	propagators and vertices. 
	\item organize the pionic effective action in a power-counting scheme 
	in terms of pion fields and pion momenta (similar to the NLSM and ChPT).
	\item discuss a ``radius of convergence'' for the low-energy expansion (w.r.t.\
	the running energy-momentum scale of the FRG).
	\item learn about subtleties and intricacies coming with the low-energy expansion 
	of the nonlinear pionic effective action and the truncation of its 
	higher-derivative interactions. 
\end{itemize}

As the major finding, we were able to simultaneously match (within errorbars) the 
experimental/ChPT values for the $S$-wave isospin-zero and isospin-two scattering lengths
$a_{0}^{0}$ and $a_{0}^{2}$.\ This was accomplished for the $\mathrm{LPA}^{\prime}$-truncation
within the FRG (and not quite for the $\mathrm{LPA}$) with the precondition of chiral 
symmetry breaking occurring at $\Lambda_{\chi} = 1.2\ \mathrm{GeV} \simeq 4\pi f_{\pi}$,
\begin{equation}
	\begin{aligned}
	\mathrm{LPA}\colon \quad a_{0}^{0} & = 0.2373 , \qquad & 
	a_{0}^{2} & = -0.0425 ; \qquad \\[0.1cm]
	\mathrm{LPA}^{\prime}\colon \quad a_{0}^{0} & = 0.2291 , \qquad & 
	a_{0}^{2} & = -0.0422 ; \qquad \\[0.1cm]
	\mathrm{experiment/ChPT}\colon 
	\quad a_{0}^{0} & = 0.2198 \pm 0.0126, \qquad & 
	a_{0}^{2} & = -0.0445 \pm 0.0023. 
	\label{eq:finalresult}
	\end{aligned}
\end{equation}
The integration of momentum modes was moreover constrained by the physical masses and the pion 
decay constant $f_{\pi}$ in the IR limit, where solely the dynamically generated isoscalar mass 
$M_{\sigma}$ was understood as a model prediction. The value of $M_{\sigma}$ is crucial for a 
``successful'' computation of the isospin-zero scattering length $a_{0}^{0}$ in the linear realization 
of chiral symmetry \cite{Black:2009bi, Divotgey:2016pst}.\ In the LPA and $\mathrm{LPA}^{\prime}$, 
we obtained $M_{\sigma}$ of the order of $500\ \mathrm{MeV}$. Contrary to the present work, 
similar phenomenological approaches oftentimes fail in simultaneously reproducing both scattering 
lengths or require for this purpose additional form factors and the value of $M_{\sigma}$, as 
well-chosen external input \cite{Kramer:1969gw, Basdevant:1970nu, Geddes:1975cf, Geddes:1976qf, 
Aouissat:1994sx, Black:2009bi, Parganlija:2010fz, Soto:2011ap, Parganlija:2012fy, Divotgey:2016pst, 
Lakaschus:2018rki}. The FRG framework outperforms numerous other calculations in that sense and 
gave new insights into the energy-momentum dependence of the pion-pion scattering lengths.\ To generate 
the desired values for the scattering lengths, the chiral-invariant mass of the nucleon and its chiral 
partner roughly lay in the range of $450\ \mathrm{MeV} \lesssim M_{0} \lesssim 825\ \mathrm{MeV}$ for the 
$\mathrm{LPA}^{\prime}$ ($824.5\ \mathrm{MeV}$ in the IR) and it was fixed to $824.5\ \mathrm{MeV}$
for the $\mathrm{LPA}$. These masses are broadly consistent with the corresponding parameters in 
Refs.\ \cite{Detar:1988kn, Jido:2001nt, Gallas:2009qp, Nemoto:1998um, *Bramon:2003xq, *Zschiesche:2006zj, 
*Wilms:2007uc, *Dexheimer:2007tn, *Dexheimer:2008cv, *Hayano:2008vn, *Sasaki:2010bp, *Sasaki:2011ff, 
*Giacosa:2011qd, *Gallas:2011qp, *Steinheimer:2011ea, *Paeng:2011hy, *Dexheimer:2012eu, *Gallas:2013ipa, 
*Heinz:2013hza, *Paeng:2013xya, *Benic:2015pia, *Motohiro:2015taa, *Olbrich:2015gln, *Mukherjee:2016nhb, 
*Mukherjee:2017jzi, *Suenaga:2017wbb, *Takeda:2017mrm, *Paeng:2017qvp, *Marczenko:2017huu, *Sasaki:2017glk,
*Marczenko:2018jui, *Takeda:2018ldi, *Yamazaki:2018stk, *Yamazaki:2019tuo, *Marczenko:2019trv, 
*Suenaga:2019urn, *Marczenko:2020jma, *Minamikawa:2020jfj, *Marczenko:2020omo, *Minamikawa:2021fln}.
The mass parameter $m_{0}$ and the influence of its variation on the FRG flows will be discussed 
in detail in a follow-up work.

In the simplest approximations (MF and one-loop), we did not reach agreement for $a_{0}^{0}$, 
which is mainly due to the relatively large value of $M_{\sigma}$ obtained in these calculations, 
$M_{\sigma} > 1\ \mathrm{GeV}$. However, such a value turns out to be necessary to meet the above 
constraints for the chiral-symmetry breaking scale and the physical masses in the IR. Interestingly, 
large $\sigma$-masses of more than $1\ \mathrm{GeV}$ were also favored by the (extended) LSMs of 
Refs.\ \cite{Parganlija:2010fz, Parganlija:2012fy}, which were fitted to experimental data like 
mesonic decay constants and decay widths.\ On these grounds, the computation of the pion-pion scattering 
lengths in Ref.\ \cite{Divotgey:2016pst} further motivated the current publication, since its findings 
were not compatible with the experimental values. Suitable to the discussion of the $\sigma$-mass, recent 
extensions of these models with tetraquarks and glueballs as additional ``exotic'' resonances in the light 
scalar sector below $1.3\ \mathrm{GeV}$ improve the scattering lengths, but still not to a satisfactory 
conclusion (regarding the experimental/ChPT values) \cite{Lakaschus:2018rki}.

In addition to the scattering lengths, we supplied a comprehensive investigation of the 
scale evolution of important low-energy parameters, which determines the ``radius of
convergence'' of the expansion in small pion momenta (w.r.t.\ the energy-momentum scale).\
Close to $\Lambda_{\chi} \simeq 4\pi f_{\pi}$, those parameters become large and signal 
the breakdown of the corresponding geometric series. There, the predictability of the pion-pion
scattering lengths within the presented formalism (with ``stereographic pions'') and its convergence 
is lost. To compute the scattering lengths in the convergent regime up to a numerical error of
less than $10^{-4}$ (as compared to the universal LSM formulas), we were urged to take pion 
self-interactions of $\smash{\mathcal{O}\!\left(p^{12}\right)}$ in the pion momentum $p$ into account.\
In general, our choice of the low-energy parameter $\epsilon = \smash{M_{\pi}^{2}/M_{\sigma}^{2}}$ 
for this expansion is not unique and might be replaced by other appropriate choices.\
The analysis of the low-energy parameters was finally accompanied by an explicit proof of 
coordinate independence of the (on-shell) four-point function.\ In summary, we have thereby 
drawn---with the FRG formalism applied to a nucleon-meson model and in an unprecedented manner---a 
direct connection between the scale of $4\pi f_{\pi} \simeq 1.2\ \mathrm{GeV}$ and the $S$-wave 
pion-pion scattering lengths as low-energy observables of QCD.

Nonetheless, in the current approximation, it is not possible to yield analogous corrections 
beyond the leading-order NLSM in the nonlinear realization of chiral symmetry, which inevitably 
demands the elimination of the $\sigma$-mode (or the ``radial excitation'') through the limit 
$M_{\sigma} \rightarrow \infty$ \cite{Bessis:1972sn, Appelquist:1980ae, Zinn-Justin:1996khx}. 
Going beyond the leading order requires the explicit treatment of higher-derivative interactions 
in the (parity-doublet) nucleon-meson model, as done in Refs.\ \cite{Eser:2018jqo, Divotgey:2019xea,
Cichutek:2020bli} for the quark-meson(-diquark) models. These terms generate low-energy couplings 
independent of the isoscalar mass $M_{\sigma}$ and will be addressed in upcoming publications.
In addition, future efforts could focus on the scattering process at nonzero temperatures and densities 
(or external fields) \cite{Schaefer:1997nd, Loewe:2008ui, *Loewe:2017kiw, *Loewe:2019zwq, *Loewe:2019xtn}, 
also involving latest fluid-dynamical perspectives on the nonperturbative FRG flows \cite{Grossi:2019urj, 
*Grossi:2021ksl, *Koenigstein:2021syz, *Koenigstein:2021rxj, *Steil:2021cbu}. It would be eventually interesting
to analyze the (tree-level) effect of vector mesons on the pion-pion scattering lengths (in the 
light of resonance saturation) \cite{Ecker:1988te, Divotgey:2016pst}.

Finally, based on the presented calculations of the $S$-wave pion-pion scattering lengths, we do not 
draw conclusions in favor of or against specific features of parity doubling in the investigated 
nucleon-meson model.\ We regard the present investigation as an interesting application 
of the parity-doublet model, in which we could combine pertinent observables, parameters, and scales 
of low-energy QCD. In general, our findings do not rule out other model settings, especially concerning
the debate about the chiral-invariant nucleon mass in the chiral-restored phase.

\begin{acknowledgments}
We are grateful to F.\ Divotgey for contributing to the early stages of this work. J.E.\ is indebted 
to J.M.\ Pawlowski, D.H.\ Rischke, B.J.\ Schaefer and L.\ von Smekal for their longstanding support. 
The authors further thank F.\ Giacosa and A.\ Koenigstein for valuable discussions. J.E.\ acknowledges 
funding by the German National Academy of Sciences Leopoldina (through the scholarship 2020-06). 
J.E.\ thanks the IPhT for its hospitality.
\end{acknowledgments}

\appendix

\section{Model construction and physical basis}
\label{sec:modelconstruction}

The parity-doublet nucleon-meson model (\ref{eq:truncation}) incorporates baryons in the 
``mirror assignment'' of chiral symmetry. They ``live'' in the three-fold tensor product of the 
Dirac-fermion representation $(2,1) \oplus (1,2)$ (w.r.t.\ Lorentz symmetry),
\begin{equation}
	\left[(2,1) \oplus (1,2)\right]^{\otimes 3} \cong 
	5 \left[(2,1) \oplus (1,2)\right]
	+ 3 \left[(3,2) \oplus (2,3)\right]
	+ (4,1) \oplus (1,4) . \label{eq:representations}
\end{equation}
Each of the direct sums on the left-hand side of the above equation symbolizes a quark 
representation. To obtain this result, we made use of the decomposition
\begin{equation}
	\left[(2,1) \oplus (1,2)\right]^{\otimes 2} \cong
	2 (1,1) \oplus 2 (2,2) \oplus (3,1) \oplus (1,3).
\end{equation}
More precisely, the baryons correspond to one of the Dirac representations on the right-hand 
side of Eq.\ (\ref{eq:representations}). In the mirror assignment of the two fermion
fields $\psi_{1}$ and $\psi_{2}$, however, the Weyl spinors are considered as twisted, i.e.,
\begin{equation}
	\begin{aligned}
	\psi_{1}\colon \ (2,1) \oplus (1,2) & = \psi_{1,l} \oplus \psi_{1,r}, \\[0.2cm]
	\psi_{2}\colon \ (1,2) \oplus (2,1) & = \psi_{2,l} \oplus \psi_{2,r} .
	\end{aligned}
\end{equation}
The transformation behavior of the ``left'' and ``right'' components thus reads
\begin{equation}
	\begin{aligned}
	\psi_{1,l} & \longrightarrow u_{l} \psi_{1,l}, \qquad &
	\psi_{1,r} & \longrightarrow u_{r} \psi_{1,r}, \\
	\psi_{2,l} & \longrightarrow u_{r} \psi_{2,l}, \qquad &
	\psi_{2,r} & \longrightarrow u_{l} \psi_{2,r},
	\end{aligned} \qquad\qquad
	(u_{l},u_{r}) \in \mathsf{SU}(2) \times \mathsf{SU}(2).
\end{equation}
Further details about the mirror assignment and a comparison to the ``naive'' assignment---where
both fields behave equally under chiral transformations---can be found in Refs.\ \cite{Detar:1988kn, 
Hatsuda:1988mv, Jido:1998av, Jido:2001nt}.\footnote{Especially, a very thorough introduction is
given in Ref.\ \cite{Jido:2001nt}.} Equipped with this setting, it is then straightforward to 
construct invariant interaction terms with mesons [in the $(2,2)$-representation of the chiral 
group] in the ``usual manner'' \cite{Giacosa:2017pos} as well as a chiral-invariant mass term,
\begin{equation}
	m_{0} \left(\bar{\psi}_{1,l}\psi_{2,r} + \bar{\psi}_{2,l}\psi_{1,r}\right)
	\equiv m_{0} \left(\bar{\psi}_{1}\psi_{2} + \bar{\psi}_{2}\psi_{1}\right).
\end{equation}
The last identity is inferred from the idempotence of chiral projection operators (not shown
here). Eventually, these considerations result in the truncation (\ref{eq:truncation}), where the
opposite minus sign in front of the term $y_{2}\bar{\psi}_{2}\sigma\psi_{2}$ is the
consequence of chiral invariance in the mirror assignment.

The fermions $\psi_{1}$ and $\psi_{2}$ are initially defined with even parity, $\mathsf{P} = 1$. 
For them to describe the nucleons and their chiral-partner fields, it is necessary to 
diagonalize the mass matrix (\ref{eq:massunphys}) and switch the parity eigenvalue of 
one of the spinors. This is achieved by the orthogonal transformation given by
\begin{IEEEeqnarray}{rCl}
	N & = & \begin{pmatrix}
	N_{+} \\ N_{-}
	\end{pmatrix}
	= \begin{pmatrix}
	\mathbbmss{1} & 0 \\
	0 & \gamma_{5}
	\end{pmatrix}
	\begin{pmatrix}
	\cos \omega & \sin \omega \\
	- \sin \omega & \cos \omega
	\end{pmatrix}
	\begin{pmatrix}
	\psi_{1} \\ \psi_{2}
	\end{pmatrix}
	\eqqcolon \mathcal{P} O \Psi
	= \mathcal{P} \Psi', \\[0.2cm]
	\bar{N} & = & \left(\bar{N}_{+},\bar{N}_{-}\right) =
	\bar{\Psi} O^{\mathsf{T}} \mathcal{P} 
	\begin{pmatrix}
	1 & 0 \\
	0 & -1
	\end{pmatrix} , \label{eq:rotation}
\end{IEEEeqnarray}
such that $N_{+} = \psi_{1}'$ and $N_{-} = \gamma_{5} \psi_{2}'$. The matrix $\mathcal{P}$ 
obviously attaches an additional Dirac matrix $\gamma_{5}$ to the rotated fermion $\psi_{2}'$, 
turning its parity to $\mathsf{P} = -1$ \cite{Jido:2001nt}. In Eq.\ (\ref{eq:rotation}), we
exploited the fact that $\mathcal{P}^{2} = \mathbbmss{1}$. For the rotation angle $\omega$, 
one finds
\begin{equation}
	\omega = \frac{1}{2} \arctan \left[
	\frac{2 m_{0}}{(y_{1} + y_{2})\sigma_{0}} \right]
	\equiv \frac{1}{2} \arctan \left[
	\frac{2 M_{0}}{(\tilde{y}_{1} + \tilde{y}_{2}) \tilde{\sigma}_{0}} \right] .
	\label{eq:rotangle}
\end{equation}
This expression is renormalization-group invariant, cf.\ the discussion around
Eq.\ (\ref{eq:npointrenorm}) in Sec.\ \ref{sec:model}. The angle $\omega$ finally converts 
the mass matrix (\ref{eq:massunphys}) to the equivalent (\ref{eq:massphys}).\ Its scale
evolution is shown in Fig.\ \ref{fig:omega}(a) in Appendix \ref{sec:FRG}. In particular,
$\omega \rightarrow \pi/4$ in the UV when $N_{+}$ and $N_{-}$ are degenerate in mass.

\section{Geometry and algebra}
\label{sec:algebra}

\begin{figure}[t]
	\centering
	\includegraphics[scale=1.0]{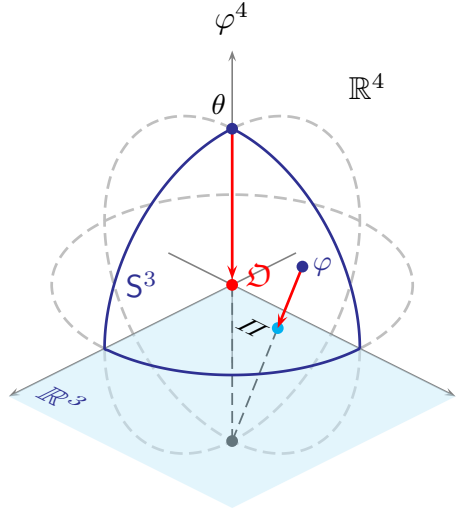}
	\caption{Vacuum manifold of the nucleon-meson model. Schematic three-dimensional sketch of 
	stereographic projections from the south pole (red arrows) of the three-sphere $\mathsf{S}^{3}$ 
	onto $\mathbb{R}^{3}$. The contours of the sphere $\mathsf{S}^{3}$ are depicted by blue solid 
	(``octant'' directed at the viewer) and gray dashed lines.}
	\label{fig:stereo}
\end{figure}

A schematic picture of stereographic projections (from the south pole) is shown in Fig.\ \ref{fig:stereo}.
The north pole $\tilde{\varphi} = (0,0,0,\theta)$ is mapped onto the origin of $\mathbb{R}^{3}$, 
$\mathfrak{O} = (0,0,0)$ (indicated as a red dot). A local dreibein on the sphere $\mathsf{S}^{3}$ 
is constructed from the Maurer-Cartan form $\alpha_{\mu}(\Pi)$,
\begin{equation}
	\alpha_{\mu}(\Pi) = \Sigma^{-1}(\Pi)\partial_{\mu}\Sigma(\Pi) ,
\end{equation}
where $\Sigma(\Pi)$ is a coset representative of the coset space $\mathsf{SO}(4)/\mathsf{SO}(3)
\cong \mathsf{S}^{3}$. The component $\alpha_{\mu}(\Pi)$ takes on values in the (six-dimensional) 
Lie algebra $\mathfrak{so}(4)$, therefore decomposing as
\begin{equation}
	\alpha_{\mu}(\Pi) = ie_{\mu}^{a}(\Pi) c_{a} + iw_{\mu}^{i}(\Pi) s_{i}, \qquad
	a,i = 1,2,3.
\end{equation}
The vectors $c_{a}$, $a = 1,2,3$, and $s_{i}$, $i = 1,2,3$, are the coset generators 
and those of the (residual) isospin subalgebra $\mathfrak{so}(3)$, respectively. The 
coefficients $e_{\mu}^{a}$ can be expanded in terms of the local coordinates $\Pi^{\bar{a}}$,\footnote{The
coefficients $w_{\mu}^{i}$ parametrize isospin transformations; they are of no relevance for the purely
pionic model of our work.}
\begin{equation}
	e_{\mu}^{a}(\Pi) = e^{a}_{\bar{a}}(\Pi)\;\! \partial_{\mu} \Pi^{\bar{a}}
	= \frac{4f_{\pi}}{4f_{\pi}^{2} + \Pi^{2}}\;\! \delta^{a}_{\bar{a}} 
	\partial_{\mu} \Pi^{\bar{a}} ,
	\qquad \bar{a} = 1,2,3. \label{eq:evectors}
\end{equation}
The objects $e_{\bar{a}}^{a}$ form the requested local frame (dreibein) on $\mathsf{S}^{3}$ 
with the related metric tensor 
\begin{equation}
	g_{\bar{a}\bar{b}}(\Pi) = \delta_{ab} 
	e^{a}_{\bar{a}}(\Pi) e^{b}_{\bar{b}}(\Pi) 
	= \frac{16 f_{\pi}^2}{(4 f_{\pi}^2 + \Pi^2)^{2}} \;\!
	\delta_{\bar{a}\bar{b}} .
\end{equation}

The generators $s_{i}$, $i = 1,2,3$, of the Lie algebra $\mathfrak{so}(3)$ and
their algebraic complement $c_{a}$, $a = 1,2,3$, which together form those of the
Lie algebra $\mathfrak{so}(4)$, are chosen as
\begin{equation}
	\begin{aligned}
	s_{1} = & \begin{pmatrix}
	\phantom{|} 0 & 0 & 0 & 0 \phantom{|}\\
	\phantom{|} 0 & 0 & -i & 0 \phantom{|}\\
	\phantom{|} 0 & i & 0 & 0 \phantom{|}\\
	\phantom{|} 0 & 0 & 0 & 0 \phantom{|}
	\end{pmatrix}, & \qquad
	s_{2} = & \begin{pmatrix}
	0 & 0 & i & 0 \phantom{|}\\
	0 & 0 & 0 & 0 \phantom{|}\\
	-i & 0 & 0 & 0 \phantom{|}\\
	0 & 0 & 0 & 0 \phantom{|}
	\end{pmatrix}, & \qquad
	s_{3} = & \begin{pmatrix}
	\phantom{|} 0 & -i & 0 & 0 \phantom{|}\\
	\phantom{|} i & 0 & 0 & 0 \phantom{|}\\
	\phantom{|} 0 & 0 & 0 & 0 \phantom{|}\\
	\phantom{|} 0 & 0 & 0 & 0 \phantom{|}
	\end{pmatrix}, \\[0.2cm]
	c_{1} = & \begin{pmatrix}
	\phantom{|} 0 & 0 & 0 & -i \phantom{|}\\
	\phantom{|} 0 & 0 & 0 & 0 \phantom{|}\\
	\phantom{|} 0 & 0 & 0 & 0 \phantom{|}\\
	\phantom{|} i & 0 & 0 & 0 \phantom{|}
	\end{pmatrix} , &
	c_{2} = & \begin{pmatrix}
	\phantom{|} 0 & 0 & 0 & 0\\
	\phantom{|} 0 & 0 & 0 & -i\\
	\phantom{|} 0 & 0 & 0 & 0\\
	\phantom{|} 0 & i & 0 & 0
	\end{pmatrix} , &
	c_{3} = & \begin{pmatrix}
	\phantom{|} 0 & 0 & 0 & 0\\
	\phantom{|} 0 & 0 & 0 & 0\\
	\phantom{|} 0 & 0 & 0 & -i\\
	\phantom{|} 0 & 0 & i & 0
	\end{pmatrix} . \\
	\end{aligned}
\end{equation}
They fulfill the commutator (Lie-bracket) and trace relations
\begin{equation}
	\begin{aligned}
	[s_{i},s_{j}] & = i\epsilon_{ij}^{\ \ k} s_{k}, & \qquad
	\tr\left(s_{i}s_{j}\right) & = 2\delta_{ij}, \\
	[s_{i},c_{a}] & = i\epsilon_{ia}^{\ \ b} c_{b}, & \qquad
	\tr\left(s_{i}c_{a}\right) & = 0, \\
	[c_{a},c_{b}] & = i\epsilon_{ab}^{\ \ i} s_{i}, & \qquad
	\tr\left(c_{a}c_{b}\right) & = 2\delta_{ab},
	\end{aligned}
\end{equation}
with the Levi-Civita symbols $\smash{\epsilon_{ij}^{\ \ k}}$ as the structure constants. An 
appropriate coset representative $\Sigma(\Pi)$, i.e., an $\mathsf{SO}(4)$-matrix parametrized
through the stereographic pions $\Pi$, and the corresponding skew-symmetric Maurer-Cartan form
$\alpha_{\mu}(\Pi)$ on the vacuum manifold $\mathsf{S}^{3}$ are given by
\begin{IEEEeqnarray}{rCl}
	\Sigma(\Pi) & = & \begin{pmatrix}
	\mathbbmss{1} - \frac{2\Pi\Pi^{\mathsf{T}}}
	{\rule[1.4ex]{0pt}{0.1ex} 4f_{\pi}^{2} + \Pi^{2}} & 
	\frac{4 f_{\pi}\Pi}
	{\rule[1.4ex]{0pt}{0.1ex}4f_{\pi}^{2} + \Pi^{2}} \\[0.25cm] 
	- \frac{4 f_{\pi}\Pi^{\mathsf{T}}}
	{\rule[1.4ex]{0pt}{0.1ex} 4f_{\pi}^{2} + \Pi^{2}} & 
	\frac{4f_{\pi}^{2} - \Pi^{2}}
	{\rule[1.4ex]{0pt}{0.1ex} 4f_{\pi}^{2} + \Pi^{2}}
	\end{pmatrix} \in \mathsf{SO}(4),
	\qquad \Pi = \left(\Pi^{1},\Pi^{2},\Pi^{3}\right)^{\mathsf{T}}, \\[0.25cm]
	\alpha_{\mu}(\Pi) & = & \Sigma^{-1}(\Pi)\partial_{\mu}\Sigma(\Pi)
	= \begin{pmatrix}
	\frac{2 \left[\Pi \partial_{\mu}\Pi^{\mathsf{T}} 
	- \left(\partial_{\mu}\Pi\right)\Pi^{\mathsf{T}}\right]}
	{\rule[1.4ex]{0pt}{0.1ex} 4f_{\pi}^{2} + \Pi^{2}} & 
	\frac{4f_{\pi}\partial_{\mu}\Pi}
	{\rule[1.4ex]{0pt}{0.1ex} 4f_{\pi}^{2} + \Pi^{2}} \\[0.25cm] 
	- \frac{4f_{\pi}\partial_{\mu}\Pi^{\mathsf{T}}}
	{\rule[1.4ex]{0pt}{0.1ex} 4f_{\pi}^{2} + \Pi^{2}} & 0
	\end{pmatrix} \in \mathfrak{so}(4) .
	\label{eq:mcform}
\end{IEEEeqnarray}
The coefficients $e_{\mu}^{a}(\Pi)$ are extracted from Eq.\ (\ref{eq:mcform}) via
\begin{equation}
	e^{a}_{\mu}(\Pi) = - \frac{i}{2} \tr\left[\alpha_{\mu}(\Pi) c_{a}\right]
	= \frac{4f_{\pi}\partial_{\mu}\Pi^{a}}{4f_{\pi}^{2} + \Pi^{2}},
\end{equation}
which was used in Eq.\ (\ref{eq:evectors}). The transformation properties of these geometric 
objects are extensively examined in Refs.\ \cite{Divotgey:2019xea, Eser:2020phd, Divotgey:2020phd} 
(which also contain a lot more details about the geometric interpretation and the nonlinear 
transformation behavior of the ``stereographic pions'').

\section{Coordinate independence of the on-shell four-point function}
\label{sec:fourpoint}

Consider the change of variable $\varphi \rightarrow \varphi'$ in the effective-action functional 
$\Gamma[\varphi]$. We assume that the transformation function $\varphi = \varphi[\varphi']$ is 
smooth and invertible, i.e., it is differentiable and its inverse $\varphi^{-1}$ exists. The 
transformed action functional is $\Gamma'[\varphi'] = \Gamma[\varphi[\varphi']]$.\ Specifically, 
we regard the Cartesian coordinates $\varphi$ as (invertible) mappings of the stereographic 
projections $\varphi'$ [cf.\ Eq.\ (\ref{eq:stereo})].\footnote{We do not print the prime in 
the main part.} Taking a solution $\varphi_{\mathrm{sol}}$ of the EOM,
\begin{equation}
	\left.\frac{\delta \Gamma}{\delta \varphi^{i}}\right|_{\varphi_{\mathrm{sol}}} 
	= 0, \qquad i = 1,\ldots,4 ,
\end{equation}
the functional derivative of $\Gamma'$ evaluated at $\varphi^{-1}[\varphi_{\mathrm{sol}}]$ 
also vanishes by the chain rule,
\begin{equation}
	\left.\frac{\delta \Gamma'[\varphi']}{\delta\varphi'^{j}(x)}\right|_{
	\varphi^{-1}\left[\varphi_{\mathrm{sol}}\right]}
	= \int_{z} \left.\frac{\delta\Gamma[\varphi]}{\delta\varphi^{i}(z)}
	\right|_{\varphi_{\mathrm{sol}}} 
	\left.\frac{\delta\varphi^{i}(z)}{\delta \varphi'^{j}(x)}
	\right|_{\varphi^{-1}\left[\varphi_{\mathrm{sol}}\right]}
	= 0, \qquad j = 1,\ldots,4 .
\end{equation}
This equation shows that $\varphi^{-1}\!\left[\varphi_{\mathrm{sol}}\right]$ is a solution to 
the EOM w.r.t.\ $\varphi'$. The map $\varphi^{-1}$ is well-defined in our case, since we 
restrict the discussion to the vicinity of the origin $\mathfrak{O}$ of $\mathbb{R}^{3}$; see once more 
Fig.\ \ref{fig:stereo}. Consequently, we directly obtain the solution $\varphi'_{\mathrm{sol}} 
= \varphi^{-1}[\varphi_{\mathrm{sol}}]$ once $\varphi_{\mathrm{sol}}$ is known, and vice versa.

Solving the EOM w.r.t.\ the radial $\theta$-field implicitly defines the function 
$\theta_{\mathrm{sol}}[\Pi]$,
\begin{equation}
	\left.\frac{\delta\Gamma'}
	{\delta\theta}\right|_{\theta_{\mathrm{sol}}[\Pi]} = 0.
	\label{eq:thetasol}
\end{equation}
The second functional derivative of Eq.\ (\ref{eq:thetasol}),
\begin{equation}
	\frac{\delta^{2}}{\delta\Pi^{a}(x_{1})\delta\Pi^{b}(x_{2})}
	\left(\left.\frac{\delta\Gamma'[\varphi']}
	{\delta\theta(x_{3})}\right|_{\theta_{\mathrm{sol}}}\right)
	= \Gamma'^{(3)\;\! ab}_{\Pi\Pi\theta}(x_{1},x_{2},x_{3})
	+ \int_{z} \Gamma'^{(2)}_{\theta\theta}(z,x_{3})\,
	\frac{\delta^{2}\theta_{\mathrm{sol}}(z)}
	{\delta\Pi^{a}(x_{1})\delta\Pi^{b}(x_{2})} = 0,
\end{equation}
yields the important relation
\begin{equation}
	\frac{\delta^{2}\theta_{\mathrm{sol}}(x_{1})}{\delta\Pi^{a}(x_{2})
	\delta\Pi^{b}(x_{3})} = - \int_{z} 
	\frac{\delta^{2}W'[J']}{\delta J'_{\theta}(x_{1})\delta J'_{\theta}(z)}\,
	\Gamma'^{(3)\!\; ab}_{\Pi\Pi\theta}(x_{2},x_{3},z)
	\equiv - \int_{z} W'^{(2)}_{\theta\theta}(x_{1},z)\,
	\Gamma'^{(3)\!\; ab}_{\Pi\Pi\theta}(x_{2},x_{3},z) .
	\label{eq:thetasolid}
\end{equation}
To derive the above Eq.\ (\ref{eq:thetasolid}), we used the fact that the first derivative 
of $\theta_{\mathrm{sol}}[\Pi]$ is zero (due to the necessary isospin symmetry of the solution)
as well as the well-known identity
\begin{equation}
	\int_{z} \frac{\delta^{2} W'[J']}{\delta J'^{i}(x_{1}) \delta J'^{k}(z)}
	\frac{\delta^{2} \Gamma'[\varphi']}{\delta \varphi'^{k}(z) 
	\delta \varphi'^{j}(x_{2})} = \delta^{ij} \delta(x_{1} - x_{2}).
\end{equation}
The functional $W'[J']$ (with the sources $J'$) is the Schwinger functional, i.e., the Legendre 
transform of $\Gamma'[\varphi']$. Defining the generating functional $\Gamma'_{\mathrm{sol}}$,
\begin{equation}
	\Gamma'_{\mathrm{sol}}[\Pi] \coloneqq \left.\Gamma'[\varphi']\right|_{\theta\,
	=\, \theta_{\mathrm{sol}}} ,
\end{equation}
one obtains the four-point function of $\Gamma'_{\mathrm{sol}}$ as
\begin{IEEEeqnarray}{rCl}
	\Gamma'^{(4)\;\! abcd}_{\mathrm{sol}\,\Pi\Pi\Pi\Pi}(x_{1},x_{2},x_{3},x_{4})
	& = & \Gamma'^{(4)\;\! abcd}_{\Pi\Pi\Pi\Pi}(x_{1},x_{2},x_{3},x_{4})
	- \int_{z_{1}} \int_{z_{2}} \Gamma'^{(3)\;\! ab}_{\Pi\Pi\theta}(x_{1},x_{2},z_{1})\,
	W'^{(2)}_{\theta\theta}(z_{1},z_{2})\,
	\Gamma'^{(3)\;\! cd}_{\Pi\Pi\theta}(x_{3},x_{4},z_{2}) \nonumber\\[0.2cm]
	& & \hspace{3cm}
	- \int_{z_{1}} \int_{z_{2}} \Gamma'^{(3)\;\! ac}_{\Pi\Pi\theta}(x_{1},x_{3},z_{1})\,
	W'^{(2)}_{\theta\theta}(z_{1},z_{2})\,
	\Gamma'^{(3)\;\! bd}_{\Pi\Pi\theta}(x_{2},x_{4},z_{2}) \nonumber\\[0.2cm]
	& & \hspace{3cm}
	- \int_{z_{1}} \int_{z_{2}} \Gamma'^{(3)\;\! ad}_{\Pi\Pi\theta}(x_{1},x_{4},z_{1})\,
	W'^{(2)}_{\theta\theta}(z_{1},z_{2})\,
	\Gamma'^{(3)\;\! bc}_{\Pi\Pi\theta}(x_{2},x_{3},z_{2}) ,
	\label{eq:solfourpoint}
\end{IEEEeqnarray}
where we again utilized that odd derivatives of $\theta_{\mathrm{sol}}$ vanish. In the language 
of Feynman diagrams, Eq.\ (\ref{eq:solfourpoint}) translates into the right-hand sides of 
Eqs.\ (\ref{eq:treeamplitude}) and (\ref{eq:eom}), with the $\theta$-propagator
\begin{equation}
	G'_{\theta\theta}(x_{1},x_{2}) = i W'^{(2)}_{\theta\theta}(x_{1},x_{2})
	\equiv \frac{i \delta^{2} W'[J']}{\delta J'_{\theta}(x_{1})\delta J'_{\theta}(x_{2})} .
\end{equation}

To demonstrate the coordinate independence of the on-shell four-point function, we
now express Eq.\ (\ref{eq:solfourpoint}) in terms of the original vertex functions 
(derivatives of $\Gamma$): As prerequisites, we therefore explicitly calculate the 
derivatives of the coordinate function $\varphi[\varphi']$,
\begin{IEEEeqnarray}{rCl}
	\frac{\delta\varphi^{i}(x_{1})}{\delta\Pi^{a}(x_{2})} & = & \frac{\theta}{f_{\pi}}
	\;\! \delta^{ia} \delta(x_{1} - x_{2}), \\[0.2cm]
	\frac{\delta\varphi^{i}(x_{1})}{\delta\theta(x_{2})} & = & \delta^{i4}
	\delta(x_{1} - x_{2}), \\[0.2cm]
	\frac{\delta^{2}\varphi^{i}(x_{1})}{\delta\Pi^{a}(x_{2})\delta\Pi^{b}(x_{3})}
	& = & - \frac{\theta}{f_{\pi}^{2}}\;\! \delta^{i4}\delta^{ab}\delta(x_{1} - x_{2})
	\delta(x_{1} - x_{3}), \\[0.2cm]
	\frac{\delta^{2}\varphi^{i}(x_{1})}{\delta\Pi^{a}(x_{2})\delta\theta(x_{3})}
	& = & \frac{1}{f_{\pi}}\;\! \delta^{ia}\delta(x_{1} - x_{2}) 
	\delta(x_{1} - x_{3}), \\[0.2cm]
	\frac{\delta^{2}\varphi^{i}(x_{1})}{\delta\Pi^{a}(x_{2})\delta\Pi^{b}(x_{3})
	\delta\Pi^{c}(x_{4})} & = & - \frac{\theta}{2 f_{\pi}^{2}}\left(\delta^{ia}\delta^{bc}
	+ \delta^{ib}\delta^{ac} + \delta^{ic}\delta^{ab}\right) \delta(x_{1} - x_{2})
	\delta(x_{1} - x_{3}) \delta(x_{1} - x_{4}),
\end{IEEEeqnarray}
and provide the chiral Ward identity between the two-point and the three-point functions
based on the (renormalized) bosonic part of the Minkowskian analogue of Eq.\ (\ref{eq:truncation}),
\begin{IEEEeqnarray}{rCl}
	\frac{\delta^{2}\Gamma[\varphi]}{\delta\varphi^{i}(x_{1})\delta\varphi^{j}(x_{2})}
	& = & - \left[\left(\Box_{x_{1}} + 2 V'\right) \delta^{ij} 
	+ 4 V'' \varphi^{i}(x_{1}) \varphi^{j}(x_{1})\right] \delta(x_{1} - x_{2}), \\[0.2cm]
	\frac{\delta^{3}\Gamma[\varphi]}{\delta\varphi^{i}(x_{1})\delta\varphi^{j}(x_{2})
	\delta\varphi^{k}(x_{3})} & = & -4\Big\lbrace V''\left[\varphi^{i}(x_{1}) \delta^{jk}
	+ \varphi^{j}(x_{1}) \delta^{ik} + \varphi^{k}(x_{1}) \delta^{ij}\right] \nonumber\\
	& & \qquad +\, 2V^{(3)}\varphi^{i}(x_{1}) \varphi^{j}(x_{1}) \varphi^{k}(x_{1})\Big\rbrace
	\, \delta(x_{1} - x_{2}) \delta(x_{1} - x_{3}), \\[0.2cm]
	\Rightarrow \qquad \Gamma^{(3)\;\! ab}_{\pi\pi\sigma}(x_{1},x_{2},x_{3})
	& = & \frac{1}{\theta} \left[\Gamma^{(2)}_{\sigma\sigma}(x_{1},x_{3})\;\! \delta^{ab}
	- \Gamma_{\pi\pi}^{(2)\;\! ab}(x_{1},x_{3})\right] \delta(x_{1} - x_{2}),
	\label{eq:wardidentity}
\end{IEEEeqnarray}
where $V' \equiv \mathrm{d}V\!\left(\varphi^{2}\right)\!/\mathrm{d}\varphi^{2}$, $V'' \equiv 
\mathrm{d}^{2}V\!\left(\varphi^{2}\right)\!/(\mathrm{d}\varphi^{2})^{2}$, and so forth (to be
distinguished from the above prime indicating the transformed effective action). By virtue 
of the chain rule and the EOM from the beginning of this Appendix, the three-point 
and four-point functions on the right of Eq.\ (\ref{eq:solfourpoint}) are rewritten as
\begin{IEEEeqnarray}{rCl}
	\Gamma'^{(3)\;\! ab}_{\Pi\Pi\theta}(x_{1},x_{2},x_{3}) & = & 
	\int_{z_{1}} \int_{z_{2}} \frac{\delta^{2}\Gamma[\varphi]}
	{\delta\varphi^{i}(z_{1})\delta\varphi^{j}(z_{2})} \bigg[
	\frac{\delta\varphi^{i}(z_{1})}{\delta\Pi^{a}(x_{1})}
	\frac{\delta^{2}\varphi^{j}(z_{2})}{\delta\Pi^{b}(x_{2})\delta\theta(x_{3})}
	+ \frac{\delta\varphi^{i}(z_{1})}{\delta\Pi^{b}(x_{2})}
	\frac{\delta^{2}\varphi^{j}(z_{2})}{\delta\Pi^{a}(x_{1})\delta\theta(x_{3})}
	\nonumber\\[0.2cm]
	& & \hspace{3.75cm} +\, \frac{\delta^{2}\varphi^{i}(z_{1})}
	{\delta\Pi^{a}(x_{1})\delta\Pi^{b}(x_{2})}
	\frac{\delta\varphi^{j}(z_{2})}{\delta\theta(x_{3})} \bigg] \nonumber\\[0.2cm]
	& & +\, \int_{z_{1}} \int_{z_{2}} \int_{z_{3}}
	\frac{\delta^{3}\Gamma[\varphi]}{\delta\varphi^{i}(z_{1})
	\delta\varphi^{j}(z_{2})\delta\varphi^{l}(z_{3})}
	\frac{\delta\varphi^{i}(z_{1})}{\delta\Pi^{a}(x_{1})}
	\frac{\delta\varphi^{j}(z_{2})}{\delta\Pi^{b}(x_{2})}
	\frac{\delta\varphi^{l}(z_{3})}{\delta\theta(x_{3})} \nonumber\\[0.2cm]
	& = & \frac{\theta}{f_{\pi}^{2}}
	\Big\lbrace \theta \Gamma^{(3)\;\! ab}_{\pi\pi\sigma}(x_{1},x_{2},x_{3})
	+ \Gamma^{(2)\;\! ab}_{\pi\pi}(x_{1},x_{2})
	\left[\delta(x_{1} - x_{3}) + \delta(x_{2} - x_{3})\right] \nonumber\\
	& & \qquad -\, \Gamma^{(2)}_{\sigma\sigma}(x_{1},x_{3})\;\! 
	\delta^{ab}\delta(x_{1} - x_{2}) \Big\rbrace , \\[0.2cm]
	\Gamma'^{(4)\;\! abcd}_{\Pi\Pi\Pi\Pi}(x_{1},x_{2},x_{3},x_{4}) & = &
	\frac{\theta^{2}}{f_{\pi}^{4}} \bigg\lbrace
	\theta^{2}\Gamma^{(4)\;\! abcd}_{\pi\pi\pi\pi}(x_{1},x_{2},x_{3},x_{4})
	- \delta^{ab}\delta^{cd} \Big[
	\Gamma^{(2)}_{\sigma\sigma}(x_{1},x_{2})\;\! 
	\delta(x_{2} - x_{3})\delta(x_{3} - x_{4}) \nonumber\\
	& & \qquad +\, \Gamma^{(2)}_{\sigma\sigma}(x_{3},x_{4}) \;\!
	\delta(x_{1} - x_{2})\delta(x_{1} - x_{4})
	- \Gamma^{(2)}_{\sigma\sigma}(x_{1},x_{3}) \;\!
	\delta(x_{1} - x_{2})\delta(x_{3} - x_{4}) \Big] \nonumber\\
	& & \qquad +\, \frac{1}{2}\;\! 
	\Gamma^{(2)\;\! ab}_{\pi\pi}(x_{1},x_{2})\;\! \delta^{cd}
	\left[\delta(x_{2} - x_{3})\delta(x_{2} - x_{4}) 
	- \delta(x_{1} - x_{3})\delta(x_{1} - x_{4}) \right] \nonumber\\[0.2cm]
	& & \qquad +\, \frac{1}{2}\;\! 
	\Gamma^{(2)\;\! cd}_{\pi\pi}(x_{3},x_{4})\;\! \delta^{ab}
	\left[\delta(x_{1} - x_{4})\delta(x_{2} - x_{4})
	- \delta(x_{1} - x_{3})\delta(x_{2} - x_{3})\right] \nonumber\\
	& & \qquad +\, \text{crossings to the previous three terms} \bigg\rbrace .
\end{IEEEeqnarray}
Plugging in these relations in Eq.\ (\ref{eq:solfourpoint}), taking advantage of the 
Ward identity (\ref{eq:wardidentity}), and transferring everything into momentum space 
along the conventions visible in Fig.\ \ref{fig:amplitude},
\begin{IEEEeqnarray}{rCl}
	\vcenter{\hbox{
	\begin{pspicture}[showgrid=false](3.0,3.0)
		\psline[linewidth=0.03,linestyle=dashed,
		dash=2pt 1pt,linecolor=Red,ArrowInside=->,ArrowInsidePos=0.33,
		arrowsize=2pt 3,arrowinset=0](0.55,2.45)(1.5,1.5)
		\psline[linewidth=0.03,linestyle=dashed,
		dash=2pt 1pt,linecolor=Red,ArrowInside=->,ArrowInsidePos=0.33,
		arrowsize=2pt 3,arrowinset=0](2.45,2.45)(1.5,1.5)
		\psline[linewidth=0.03,linestyle=dashed,
		dash=2pt 1pt,linecolor=Red,ArrowInside=->,ArrowInsidePos=0.33,
		arrowsize=2pt 3,arrowinset=0](0.55,0.55)(1.5,1.5)
		\psline[linewidth=0.03,linestyle=dashed,
		dash=2pt 1pt,linecolor=Red,ArrowInside=->,ArrowInsidePos=0.33,
		arrowsize=2pt 3,arrowinset=0](2.45,0.55)(1.5,1.5)
		\rput[b]{*0}(0.35,2.55){\small{$\Pi(x_{1})$}}
		\rput[b]{*0}(2.65,2.55){\small{$\Pi(x_{3})$}}
		\rput[b]{*0}(0.35,0.10){\small{$\Pi(x_{2})$}}
		\rput[b]{*0}(2.65,0.10){\small{$\Pi(x_{4})$}}
		\rput(0.55,1.8){\small{$p_{1},a$}}
		\rput(2.45,1.8){\small{$p_{3},c$}}
		\rput(0.55,1.2){\small{$p_{2},b$}}
		\rput(2.45,1.2){\small{$p_{4},d$}}
		\pscircle[linewidth=0.03,fillstyle=hlines*,hatchwidth=0.02,
		fillcolor=lightgray!25,hatchsep=0.05,hatchcolor=gray](1.5,1.5){0.41}
	\end{pspicture}
	}}
	\qquad & \Rightarrow & \qquad
	\int_{x_{1}} \int_{x_{2}} \int_{x_{3}} \int_{x_{4}}
	\exp\left(i\sum_{n\, =\, 1}^{4} x_{n} \cdot p_{n}\right) \;\!
	i\Gamma'^{(4)\;\! abcd}_{\mathrm{sol}\,\Pi\Pi\Pi\Pi}(x_{1},x_{2},x_{3},x_{4})
	\nonumber\\[-0.75cm]
	& & \hspace{3cm}
	\equiv (2\pi)^{4} \delta\!\left(\sum_{n\, =\, 1}^{4} x_{n} \cdot p_{n}\right) \;\!
	i\Gamma'^{(4)\;\! abcd}_{\mathrm{sol}\,\Pi\Pi\Pi\Pi}(p_{1},p_{2},p_{3},p_{4}) ,
	\qquad \\[0.2cm]
	\delta(x_{1} - x_{2}) & = & \int\frac{\mathrm{d}^{4}q}{(2\pi)^{4}}
	\;\! e^{-i q\cdot (x_{1} - x_{2})} \equiv \int_{q}
	\;\! e^{-i q\cdot (x_{1} - x_{2})}, \\[0.2cm]
	\Gamma^{(2)\;\! ab}_{\pi\pi}(x_{1},x_{2}) & = & 
	\int_{q} \left(q^{2} - M_{\pi}^{2}\right)
	\delta^{ab} e^{-i q\cdot (x_{1} - x_{2})}, \qquad
	W^{(2)}_{\sigma\sigma}(x_{1},x_{2}) = \int_{q} \frac{1}{q^{2} - M_{\sigma}^{2}}
	\;\! e^{-i q\cdot (x_{1} - x_{2})}, \qquad\quad
\end{IEEEeqnarray}
we find for the amplitude $\mathfrak{A}(s,t,u)$ in Eq.\ (\ref{eq:amplitude}) the expression
\begin{equation}
	\mathfrak{A}(s,t,u) = \frac{1}{f_{\pi}^{2}} \left\lbrace 
	\left(M_{\pi}^{2} - M_{\sigma}^{2}\right)
	\frac{s - M_{\pi}^{2}}{s - M_{\sigma}^{2}}
	- \left[\left(\frac{1}{2} - \frac{s - M_{\pi}^{2}}{s - M_{\sigma}^{2}}
	\right) \sum_{i\, =\, 1}^{4} \mathbb{P}_{i}
	+ \frac{\mathbb{P}_{1} \mathbb{P}_{3} + \mathbb{P}_{1} \mathbb{P}_{4} 
	+ \mathbb{P}_{2} \mathbb{P}_{3} + \mathbb{P}_{2} \mathbb{P}_{4}}
	{s - M_{\sigma}^{2}} \right] \right\rbrace ,
\end{equation}
with $\mathbb{P}_{n} \coloneqq p_{n}^{2} - M_{\pi}^{2}$, $n = 1, \ldots, 4$. The momentum-space
Feynman rules, needed for the last step, read for the relevant $s$-channel contributions:
\begin{IEEEeqnarray}{rCl}
	\vcenter{\hbox{
	\begin{pspicture}[showgrid=false](2.0,2.0)
		\psline[linewidth=0.03,linestyle=dashed,
		dash=2pt 1pt,linecolor=Red](0.45,1.55)(1.0,1.0)
		\psline[linewidth=0.03,linestyle=dashed,
		dash=2pt 1pt,linecolor=Red](1.55,1.55)(1.0,1.0)
		\psline[linewidth=0.03,linestyle=dashed,
		dash=2pt 1pt,linecolor=Red](0.45,0.45)(1.0,1.0)
		\psline[linewidth=0.03,linestyle=dashed,
		dash=2pt 1pt,linecolor=Red](1.55,0.45)(1.0,1.0)
		\rput[b]{*0}(0.25,1.65){\small{$\pi$}}
		\rput[b]{*0}(1.75,1.65){\small{$\pi$}}
		\rput[b]{*0}(0.25,0.15){\small{$\pi$}}
		\rput[b]{*0}(1.75,0.15){\small{$\pi$}}
		\pscircle[linewidth=0.03,fillstyle=solid,
		fillcolor=lightgray!25](1.0,1.0){0.20}
		\rput[b]{*0}(0.99,0.92){\fontsize{7pt}{0pt}{$4$}\selectfont}
	\end{pspicture}
	}} & \quad \Rightarrow \quad & - 4 i V'' \delta^{ab}\delta^{cd}
	\equiv i\;\! \frac{M_{\pi}^{2} - M_{\sigma}^{2}}{f_{\pi}^{2}} \!\;
	\delta^{ab}\delta^{cd}, \\
	\vcenter{\hbox{
	\begin{pspicture}[showgrid=false](3.0,2.0)
		\psline[linewidth=0.03,linestyle=dashed,
		dash=2pt 1pt,linecolor=Red](0.45,1.55)(1.0,1.0)
		\psline[linewidth=0.03,linestyle=dashed,
		dash=2pt 1pt,linecolor=Red](2.55,1.55)(2.0,1.0)
		\psline[linewidth=0.03,linestyle=dashed,
		dash=2pt 1pt,linecolor=Red](0.45,0.45)(1.0,1.0)
		\psline[linewidth=0.03,linestyle=dashed,
		dash=2pt 1pt,linecolor=Red](2.55,0.45)(2.0,1.0)
		\psline[linewidth=0.03,linestyle=dashed,
		dash=2pt 1pt,linecolor=Blue](1.0,1.0)(2.0,1.0)
		\rput[b]{*0}(0.25,1.65){\small{$\pi$}}
		\rput[b]{*0}(2.75,1.65){\small{$\pi$}}
		\rput[b]{*0}(0.25,0.15){\small{$\pi$}}
		\rput[b]{*0}(2.75,0.15){\small{$\pi$}}
		\rput[b]{*0}(1.5,0.7){\small{$\sigma$}}
		\pscircle[linewidth=0.03,fillstyle=solid,
		fillcolor=lightgray!25](1.0,1.0){0.20}
		\rput[b]{*0}(1.0,0.92){\fontsize{7pt}{0pt}{$3$}\selectfont}
		\pscircle[linewidth=0.03,fillstyle=solid,
		fillcolor=lightgray!25](2.0,1.0){0.20}
		\rput[b]{*0}(2.0,0.92){\fontsize{7pt}{0pt}{$3$}\selectfont}
	\end{pspicture}
	}} & \quad \Rightarrow \quad & 
	(-4 iV'')^{2} f_{\pi}^{2}\;\! \frac{i}{s - M_{\sigma}^{2}}\;\!
	\delta^{ab} \delta^{cd} \equiv -\frac{(M_{\pi}^{2} - M_{\sigma}^{2})^{2}}
	{f_{\pi}^{2}} \frac{i}{s - M_{\sigma}^{2}}
	\;\! \delta^{ab} \delta^{cd} ,
\end{IEEEeqnarray}
where we used the chiral Ward identity (\ref{eq:wardidentity}). On the mass shell, 
$\mathbb{P}_{n} = 0\ \forall\, n$, the result simplifies to
\begin{equation}
	\mathfrak{A}(s,t,u) = \frac{M_{\pi}^{2} - M_{\sigma}^{2}}{f_{\pi}^{2}}
	\frac{s - M_{\pi}^{2}}{s - M_{\sigma}^{2}}, \label{eq:universal}
\end{equation}
which is the amplitude in the (original) coordinates $\varphi = (\vec{\pi},\sigma)$
\cite{Bessis:1972sn, Appelquist:1980ae}. For $s = M_{\pi}^{2}$, the amplitude $\mathfrak{A}(s,t,u)$ 
vanishes, since the four-pion vertex and the $s$-channel $\sigma$-exchange diagram cancel each other. 
This cancelation phenomenon is known as the ``Adler zero'' \cite{Adler:1964um, *Adler:1965ga, 
Weinberg:1966kf, Scadron:2006mq}. Equation (\ref{eq:universal}) is the universal tree-level 
amplitude of the $\mathsf{SO}(4)$-symmetric LSM (with pions and the $\sigma$-meson).

\section{FRG flow of scale-dependent quantities}
\label{sec:FRG}

The flow equations for the scale-dependent couplings in Eq.\ (\ref{eq:truncation}) and the 
corresponding vertex functions are obtained by calculating the functional derivatives of 
Eq.\ (\ref{eq:Wetterich}). The Taylor coefficients of the effective potential are determined
by Eq.\ (\ref{eq:potential}). All differential equations are evaluated at the $k$-independent 
expansion point $\varphi_{0}^{2} = \mathrm{const.}$ of the potential, which is chosen to be 
slightly larger than the IR minimum ($\varphi_{0}^{2} \gtrsim f_{\pi}^{2}$). The latter is the 
physical vacuum configuration of the model, where the masses and other observables are defined. 
The integration of momentum modes is initialized at $\Lambda_{\chi} = 1.2\ \mathrm{GeV} \simeq 
4\pi f_{\pi}$.\ At this scale, the potential $V\!\left(\varphi^{2}\right)$ is assumed to 
be of the form given in Eq.\ (\ref{eq:potential}), with $\alpha_{1}, \alpha_{2} > 0$ and $\alpha_{n} 
= 0$, $n > 2$. The other scale-dependent quantities are initialized as listed in Table \ref{tab:UV}. 
The fixing of the chiral-invariant mass $m_{0}$ to a common value among all approximations 
determines the Yukawa couplings $y_{1}$ and $y_{2}$ in the MF and one-loop approximations
and the LPA, cf.\ Fig.\ \ref{fig:Msigma_MF}(a) in Appendix \ref{sec:analytic}. It leaves only
two adjustable model parameters (for the MF, one-loop, and LPA cases), namely, the Taylor coefficients 
$\alpha_{1}$ and $\alpha_{2}$, to account for the breaking scale $\Lambda_{\chi}$ and the IR observables 
(including the pion-pion scattering lengths).\ The expansion point of the effective potential 
and the $h$-parameter are indirectly fixed by the pion decay constant $f_{\pi}$ and the pion 
mass $M_{\pi}$, respectively. The integration is then carried out down to $k = 0$, and additionally 
up to $k = 1.5\ \mathrm{GeV}$ in order to verify chiral symmetry restoration above $\Lambda_{\chi}$. 
This upper limit of the considered $k$-range is set to the largest mass scale of the model, 
$M_{N(1535)} \simeq 1.5\ \mathrm{GeV}$.

Employing the language of Feynman diagrams, the flow equations for the effective potential $V$, 
the chiral-invariant mass $m_{0}$, the Yukawa couplings $y_{1}$ and $y_{2}$, as well as the
wave-function renormalizations $Z$ and $Z^{\psi}$ read
\vspace*{-0.2cm}
\begin{IEEEeqnarray}{rCl}
	\partial_{k}V_{k} & = & \mathcal{V}^{-1} 
	\partial_{k}\Gamma_{k} = \mathcal{V}^{-1} \biggg( 
	\frac{1}{2} \! \! \vcenter{\hbox{
	\begin{pspicture}[showgrid=false](1.5,2.4)
		\psarc[linewidth=0.03,linestyle=dashed,
		dash=2pt 1pt,linecolor=Blue](0.75,1.2){0.6}{115}{65}
		\pscircle[linewidth=0.03,fillstyle=solid,
		fillcolor=RoyalBlue](0.75,1.8){0.25}
		\psline[linewidth=0.03](0.75,1.8)(0.92,1.97)
		\psline[linewidth=0.03](0.75,1.8)(0.58,1.97)
		\psline[linewidth=0.03](0.75,1.8)(0.58,1.63)
		\psline[linewidth=0.03](0.75,1.8)(0.92,1.63)
		\rput[b]{*0}(0.75,0.3){\small $\sigma$}
	\end{pspicture}
	}} \! \! \!
	+ \frac{1}{2} \! \!
	\vcenter{\hbox{
	\begin{pspicture}(1.5,2.4)
		\psarc[linewidth=0.03,linestyle=dashed,
		dash=2pt 1pt,linecolor=Red](0.75,1.2){0.6}{115}{65}
		\pscircle[linewidth=0.03,fillstyle=solid,
		fillcolor=RedOrange](0.75,1.8){0.25}
		\psline[linewidth=0.03](0.75,1.8)(0.92,1.97)
		\psline[linewidth=0.03](0.75,1.8)(0.58,1.97)
		\psline[linewidth=0.03](0.75,1.8)(0.58,1.63)
		\psline[linewidth=0.03](0.75,1.8)(0.92,1.63)
		\rput[b]{*0}(0.75,0.3){\small $\pi$}
	\end{pspicture}
	}} \! \! \!
	- \! \! \!
	\vcenter{\hbox{
	\begin{pspicture}(1.5,2.4)
		\psarc[linewidth=0.02,arrowsize=2pt 3,
		arrowinset=0]{->}(0.75,1.2){0.6}{115}{189}
		\psarc[linewidth=0.02](0.75,1.2){0.6}{180}{65}
		\pscircle[linewidth=0.03,fillstyle=solid,
		fillcolor=lightgray](0.75,1.8){0.25}
		\psline[linewidth=0.03](0.75,1.8)(0.92,1.97)
		\psline[linewidth=0.03](0.75,1.8)(0.58,1.97)
		\psline[linewidth=0.03](0.75,1.8)(0.58,1.63)
		\psline[linewidth=0.03](0.75,1.8)(0.92,1.63)
		\rput[b]{*0}(0.75,0.25){\small $\Psi$}
	\end{pspicture}
	}} \! \! \biggg), \label{eq:u} \\[0.2cm]
	\partial_{k} m_{0,k} & = & -\frac{1}{4} \mathcal{V}^{-1}
	\tr_{\gamma} \left[\left.\frac{\delta^{2}\partial_{k}\Gamma_{k}}
	{\rule[0.20cm]{0pt}{1ex}\delta \bar{\Psi}_{f_{1},\mathsf{p}_{1}}(0) 
	\delta\Psi_{f_{2},\mathsf{p}_{2}}(0)}
	\right|_{f_{1}\, =\, f_{2}\, =\, 1,\, \mathsf{p}_{1}\, =\, 1,\, 
	\mathsf{p}_{2}\, =\, 2}\right] \nonumber\\[0.1cm]
	& = & -\frac{1}{4} \mathcal{V}^{-1}
	\tr_{\gamma}\biggg(
	\frac{1}{2} \! \! \vcenter{\hbox{
	\begin{pspicture}(3.0,2.0)
		\psarc[linewidth=0.03,linestyle=dashed,
		dash=2pt 1pt,linecolor=Blue](1.5,1.0){0.6}{20}{65}
		\psarc[linewidth=0.03,linestyle=dashed,
		dash=2pt 1pt,linecolor=Blue](1.5,1.0){0.6}{115}{160}
		\psarc[linewidth=0.02,arrowsize=2pt 3,
		arrowinset=0]{->}(1.5,1.0){0.6}{200}{279}
		\psarc[linewidth=0.02](1.5,1.0){0.6}{270}{340}
		\psline[linewidth=0.02,ArrowInside=->,
		ArrowInsidePos=0.55,arrowsize=2pt 3,
		arrowinset=0](0.1,1.0)(0.7,1.0)
		\psline[linewidth=0.02,ArrowInside=->,
		ArrowInsidePos=0.55,arrowsize=2pt 3,
		arrowinset=0](2.3,1.0)(2.9,1.0)
		\pscircle[linewidth=0.03,fillstyle=solid,
		fillcolor=RoyalBlue](1.5,1.6){0.25}
		\psline[linewidth=0.03](1.5,1.6)(1.67,1.77)
		\psline[linewidth=0.03](1.5,1.6)(1.33,1.77)
		\psline[linewidth=0.03](1.5,1.6)(1.33,1.43)
		\psline[linewidth=0.03](1.5,1.6)(1.67,1.43)
		\pscircle[linewidth=0.03,fillstyle=solid,
		fillcolor=lightgray!25](0.9,1.0){0.20}
		\pscircle[linewidth=0.03,fillstyle=solid,
		fillcolor=lightgray!25](2.1,1.0){0.20}
		\rput[b]{*0}(2.12,1.45){\small $\sigma$}
		\rput[b]{*0}(0.88,1.45){\small $\sigma$}
		\rput[b]{*0}(0.4,0.6){\small $\Psi$}
		\rput[b]{*0}(2.6,0.6){\small $\Psi$}
		\rput[b]{*0}(1.5,0.0){\small $\Psi$}
		\rput[b]{*0}(0.9,0.92){\fontsize{7pt}{0pt}{$3$}\selectfont}
		\rput[b]{*0}(2.1,0.92){\fontsize{7pt}{0pt}{$3$}\selectfont}
	\end{pspicture}
	}} \! \! \! 
	+ \frac{1}{2} \! \! \vcenter{\hbox{
	\begin{pspicture}(3.0,2.0)
		\psarc[linewidth=0.03,linestyle=dashed,
		dash=2pt 1pt,linecolor=Red](1.5,1.0){0.6}{20}{65}
		\psarc[linewidth=0.03,linestyle=dashed,
		dash=2pt 1pt,linecolor=Red](1.5,1.0){0.6}{115}{160}
		\psarc[linewidth=0.02,arrowsize=2pt 3,
		arrowinset=0]{->}(1.5,1.0){0.6}{200}{279}
		\psarc[linewidth=0.02](1.5,1.0){0.6}{270}{340}
		\psline[linewidth=0.02,ArrowInside=->,
		ArrowInsidePos=0.55,arrowsize=2pt 3,
		arrowinset=0](0.1,1.0)(0.7,1.0)
		\psline[linewidth=0.02,ArrowInside=->,
		ArrowInsidePos=0.55,arrowsize=2pt 3,
		arrowinset=0](2.3,1.0)(2.9,1.0)
		\pscircle[linewidth=0.03,fillstyle=solid,
		fillcolor=RedOrange](1.5,1.6){0.25}
		\psline[linewidth=0.03](1.5,1.6)(1.67,1.77)
		\psline[linewidth=0.03](1.5,1.6)(1.33,1.77)
		\psline[linewidth=0.03](1.5,1.6)(1.33,1.43)
		\psline[linewidth=0.03](1.5,1.6)(1.67,1.43)
		\pscircle[linewidth=0.03,fillstyle=solid,
		fillcolor=lightgray!25](0.9,1.0){0.20}
		\pscircle[linewidth=0.03,fillstyle=solid,
		fillcolor=lightgray!25](2.1,1.0){0.20}
		\rput[b]{*0}(2.12,1.45){\small $\pi$}
		\rput[b]{*0}(0.88,1.45){\small $\pi$}
		\rput[b]{*0}(0.4,0.6){\small $\Psi$}
		\rput[b]{*0}(2.6,0.6){\small $\Psi$}
		\rput[b]{*0}(1.5,0.0){\small $\Psi$}
		\rput[b]{*0}(0.9,0.92){\fontsize{7pt}{0pt}{$3$}\selectfont}
		\rput[b]{*0}(2.1,0.92){\fontsize{7pt}{0pt}{$3$}\selectfont}
	\end{pspicture}
	}} \! \! \! \nonumber\\[0.1cm]
	& & \qquad \qquad \qquad
	- \! \! \vcenter{\hbox{
	\begin{pspicture}(3.0,2.0)
		\psarc[linewidth=0.02,arrowsize=2pt 3,
		arrowinset=0]{->}(1.5,1.0){0.6}{20}{55}
		\psarc[linewidth=0.02](1.5,1.0){0.6}{46}{65}
		\psarc[linewidth=0.02,arrowsize=2pt 3,
		arrowinset=0]{->}(1.5,1.0){0.6}{115}{150}
		\psarc[linewidth=0.02](1.5,1.0){0.6}{141}{160}
		\psarc[linewidth=0.03,linestyle=dashed,
		dash=2pt 1pt,linecolor=Blue](1.5,1.0){0.6}{200}{340}
		\psline[linewidth=0.02,ArrowInside=->,
		ArrowInsidePos=0.55,arrowsize=2pt 3,
		arrowinset=0](0.7,1.0)(0.1,1.0)
		\psline[linewidth=0.02,ArrowInside=->,
		ArrowInsidePos=0.55,arrowsize=2pt 3,
		arrowinset=0](2.9,1.0)(2.3,1.0)
		\pscircle[linewidth=0.03,fillstyle=solid,
		fillcolor=lightgray](1.5,1.6){0.25}
		\psline[linewidth=0.03](1.5,1.6)(1.67,1.77)
		\psline[linewidth=0.03](1.5,1.6)(1.33,1.77)
		\psline[linewidth=0.03](1.5,1.6)(1.33,1.43)
		\psline[linewidth=0.03](1.5,1.6)(1.67,1.43)
		\pscircle[linewidth=0.03,fillstyle=solid,
		fillcolor=lightgray!25](0.9,1.0){0.20}
		\pscircle[linewidth=0.03,fillstyle=solid,
		fillcolor=lightgray!25](2.1,1.0){0.20}
		\rput[b]{*0}(2.12,1.45){\small $\Psi$}
		\rput[b]{*0}(0.88,1.45){\small $\Psi$}
		\rput[b]{*0}(0.4,0.6){\small $\Psi$}
		\rput[b]{*0}(2.6,0.6){\small $\Psi$}
		\rput[b]{*0}(1.5,0.1){\small $\sigma$}
		\rput[b]{*0}(0.9,0.92){\fontsize{7pt}{0pt}{$3$}\selectfont}
		\rput[b]{*0}(2.1,0.92){\fontsize{7pt}{0pt}{$3$}\selectfont}
	\end{pspicture}
	}} \! \! \! 
	- \! \! \! \vcenter{\hbox{
	\begin{pspicture}(3.0,2.0)
		\psarc[linewidth=0.02,arrowsize=2pt 3,
		arrowinset=0]{->}(1.5,1.0){0.6}{20}{55}
		\psarc[linewidth=0.02](1.5,1.0){0.6}{46}{65}
		\psarc[linewidth=0.02,arrowsize=2pt 3,
		arrowinset=0]{->}(1.5,1.0){0.6}{115}{150}
		\psarc[linewidth=0.02](1.5,1.0){0.6}{141}{160}
		\psarc[linewidth=0.03,linestyle=dashed,
		dash=2pt 1pt,linecolor=Red](1.5,1.0){0.6}{200}{340}
		\psline[linewidth=0.02,ArrowInside=->,
		ArrowInsidePos=0.55,arrowsize=2pt 3,
		arrowinset=0](0.7,1.0)(0.1,1.0)
		\psline[linewidth=0.02,ArrowInside=->,
		ArrowInsidePos=0.55,arrowsize=2pt 3,
		arrowinset=0](2.9,1.0)(2.3,1.0)
		\pscircle[linewidth=0.03,fillstyle=solid,
		fillcolor=lightgray](1.5,1.6){0.25}
		\psline[linewidth=0.03](1.5,1.6)(1.67,1.77)
		\psline[linewidth=0.03](1.5,1.6)(1.33,1.77)
		\psline[linewidth=0.03](1.5,1.6)(1.33,1.43)
		\psline[linewidth=0.03](1.5,1.6)(1.67,1.43)
		\pscircle[linewidth=0.03,fillstyle=solid,
		fillcolor=lightgray!25](0.9,1.0){0.20}
		\pscircle[linewidth=0.03,fillstyle=solid,
		fillcolor=lightgray!25](2.1,1.0){0.20}
		\rput[b]{*0}(2.12,1.45){\small $\Psi$}
		\rput[b]{*0}(0.88,1.45){\small $\Psi$}
		\rput[b]{*0}(0.4,0.6){\small $\Psi$}
		\rput[b]{*0}(2.6,0.6){\small $\Psi$}
		\rput[b]{*0}(1.5,0.1){\small $\pi$}
		\rput[b]{*0}(0.9,0.92){\fontsize{7pt}{0pt}{$3$}\selectfont}
		\rput[b]{*0}(2.1,0.92){\fontsize{7pt}{0pt}{$3$}\selectfont}
	\end{pspicture}
	}} \! \! \biggg) \! ,\label{eq:m0} \\[0.4cm]
	\partial_{k} y_{1,k} & = & \frac{i}{24} \mathcal{V}^{-1}
	\tr_{\gamma,\mathrm{flavor}} 
	\left[\left.\frac{\delta^{3}\partial_{k}\Gamma_{k}}
	{\rule[0.20cm]{0pt}{1ex}\delta\pi^{a}(0)\delta \bar{\Psi}_{\mathsf{p}_{1}}(0)
	\delta\Psi_{\mathsf{p}_{2}}(0)} \right|_{\mathsf{p}_{1}\, =\, 
	\mathsf{p}_{2}\, =\, 1} \gamma_{5} \tau^{a}\right] \nonumber\\
	& = & \frac{i}{24} \mathcal{V}^{-1}
	\tr_{\gamma,\mathrm{flavor}} \Biggg[ \Biggg(
	\! \! - \frac{1}{2} \! \! \vcenter{\hbox{
	\begin{pspicture}[showgrid=false](3.0,3.2)
		\psarc[linewidth=0.03,linestyle=dashed,
		dash=2pt 1pt,linecolor=Blue](1.5,1.6){0.6}{20}{65}
		\psarc[linewidth=0.03,linestyle=dashed,
		dash=2pt 1pt,linecolor=Blue](1.5,1.6){0.6}{115}{160}
		\psarc[linewidth=0.03,linestyle=dashed,
		dash=2pt 1pt,linecolor=Red](1.5,1.6){0.6}{290}{340}
		\psarc[linewidth=0.02,arrowsize=2pt 3,
		arrowinset=0]{->}(1.5,1.6){0.6}{200}{235}
		\psarc[linewidth=0.02](1.5,1.6){0.6}{226}{250}
		\psline[linewidth=0.02,ArrowInside=->,
		ArrowInsidePos=0.55,arrowsize=2pt 3,
		arrowinset=0](0.1,1.6)(0.7,1.6)
		\psline[linewidth=0.03,linestyle=dashed,
		dash=2pt 1pt,linecolor=Red](2.3,1.6)(2.9,1.6)
		\psline[linewidth=0.02,ArrowInside=->,
		ArrowInsidePos=0.55,arrowsize=2pt 3,
		arrowinset=0](1.5,0.8)(1.5,0.2)
		\pscircle[linewidth=0.03,fillstyle=solid,
		fillcolor=RoyalBlue](1.5,2.2){0.25}
		\psline[linewidth=0.03](1.5,2.2)(1.67,2.37)
		\psline[linewidth=0.03](1.5,2.2)(1.33,2.37)
		\psline[linewidth=0.03](1.5,2.2)(1.33,2.03)
		\psline[linewidth=0.03](1.5,2.2)(1.67,2.03)
		\pscircle[linewidth=0.03,fillstyle=solid,
		fillcolor=lightgray!25](1.5,1.0){0.20}
		\rput[b]{*0}(1.5,0.92){\scriptsize{$3$}}
		\pscircle[linewidth=0.03,fillstyle=solid,
		fillcolor=lightgray!25](0.9,1.6){0.20}
		\pscircle[linewidth=0.03,fillstyle=solid,
		fillcolor=lightgray!25](2.1,1.6){0.20}
		\rput[b]{*0}(0.9,1.52){\scriptsize{$3$}}
		\rput[b]{*0}(2.1,1.52){\scriptsize{$3$}}
		\rput[b]{*0}(2.6,1.3){\small $\pi$}
		\rput[b]{*0}(1.27,0.4){\small $\Psi$}
		\rput[b]{*0}(2.12,2.05){\small $\sigma$}
		\rput[b]{*0}(0.88,2.05){\small $\sigma$}
		\rput[t]{*0}(2.12,1.12){\small $\pi$}
		\rput[t]{*0}(0.88,1.12){\small $\Psi$}
		\rput[b]{*0}(0.4,1.2){\small $\Psi$}
	\end{pspicture}
	}} \! \! \!
	- \frac{1}{2} \! \! \vcenter{\hbox{
	\begin{pspicture}[showgrid=false](3.0,3.2)
		\psarc[linewidth=0.03,linestyle=dashed,
		dash=2pt 1pt,linecolor=Red](1.5,1.6){0.6}{20}{65}
		\psarc[linewidth=0.03,linestyle=dashed,
		dash=2pt 1pt,linecolor=Red](1.5,1.6){0.6}{115}{160}
		\psarc[linewidth=0.03,linestyle=dashed,
		dash=2pt 1pt,linecolor=Blue](1.5,1.6){0.6}{290}{340}
		\psarc[linewidth=0.02,arrowsize=2pt 3,
		arrowinset=0]{->}(1.5,1.6){0.6}{200}{235}
		\psarc[linewidth=0.02](1.5,1.6){0.6}{226}{250}
		\psline[linewidth=0.02,ArrowInside=->,
		ArrowInsidePos=0.55,arrowsize=2pt 3,
		arrowinset=0](0.1,1.6)(0.7,1.6)
		\psline[linewidth=0.03,linestyle=dashed,
		dash=2pt 1pt,linecolor=Red](2.3,1.6)(2.9,1.6)
		\psline[linewidth=0.02,ArrowInside=->,
		ArrowInsidePos=0.55,arrowsize=2pt 3,
		arrowinset=0](1.5,0.8)(1.5,0.2)
		\pscircle[linewidth=0.03,fillstyle=solid,
		fillcolor=RedOrange](1.5,2.2){0.25}
		\psline[linewidth=0.03](1.5,2.2)(1.67,2.37)
		\psline[linewidth=0.03](1.5,2.2)(1.33,2.37)
		\psline[linewidth=0.03](1.5,2.2)(1.33,2.03)
		\psline[linewidth=0.03](1.5,2.2)(1.67,2.03)
		\pscircle[linewidth=0.03,fillstyle=solid,
		fillcolor=lightgray!25](1.5,1.0){0.20}
		\rput[b]{*0}(1.5,0.92){\scriptsize{$3$}}
		\pscircle[linewidth=0.03,fillstyle=solid,
		fillcolor=lightgray!25](0.9,1.6){0.20}
		\pscircle[linewidth=0.03,fillstyle=solid,
		fillcolor=lightgray!25](2.1,1.6){0.20}
		\rput[b]{*0}(0.9,1.52){\scriptsize{$3$}}
		\rput[b]{*0}(2.1,1.52){\scriptsize{$3$}}
		\rput[b]{*0}(2.6,1.3){\small $\pi$}
		\rput[b]{*0}(1.27,0.4){\small $\Psi$}
		\rput[b]{*0}(2.12,2.05){\small $\pi$}
		\rput[b]{*0}(0.88,2.05){\small $\pi$}
		\rput[t]{*0}(2.12,1.12){\small $\sigma$}
		\rput[t]{*0}(0.88,1.12){\small $\Psi$}
		\rput[b]{*0}(0.4,1.2){\small $\Psi$}
	\end{pspicture}
	}} \! \! \!
	- \frac{1}{2} \! \! \vcenter{\hbox{
	\begin{pspicture}[showgrid=false](3.0,3.2)
		\psarc[linewidth=0.03,linestyle=dashed,
		dash=2pt 1pt,linecolor=Blue](1.5,1.6){0.6}{20}{65}
		\psarc[linewidth=0.03,linestyle=dashed,
		dash=2pt 1pt,linecolor=Blue](1.5,1.6){0.6}{115}{160}
		\psarc[linewidth=0.02,arrowsize=2pt 3,
		arrowinset=0]{->}(1.5,1.6){0.6}{200}{235}
		\psarc[linewidth=0.02](1.5,1.6){0.6}{226}{250}
		\psarc[linewidth=0.02,arrowsize=2pt 3,
		arrowinset=0]{->}(1.5,1.6){0.6}{290}{330}
		\psarc[linewidth=0.02](1.5,1.6){0.6}{321}{340}
		\psline[linewidth=0.02,ArrowInside=->,
		ArrowInsidePos=0.55,arrowsize=2pt 3,
		arrowinset=0](0.1,1.6)(0.7,1.6)
		\psline[linewidth=0.02,ArrowInside=->,
		ArrowInsidePos=0.55,arrowsize=2pt 3,
		arrowinset=0](2.3,1.6)(2.9,1.6)
		\psline[linewidth=0.03,linestyle=dashed,
		dash=2pt 1pt,linecolor=Red](1.5,0.8)(1.5,0.2)
		\pscircle[linewidth=0.03,fillstyle=solid,
		fillcolor=RoyalBlue](1.5,2.2){0.25}
		\psline[linewidth=0.03](1.5,2.2)(1.67,2.37)
		\psline[linewidth=0.03](1.5,2.2)(1.33,2.37)
		\psline[linewidth=0.03](1.5,2.2)(1.33,2.03)
		\psline[linewidth=0.03](1.5,2.2)(1.67,2.03)
		\pscircle[linewidth=0.03,fillstyle=solid,
		fillcolor=lightgray!25](0.9,1.6){0.20}
		\pscircle[linewidth=0.03,fillstyle=solid,
		fillcolor=lightgray!25](2.1,1.6){0.20}
		\pscircle[linewidth=0.03,fillstyle=solid,
		fillcolor=lightgray!25](1.5,1.0){0.20}
		\rput[b]{*0}(1.5,0.92){\scriptsize{$3$}}
		\rput[b]{*0}(2.12,2.05){\small $\sigma$}
		\rput[b]{*0}(0.88,2.05){\small $\sigma$}
		\rput[b]{*0}(0.4,1.2){\small $\Psi$}
		\rput[b]{*0}(2.6,1.2){\small $\Psi$}
		\rput[b]{*0}(1.27,0.4){\small $\pi$}
		\rput[t]{*0}(2.12,1.12){\small $\Psi$}
		\rput[t]{*0}(0.88,1.12){\small $\Psi$}
		\rput[b]{*0}(0.9,1.52){\fontsize{7pt}{0pt}{$3$}\selectfont}
		\rput[b]{*0}(2.1,1.52){\fontsize{7pt}{0pt}{$3$}\selectfont}
	\end{pspicture}
	}} \! \! \! \nonumber\\[-0.5cm]
	& & \qquad \qquad \qquad \qquad \ \ 
	- \, \frac{1}{2} \! \! \vcenter{\hbox{
	\begin{pspicture}[showgrid=false](3.0,3.2)
		\psarc[linewidth=0.03,linestyle=dashed,
		dash=2pt 1pt,linecolor=Red](1.5,1.6){0.6}{20}{65}
		\psarc[linewidth=0.03,linestyle=dashed,
		dash=2pt 1pt,linecolor=Red](1.5,1.6){0.6}{115}{160}
		\psarc[linewidth=0.02,arrowsize=2pt 3,
		arrowinset=0]{->}(1.5,1.6){0.6}{200}{235}
		\psarc[linewidth=0.02](1.5,1.6){0.6}{226}{250}
		\psarc[linewidth=0.02,arrowsize=2pt 3,
		arrowinset=0]{->}(1.5,1.6){0.6}{200}{330}
		\psarc[linewidth=0.02](1.5,1.6){0.6}{321}{340}
		\psline[linewidth=0.02,ArrowInside=->,
		ArrowInsidePos=0.55,arrowsize=2pt 3,
		arrowinset=0](0.1,1.6)(0.7,1.6)
		\psline[linewidth=0.02,ArrowInside=->,
		ArrowInsidePos=0.55,arrowsize=2pt 3,
		arrowinset=0](2.3,1.6)(2.9,1.6)
		\psline[linewidth=0.03,linestyle=dashed,
		dash=2pt 1pt,linecolor=Red](1.5,0.8)(1.5,0.2)
		\pscircle[linewidth=0.03,fillstyle=solid,
		fillcolor=RedOrange](1.5,2.2){0.25}
		\psline[linewidth=0.03](1.5,2.2)(1.67,2.37)
		\psline[linewidth=0.03](1.5,2.2)(1.33,2.37)
		\psline[linewidth=0.03](1.5,2.2)(1.33,2.03)
		\psline[linewidth=0.03](1.5,2.2)(1.67,2.03)
		\pscircle[linewidth=0.03,fillstyle=solid,
		fillcolor=lightgray!25](0.9,1.6){0.20}
		\pscircle[linewidth=0.03,fillstyle=solid,
		fillcolor=lightgray!25](2.1,1.6){0.20}
		\pscircle[linewidth=0.03,fillstyle=solid,
		fillcolor=lightgray!25](1.5,1.0){0.20}
		\rput[b]{*0}(1.5,0.92){\scriptsize{$3$}}
		\rput[b]{*0}(2.12,2.05){\small $\pi$}
		\rput[b]{*0}(0.88,2.05){\small $\pi$}
		\rput[b]{*0}(0.4,1.2){\small $\Psi$}
		\rput[b]{*0}(2.6,1.2){\small $\Psi$}
		\rput[b]{*0}(1.27,0.4){\small $\pi$}
		\rput[t]{*0}(2.12,1.12){\small $\Psi$}
		\rput[t]{*0}(0.88,1.12){\small $\Psi$}
		\rput[b]{*0}(0.9,1.52){\fontsize{7pt}{0pt}{$3$}\selectfont}
		\rput[b]{*0}(2.1,1.52){\fontsize{7pt}{0pt}{$3$}\selectfont}
	\end{pspicture}
	}} \! \! \! 
	- \frac{1}{2} \! \! \vcenter{\hbox{
	\begin{pspicture}[showgrid=false](3.0,3.2)
		\psarc[linewidth=0.03,linestyle=dashed,
		dash=2pt 1pt,linecolor=Blue](1.5,1.6){0.6}{20}{65}
		\psarc[linewidth=0.03,linestyle=dashed,
		dash=2pt 1pt,linecolor=Blue](1.5,1.6){0.6}{115}{160}
		\psarc[linewidth=0.03,linestyle=dashed,
		dash=2pt 1pt,linecolor=Red](1.5,1.6){0.6}{290}{340}
		\psarc[linewidth=0.02,arrowsize=2pt 3,
		arrowinset=0]{<-}(1.5,1.6){0.6}{215}{250}
		\psarc[linewidth=0.02](1.5,1.6){0.6}{200}{224}
		\psline[linewidth=0.02,ArrowInside=->,
		ArrowInsidePos=0.55,arrowsize=2pt 3,
		arrowinset=0](0.7,1.6)(0.1,1.6)
		\psline[linewidth=0.03,linestyle=dashed,
		dash=2pt 1pt,linecolor=Red](2.3,1.6)(2.9,1.6)
		\psline[linewidth=0.02,ArrowInside=->,
		ArrowInsidePos=0.55,arrowsize=2pt 3,
		arrowinset=0](1.5,0.2)(1.5,0.8)
		\pscircle[linewidth=0.03,fillstyle=solid,
		fillcolor=RoyalBlue](1.5,2.2){0.25}
		\psline[linewidth=0.03](1.5,2.2)(1.67,2.37)
		\psline[linewidth=0.03](1.5,2.2)(1.33,2.37)
		\psline[linewidth=0.03](1.5,2.2)(1.33,2.03)
		\psline[linewidth=0.03](1.5,2.2)(1.67,2.03)
		\pscircle[linewidth=0.03,fillstyle=solid,
		fillcolor=lightgray!25](1.5,1.0){0.20}
		\rput[b]{*0}(1.5,0.92){\scriptsize{$3$}}
		\pscircle[linewidth=0.03,fillstyle=solid,
		fillcolor=lightgray!25](0.9,1.6){0.20}
		\pscircle[linewidth=0.03,fillstyle=solid,
		fillcolor=lightgray!25](2.1,1.6){0.20}
		\rput[b]{*0}(0.9,1.52){\scriptsize{$3$}}
		\rput[b]{*0}(2.1,1.52){\scriptsize{$3$}}
		\rput[b]{*0}(2.6,1.3){\small $\pi$}
		\rput[b]{*0}(1.27,0.4){\small $\Psi$}
		\rput[b]{*0}(2.12,2.05){\small $\sigma$}
		\rput[b]{*0}(0.88,2.05){\small $\sigma$}
		\rput[t]{*0}(2.12,1.12){\small $\pi$}
		\rput[t]{*0}(0.88,1.12){\small $\Psi$}
		\rput[b]{*0}(0.4,1.2){\small $\Psi$}
	\end{pspicture}
	}} \! \! \!
	- \frac{1}{2} \! \! \vcenter{\hbox{
	\begin{pspicture}[showgrid=false](3.0,3.2)
		\psarc[linewidth=0.03,linestyle=dashed,
		dash=2pt 1pt,linecolor=Red](1.5,1.6){0.6}{20}{65}
		\psarc[linewidth=0.03,linestyle=dashed,
		dash=2pt 1pt,linecolor=Red](1.5,1.6){0.6}{115}{160}
		\psarc[linewidth=0.03,linestyle=dashed,
		dash=2pt 1pt,linecolor=Blue](1.5,1.6){0.6}{290}{340}
		\psarc[linewidth=0.02,arrowsize=2pt 3,
		arrowinset=0]{<-}(1.5,1.6){0.6}{215}{250}
		\psarc[linewidth=0.02](1.5,1.6){0.6}{200}{224}
		\psline[linewidth=0.02,ArrowInside=->,
		ArrowInsidePos=0.55,arrowsize=2pt 3,
		arrowinset=0](0.7,1.6)(0.1,1.6)
		\psline[linewidth=0.03,linestyle=dashed,
		dash=2pt 1pt,linecolor=Red](2.3,1.6)(2.9,1.6)
		\psline[linewidth=0.02,ArrowInside=->,
		ArrowInsidePos=0.55,arrowsize=2pt 3,
		arrowinset=0](1.5,0.2)(1.5,0.8)
		\pscircle[linewidth=0.03,fillstyle=solid,
		fillcolor=RedOrange](1.5,2.2){0.25}
		\psline[linewidth=0.03](1.5,2.2)(1.67,2.37)
		\psline[linewidth=0.03](1.5,2.2)(1.33,2.37)
		\psline[linewidth=0.03](1.5,2.2)(1.33,2.03)
		\psline[linewidth=0.03](1.5,2.2)(1.67,2.03)
		\pscircle[linewidth=0.03,fillstyle=solid,
		fillcolor=lightgray!25](1.5,1.0){0.20}
		\rput[b]{*0}(1.5,0.92){\scriptsize{$3$}}
		\pscircle[linewidth=0.03,fillstyle=solid,
		fillcolor=lightgray!25](0.9,1.6){0.20}
		\pscircle[linewidth=0.03,fillstyle=solid,
		fillcolor=lightgray!25](2.1,1.6){0.20}
		\rput[b]{*0}(0.9,1.52){\scriptsize{$3$}}
		\rput[b]{*0}(2.1,1.52){\scriptsize{$3$}}
		\rput[b]{*0}(2.6,1.3){\small $\pi$}
		\rput[b]{*0}(1.27,0.4){\small $\Psi$}
		\rput[b]{*0}(2.12,2.05){\small $\pi$}
		\rput[b]{*0}(0.88,2.05){\small $\pi$}
		\rput[t]{*0}(2.12,1.12){\small $\sigma$}
		\rput[t]{*0}(0.88,1.12){\small $\Psi$}
		\rput[b]{*0}(0.4,1.2){\small $\Psi$}
	\end{pspicture}
	}} \! \! \! \nonumber\\[-0.5cm]
	& & \qquad \qquad \qquad \qquad \ \ 
	+ \! \! \vcenter{\hbox{
	\begin{pspicture}[showgrid=false](3.0,3.2)
		\psarc[linewidth=0.02,arrowsize=2pt 3,
		arrowinset=0]{->}(1.5,1.6){0.6}{20}{55}
		\psarc[linewidth=0.02](1.5,1.6){0.6}{46}{65}
		\psarc[linewidth=0.02,arrowsize=2pt 3,
		arrowinset=0]{->}(1.5,1.6){0.6}{115}{150}
		\psarc[linewidth=0.02](1.5,1.6){0.6}{141}{160}
		\psarc[linewidth=0.03,linestyle=dashed,
		dash=2pt 1pt,linecolor=Red](1.5,1.6){0.6}{200}{250}
		\psarc[linewidth=0.03,linestyle=dashed,
		dash=2pt 1pt,linecolor=Blue](1.5,1.6){0.6}{290}{340}
		\psline[linewidth=0.02,ArrowInside=->,
		ArrowInsidePos=0.55,arrowsize=2pt 3,
		arrowinset=0](0.7,1.6)(0.1,1.6)
		\psline[linewidth=0.02,ArrowInside=->,
		ArrowInsidePos=0.55,arrowsize=2pt 3,
		arrowinset=0](2.9,1.6)(2.3,1.6)
		\psline[linewidth=0.03,linestyle=dashed,
		dash=2pt 1pt,linecolor=Red](1.5,0.8)(1.5,0.2)
		\pscircle[linewidth=0.03,fillstyle=solid,
		fillcolor=lightgray](1.5,2.2){0.25}
		\psline[linewidth=0.03](1.5,2.2)(1.67,2.37)
		\psline[linewidth=0.03](1.5,2.2)(1.33,2.37)
		\psline[linewidth=0.03](1.5,2.2)(1.33,2.03)
		\psline[linewidth=0.03](1.5,2.2)(1.67,2.03)
		\pscircle[linewidth=0.03,fillstyle=solid,
		fillcolor=lightgray!25](0.9,1.6){0.20}
		\pscircle[linewidth=0.03,fillstyle=solid,
		fillcolor=lightgray!25](2.1,1.6){0.20}
		\pscircle[linewidth=0.03,fillstyle=solid,
		fillcolor=lightgray!25](1.5,1.0){0.20}
		\rput[b]{*0}(1.5,0.92){\scriptsize{$3$}}
		\rput[b]{*0}(2.12,2.05){\small $\Psi$}
		\rput[b]{*0}(0.88,2.05){\small $\Psi$}
		\rput[b]{*0}(0.4,1.2){\small $\Psi$}
		\rput[b]{*0}(2.6,1.2){\small $\Psi$}
		\rput[b]{*0}(1.27,0.4){\small $\pi$}
		\rput[t]{*0}(2.12,1.12){\small $\sigma$}
		\rput[t]{*0}(0.88,1.12){\small $\pi$}
		\rput[b]{*0}(0.9,1.52){\fontsize{7pt}{0pt}{$3$}\selectfont}
		\rput[b]{*0}(2.1,1.52){\fontsize{7pt}{0pt}{$3$}\selectfont}
	\end{pspicture}
	}} \! \! \! 
	+ \! \! \! \vcenter{\hbox{
	\begin{pspicture}[showgrid=false](3.0,3.2)
		\psarc[linewidth=0.02,arrowsize=2pt 3,
		arrowinset=0]{->}(1.5,1.6){0.6}{20}{55}
		\psarc[linewidth=0.02](1.5,1.6){0.6}{46}{65}
		\psarc[linewidth=0.02,arrowsize=2pt 3,
		arrowinset=0]{->}(1.5,1.6){0.6}{115}{150}
		\psarc[linewidth=0.02](1.5,1.6){0.6}{141}{160}
		\psarc[linewidth=0.03,linestyle=dashed,
		dash=2pt 1pt,linecolor=Blue](1.5,1.6){0.6}{200}{250}
		\psarc[linewidth=0.03,linestyle=dashed,
		dash=2pt 1pt,linecolor=Red](1.5,1.6){0.6}{290}{340}
		\psline[linewidth=0.02,ArrowInside=->,
		ArrowInsidePos=0.55,arrowsize=2pt 3,
		arrowinset=0](0.7,1.6)(0.1,1.6)
		\psline[linewidth=0.02,ArrowInside=->,
		ArrowInsidePos=0.55,arrowsize=2pt 3,
		arrowinset=0](2.9,1.6)(2.3,1.6)
		\psline[linewidth=0.03,linestyle=dashed,
		dash=2pt 1pt,linecolor=Red](1.5,0.8)(1.5,0.2)
		\pscircle[linewidth=0.03,fillstyle=solid,
		fillcolor=lightgray](1.5,2.2){0.25}
		\psline[linewidth=0.03](1.5,2.2)(1.67,2.37)
		\psline[linewidth=0.03](1.5,2.2)(1.33,2.37)
		\psline[linewidth=0.03](1.5,2.2)(1.33,2.03)
		\psline[linewidth=0.03](1.5,2.2)(1.67,2.03)
		\pscircle[linewidth=0.03,fillstyle=solid,
		fillcolor=lightgray!25](0.9,1.6){0.20}
		\pscircle[linewidth=0.03,fillstyle=solid,
		fillcolor=lightgray!25](2.1,1.6){0.20}
		\pscircle[linewidth=0.03,fillstyle=solid,
		fillcolor=lightgray!25](1.5,1.0){0.20}
		\rput[b]{*0}(1.5,0.92){\scriptsize{$3$}}
		\rput[b]{*0}(2.12,2.05){\small $\Psi$}
		\rput[b]{*0}(0.88,2.05){\small $\Psi$}
		\rput[b]{*0}(0.4,1.2){\small $\Psi$}
		\rput[b]{*0}(2.6,1.2){\small $\Psi$}
		\rput[b]{*0}(1.27,0.4){\small $\pi$}
		\rput[t]{*0}(2.12,1.12){\small $\pi$}
		\rput[t]{*0}(0.88,1.12){\small $\sigma$}
		\rput[b]{*0}(0.9,1.52){\fontsize{7pt}{0pt}{$3$}\selectfont}
		\rput[b]{*0}(2.1,1.52){\fontsize{7pt}{0pt}{$3$}\selectfont}
	\end{pspicture}
	}} \! \! \!
	+ \! \! \! \vcenter{\hbox{
	\begin{pspicture}[showgrid=false](3.0,3.2)
		\psarc[linewidth=0.02,arrowsize=2pt 3,
		arrowinset=0]{->}(1.5,1.6){0.6}{20}{55}
		\psarc[linewidth=0.02](1.5,1.6){0.6}{46}{65}
		\psarc[linewidth=0.02,arrowsize=2pt 3,
		arrowinset=0]{->}(1.5,1.6){0.6}{115}{150}
		\psarc[linewidth=0.02](1.5,1.6){0.6}{141}{160}
		\psarc[linewidth=0.03,linestyle=dashed,
		dash=2pt 1pt,linecolor=Blue](1.5,1.6){0.6}{200}{250}
		\psarc[linewidth=0.02,arrowsize=2pt 3,
		arrowinset=0]{->}(1.5,1.6){0.6}{290}{325}
		\psarc[linewidth=0.02](1.5,1.6){0.6}{316}{340}
		\psline[linewidth=0.02,ArrowInside=->,
		ArrowInsidePos=0.55,arrowsize=2pt 3,
		arrowinset=0](0.7,1.6)(0.1,1.6)
		\psline[linewidth=0.03,linestyle=dashed,
		dash=2pt 1pt,linecolor=Red](2.9,1.6)(2.3,1.6)
		\psline[linewidth=0.02,ArrowInside=->,
		ArrowInsidePos=0.55,arrowsize=2pt 3,
		arrowinset=0](1.5,0.2)(1.5,0.8)
		\pscircle[linewidth=0.03,fillstyle=solid,
		fillcolor=lightgray](1.5,2.2){0.25}
		\psline[linewidth=0.03](1.5,2.2)(1.67,2.37)
		\psline[linewidth=0.03](1.5,2.2)(1.33,2.37)
		\psline[linewidth=0.03](1.5,2.2)(1.33,2.03)
		\psline[linewidth=0.03](1.5,2.2)(1.67,2.03)
		\pscircle[linewidth=0.03,fillstyle=solid,
		fillcolor=lightgray!25](0.9,1.6){0.20}
		\pscircle[linewidth=0.03,fillstyle=solid,
		fillcolor=lightgray!25](2.1,1.6){0.20}
		\pscircle[linewidth=0.03,fillstyle=solid,
		fillcolor=lightgray!25](1.5,1.0){0.20}
		\rput[b]{*0}(1.5,0.92){\scriptsize{$3$}}
		\rput[b]{*0}(2.12,2.05){\small $\Psi$}
		\rput[b]{*0}(0.88,2.05){\small $\Psi$}
		\rput[b]{*0}(0.4,1.2){\small $\Psi$}
		\rput[b]{*0}(2.6,1.3){\small $\pi$}
		\rput[b]{*0}(1.27,0.4){\small $\Psi$}
		\rput[t]{*0}(2.12,1.12){\small $\Psi$}
		\rput[t]{*0}(0.88,1.12){\small $\sigma$}
		\rput[b]{*0}(0.9,1.52){\fontsize{7pt}{0pt}{$3$}\selectfont}
		\rput[b]{*0}(2.1,1.52){\fontsize{7pt}{0pt}{$3$}\selectfont}
	\end{pspicture}
	}} \! \! \! \vspace*{1.2cm} \nonumber\\[-0.4cm]
	& & \qquad \qquad \qquad \qquad \ \ 
	+ \! \! \vcenter{\hbox{
	\begin{pspicture}[showgrid=false](3.0,3.2)
		\psarc[linewidth=0.02,arrowsize=2pt 3,
		arrowinset=0]{->}(1.5,1.6){0.6}{20}{55}
		\psarc[linewidth=0.02](1.5,1.6){0.6}{46}{65}
		\psarc[linewidth=0.02,arrowsize=2pt 3,
		arrowinset=0]{->}(1.5,1.6){0.6}{115}{150}
		\psarc[linewidth=0.02](1.5,1.6){0.6}{141}{160}
		\psarc[linewidth=0.03,linestyle=dashed,
		dash=2pt 1pt,linecolor=Red](1.5,1.6){0.6}{200}{250}
		\psarc[linewidth=0.02,arrowsize=2pt 3,
		arrowinset=0]{->}(1.5,1.6){0.6}{290}{325}
		\psarc[linewidth=0.02](1.5,1.6){0.6}{316}{340}
		\psline[linewidth=0.02,ArrowInside=->,
		ArrowInsidePos=0.55,arrowsize=2pt 3,
		arrowinset=0](0.7,1.6)(0.1,1.6)
		\psline[linewidth=0.03,linestyle=dashed,
		dash=2pt 1pt,linecolor=Red](2.9,1.6)(2.3,1.6)
		\psline[linewidth=0.02,ArrowInside=->,
		ArrowInsidePos=0.55,arrowsize=2pt 3,
		arrowinset=0](1.5,0.2)(1.5,0.8)
		\pscircle[linewidth=0.03,fillstyle=solid,
		fillcolor=lightgray](1.5,2.2){0.25}
		\psline[linewidth=0.03](1.5,2.2)(1.67,2.37)
		\psline[linewidth=0.03](1.5,2.2)(1.33,2.37)
		\psline[linewidth=0.03](1.5,2.2)(1.33,2.03)
		\psline[linewidth=0.03](1.5,2.2)(1.67,2.03)
		\pscircle[linewidth=0.03,fillstyle=solid,
		fillcolor=lightgray!25](0.9,1.6){0.20}
		\pscircle[linewidth=0.03,fillstyle=solid,
		fillcolor=lightgray!25](2.1,1.6){0.20}
		\pscircle[linewidth=0.03,fillstyle=solid,
		fillcolor=lightgray!25](1.5,1.0){0.20}
		\rput[b]{*0}(1.5,0.92){\scriptsize{$3$}}
		\rput[b]{*0}(2.12,2.05){\small $\Psi$}
		\rput[b]{*0}(0.88,2.05){\small $\Psi$}
		\rput[b]{*0}(0.4,1.2){\small $\Psi$}
		\rput[b]{*0}(2.6,1.3){\small $\pi$}
		\rput[b]{*0}(1.27,0.4){\small $\Psi$}
		\rput[t]{*0}(2.12,1.12){\small $\Psi$}
		\rput[t]{*0}(0.88,1.12){\small $\pi$}
		\rput[b]{*0}(0.9,1.52){\fontsize{7pt}{0pt}{$3$}\selectfont}
		\rput[b]{*0}(2.1,1.52){\fontsize{7pt}{0pt}{$3$}\selectfont}
	\end{pspicture}
	}} \! \! \!
	+ \! \! \! \vcenter{\hbox{
	\begin{pspicture}[showgrid=false](3.0,3.2)
		\psarc[linewidth=0.02,arrowsize=2pt 3,
		arrowinset=0]{<-}(1.5,1.6){0.6}{30}{65}
		\psarc[linewidth=0.02](1.5,1.6){0.6}{20}{39}
		\psarc[linewidth=0.02,arrowsize=2pt 3,
		arrowinset=0]{<-}(1.5,1.6){0.6}{125}{160}
		\psarc[linewidth=0.02](1.5,1.6){0.6}{110}{134}
		\psarc[linewidth=0.02,arrowsize=2pt 3,
		arrowinset=0]{<-}(1.5,1.6){0.6}{215}{250}
		\psarc[linewidth=0.02](1.5,1.6){0.6}{200}{224}
		\psarc[linewidth=0.03,linestyle=dashed,
		dash=2pt 1pt,linecolor=Blue](1.5,1.6){0.6}{290}{340}
		\psline[linewidth=0.02,ArrowInside=->,
		ArrowInsidePos=0.55,arrowsize=2pt 3,
		arrowinset=0](2.3,1.6)(2.9,1.6)
		\psline[linewidth=0.03,linestyle=dashed,
		dash=2pt 1pt,linecolor=Red](0.1,1.6)(0.7,1.6)
		\psline[linewidth=0.02,ArrowInside=->,
		ArrowInsidePos=0.55,arrowsize=2pt 3,
		arrowinset=0](1.5,0.2)(1.5,0.8)
		\pscircle[linewidth=0.03,fillstyle=solid,
		fillcolor=lightgray](1.5,2.2){0.25}
		\psline[linewidth=0.03](1.5,2.2)(1.67,2.37)
		\psline[linewidth=0.03](1.5,2.2)(1.33,2.37)
		\psline[linewidth=0.03](1.5,2.2)(1.33,2.03)
		\psline[linewidth=0.03](1.5,2.2)(1.67,2.03)
		\pscircle[linewidth=0.03,fillstyle=solid,
		fillcolor=lightgray!25](0.9,1.6){0.20}
		\pscircle[linewidth=0.03,fillstyle=solid,
		fillcolor=lightgray!25](2.1,1.6){0.20}
		\pscircle[linewidth=0.03,fillstyle=solid,
		fillcolor=lightgray!25](1.5,1.0){0.20}
		\rput[b]{*0}(1.5,0.92){\scriptsize{$3$}}
		\rput[b]{*0}(2.12,2.05){\small $\Psi$}
		\rput[b]{*0}(0.88,2.05){\small $\Psi$}
		\rput[b]{*0}(0.4,1.3){\small $\pi$}
		\rput[b]{*0}(2.6,1.2){\small $\Psi$}
		\rput[b]{*0}(1.27,0.4){\small $\Psi$}
		\rput[t]{*0}(2.12,1.12){\small $\sigma$}
		\rput[t]{*0}(0.88,1.12){\small $\Psi$}
		\rput[b]{*0}(0.9,1.52){\fontsize{7pt}{0pt}{$3$}\selectfont}
		\rput[b]{*0}(2.1,1.52){\fontsize{7pt}{0pt}{$3$}\selectfont}
	\end{pspicture}
	}} \! \! \!
	+ \! \! \! \vcenter{\hbox{
	\begin{pspicture}[showgrid=false](3.0,3.2)
		\psarc[linewidth=0.02,arrowsize=2pt 3,
		arrowinset=0]{<-}(1.5,1.6){0.6}{30}{65}
		\psarc[linewidth=0.02](1.5,1.6){0.6}{20}{39}
		\psarc[linewidth=0.02,arrowsize=2pt 3,
		arrowinset=0]{<-}(1.5,1.6){0.6}{125}{160}
		\psarc[linewidth=0.02](1.5,1.6){0.6}{110}{134}
		\psarc[linewidth=0.02,arrowsize=2pt 3,
		arrowinset=0]{<-}(1.5,1.6){0.6}{215}{250}
		\psarc[linewidth=0.02](1.5,1.6){0.6}{200}{224}
		\psarc[linewidth=0.03,linestyle=dashed,
		dash=2pt 1pt,linecolor=Red](1.5,1.6){0.6}{290}{340}
		\psline[linewidth=0.02,ArrowInside=->,
		ArrowInsidePos=0.55,arrowsize=2pt 3,
		arrowinset=0](2.3,1.6)(2.9,1.6)
		\psline[linewidth=0.03,linestyle=dashed,
		dash=2pt 1pt,linecolor=Red](0.1,1.6)(0.7,1.6)
		\psline[linewidth=0.02,ArrowInside=->,
		ArrowInsidePos=0.55,arrowsize=2pt 3,
		arrowinset=0](1.5,0.2)(1.5,0.8)
		\pscircle[linewidth=0.03,fillstyle=solid,
		fillcolor=lightgray](1.5,2.2){0.25}
		\psline[linewidth=0.03](1.5,2.2)(1.67,2.37)
		\psline[linewidth=0.03](1.5,2.2)(1.33,2.37)
		\psline[linewidth=0.03](1.5,2.2)(1.33,2.03)
		\psline[linewidth=0.03](1.5,2.2)(1.67,2.03)
		\pscircle[linewidth=0.03,fillstyle=solid,
		fillcolor=lightgray!25](0.9,1.6){0.20}
		\pscircle[linewidth=0.03,fillstyle=solid,
		fillcolor=lightgray!25](2.1,1.6){0.20}
		\pscircle[linewidth=0.03,fillstyle=solid,
		fillcolor=lightgray!25](1.5,1.0){0.20}
		\rput[b]{*0}(1.5,0.92){\scriptsize{$3$}}
		\rput[b]{*0}(2.12,2.05){\small $\Psi$}
		\rput[b]{*0}(0.88,2.05){\small $\Psi$}
		\rput[b]{*0}(0.4,1.3){\small $\pi$}
		\rput[b]{*0}(2.6,1.2){\small $\Psi$}
		\rput[b]{*0}(1.27,0.4){\small $\Psi$}
		\rput[t]{*0}(2.12,1.12){\small $\pi$}
		\rput[t]{*0}(0.88,1.12){\small $\Psi$}
		\rput[b]{*0}(0.9,1.52){\fontsize{7pt}{0pt}{$3$}\selectfont}
		\rput[b]{*0}(2.1,1.52){\fontsize{7pt}{0pt}{$3$}\selectfont}
	\end{pspicture}
	}} \! \! \!
	\Biggg) \gamma_{5} \tau^{a}\Biggg] , \quad \qquad \label{eq:yc1} \\[0.4cm]
	\partial_{k} y_{2,k} & = & \frac{i}{24} \mathcal{V}^{-1}
	\tr_{\gamma,\mathrm{flavor}} 
	\left[\left.\frac{\delta^{3}\partial_{k}\Gamma_{k}}
	{\rule[0.20cm]{0pt}{1ex}\delta\pi^{a}(0)\delta \bar{\Psi}_{\mathsf{p}_{1}}(0)
	\delta\Psi_{\mathsf{p}_{2}}(0)} \right|_{\mathsf{p}_{1}\, =\, \mathsf{p}_{2}\, =\, 2}
	\gamma_{5} \tau^{a}\right], \label{eq:yc2} \\[0.4cm]
	\partial_{k} Z_{k} & = & \mathcal{V}^{-1}
	\left.\frac{\mathrm{d}}{\mathrm{d}q^{2}}\right|_{q^{2}\, =\, 0}
	\frac{\delta^{2}\partial_{k}\Gamma_{k}}
	{\delta\pi^{1}(-q)\delta\pi^{1}(q)} \nonumber\\
    & = & \mathcal{V}^{-1} \! \left.\frac{\mathrm{d}}
    {\mathrm{d}q^{2}}\right|_{q^{2}\, =\, 0}\biggg(
	\frac{1}{2} \! \! \vcenter{\hbox{
	\begin{pspicture}(3.0,2.0)
		\psarc[linewidth=0.03,linestyle=dashed,
		dash=2pt 1pt,linecolor=Blue](1.5,1.0){0.6}{20}{65}
		\psarc[linewidth=0.03,linestyle=dashed,
		dash=2pt 1pt,linecolor=Blue](1.5,1.0){0.6}{115}{160}
		\psarc[linewidth=0.03,linestyle=dashed,
		dash=2pt 1pt,linecolor=Red](1.5,1.0){0.6}{200}{340}
		\psline[linewidth=0.03,linestyle=dashed,
		dash=2pt 1pt,linecolor=Red](0.1,1.0)(0.7,1.0)
		\psline[linewidth=0.03,linestyle=dashed,
		dash=2pt 1pt,linecolor=Red](2.3,1.0)(2.9,1.0)
		\pscircle[linewidth=0.03,fillstyle=solid,
		fillcolor=RoyalBlue](1.5,1.6){0.25}
		\psline[linewidth=0.03](1.5,1.6)(1.67,1.77)
		\psline[linewidth=0.03](1.5,1.6)(1.33,1.77)
		\psline[linewidth=0.03](1.5,1.6)(1.33,1.43)
		\psline[linewidth=0.03](1.5,1.6)(1.67,1.43)
		\pscircle[linewidth=0.03,fillstyle=solid,
		fillcolor=lightgray!25](0.9,1.0){0.20}
		\pscircle[linewidth=0.03,fillstyle=solid,
		fillcolor=lightgray!25](2.1,1.0){0.20}
		\rput[b]{*0}(2.12,1.45){\small $\sigma$}
		\rput[b]{*0}(0.88,1.45){\small $\sigma$}
		\rput[b]{*0}(0.4,0.7){\small $\pi$}
		\rput[b]{*0}(2.6,0.7){\small $\pi$}
		\rput[b]{*0}(1.5,0.1){\small $\pi$}
		\rput[b]{*0}(0.9,0.92){\fontsize{7pt}{0pt}{$3$}\selectfont}
		\rput[b]{*0}(2.1,0.92){\fontsize{7pt}{0pt}{$3$}\selectfont}
	\end{pspicture}
	}} \! \! \! 
	 + \frac{1}{2} \! \! \vcenter{\hbox{
	\begin{pspicture}(3.0,2.0)
		\psarc[linewidth=0.03,linestyle=dashed,
		dash=2pt 1pt,linecolor=Red](1.5,1.0){0.6}{20}{65}
		\psarc[linewidth=0.03,linestyle=dashed,
		dash=2pt 1pt,linecolor=Red](1.5,1.0){0.6}{115}{160}
		\psarc[linewidth=0.03,linestyle=dashed,
		dash=2pt 1pt,linecolor=Blue](1.5,1.0){0.6}{200}{340}
		\psline[linewidth=0.03,linestyle=dashed,
		dash=2pt 1pt,linecolor=Red](0.1,1.0)(0.7,1.0)
		\psline[linewidth=0.03,linestyle=dashed,
		dash=2pt 1pt,linecolor=Red](2.3,1.0)(2.9,1.0)
		\pscircle[linewidth=0.03,fillstyle=solid,
		fillcolor=RedOrange](1.5,1.6){0.25}
		\psline[linewidth=0.03](1.5,1.6)(1.67,1.77)
		\psline[linewidth=0.03](1.5,1.6)(1.33,1.77)
		\psline[linewidth=0.03](1.5,1.6)(1.33,1.43)
		\psline[linewidth=0.03](1.5,1.6)(1.67,1.43)
		\pscircle[linewidth=0.03,fillstyle=solid,
		fillcolor=lightgray!25](0.9,1.0){0.20}
		\pscircle[linewidth=0.03,fillstyle=solid,
		fillcolor=lightgray!25](2.1,1.0){0.20}
		\rput[b]{*0}(2.12,1.45){\small $\pi$}
		\rput[b]{*0}(0.88,1.45){\small $\pi$}
		\rput[b]{*0}(0.4,0.7){\small $\pi$}
		\rput[b]{*0}(2.6,0.7){\small $\pi$}
		\rput[b]{*0}(1.5,0.1){\small $\sigma$}
		\rput[b]{*0}(0.9,0.92){\fontsize{7pt}{0pt}{$3$}\selectfont}
		\rput[b]{*0}(2.1,0.92){\fontsize{7pt}{0pt}{$3$}\selectfont}
	\end{pspicture}
	}} \! \! \!
	 - \! \! \! \vcenter{\hbox{
	\begin{pspicture}(3.0,2.0)
		\psarc[linewidth=0.02,arrowsize=2pt 3,
		arrowinset=0]{->}(1.5,1.0){0.6}{20}{55}
		\psarc[linewidth=0.02](1.5,1.0){0.6}{46}{65}
		\psarc[linewidth=0.02,arrowsize=2pt 3,
		arrowinset=0]{->}(1.5,1.0){0.6}{115}{150}
		\psarc[linewidth=0.02](1.5,1.0){0.6}{141}{160}
		\psarc[linewidth=0.02,arrowsize=2pt 3,
		arrowinset=0]{->}(1.5,1.0){0.6}{200}{279}
		\psarc[linewidth=0.02](1.5,1.0){0.6}{270}{340}
		\psline[linewidth=0.03,linestyle=dashed,
		dash=2pt 1pt,linecolor=Red](0.1,1.0)(0.7,1.0)
		\psline[linewidth=0.03,linestyle=dashed,
		dash=2pt 1pt,linecolor=Red](2.3,1.0)(2.9,1.0)
		\pscircle[linewidth=0.03,fillstyle=solid,
		fillcolor=lightgray](1.5,1.6){0.25}
		\psline[linewidth=0.03](1.5,1.6)(1.67,1.77)
		\psline[linewidth=0.03](1.5,1.6)(1.33,1.77)
		\psline[linewidth=0.03](1.5,1.6)(1.33,1.43)
		\psline[linewidth=0.03](1.5,1.6)(1.67,1.43)
		\pscircle[linewidth=0.03,fillstyle=solid,
		fillcolor=lightgray!25](0.9,1.0){0.20}
		\pscircle[linewidth=0.03,fillstyle=solid,
		fillcolor=lightgray!25](2.1,1.0){0.20}
		\rput[b]{*0}(2.12,1.45){\small $\Psi$}
		\rput[b]{*0}(0.88,1.45){\small $\Psi$}
		\rput[b]{*0}(0.4,0.7){\small $\pi$}
		\rput[b]{*0}(2.6,0.7){\small $\pi$}
		\rput[b]{*0}(1.5,0.0){\small $\Psi$}
		\rput[b]{*0}(0.9,0.92){\fontsize{7pt}{0pt}{$3$}\selectfont}
		\rput[b]{*0}(2.1,0.92){\fontsize{7pt}{0pt}{$3$}\selectfont}
	\end{pspicture}
	}} \nonumber\\
	& & \qquad \qquad \qquad \quad \ -\, \frac{1}{2} \! \!
	\vcenter{\hbox{
	\begin{pspicture}[showgrid=false](2.0,2.0)
		\psarc[linewidth=0.03,linestyle=dashed,
		dash=2pt 1pt,linecolor=Blue](1.0,1.0){0.6}{290}{65}
		\psarc[linewidth=0.03,linestyle=dashed,
		dash=2pt 1pt,linecolor=Blue](1.0,1.0){0.6}{115}{250}
		\psline[linewidth=0.03,linestyle=dashed,
		dash=2pt 1pt,linecolor=Red](0.35,0.1)(0.818408,0.316188)
		\psline[linewidth=0.03,linestyle=dashed,
		dash=2pt 1pt,linecolor=Red](1.181592,0.316188)(1.65,0.1)
		\pscircle[linewidth=0.03,fillstyle=solid,
		fillcolor=RoyalBlue](1.0,1.6){0.25}
		\psline[linewidth=0.03](1.0,1.6)(1.17,1.77)
		\psline[linewidth=0.03](1.0,1.6)(0.83,1.77)
		\psline[linewidth=0.03](1.0,1.6)(0.83,1.43)
		\psline[linewidth=0.03](1.0,1.6)(1.17,1.43)
		\pscircle[linewidth=0.03,fillstyle=solid,
		fillcolor=lightgray!25](1.0,0.4){0.20}
		\rput[b]{*0}(0.99,0.32){\fontsize{7pt}{0pt}{$4$}\selectfont}
		\rput[b]{*0}(0.2,0.9){\small $\sigma$}
		\rput[b]{*0}(1.8,0.9){\small $\sigma$}
		\rput[b]{*0}(0.15,0.0){\small $\pi$}
		\rput[b]{*0}(1.85,0.0){\small $\pi$}
	\end{pspicture}
	}} \! \! \!
	- \frac{1}{2} \! \!
	\vcenter{\hbox{
	\begin{pspicture}[showgrid=false](2.0,2.0)
		\psarc[linewidth=0.03,linestyle=dashed,
		dash=2pt 1pt,linecolor=Red](1.0,1.0){0.6}{290}{65}
		\psarc[linewidth=0.03,linestyle=dashed,
		dash=2pt 1pt,linecolor=Red](1.0,1.0){0.6}{115}{250}
		\psline[linewidth=0.03,linestyle=dashed,
		dash=2pt 1pt,linecolor=Red](0.35,0.1)(0.818408,0.316188)
		\psline[linewidth=0.03,linestyle=dashed,
		dash=2pt 1pt,linecolor=Red](1.181592,0.316188)(1.65,0.1)
		\pscircle[linewidth=0.03,fillstyle=solid,
		fillcolor=RedOrange](1.0,1.6){0.25}
		\psline[linewidth=0.03](1.0,1.6)(1.17,1.77)
		\psline[linewidth=0.03](1.0,1.6)(0.83,1.77)
		\psline[linewidth=0.03](1.0,1.6)(0.83,1.43)
		\psline[linewidth=0.03](1.0,1.6)(1.17,1.43)
		\pscircle[linewidth=0.03,fillstyle=solid,
		fillcolor=lightgray!25](1.0,0.4){0.20}
		\rput[b]{*0}(0.99,0.32){\fontsize{7pt}{0pt}{$4$}\selectfont}
		\rput[b]{*0}(0.2,0.9){\small $\pi$}
		\rput[b]{*0}(1.8,0.9){\small $\pi$}
		\rput[b]{*0}(0.15,0.0){\small $\pi$}
		\rput[b]{*0}(1.85,0.0){\small $\pi$}
	\end{pspicture}
	}} \, \, \biggg) \! ,\label{eq:zp} \\[0.4cm]
	\partial_{k}Z_{k}^{\psi} & = & - \frac{i}{4} \mathcal{V}^{-1}
	\left.\frac{\mathrm{d}}{\mathrm{d}q^2}\right|_{q^{2}\, =\, 0}
	\tr_{\gamma} \left[\left.\frac{\delta^{2}\partial_{k}\Gamma_{k}}
	{\rule[0.20cm]{0pt}{1ex}\delta \bar{\Psi}_{f_{1},\mathsf{p}_{1}}(q)
	\delta\Psi_{f_{2},\mathsf{p}_{2}}(q)} 
	\right|_{f_{1}\, =\, f_{2}\, =\, 1,\, \mathsf{p}_{1}\, =\, 
	\mathsf{p}_{2}\, =\, 1}\gamma_{\mu}q_{\mu}\right] . \label{eq:zf}
\end{IEEEeqnarray}
In the graphical translation, we left out particular leg specifications and intrinsic multiplicity/symmetry 
or sign factors. For the flow equations (\ref{eq:yc2}) and (\ref{eq:zf}), we have the same Feynman diagrams 
as in Eqs.\ (\ref{eq:yc1}) and (\ref{eq:m0}), respectively. The object $\Psi$ represents the vector 
$\smash{(\psi_{1},\psi_{2})^{\mathsf{T}}}$ in Eq.\ (\ref{eq:rotation}) of Appendix \ref{sec:modelconstruction}; 
it carries the $\mathsf{SU}(2)$ flavor indices $f_{1}$, $f_{2}$ and the $\mathsf{SO}(2)$ parity (or 
rotation) indices $\mathsf{p}_{1}$, $\mathsf{p}_{2}$ (in addition to its spin indices, which are not shown). 
The different fields of the nucleon-meson model are ``highlighted'' in red (pions), blue ($\sigma$-field), 
and black/gray (fermions). Finally, the spacetime volume is written as $\mathcal{V}$. The regulator 
insertions, which are colored equally as the propagator lines, are proportional to the wave-function 
renormalizations,
\begin{IEEEeqnarray}{rCl}
	& R_{k}^{\pi}\!\left(q^{2}\right) \equiv R_{k}^{\sigma}\!\left(q^{2}\right)
	= Z_{k}\!\; q^{2}\!\; r_{k}\!\left(q^{2}\right), & \\[0.2cm]
	& R_{k}^{\psi}\!\left(q^{2}\right) 
	= i Z_{k}^{\psi}\!\; \gamma_{\mu}q_{\mu} 
	r_{k}^{\psi} \!\left(q^{2}\right), \qquad
	R_{k,f_{1}f_{2},\mathsf{p}_{1}\mathsf{p}_{2}}^{\Psi}\!\left(q^{2}\right) 
	= R_{k}^{\psi}\!\left(q^{2}\right)
	\delta_{f_{1}f_{2}}\delta_{\mathsf{p}_{1}\mathsf{p}_{2}} , &
\end{IEEEeqnarray}
where we take the optimized Litim regulators as the shape functions \cite{Litim:2001up},
\begin{IEEEeqnarray}{rCl}
	r_{k}\!\left(q^{2}\right) & \equiv & r \!\left(\frac{q^{2}}{k^{2}}\right) 
	= \left(\frac{k^{2}}{q^{2}} - 1\right)
	\Theta\!\left(k^{2} - q^{2}\right), \\[0.25cm]
	r_{k}^{\psi}\!\left(q^{2}\right) & \equiv & 
	r^{\psi}\!\left(\frac{q^{2}}{k^{2}}\right) = 
	\left(\sqrt{\frac{k^{2}}{q^{2}}} - 1\right)
	\Theta\!\left(k^{2} - q^{2}\right).
\end{IEEEeqnarray}
The symbol $\Theta(\,\boldsymbol{\cdot}\,)$ denotes the Heaviside step function. The regulators obey
the limits
\begin{equation}
	\lim_{k\, \rightarrow\, 0} R_{k} = 0, \qquad
	\lim_{k\, \rightarrow\, \Lambda\, \rightarrow\, \infty} R_{k} = \infty , \qquad
	\lim_{q\, \rightarrow\, 0} R_{k} > 0, \qquad
	\lim_{q\, \rightarrow\, \infty} R_{k} = 0.
\end{equation}
In the first limit, one recovers the (quantum) effective action $\Gamma$.\ The second leads
to $\Gamma_{k} \rightarrow S$ (for $k \rightarrow \Lambda \rightarrow \infty$).\ The last
two limits imply the massive suppression of soft modes only during the integration process.\ 
To calculate the diagrams, we used the dedicated algebra tools \texttt{DoFun} \cite{Huber:2011qr, 
*Huber:2019dkb}, \texttt{FormTracer} \cite{Cyrol:2016zqb}, and \texttt{FeynCalc} \cite{Mertig:1990an, 
*Shtabovenko:2016sxi, *Shtabovenko:2020gxv}.
\begin{table}
	\caption{\label{tab:UV}Initialization parameters for the integration
	of momentum modes.}
	\begin{ruledtabular}
		\begin{tabular}{lccccccccc}
		& $\varphi_{0}$ & $\alpha_{1}$ & & $m_{0}$ & & & & & $h$ \\[0.075cm]
		Approximation & $\left[\mathrm{MeV}\right]$ & $\left[\mathrm{MeV}^{2}\right]$ &
		$\alpha_{2}$ & $\left[\mathrm{MeV}\right]$ & $y_{1}$ & $y_{2}$ & 
		$Z$ & $Z^{\psi}$ & $\left[\mathrm{MeV}^{3}\right]$
		\\[0.075cm]
		\colrule\\[-0.3cm]
		FRG $\left(\mathrm{LPA}^{\prime}\right)$ & $60.74$ & $510^{2}$ & $58.2$ & $770$ 
		& $12.0$ & $19.4$ & $1.6$ & $1.4$ & $2.62 \times 10^{6}$\\
		FRG (LPA) & $91.92$ & $330^{2}$ & $10.5$ & $824.45$ 
		& $6.765$ & $13.06$ & $1.0$ & $1.0$ & $1.80 \times 10^{6}$\\
		One-loop & $91.57$ & $460^{2}$ & $21.3$ & $824.45$ 
		& $6.765$ & $13.06$ & $1.0$ & $1.0$ & $1.80 \times 10^{6}$\\
		MF & $91.80$ & $598^{2}$ & $37.5$ & $824.45$ 
		& $6.765$ & $13.06$ & $1.0$ & $1.0$ & $1.80 \times 10^{6}$
		\end{tabular}
	\end{ruledtabular}
\end{table}

The fermion propagator has off-diagonal elements in the basis of $\Psi$. The two-point function
(plus regulator) on the right-hand side of Eq.\ (\ref{eq:Wetterich}) has in momentum space the 
structure
\begin{equation}
	\mathcal{V}^{-1} \Gamma_{k}^{(2)\;\!\bar{\Psi}\Psi} 
	+ R_{k}^{\Psi} \sim
	\begin{pmatrix}
	i Z^{\psi} \gamma_{\mu} q_{\mu} 
	\left[1 + r_{k}^{\psi}\!\left(q^2\right)\right]
	- y_{1}\sigma & -m_{0} \\
	-m_{0} & iZ^{\psi} \gamma_{\mu} q_{\mu}
	\left[1 + r_{k}^{\psi}\!\left(q^2\right)\right]
	+ y_{2}\sigma 
	\end{pmatrix} .
\end{equation}
Its inverse, i.e., the fermionic propagator tensor $G_{k}^{\Psi\bar{\Psi}}$, is given by
\begin{IEEEeqnarray}{rCl}
	G_{k,d_{1}d_{2},f_{1}f_{2},\mathsf{p}_{1}\mathsf{p}_{2}}^{\Psi\bar{\Psi}}(q,q')
	& = - \bigg\lbrace & \left[f_{k}^{++}\!\left(q^{2}\right) \delta_{d_{1}d_{2}}
	+ ig_{k,\mu}^{++}\!\left(q\right) \left(\gamma_{\mu}\right)_{d_{1}d_{2}}\right]
	\delta_{\mathsf{p}_{1}1} \delta_{\mathsf{p}_{2}1} \nonumber\\
	& & \left[f_{k}^{+-}\!\left(q^{2}\right) \delta_{d_{1}d_{2}}
	+ ig_{k,\mu}^{+-}\!\left(q\right) \left(\gamma_{\mu}\right)_{d_{1}d_{2}}\right]
	\delta_{\mathsf{p}_{1}1} \delta_{\mathsf{p}_{2}2} \nonumber\\[0.1cm]
	& & \left[f_{k}^{-+}\!\left(q^{2}\right) \delta_{d_{1}d_{2}}
	+ ig_{k,\mu}^{-+}\!\left(q\right) \left(\gamma_{\mu}\right)_{d_{1}d_{2}}\right]
	\delta_{\mathsf{p}_{1}2} \delta_{\mathsf{p}_{2}1} \nonumber\\
	& & \left[f_{k}^{--}\!\left(q^{2}\right) \delta_{d_{1}d_{2}}
	+ ig_{k,\mu}^{--}\!\left(q\right) \left(\gamma_{\mu}\right)_{d_{1}d_{2}}\right]
	\delta_{\mathsf{p}_{1}2} \delta_{\mathsf{p}_{2}2}  
	\bigg\rbrace\, \delta_{f_{1}f_{2}}\delta(q - q') .
\end{IEEEeqnarray}
The indices $d_{1}$, $d_{2}$ are those of the Dirac (spin) space. The functions $f_{k}$ (even
in the momentum $q$) and $g_{k,\mu}$ (odd in the momentum) are deduced as
\begin{IEEEeqnarray}{rCl}
	f_{k}^{\pm\pm}\!\left(q^{2}\right) & = & \pm\, \mathcal{N}_{k}\!\left(q^{2}\right)
	\left[(Z^{\psi})^{2} q^{2} (1 + r_{k}^{\psi})^{2}
	y_{1/2}\sigma + m^{+}m^{-} y_{2/1} \sigma\right], \\[0.2cm]
	f_{k}^{\pm\mp}\!\left(q^{2}\right) & = & \mathcal{N}_{k}\!\left(q^{2}\right)
	\left[(Z^{\psi})^{2} q^{2} (1 + r_{k}^{\psi})^{2}
	+ m^{+}m^{-}\right] m_{0}, \\[0.2cm]
	g_{k,\mu}^{\pm\pm}\!\left(q\right) & = & \mathcal{N}_{k}\!\left(q^{2}\right)
	Z^{\psi} q_{\mu} (1 + r_{k}^{\psi})
	\left[(Z^{\psi})^{2} q^{2} (1 + r_{k}^{\psi})^{2}
	+ (m^{+})^{2} + (m^{-})^{2} - m_{0}^{2}
	- \left(y_{1/2}\sigma\right)^{2}\right] , \\[0.2cm]
	g_{k,\mu}^{\pm\mp}\!\left(q\right) & = & -\, \mathcal{N}_{k}\!\left(q^{2}\right)
	Z^{\psi} q_{\mu} (1 + r_{k}^{\psi})
	(m^{+} - m^{-}) \!\; m_{0} ,
\end{IEEEeqnarray}
with the normalization factor 
\begin{equation}
	\mathcal{N}_{k}\!\left(q^{2}\right) = \left\lbrace \left[(Z^{\psi})^{2} 
	q^{2} (1 + r_{k}^{\psi})^2 + (m^{+})^{2} \right] \left[(Z^{\psi})^{2} 
	q^{2}(1 + r_{k}^{\psi})^2 + (m^{-})^{2} \right] \right\rbrace^{-1} .
\end{equation}
For $m_{0} = 0$, the off-diagonal elements vanish, $f_{k}^{\pm\mp} = g_{k,\mu}^{\pm\mp}
= 0$, and the diagonal elements become
\begin{IEEEeqnarray}{rCl}
	f_{k}^{\pm\pm}\!\left(q^{2}\right) = \pm 
	\frac{y_{1/2}\sigma}{(Z^{\psi})^{2} q^{2}
	(1 + r_{k}^{\psi})^{2} + (y_{1/2}\sigma)^{2}} , \qquad
	g_{k,\mu}^{\pm\pm}(q) = \frac{Z^{\psi}q_{\mu}(1 + r_{k}^{\psi})}
	{(Z^{\psi})^{2} q^{2}\bigl(1 + r_{k}^{\psi})^{2} + (y_{1/2}\sigma)^{2}},
\end{IEEEeqnarray}
which are the well-known propagator functions of quark-meson-type models with the fermion 
masses $y_{1/2}\sigma$ \cite{Eser:2018jqo, Divotgey:2019xea}. As an example, the fermionic 
loop in the flow equation (\ref{eq:u}) of the effective potential is calculated as
\vspace{-0.3cm}
\begin{IEEEeqnarray}{rCl}
	- \!
	\vcenter{\hbox{
	\begin{pspicture}(1.5,2.4)
		\psarc[linewidth=0.02,arrowsize=2pt 3,
		arrowinset=0]{->}(0.75,1.2){0.6}{115}{189}
		\psarc[linewidth=0.02](0.75,1.2){0.6}{180}{65}
		\pscircle[linewidth=0.03,fillstyle=solid,
		fillcolor=lightgray](0.75,1.8){0.25}
		\psline[linewidth=0.03](0.75,1.8)(0.92,1.97)
		\psline[linewidth=0.03](0.75,1.8)(0.58,1.97)
		\psline[linewidth=0.03](0.75,1.8)(0.58,1.63)
		\psline[linewidth=0.03](0.75,1.8)(0.92,1.63)
		\rput[b]{*0}(0.75,0.25){\small $\Psi$}
	\end{pspicture}
	}} \! \!  
	& = & - 8 \mathcal{V} \int_{q} \left[g_{k,\mu}^{++}(q) h_{k,\mu}(q)
	+ g_{k,\mu}^{--}(q) h_{k,\mu}(q) \right] , \\[-0.2cm]
	h_{k,\mu}(q) & = & Z^{\psi} q_{\mu} 
	\left[\partial_{k} r_{k}^{\psi}\!\left(q^{2}\right)
	- \frac{\eta^{\psi}}{k}\;\! r_{k}^{\psi}\!\left(q^{2}\right)\right], 
	\label{eq:h} \\[0.2cm]
	\Rightarrow \qquad \left.\partial_{k}V_{k}\right|_{\mathrm{fermions}} 
	& = & - \left(5 - \eta^{\psi}\right) \frac{k^{5}}{20\pi^{2}}
	\left[\frac{1}{k^{2} + (M^{+})^{2}} + 
	\frac{1}{k^{2} + (M^{-})^{2}}\right] . \label{eq:ufermion}
\end{IEEEeqnarray}
As expected, the trace operation of the loop extracts the diagonal propagator parts, 
i.e., the physical eigenvalues of the fermions with the (squared) particle energies 
$\smash{k^{2} + (M^{\pm})^{2}}$ and the physical masses $\smash{M^{\pm}}$. In Eq.\ (\ref{eq:h}),
we introduced the anomalous dimension $\smash{\eta^{\psi}}$. The bosonic and fermionic
anomalous dimensions $\smash{\eta}$ and $\smash{\eta^{\psi}}$ parametrize the scaling 
of the wave-function renormalization factors,
\begin{equation}
	\eta_{k} = - k\!\; \partial_{k}\ln Z_{k}, \qquad
	\eta_{k}^{\psi} = - k\!\; \partial_{k}\ln Z_{k}^{\psi}. \label{eq:anomalous}
\end{equation}
The flow equations in terms of renormalized fields and couplings solely depend on the 
anomalous dimensions, see e.g.\ Eq.\ (\ref{eq:ufermion}), such that the absolute values 
of the factors $Z$ and $Z^{\psi}$ actually do not explicitly enter the physical solution 
of the equations.

The numeric solution of the differential equations for the Taylor coefficients $\alpha_{n}$,
$n = 1, \ldots , N_{\alpha}$, is already (indirectly) plotted in Fig.\ \ref{fig:masses} of
Sec.\ \ref{sec:results}, since the masses (\ref{eq:renormmasses}) can be expressed through 
the coefficients as follows:
\begin{IEEEeqnarray}{rCl}
	M_{\pi}^{2} & = & 2\tilde{V}'\!\left(f_{\pi}^{2}\right)
	\equiv 2 \sum_{n\, =\, 1}^{N_{\alpha}} \frac{\tilde{\alpha}_{n}}{(n - 1)!} 
	\left(f_{\pi}^{2} - \tilde{\varphi}_{0}^{2}\right)^{n -  1} , \\[0.2cm]
	M_{\sigma}^{2} & = & 2\tilde{V}'\!\left(f_{\pi}^{2}\right) 
	+ 4 f_{\pi}^{2} \tilde{V}''\!\left(f_{\pi}^{2}\right)
	\equiv 2 \left[\sum_{n\, =\, 1}^{N_{\alpha} - 1} \frac{\tilde{\alpha}_{n}
	+ 2 f_{\pi}^{2} \tilde{\alpha}_{n+1}}{(n - 1)!} 
	\left(f_{\pi}^{2} - \tilde{\varphi}_{0}^{2}\right)^{n -  1}
	+ \frac{\tilde{\alpha}_{N_{\alpha}}}{(N_{\alpha} - 1)!} \left(f_{\pi}^{2}
	- \tilde{\varphi}_{0}^{2}\right)^{N_{\alpha} -  1} \right] , \qquad
\end{IEEEeqnarray}
where we utilized the renormalized coefficients $\tilde{\alpha}_{n} = \alpha_{n}/Z^{n}$
and the renormalized expansion point $\tilde{\varphi}_{0}^{2} = Z \varphi_{0}^{2}$. Instead 
of deriving the masses from the scale-dependent minimum $\tilde{\sigma}_{0}$ of the effective 
potential, we alternatively compute their flow equations,
\begin{IEEEeqnarray}{rCl}
	\partial_{k} \tilde{\sigma}_{0} & = & - \tilde{\sigma}_{0} \left[\frac{1}{2} 
	\frac{\eta}{k} + \frac{2}{M_{\sigma}^{2}} \left.
	\frac{\partial}{\partial\tilde{\varphi}^{2}}\right|_{\tilde{\sigma}_{0}^{2}}\partial_{k} 
	\tilde{V}_{k} \!\left(\tilde{\varphi}^{2}\right) \right] , \\[0.2cm]
	\partial_{k} M_{\pi} & = & M_{\pi} \left[
	\frac{1}{2} \frac{\eta}{k} + \frac{1}{M_{\sigma}^{2}} 
	\left. \frac{\partial}{\partial\tilde{\varphi}^{2}}\right|_{\tilde{\sigma}_{0}^{2}}
	\partial_{k} \tilde{V}_{k} \!\left(\tilde{\varphi}^{2}\right) \right], \\[0.2cm]
	\partial_{k} M_{\sigma} & = & M_{\sigma} \left\lbrace \frac{1}{2} 
	\frac{\eta}{k} + \frac{1}{M_{\sigma}^{2}}\left[ \frac{M_{\pi}^{2} 
	- 8 \tilde{\sigma}_{0}^{2} \tilde{V}''(\tilde{\sigma}_{0}^{2}) - 8 \tilde{\sigma}_{0}^{4} 
	\tilde{V}^{(3)}(\tilde{\sigma}_{0}^{2})}{M_{\sigma}^{2}}
	\left.\frac{\partial}{\partial\tilde{\varphi}^{2}}\right|_{\tilde{\sigma}_{0}^{2}}
	+ 2 \tilde{\sigma}_{0}^{2} \left.\frac{\partial^{2}}
	{(\partial\tilde{\varphi}^{2})^{2}} \right|_{\tilde{\sigma}_{0}^{2}} \right]
	\partial_{k} \tilde{V}_{k} \!\left(\tilde{\varphi}^{2}\right)\right\rbrace ,
	\qquad \ \ \label{eq:flowmsigma}
\end{IEEEeqnarray}
and evaluate them on the solution for the Taylor coefficients. In the UV, i.e., $\tilde{\sigma}_{0} 
\rightarrow 0$ and $M_{\sigma} \rightarrow M_{\pi}$, the flow equation for $M_{\sigma}$ 
reduces to
\begin{equation}
	\partial_{k} M_{\sigma} = M_{\pi} 
	\left[\frac{1}{2} \frac{\eta}{k} + \frac{1}{M_{\pi}^{2}} 
	\left. \frac{\partial}{\partial\tilde{\varphi}^{2}}\right|_{\tilde{\sigma}_{0}^{2}}
	\partial_{k} \tilde{V}_{k} \!\left(\tilde{\varphi}^{2}\right) \right]
	\equiv \partial_{k} M_{\pi} . \label{eq:asymptotic}
\end{equation}
\begin{figure*}
	\centering
	\includegraphics[scale=1.0]{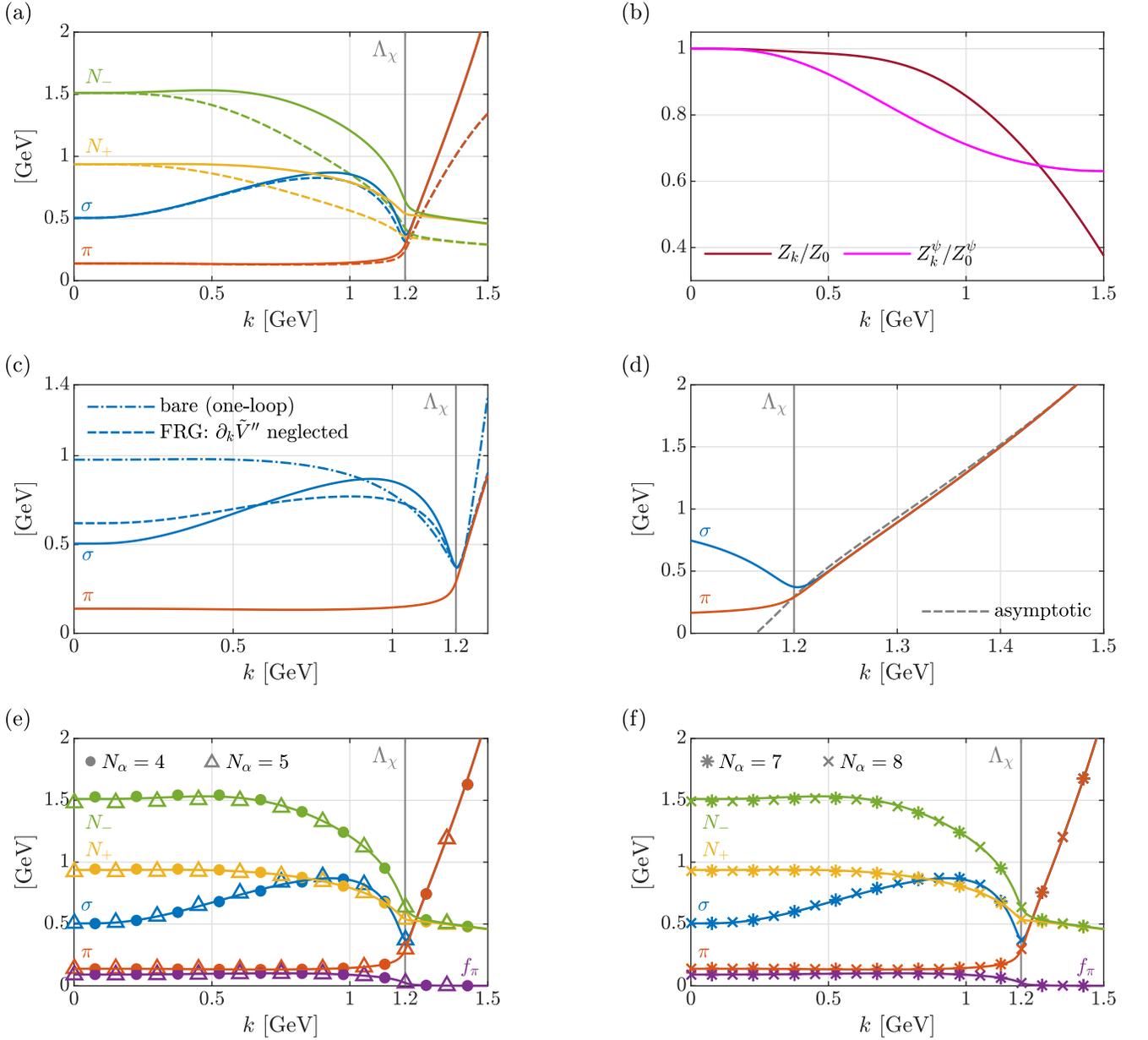}
	\caption{Scale-dependent (renormalized and bare) masses, scale evolution of the 
	wave-function renormalization, and numeric analysis of the $\smash{\mathrm{LPA}^{\prime}}$ 
	flow. (a) Boson and fermion masses [solid lines; identical to Fig.\ \ref{fig:masses}(a)] 
	in comparison to the bare quantities (dashed lines). (b) Scale evolution of the
	factors $Z$ and $\smash{Z^{\psi}}$ (normalized to one in the IR). (c) Numeric 
	analysis of the flow of the $\sigma$-mass (detailed explanations are given in the main 
	text). (d) Asymptotic behavior of the boson masses for large $k$-scales.
	(e) and (f) Variations of the $N_{\alpha}$-parameter of the effective potential
	(\ref{eq:potential}) and the effect on the numeric results (solid lines:
	$N_{\alpha} = 6$).}
	\label{fig:LPAprime}
\end{figure*}

Since the above equations are in their renormalization-group-invariant form, meaning that 
they neither explicitly depend on $Z$ nor $\smash{Z^{\psi}}$ [as discussed below Eq.\ 
(\ref{eq:anomalous})], one may rescale the values of the wave-function renormalizations 
in the definition of the bare boson and fermion masses,
\begin{equation}
	m_{\pi/\sigma}^{\mathrm{bare}} \coloneqq \sqrt{\frac{Z_{k}}{Z_{k_{\mathrm{IR}}}}} 
	\;\! M_{\pi/\sigma} \equiv \frac{m_{\pi/\sigma}}{Z_{0}^{1/2}},
	\qquad m^{\pm,\mathrm{bare}} \coloneqq 
	\frac{Z_{k}^{\psi}}{Z^{\psi}_{k_{\mathrm{IR}}}}
	\;\! M^{\pm} \equiv \frac{m^{\pm}}{Z^{\psi}_{0}}.
	\label{eq:bare}
\end{equation}
This leads (in the $\mathrm{LPA}^{\prime}$) to the coincidence of bare and renormalized 
masses at $k = 0$, cf.\ Fig.\ \ref{fig:LPAprime}(a), which combines the curves 
from Fig.\ \ref{fig:masses}(a) (solid lines) with the bare ones (dashed lines in the respective
colors). It reveals that the bare fermion masses are notably smaller over the almost entire
$k$-range, while the bosonic masses mainly differ for scales above $\Lambda_{\chi}$. For
the sake of completeness, Fig.\ \ref{fig:LPAprime}(b) shows the (normalized) flows of the
wave-function renormalization factors $\smash{Z}$ and $\smash{Z^{\psi}}$. Interestingly,
the flow of the fermionic wave-function renormalization $\smash{Z^{\psi}}$ flattens out 
in the UV, which is not observed for its bosonic counterpart $\smash{Z}$; the scaling 
behavior of the renormalized and bare fermion masses is nearly ``parallel'' above the 
chiral-symmetry breaking scale, see once more Fig.\ \ref{fig:LPAprime}(a).
\begin{figure*}
	\centering
	\includegraphics[scale=1.0]{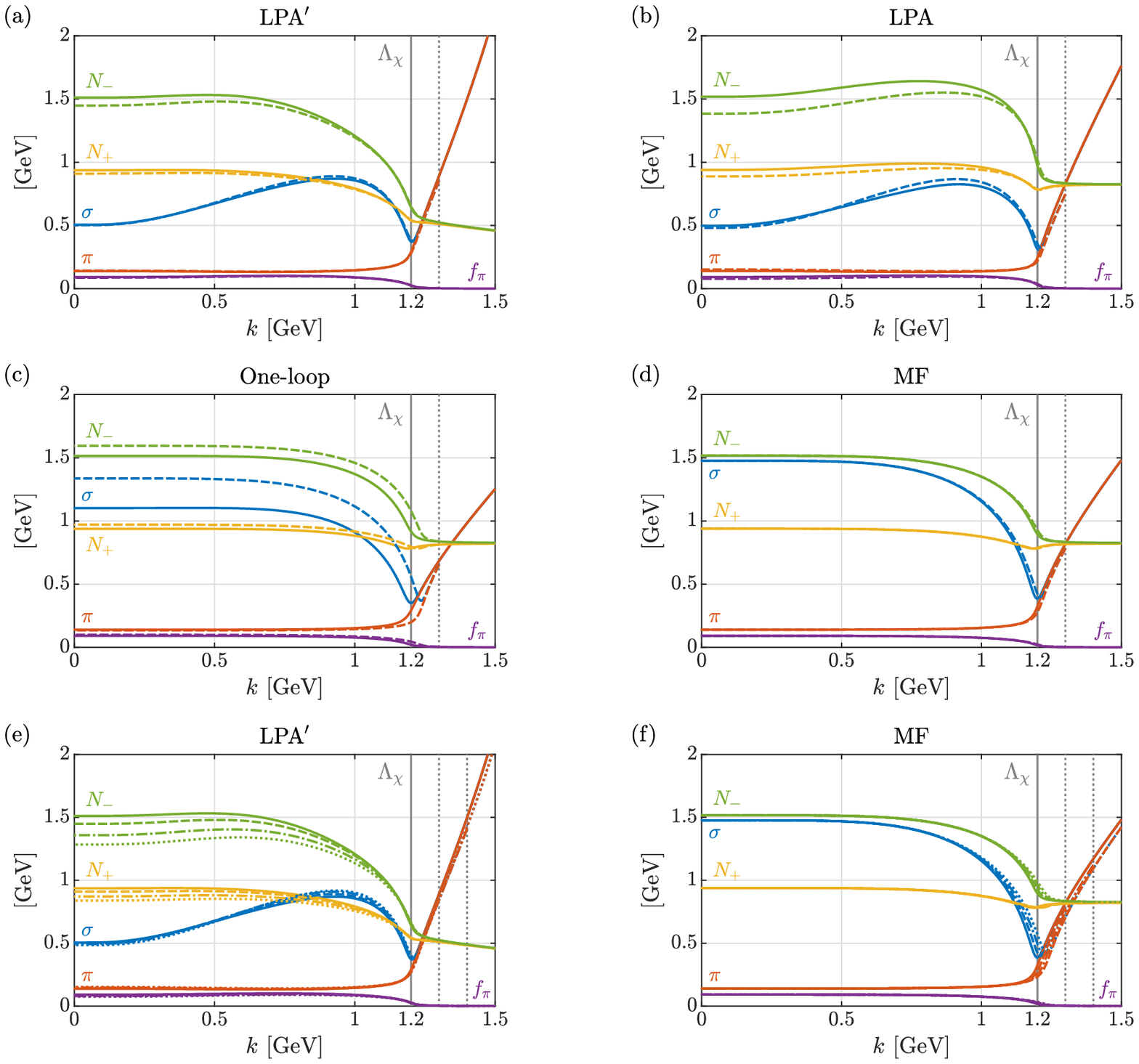}
	\caption{Influence of higher Taylor coefficients ($\alpha_{n}$, $n > 2$). 
	(a) to (d) Shift of the initialization scale $\Lambda_{\chi} = 1.2\ \mathrm{GeV}$
	to $1.3\ \mathrm{GeV}$ (vertical gray-dotted line) in all approximations
	($\mathrm{LPA}^{\prime}$, LPA, one-loop, and MF). The integration towards the IR 
	starting from the cutoff $1.3\ \mathrm{GeV}$ (dashed lines) is initialized as a cubic 
	polynomial in $\varphi^{2}$, cf.\ Eq.\ (\ref{eq:potential}). This is compared to the 
	corresponding original flows (solid lines; quadratic polynomials tuned at $\Lambda_{\chi} 
	= 1.2\ \mathrm{GeV}$). The respective initial values of the coefficients and couplings 
	at $1.3\ \mathrm{GeV}$ are obtained by integrating the original flow ``upwards'' from 
	$\Lambda_{\chi}$ to $1.3\ \mathrm{GeV}$ and then dropping the Taylor coefficients 
	$\alpha_{n\, >\, 3}$. (e) Initialization of the $\mathrm{LPA}^{\prime}$-flow at
	$1.3\ \mathrm{GeV}$, $1.4\ \mathrm{GeV}$, and $1.5\ \mathrm{GeV}$ (dashed, dash-dotted,
	and dotted lines, respectively). The initial values of the Taylor coefficients and 
	couplings are determined as in panel (a). (f) Analogous figure to panel (e) for the 
	MF approximation, demonstrating the cutoff independence of the IR observables.}
	\label{fig:taylor}
\end{figure*}

In Fig.\ \ref{fig:LPAprime}(c), we analyzed the characteristic ``down-bending'' of the $\sigma$-mass
in the FRG truncations (LPA and $\mathrm{LPA}^{\prime}$) and isolated its triggers. The dash-dotted
line, which we named ``bare,'' is the scale evolution of the $\sigma$-mass for taking the bare two-point 
function $\smash{S^{(2)}}$ as input for the flow equations---as it is done in the MF and one-loop 
approximations. The absence of interactions beyond quartic order in $\smash{S^{(2)}}$ leads to 
the rather monotonous behavior below the scale $\Lambda_{\chi}$, where all lines are fixed. Secondly, 
the dashed line stands for the FRG flow of $M_{\sigma}$, which obeys Eq.\ (\ref{eq:flowmsigma}), 
but where we omitted the last contribution proportional to $\smash{\partial_{k}\tilde{V}''}$.\ The 
resulting curve is ``damped'' as compared to the total flow (solid line) and the IR mass shrinks 
to smaller values. The appearing non-monotonicity then gets amplified by the inclusion of the flow 
of $\tilde{V}''$, which reinforces the influence of the higher Taylor coefficients. The IR mass 
$M_{\sigma}$, in summary, decreases by a factor of almost $2$. The asymptotic scale evolution 
of the bosonic masses in the UV, following Eq.\ (\ref{eq:asymptotic}), is verified in Fig.\ \ref{fig:LPAprime}(d).

Against variations of the $N_{\alpha}$-parameter of the effective potential (\ref{eq:potential}), the 
numeric masses and the pion decay constant are relatively robust in the $\mathrm{LPA}^{\prime}$, 
i.e., the flows and the physical IR values are not substantially affected; see Figs.\ \ref{fig:LPAprime}(e) 
and \ref{fig:LPAprime}(f) for changes of $N_{\alpha}$ to smaller ($N_{\alpha} = 4,5$) and larger integers 
($N_{\alpha} = 7,8$), respectively. The solid lines in both subfigures indicate the flows of Figs.\ 
\ref{fig:masses}(a) and \ref{fig:masses}(b) ($N_{\alpha} = 6$). Clearly, the effect of increasing 
$N_{\alpha}$, which amounts to less than $2\%$ in the (most relevant) quantities $f_{\pi}$, $M_{\pi}$, 
and $M_{\sigma}$, can be smoothed out by slightly readjusting the UV parameters. Thus choosing 
$N_{\alpha} = 6$ [as also argued by Refs.\ \cite{Divotgey:2019xea, Cichutek:2020bli, Pawlowski:2014zaa}] 
does not a priori distort the main conclusion of our work. In particular, the crucial value of 
$M_{\sigma}$ changes by less than $0.5\%$ for smaller or larger values of $N_{\alpha}$.

To numerically support the ansatz for the effective potential as a quadratic polynomial (in $\varphi^{2}$) 
at $\Lambda_{\chi} = 1.2\ \mathrm{GeV}$, we estimate the influence of the higher Taylor coefficients
($\alpha_{n}$, $n > 2$) that could be present at $\Lambda_{\chi}$ by increasing the initialization
scale to $1.3\ \mathrm{GeV}$. This is illustrated in Figs.\ \ref{fig:taylor}(a) to \ref{fig:taylor}(d)
(see the vertical gray-dotted line as compared to the gray line $\Lambda_{\chi}$); the dashed curves 
represent the integration of the flow equations starting with a cubic potential ($\alpha_{n} > 0$, 
$n \le 3$, and $\alpha_{n} = 0$, $n > 3$) initialized at $1.3\ \mathrm{GeV}$ in the chiral-restored
phase, such that all higher coefficients already acquire nonzero values at $\Lambda_{\chi}$.\ The 
respective initial values of the coefficients and couplings are obtained by integrating the original
quadratic potential ``upwards'' from $\Lambda_{\chi}$ to $1.3\ \mathrm{GeV}$ and then dropping
the coefficients $\alpha_{n\, >\, 3}$. Let us remark that it was necessary to involve (at least)
the first three Taylor coefficients in this upwards-integrating procedure in order to find stable 
solutions for the subsequent ``downwards'' integration starting from $1.3\ \mathrm{GeV}$ (which 
was e.g.\ not possible for a similar procedure solely involving the first two coefficients).
Hence, the dashed curves can be interpreted as alternative UV tuning of the flows in the 
chiral-restored phase above $\Lambda_{\chi}$. It obviously turns out that the overall picture
does not change drastically, especially, the boson masses and the pion decay constant appear 
to be only mildly affected in the $\mathrm{LPA}^{\prime}$, $\mathrm{LPA}$, and the MF approximation
(which is the most important criterion regarding the computation of the pion-pion scattering lengths 
and the $\sigma$-mass prediction). In the MF case, the IR values of the observables are actually not 
affected at all. Only the one-loop integration in Fig.\ \ref{fig:taylor}(c) exhibits a substantial 
difference in the chiral breaking scale and the IR value of the isoscalar mass (but it still produces 
a $\sigma$-mass larger than $1\ \mathrm{GeV}$).

Figure \ref{fig:taylor}(e) shows the $\mathrm{LPA}^{\prime}$-flows initialized at UV scales 
even larger than $1.3\ \mathrm{GeV}$ ($1.4\ \mathrm{GeV}$ and $1.5\ \mathrm{GeV}$), compare 
the additionally added gray-dotted lines. Again, the respective starting values were computed 
by upwards integration with a cubic potential as described in the context of Fig.\ \ref{fig:taylor}(a).
Once more, the bosonic observables are rather robust and to some extent almost independent
of the initalization scale as well as the presence of the higher Taylor coefficients already 
at $\Lambda_{\chi}$. The largest effects are found in the IR fermion masses $M^{\pm}$. Finally,
Fig.\ \ref{fig:taylor}(e) demonstrates the cutoff independence of the IR results in the MF 
approximation also for the initialization scales of $1.4\ \mathrm{GeV}$ and $1.5\ \mathrm{GeV}$.
The reason for this cutoff independence lies in the fact that the integration of each of the Taylor
coefficients itself is independent of the others and exclusively depends on the fermionic loop
contribution, see also Appendix \ref{sec:analytic}.

\begin{figure*}
	\centering
	\includegraphics[scale=1.0]{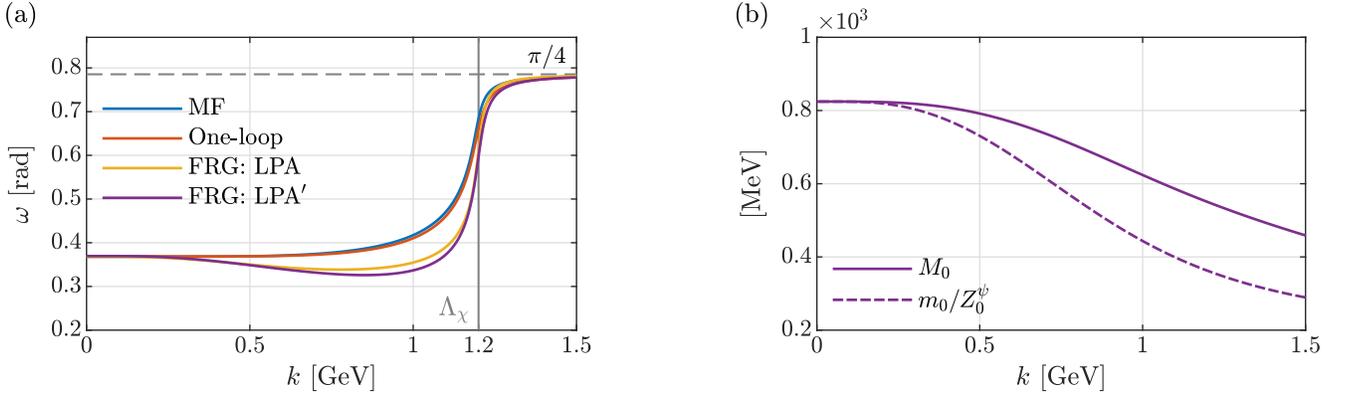}
	\caption{Scale evolution of the rotation angle $\omega$ and the (renormalized 
	and bare) chiral-invariant mass. (a) The rotation angle of Eq.\ (\ref{eq:rotangle}) 
	is plotted against the $k$-scale for all four approximations, which are considered 
	in the main part of this publication. The angle $\omega$ approaches the limit of $\pi/4\ 
	\mathrm{rad}$ (horizontal dashed line) in the UV. (b) Renormalized and bare 
	chiral-invariant masses $M_{0}$ and $\smash{m_{0}/Z_{0}^{\psi}}$, 
	respectively, as functions of the scale $k$ (in the $\mathrm{LPA}^{\prime}$); the 
	bare mass is normalized according to Eq.\ (\ref{eq:bare}).}
	\label{fig:omega}
\end{figure*}

The angle $\omega$ of Eq.\ (\ref{eq:rotangle}) quantifies the rotation from the unphysical
parity-even basis towards the physical parity-odd case; it is presented in Fig.\ \ref{fig:omega}(a)
and basically comprises the scale dependence of the chiral-invariant mass $M_{0}$ and the 
Yukawa couplings $\tilde{y}_{1}$ and $\tilde{y}_{2}$ (besides the condensate $\tilde{\sigma}_{0}$).\ 
The numeric angle $\omega$ roughly varies between $0.3\ \mathrm{rad}$ and $0.8\ \mathrm{rad}$, where 
it approaches the limit 
\begin{equation}
	\lim_{\tilde{\sigma}_{0}\, \rightarrow\, 0} \omega = \frac{\pi}{4}\ \mathrm{rad}
\end{equation}
in the chiral-restored phase ($k > \Lambda_{\chi}$). The latter is dominated by the asymptotic 
behavior of $\tilde{\sigma}_{0} \rightarrow 0$. In the broken phase ($k < \Lambda_{\chi}$), 
the curves for the different approximations split and eventually merge in the IR limit, 
$k \rightarrow 0$; they exhibit again the same (non-)monotonicity as observed in 
Figs.\ \ref{fig:masses} and \ref{fig:scattering_evolution} of Sec.\ \ref{sec:results},
ending up around $\omega = 0.37\ \mathrm{rad}$. The merging of the lines for
$k \rightarrow 0$ is caused by the common IR values of the chiral-invariant mass and the 
Yukawa couplings among all approximations, cf.\ the discussion below.

\begin{figure*}
	\centering
	\includegraphics[scale=1.0]{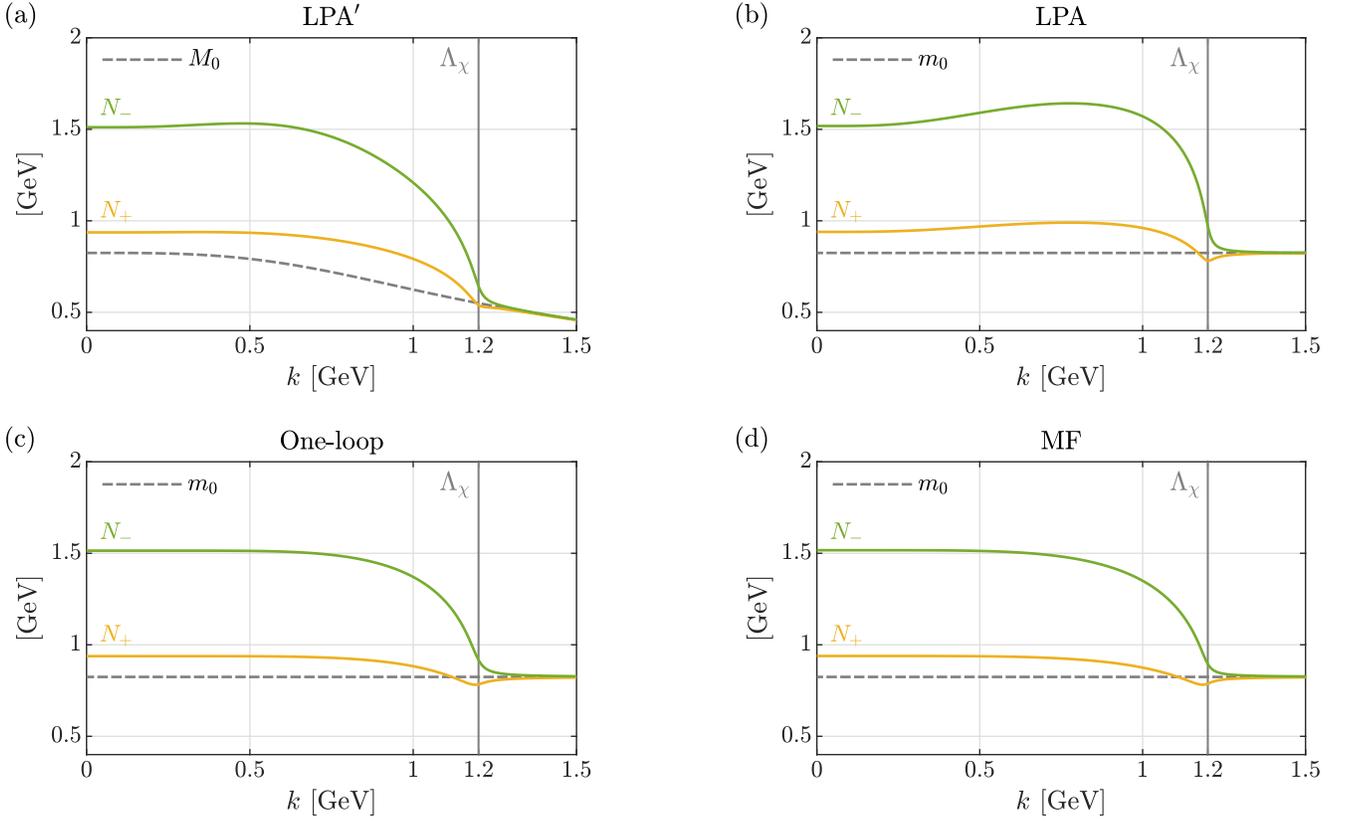}
	\caption{Scale evolution of the fermion masses $M^{\pm}$ in comparison to the chiral-invariant
	mass. (a) The mass parameters $M_{0}$ and $\smash{m_{0} = Z^{\psi} M_{0}}$ are both considered 
	as scale dependent in the $\mathrm{LPA^{\prime}}$. (b), (c), and (d) The mass parameters are 
	taken as scale independent in the LPA, one-loop, and MF approximations, respectively.}
	\label{fig:fermions}
\end{figure*}

Figure \ref{fig:omega}(b) lets us furthermore comment on the value of the renormalized and 
bare chiral-invariant mass parameters $M_{0}$ and $\smash{m_{0}^{\mathrm{bare}} \coloneqq 
m_{0}/Z_{0}^{\psi}}$ (in the $\mathrm{LPA}^{\prime}$). The (renormalized) 
physical mass matches the range of $M_{0} \simeq 450\ \mathrm{MeV}$ to $M_{0} \simeq 
825\ \mathrm{MeV}$ along its evolution (w.r.t.\ the energy-momentum scale $k$). It therefore 
principally conforms with the values cited in the Introduction. The bare mass $m_{0}^{\mathrm{bare}}$ 
turns out to be smaller, but it is still situated in the somewhat wider range of $\smash{200\ \mathrm{MeV} 
\le m_{0}^{\mathrm{bare}} \le 900\ \mathrm{MeV}}$, as also mentioned at the very beginning 
of this work. In the IR, we find a chiral-invariant mass of $M_{0} \simeq 824.5\ \mathrm{MeV}$ and
the degeneracy of the (physical) fermion masses $M^{\pm}$ with $M_{0}$ in the chiral-restored
phase is demonstrated in Fig.\ \ref{fig:fermions}(a).

Concerning the simpler approximations (LPA, one-loop, and MF), the (scale-independent) mass 
parameter $m_{0}$ is consequently set to the $\mathrm{LPA}^{\prime}$-value of $824.5\ \mathrm{MeV}$. 
In all three truncations, the fermion masses $m^{\pm}$ consistently fall into line with $m_{0}$ 
around the scale of $\Lambda_{\chi}$ (and beyond), cf.\ Figs.\ \ref{fig:fermions}(b) to 
\ref{fig:fermions}(d). Note that these results were generated under the constraints of chiral 
symmetry breaking occurring at $\Lambda_{\chi}$ and the ``successful'' reproduction of the pion-pion 
scattering lengths as IR observables within the nucleon-meson model.

\begin{figure*}
	\centering
	\includegraphics[scale=1.0]{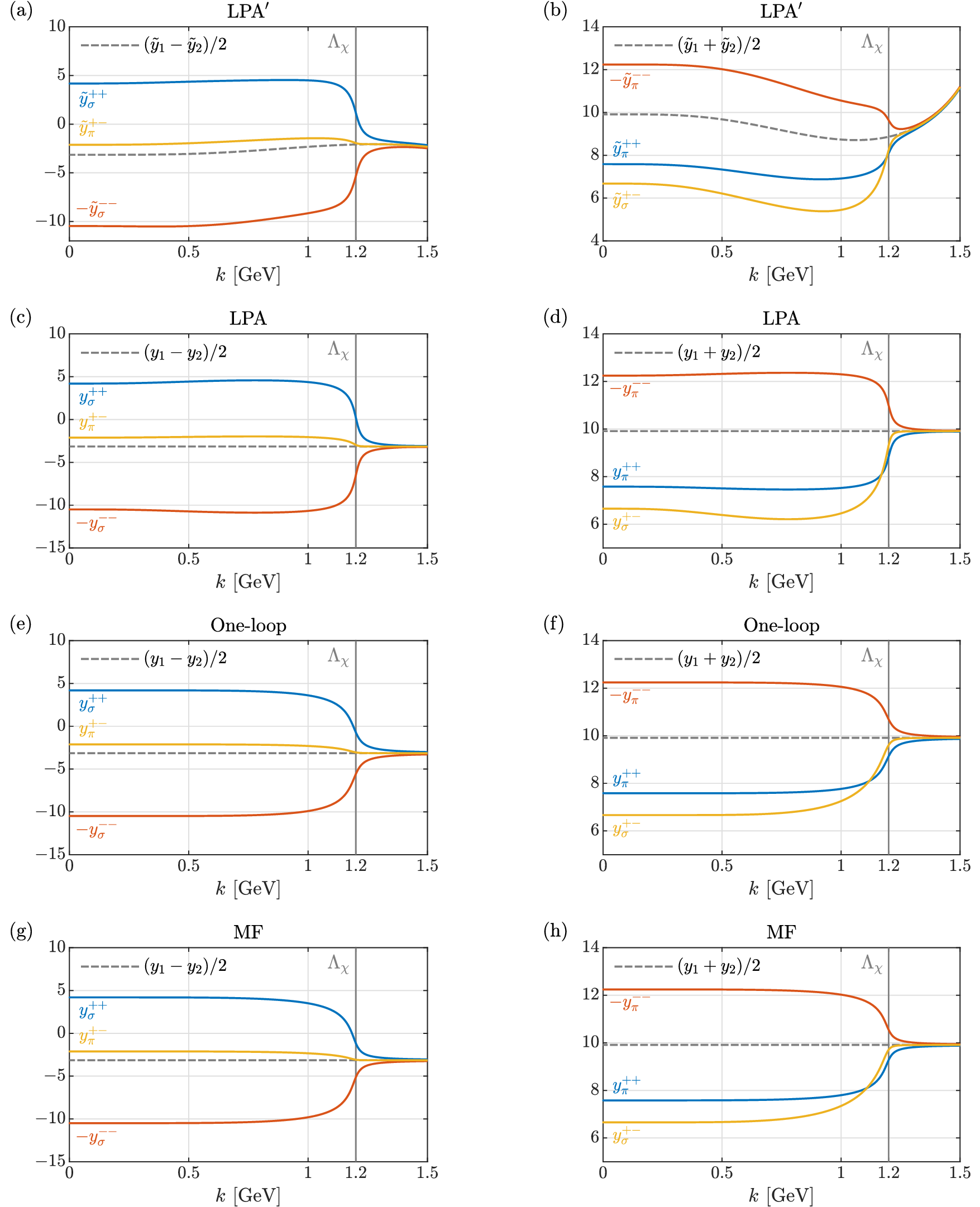}
	\caption{Scale evolution of the Yukawa couplings (in the physical basis) and
	for the four different approximations/truncations. In the left column, we
	collected the couplings that tend to the expression $\smash{(\tilde{y}_{1} -
	\tilde{y}_{2})/2}$ [or $\smash{(y_{1} - y_{2})/2}$] for large scales in the UV. 
	In the right column, the corresponding couplings converge to the numeric
	value of $\smash{(\tilde{y}_{1} + \tilde{y}_{2})/2}$ [or $\smash{(y_{1} 
	+ y_{2})/2}$].}
	\label{fig:yukawas}
\end{figure*}

Similarly to the fermion masses, the Yukawa couplings of Eqs.\ (\ref{eq:yukawa1}) to (\ref{eq:yukawa3}) 
(in the physical basis) become degenerate in the symmetric phase, $\tilde{\sigma}_{0} \rightarrow 0$, 
as it was outlined below Eq.\ (\ref{eq:yukawa3}) in Sec.\ \ref{sec:model}. In the $\mathrm{LPA}^{\prime}$, 
the expressions $(\tilde{y}_{1} \pm \tilde{y}_{2})/2$ are scale dependent themselves [cf.\ Figs.\ 
\ref{fig:yukawas}(a) and \ref{fig:yukawas}(b)], whereas the respective terms are assumed to be
constant in the other approximations [cf.\ Figs.\ \ref{fig:yukawas}(c) to \ref{fig:yukawas}(h)].
The scale dependence of the Yukawa couplings appears to be rather mild, except for scales close to
the crossover transition.\ The fixing of the chiral-invariant mass parameter to the
$\mathrm{LPA}^{\prime}$-value guarantees that the Yukawa couplings attain the same IR values 
in all approximations.

\section{Analytic one-loop integration and isoscalar-mass prediction}
\label{sec:analytic}

\subsection{Analytic integration and renormalized effective potential}

The flow equation for the effective potential $V$ is given by
\begin{equation}
	\partial_{k} V_{k} = \frac{k^{5}}{32 \pi^{2}}
	\left\lbrace \frac{3}{k^{2} + m_{\pi}^{2}}
	+ \frac{1}{k^{2} + m_{\sigma}^{2}} - 8 \left[\frac{1}{k^{2} + (m^{+})^{2}} 
	+ \frac{1}{k^{2} + (m^{-})^{2}}\right] \right\rbrace ,
\end{equation}
where we have replaced $\smash{\Gamma^{(2)}_{k}}$ by $\smash{\Gamma^{(2)}|_{\Lambda} 
\equiv S^{(2)}}$ with the UV cutoff $\Lambda$ on the right of the flow equation, 
cf.\ once again Eq.\ (\ref{eq:oneloop}) in Sec.\ \ref{sec:results}. This means that the squared 
masses $m_{\pi}^{2}$, $m_{\sigma}^{2}$, and $(m^{\pm})^{2}$ are independent of the integration 
variable $k$. These are given by
\begin{IEEEeqnarray}{rCl}
	m_{\pi}^{2}\!\left(\sigma^{2}\right) & = & 2V_{\Lambda}'\!\left(\sigma^{2}\right)
	\equiv 2\left[\alpha_{1,\Lambda} + \alpha_{2,\Lambda}
	\left(\sigma^{2} - \varphi_{0}^{2}\right) \right], \label{eq:pionmass} \\[0.2cm]
	m_{\sigma}^{2}\!\left(\sigma^{2}\right) & = & 2V_{\Lambda}'\!\left(\sigma^{2}\right)
	+ 4 \sigma^{2} V_{\Lambda}''\!\left(\sigma^{2}\right)
	\equiv 2\left[\alpha_{1,\Lambda} + \alpha_{2,\Lambda}
	\left(3 \sigma^{2} - \varphi_{0}^{2}\right)\right] , \label{eq:sigmamass} \\[0.2cm]
	m^{\pm}(\sigma) & = & \frac{1}{2} \left[\pm\sigma (y_{1} - y_{2})
	+ \sqrt{\sigma^{2}(y_{1} + y_{2})^{2} + 4 m_{0}^{2}}\right] ,
	\label{eq:fermionmasses} 
\end{IEEEeqnarray}
where we used
\begin{equation}
	V_{\Lambda}\!\left(\sigma^{2}\right) = \alpha_{1,\Lambda} 
	\left(\sigma^{2} - \varphi_{0}^{2}\right)
	+ \frac{\alpha_{2,\Lambda}}{2} \left(\sigma^{2} - \varphi_{0}^{2}\right)^{2}.
	\label{eq:UVpotential}
\end{equation}
The integration of the flow equation for the effective potential can be carried out analytically.
This yields the cutoff-regularized IR potential
\begin{IEEEeqnarray}{rCl}
	V_{0}^{\text{one-loop}} & = & V_{0}^{\mathrm{MF}} + \frac{1}{64 \pi^{2}} \left\lbrace
	3 m_{\pi}^{4} \left[\frac{1}{l^{\pi}} + \ln l^{\pi} - \ln \left(1 + l^{\pi}\right)\right]
	+ m_{\sigma}^{4} \left[\frac{1}{l^{\sigma}} + \ln l^{\sigma} -
	\ln \left(1 + l^{\sigma}\right)\right] \right\rbrace , \label{eq:IRpotential} \\[0.2cm]
	V_{0}^{\mathrm{MF}} & = & V_{\Lambda} 
	- \frac{1}{8 \pi^{2}} \left\lbrace
	(m^{+})^{4} \left[\frac{1}{l^{+}} + \ln l^{+} -
	\ln \left(1 + l^{+}\right)\right]
	+ (m^{-})^{4} \left[\frac{1}{l^{-}} + \ln l^{-} -
	\ln \left(1 + l^{-}\right)\right]
	\right\rbrace , \label{eq:IRpotentialMF}
\end{IEEEeqnarray}
with $\smash{l^{\pi/\sigma} = (m_{\pi/\sigma}/\Lambda)^{2}}$ and $\smash{l^{\pm} = 
(m^{\pm}/\Lambda)^{2}}$.\ We have dropped irrelevant terms proportional to 
$\Lambda^{4}$.\ The first two terms in each of the square brackets are divergent 
for $\smash{l^{\pi/\sigma} \rightarrow 0}$ and $\smash{l^{\pm} \rightarrow 0}$, which corresponds 
to sending the cutoff to infinity, i.e., $\Lambda \rightarrow \infty$, keeping the masses
constant. The MF approximation of the effective potential is obtained by neglecting the bosonic 
loop contributions in Eq.\ (\ref{eq:IRpotential}), keeping only the one fermion loop.

Using the minimum condition of the IR potential $V_{0}^{\text{one-loop}}$, we compute the squared 
pion mass from the relation
\begin{IEEEeqnarray}{rCl}
	h & = & M_{\pi}^{2} f_{\pi} = 2 f_{\pi} \left.V_{0}^{\text{one-loop}\!\;\prime}
	\right|_{\sigma^{2}\, =\, f_{\pi}^{2}} \nonumber\\[0.1cm]
	& = & 2 f_{\pi} \left[\alpha_{1,\Lambda} + \alpha_{2,\Lambda}
	\left(f_{\pi}^{2} - \varphi_{0}^{2}\right)\right]
	+ \frac{1}{64 \pi^{2}} \left(3\left.\frac{\mathrm{d}m_{\pi}^{4}}
	{\mathrm{d\sigma}} \left[\frac{1}{l^{\pi}} 
	\frac{1}{2}\frac{1 + 2 l^{\pi}}{1 + l^{\pi}} + \ln l^{\pi} 
	- \ln \left(1 + l^{\pi}\right)\right] \right|_{f_{\pi}} \right. \nonumber\\[0.2cm]
	& & + \left.\frac{\mathrm{d}m_{\sigma}^{4}}
	{\mathrm{d\sigma}} \left[\frac{1}{l^{\sigma}} 
	\frac{1}{2}\frac{1 + 2 l^{\sigma}}{1 + l^{\sigma}} + \ln l^{\sigma} 
	- \ln \left(1 + l^{\sigma}\right)\right] \right|_{f_{\pi}}
	- 8 \left\lbrace \left. \frac{\mathrm{d}(m^{+})^{4}}{\mathrm{d\sigma}}
	\left[\frac{1}{l^{+}} \frac{1}{2}\frac{1 + 2 l^{+}}{1 + l^{+}} 
	+ \ln l^{+} - \ln \left(1 + l^{+}\right)\right] 
	\right|_{f_{\pi}} \right. \nonumber\\[0.2cm]
	& & + \!\! \left. \left. \left. 
	\frac{\mathrm{d}(m^{-})^{4}}{\mathrm{d\sigma}}
	\left[\frac{1}{l^{-}} \frac{1}{2}\frac{1 + 2 l^{-}}{1 + l^{-}} 
	+ \ln l^{-} - \ln \left(1 + l^{-}\right)\right] \right|_{f_{\pi}} \right\rbrace
	\right). \label{eq:hminimum}
\end{IEEEeqnarray}
The squared $\sigma$-mass is analogously computed as
\begin{IEEEeqnarray}{rCl}
	M_{\sigma}^{2} & = & 2 V_{0}^{\text{one-loop}\;\!\prime} 
	+ 4 f_{\pi}^{2} \left. V_{0}^{\text{one-loop}\;\!\prime\prime} 
	\right|_{f_{\pi}^{2}} \nonumber\\[0.2cm]
	& = & 2 \left[\alpha_{1,\Lambda} + \alpha_{2,\Lambda}
	\left(3 f_{\pi}^{2} - \varphi_{0}^{2}\right)\right] \nonumber\\[0.2cm]
	& & +\, \frac{1}{64\pi^{2}} \left(3 \left\lbrace \left. 
	\frac{\mathrm{d}^{2}m_{\pi}^{4}}{\mathrm{d\sigma^{2}}}
	\left[\frac{1}{l^{\pi}} \frac{1}{2}\frac{1 + 2 l^{\pi}}{1 + l^{\pi}} 
	+ \ln l^{\pi} - \ln \left(1 + l^{\pi}\right)\right] \right|_{f_{\pi}}
	- \left. \left(\frac{\mathrm{d}m_{\pi}^{2}}{\mathrm{d\sigma}}\right)^{2} 
	\frac{1}{l^{\pi}} \frac{1}{(1 + l^{\pi})^{2}} \right|_{f_{\pi}} 
	\right\rbrace \right. \nonumber\\[0.2cm]
	& & \hspace{1.5cm} + \left. \frac{\mathrm{d}^{2}m_{\sigma}^{4}}{\mathrm{d\sigma^{2}}}
	\left[\frac{1}{l^{\sigma}} \frac{1}{2}\frac{1 + 2 l^{\sigma}}{1 + l^{\sigma}} 
	+ \ln l^{\sigma} - \ln \left(1 + l^{\sigma}\right)\right] \right|_{f_{\pi}}
	- \left. \left(\frac{\mathrm{d}m_{\sigma}^{2}}{\mathrm{d\sigma}}\right)^{2} 
	\frac{1}{l^{\sigma}} \frac{1}{(1 + l^{\sigma})^{2}} \right|_{f_{\pi}} \nonumber\\[0.2cm]
	& & \hspace{1.5cm} -\, 8 \left\lbrace 
	\left. \frac{\mathrm{d}^{2}(m^{+})^{4}}{\mathrm{d\sigma^{2}}}
	\left[\frac{1}{l^{+}} \frac{1}{2}\frac{1 + 2 l^{+}}{1 + l^{+}} 
	+ \ln l^{+} - \ln \left(1 + l^{+}\right)\right] \right|_{f_{\pi}}
	- \left. \left[\frac{\mathrm{d}(m^{+})^{2}}{\mathrm{d\sigma}}\right]^{2} 
	\frac{1}{l^{+}} \frac{1}{(1 + l^{+})^{2}} \right|_{f_{\pi}} \right. \nonumber\\[0.2cm]
	& & \hspace{2.0cm} \left. \left. + \left. 
	\frac{\mathrm{d}^{2}(m^{-})^{4}}{\mathrm{d\sigma}^{2}}
	\left[\frac{1}{l^{-}} \frac{1}{2}\frac{1 + 2 l^{-}}{1 + l^{-}} 
	+ \ln l^{-} - \ln \left(1 + l^{-}\right)\right] \right|_{f_{\pi}}
	- \left. \left[\frac{\mathrm{d}(m^{-})^{2}}{\mathrm{d\sigma}}\right]^{2} 
	\frac{1}{l^{-}} \frac{1}{(1 + l^{-})^{2}} \right|_{f_{\pi}} 
	\right\rbrace \right) . \quad \label{eq:ms}
\end{IEEEeqnarray}
If we now plug in the numeric values of Table \ref{tab:UV}, we find that $M_{\pi} = 140.2\ \mathrm{MeV}$ 
and $M_{\sigma} = 1102\ \mathrm{MeV}$ for the one-loop approximation as well as $M_{\pi} = 140.0\ 
\mathrm{MeV}$ and $M_{\sigma} = 1477\ \mathrm{MeV}$ for the MF approximation, as quoted in Table 
\ref{tab:IR}. In addition, Eq.\ (\ref{eq:hminimum}) can be used to numerically determine the pion 
decay constant $f_{\pi}$.

Writing the loop contributions in Eqs.\ (\ref{eq:IRpotential}) and (\ref{eq:IRpotentialMF}) 
in terms of the ``loop function'' $F_{\Lambda}(\sigma^{2})$,
\begin{equation}
	V_{0}^{\text{one-loop}}\!\left(\sigma^{2}\right) 
	\equiv V_{\Lambda}\!\left(\sigma^{2}\right) 
	+ F_{\Lambda}\!\left(\sigma^{2}\right) \label{eq:F},
\end{equation}
with
\begin{IEEEeqnarray}{rCl}
	F_{\Lambda}\!\left(\sigma^{2}\right) & = & F_{\Lambda}^{\mathrm{MF}}
	\!\left(\sigma^{2}\right) + \frac{1}{64 \pi^{2}} \left\lbrace
	3 m_{\pi}^{4} \left[\frac{1}{l^{\pi}} + \ln l^{\pi} - \ln \left(1 + l^{\pi}\right)\right]
	+ m_{\sigma}^{4} \left[\frac{1}{l^{\sigma}} + \ln l^{\sigma} -
	\ln \left(1 + l^{\sigma}\right)\right] \right\rbrace , \\[0.2cm]
	F_{\Lambda}^{\mathrm{MF}}\!\left(\sigma^{2}\right) 
	& = & - \frac{1}{8 \pi^{2}} \left\lbrace
	(m^{+})^{4} \left[\frac{1}{l^{+}} + \ln l^{+} -
	\ln \left(1 + l^{+}\right)\right]
	+ (m^{-})^{4} \left[\frac{1}{l^{-}} + \ln l^{-} -
	\ln \left(1 + l^{-}\right)\right]
	\right\rbrace , \label{eq:FMF}
\end{IEEEeqnarray}
we reorganize Eqs.\ (\ref{eq:hminimum}) and (\ref{eq:ms}) (with the identification $\varphi_{0}^{2} = 
f_{\pi}^{2}$) as
\begin{IEEEeqnarray}{rCl}
	\alpha_{1,\Lambda} & = & \frac{h}{2 f_{\pi}} 
	- F_{\Lambda}'\!\left(f_{\pi}^{2}\right)
	\equiv \frac{M_{\pi}^{2}}{2} - F_{\Lambda}'\!\left(f_{\pi}^{2}\right) ,
	\label{eq:deta1} \\[0.2cm]
	\alpha_{2,\Lambda} & = & \frac{M_{\sigma}^{2} - M_{\pi}^{2}}
	{4 f_{\pi}^{2}} - F_{\Lambda}''\!\left(f_{\pi}^{2}\right).
	\label{eq:deta2}
\end{IEEEeqnarray}
The loop function $F_{\Lambda}$ depends on the Taylor coefficients $\alpha_{1,\Lambda}$ 
and $\alpha_{2,\Lambda}$ in the one-loop case, whereas it is independent of these two parameters 
in the MF approximation (the MF loop function $\smash{F_{\Lambda}^{\mathrm{MF}}}$ does 
not involve the bosonic masses). Expanding the function $F_{\Lambda}$ around the vacuum state,
\begin{equation}
	F_{\Lambda}\!\left(\sigma^{2}\right) = F_{\Lambda}\!\left(f_{\pi}^{2}\right) 
	+ F_{\Lambda}'\!\left(f_{\pi}^{2}\right) \left(\sigma^{2} - f_{\pi}^{2}\right)
	+ \frac{1}{2} F_{\Lambda}''\!\left(f_{\pi}^{2}\right) \left(\sigma^{2} 
	- f_{\pi}^{2}\right)^{2} + \cdots ,
\end{equation}
one rewrites the renormalized IR potential (\ref{eq:F}) as
\begin{IEEEeqnarray}{rCl}
	V_{0}^{\text{one-loop}}\!\left(\sigma^{2}\right) & = & 
	\left[\frac{M_{\pi}^{2}}{2} - F_{\Lambda}'\!\left(f_{\pi}^{2}\right)\right] 
	\left(\sigma^{2} - f_{\pi}^{2}\right)
	+ \frac{1}{2} \left[\frac{M_{\sigma}^{2} - M_{\pi}^{2}}{4 f_{\pi}^{2}}
	- F_{\Lambda}''\!\left(f_{\pi}^{2}\right)\right]
	\left(\sigma^{2} - f_{\pi}^{2}\right)^{2} \nonumber\\[0.2cm]
	& & +\, F_{\Lambda}\!\left(f_{\pi}^{2}\right) 
	+ F_{\Lambda}'\!\left(f_{\pi}^{2}\right) \left(\sigma^{2} - f_{\pi}^{2}\right)
	+ \frac{1}{2} F_{\Lambda}''\!\left(f_{\pi}^{2}\right) 
	\left(\sigma^{2} - f_{\pi}^{2}\right)^{2}
	+ \mathcal{O}\!\left[\left(\sigma^{2} - f_{\pi}^{2}\right)^{3}\right] .
	\label{eq:potentialexpand}
\end{IEEEeqnarray}
Thus the divergent contributions to the potential cancel out to order $(\sigma^{2} - f_{\pi}^{2})^{2}$.
The infinite constant $\smash{F_{\Lambda}(f_{\pi}^{2})}$ does not affect derivatives of 
$\smash{V_{0}^{\text{one-loop}}}$. Terms higher than $\smash{(\sigma^{2} - f_{\pi}^{2})^{2}}$ 
likewise do not contribute to the first and second derivatives evaluated at the vacuum state, 
$\sigma^{2} = f_{\pi}^{2}$. This means that we find from Eq.\ (\ref{eq:potentialexpand}) 
finite observables $M_{\sigma}$ and $M_{\pi}$ even if we would take the infinite-cutoff limit, 
$\Lambda \rightarrow \infty$ ($l^{\pi/\sigma} \rightarrow 0$ and $l^{\pm} \rightarrow 0$).

The renormalized IR potential can therefore be defined as
\begin{equation}
	V_{\mathrm{ren.}}^{\text{one-loop}}\! \left(\sigma^{2}\right) 
	= \frac{M_{\pi}^{2}}{2} \left(\sigma^{2} - f_{\pi}^{2}\right)
	+ \frac{M_{\sigma}^{2} - M_{\pi}^{2}}{8f_{\pi}^{2}} 
	\left(\sigma^{2} - f_{\pi}^{2}\right)^{2} .
	\label{eq:Vren}
\end{equation}
Once we fix the physical IR values of $f_{\pi}$, $M_{\pi}$, and $M_{\sigma}$, we can determine 
at any cutoff scale the parameters $\alpha_{1}$ and $\alpha_{2}$ of the UV potential such that
the one-loop integration generates exactly those observables in the IR. The coefficient
$\alpha_{1}$ renormalizes the ``mass term'' of the effective potential and the coefficient 
$\alpha_{2}$ renormalizes the quartic interaction. We thus expect unequal sensitivity of these 
couplings on $\Lambda$, with $\alpha_{1}$ being more sensitive than $\alpha_{2}$ due
to their respective mass dimensions.

We adapt the above ``standard'' renormalization formalism in the next Section in order 
to determine initial values of the Taylor coefficients $\alpha_{1}$ and $\alpha_{2}$ at the 
chiral-symmetry breaking scale $\Lambda_{\chi} \simeq 4\pi f_{\pi}$. We emphasize that, in 
general, the breaking scale $\Lambda_{\chi}$ and the actual UV cutoff $\Lambda$ (as discussed
above) as well as the initialization scale of the integration of momentum modes do not coincide, 
especially in their precise physical meaning. It is hence important to notice that the discussion 
below is part of the approximations of our work and the findings might contain corresponding
artifacts. Being aware of this issue, we have already done numeric tests concerning the
presence of higher Taylor coefficients at $\Lambda_{\chi}$, cf.\ again Fig.\ \ref{fig:taylor}. 
In order to justify our approximations, further quantification of numeric errors induced by 
using $\Lambda_{\chi}$ in place of $\Lambda$ in the above formalism (as well as using 
$\Lambda_{\chi}$ as the initialization scale) is given below.

\subsection{Isoscalar-mass prediction}

In the MF approximation, Eqs.\ (\ref{eq:deta1}) and (\ref{eq:deta2}) unambiguously determine 
the Taylor coefficients $\alpha_{1,\Lambda}$ and $\alpha_{2,\Lambda}$ at $\Lambda = \Lambda_{\chi}$
as functions of the bosonic masses and the fermionic loop function (\ref{eq:FMF}). While the first 
coefficient $\alpha_{1}$ is set by the IR pion mass $M_{\pi}$, the second coefficient $\alpha_{2}$
controls the mass splitting between the bosons.\ Conversely, the IR isoscalar $\sigma$-mass
can be predicted after fixing all other parameters in Eq.\ (\ref{eq:deta2}) by adjusting the
coefficient $\alpha_{2}$ in such a way that chiral symmetry breaking occurs at $\Lambda_{\chi} 
= 1.2\ \mathrm{GeV} \simeq 4\pi f_{\pi}$.
\begin{figure}[t]
	\centering
	\includegraphics[scale=1.0]{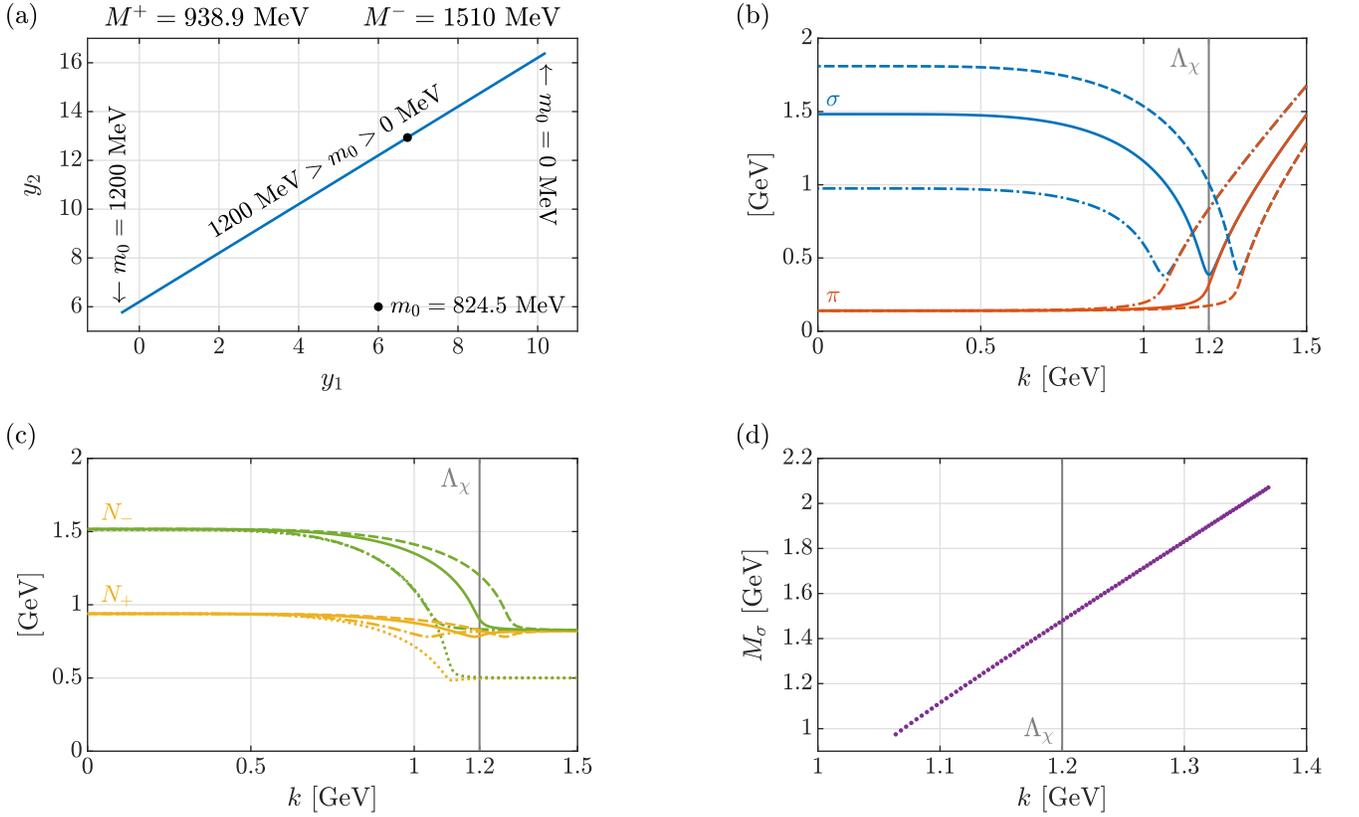}
	\caption{Prediction of the isoscalar $\sigma$-mass in the MF approximation.
	(a) Physical IR fermion masses are produced for various choices of the 
	chiral-invariant mass $m_{0}$ (parametric plot of the corresponding Yukawa
	couplings in the range of $0\ \mathrm{MeV} \le m_{0} \le 1200\ \mathrm{MeV}$); 
	the black dot indicates the IR (renormalized) $M_{0}$-value of the 
	$\mathrm{LPA}^{\prime}$-truncation. (b) Determination of the $\sigma$-mass
	by adjusting the Taylor coefficient $\alpha_{2}$ such that chiral symmetry 
	breaking occurs at the scale $\Lambda_{\chi}$ (solid: $\alpha_{2} = 38$; 
	dashed: $\alpha_{2} = 70$; dash-dotted: $\alpha_{2} = 1$). (c) Scale evolution
	of the fermion masses corresponding to the parameter scenarios of panel (b);
	the IR fermion masses are fixed and the dotted lines show a scenario with 
	varying $m_{0}$ (dotted: $m_{0} = 500\ \mathrm{MeV}$; else: $m_{0} = 824.5\ 
	\mathrm{MeV}$). (d) Scale ($k$-value; x-axis) of chiral symmetry breaking as
	a function of the IR $\sigma$-mass ($M_{\sigma}$; y-axis).}
	\label{fig:Msigma_MF}
\end{figure}

In general, requiring that the isoscalar mass $M_{\sigma}$ be larger than the pion mass
$M_{\pi}$ at the chiral breaking scale $\Lambda_{\chi}$ (and below), which is the natural 
ordering of the bosonic masses, implies a positive Taylor coefficient $\alpha_{2}$,
\begin{equation}
	\left. M_{\sigma}^{2} - M_{\pi}^{2} \right|_{\Lambda_{\chi}}
	= \left. 4 \sigma_{0}^{2} \alpha_{2} \right|_{\Lambda_{\chi}}
	> 0 \quad \Leftrightarrow \quad \alpha_{2,\Lambda_{\chi}} > 0,
\end{equation}
with the (scale-dependent) minimum $\sigma_{0}$ of the effective potential. Moreover, the 
coefficient $\alpha_{1}$ must also be positive in order to fulfill the minimum condition 
at $\Lambda_{\chi}$,
\begin{equation}
	2 \sigma_{0} V_{\Lambda_{\chi}}'\!\left(\sigma_{0}^{2}\right) 
	- h = 0 \quad \Leftrightarrow \quad 
	\alpha_{1,\Lambda_{\chi}} + \alpha_{2,\Lambda_{\chi}}
	\left(\sigma_{0}^{2} - \varphi_{0}^{2}\right)
	- \frac{h}{2 \sigma_{0}} = 0 \quad \Rightarrow
	\quad \alpha_{1,\Lambda_{\chi}} > 0,
\end{equation}
where we used that the value of $\sigma_{0}$ increases as the scale $k$ decreases, so that 
$\sigma_{0}^{2} = f_{\pi}^{2} \lesssim \varphi_{0}^{2}$ in the IR (cf.\ again Fig.\ \ref{fig:masses}).

The prediction of the isoscalar mass is only possible if one fixes the chiral-invariant 
mass parameter $m_{0}$ to a common value in all investigated approximations.\ This is 
due to the fact that the three parameters $y_{1}$, $y_{2}$, and $m_{0}$ are not 
uniquely determined by the two fermion masses (\ref{eq:fermionmasses}). The Yukawa 
couplings $y_{1}$ and $y_{2}$ are correlated. Figure \ref{fig:Msigma_MF}(a) demonstrates 
this correlation in terms of a parametric plot of the Yukawa couplings depending on 
the chosen value of $m_{0}$ for given physical fermion masses $m^{\pm} \equiv M^{\pm}$.\
As discussed in the main part of this publication, we decided to take the IR (renormalized) 
$M_{0}$-value of the $\mathrm{LPA}^{\prime}$-truncation for this purpose ($M_{0} = 
m_{0}/Z^{\psi}$ in the $\mathrm{LPA}^{\prime}$, whereas $M_{0} \equiv m_{0}$ for the 
other approximations), which yields approximately the same chiral-invariant mass and 
Yukawa couplings in the IR (along with roughly the same IR fermion masses, see Table 
\ref{tab:IR}).

In Fig.\ \ref{fig:Msigma_MF}(b), it is shown how we predict the isoscalar mass
$M_{\sigma}$ in the MF approximation by utilizing Eqs.\ (\ref{eq:deta1}) and (\ref{eq:deta2}). 
The first Taylor coefficient $\alpha_{1,\Lambda_{\chi}}$ of the potential (\ref{eq:UVpotential}) 
is obtained by fixing the pion mass in the IR to the physical value of $M_{\pi} \simeq 
138\ \mathrm{MeV}$; this coefficient is independent of the $\sigma$-mass. The second Taylor 
coefficient $\alpha_{2,\Lambda_{\chi}}$, which regulates the isoscalar mass in the IR, 
is then adjusted in order to break chiral symmetry at $\Lambda_{\chi}$ (see the solid line).
Figure \ref{fig:Msigma_MF}(c) exhibits the scale evolution of the fermion masses corresponding 
to the three parameter scenarios of Fig.\ \ref{fig:Msigma_MF}(b). The IR masses are not 
affected by the different choices for $M_{\sigma}$; they are determined by the parameter 
$m_{0} = 824.5\ \mathrm{MeV}$, the Yukawa couplings according to Fig.\ \ref{fig:Msigma_MF}(a),
and the physical value of the pion decay constant $f_{\pi}$. The additional dotted curves
exemplify a change of the chiral-invariant mass (to $m_{0} = 500\ \mathrm{MeV}$). In this
setting, the Yukawa couplings (as functions of the parameter $m_{0}$) compensate for the
change in the mass $m_{0}$ in the sense that the physical fermion masses are still attained
in the IR.

The $k$-dependence of the isoscalar mass $M_{\sigma}$ in Fig.\ \ref{fig:Msigma_MF}(b) features 
a minimum at $k > 1\ \mathrm{GeV}$ in all three cases, when chiral symmetry breaking occurs
(in fact, we adjusted the parameter $\alpha_{2}$ such that this minimum lies at $\Lambda_{\chi}
= 1.2\ \mathrm{GeV}$, cf.\ once more the solid line).\ Figure \ref{fig:Msigma_MF}(d) presents 
a scatter plot of the location of this minimum on the $k$-scale (x-axis) as a function 
of the IR value of $M_{\sigma}$ (y-axis). Manifestly, the location crosses the scale 
$\Lambda_{\chi}$ for an IR isoscalar mass of the order of $1.5\ \mathrm{GeV}$, giving the 
$\sigma$-mass prediction in the MF approximation ($M_{\sigma} = 1477\ \mathrm{MeV}$).

\begin{figure}
	\centering
	\includegraphics[scale=1.0]{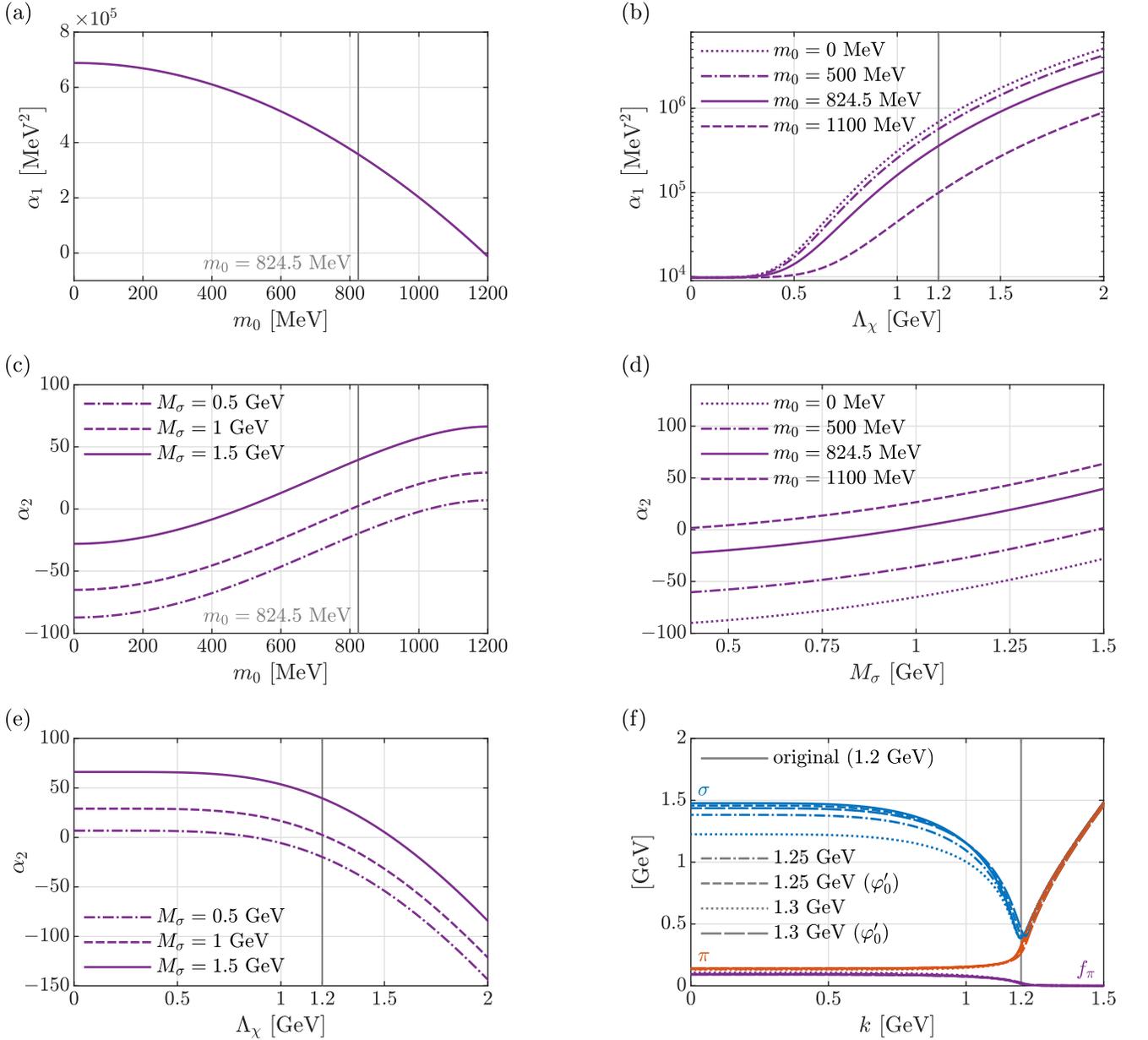}
	\caption{Analysis of the initial Taylor coefficients in the MF approximation. 
	(a) and (b) Taylor coefficient $\alpha_{1}$ as a function of the chiral-invariant
	mass $m_{0}$ and the scale $\Lambda_{\chi}$, respectively. (c) to (e) Taylor 
	coefficient $\alpha_{2}$ as a function of the chiral-invariant mass $m_{0}$, the 
	IR isoscalar mass $M_{\sigma}$, as well as the scale $\Lambda_{\chi}$, respectively.
	These findings rule out light IR $\sigma$-masses below $1\ \mathrm{GeV}$. 
	(f) Effect of changes in the initialization scale $\Lambda_{\chi}$ and the IR reference 
	scale for the bosonic masses (as well as the pion decay constant): the denotation 
	``original'' means the determination of $\alpha_{1}$ and $\alpha_{2}$ at 
	$\Lambda_{\chi} = 1.2\ \mathrm{GeV}$ using the IR bosonic masses at $k = 0$ as the 
	reference scale [described by Eqs.\ (\ref{eq:deta1}) and (\ref{eq:deta2})]; the other 
	curves are obtained by increasing $\Lambda_{\chi}$ to $1.25\ \mathrm{GeV}$ or $\mathrm{1.3}\ 
	\mathrm{GeV}$. For each of these two scales the respective line without any further
	specification stands for the case of taking the bosonic masses at $k = 1.2\ \mathrm{GeV}$ 
	as the reference scale (instead of the physical IR masses at $k = 0$), whereas the lines 
	with the specification ``$(\varphi_{0}')$'' represent those with the reference scale lying 
	at $k = 1\ \mathrm{GeV}$ and an additionally shifted expansion point $\varphi_{0} \rightarrow 
	\varphi_{0}'$. The curves for the determination scales of $1.25\ \mathrm{GeV}$ or 
	$\mathrm{1.3}\ \mathrm{GeV}$ are computed using the equivalents of Eqs.\ (\ref{eq:deta1}) 
	and (\ref{eq:deta2}) for nonzero $k$ [for even larger determination scales of e.g.\ 
	$1.4\ \mathrm{GeV}$ or $1.5\ \mathrm{GeV}$, we already find negative values for 
	the second coefficient $\alpha_{2}$, similar to panel (e)]. The results in subfigure 
	(f) give an intuition about the accuracy of the employed Taylor approximation and
	the importance of higher Taylor coefficients.}
	\label{fig:alpha}
\end{figure}

A more detailed analysis of the initial Taylor coefficients $\alpha_{1,\Lambda_{\chi}}$ and 
$\alpha_{2,\Lambda_{\chi}}$ of the potential (\ref{eq:UVpotential}) is given in Fig.\ \ref{fig:alpha}. 
As stated above, it is mandatory to have positive values for the two coefficients, which leads to restrictions 
regarding the choice of initial parameters.\ In Figs.\ \ref{fig:alpha}(a) and \ref{fig:alpha}(b), 
we illustrate the variation of $\alpha_{1}$ with the parameters $m_{0}$ and $\Lambda_{\chi}$, 
respectively.\ This coefficient is positive for $m_{0} < 1.2\ \mathrm{GeV}$.\ Its value 
generally shrinks with increasing $m_{0}$ and decreasing $\Lambda_{\chi}$.\ Concerning the coefficient 
$\alpha_{2}$, Figs.\ \ref{fig:alpha}(c) and \ref{fig:alpha}(d) reveal that, for fixed $m_{0} = 824.5\ 
\mathrm{MeV}$, the IR $\sigma$-mass has to be larger than $1\ \mathrm{GeV}$ in order to find positive 
values.\ Varying the $m_{0}$-parameter, large values of the $\sigma$-mass above $1\ \mathrm{GeV}$ are 
obviously favored as they generate more ``overlap'' with the range of $200\ \mathrm{MeV} 
\lesssim m_{0} \lesssim 900\ \mathrm{MeV}$ suggested by other phenomenological studies [cf.\ 
Fig.\ \ref{fig:alpha}(c)].\ Furthermore, light isoscalar masses ($M_{\sigma} \sim 500\ \mathrm{MeV}$) 
are only reachable for $m_{0} > 1.1\ \mathrm{GeV}$ [cf.\ Fig.\ \ref{fig:alpha}(d)].\ The dependence 
of $\alpha_{2}$ on $\Lambda_{\chi}$ again rules out IR isoscalar masses below $1\ \mathrm{GeV}$, 
see Fig.\ \ref{fig:alpha}(e). A variation of $\Lambda_{\chi}$ itself is also restricted by the 
condition $\alpha_{2} > 0$ to a rough maximum of $1.5\ \mathrm{GeV}$, given a $\sigma$-mass of 
the same order (and even to smaller values for decreasing $M_{\sigma}$). The overall sensitivity
of the coefficients $\alpha_{1}$ and $\alpha_{2}$ on $\Lambda_{\chi}$ is estimated by the difference
quotients
\begin{IEEEeqnarray}{rCl}
	\left| \frac{\Delta \alpha_{1}/f_{\pi}^{2}}{\Delta \Lambda_{\chi}} \right| 
	& = & \frac{\alpha_{1}(1.5\, \mathrm{GeV}) - \alpha_{1}(1\, \mathrm{GeV})}
	{f_{\pi}^{2} \left(1.5\, \mathrm{GeV} - 1\, \mathrm{GeV}\right)}
	\simeq 0.178\ \mathrm{MeV}^{-1}, \\[0.2cm]
	\left| \frac{\Delta \alpha_{2}}{\Delta \Lambda_{\chi}} \right| 
	& = & \frac{\alpha_{2}(1\, \mathrm{GeV}) - \alpha_{2}(1.5\, \mathrm{GeV})}
	{1.5\, \mathrm{GeV} - 1\, \mathrm{GeV}}
	\simeq 0.097\ \mathrm{MeV}^{-1},
\end{IEEEeqnarray}
reflecting a larger sensitivity of the dimensionless coefficient $\alpha_{1}/f_{\pi}^{2}$ 
w.r.t.\ changes in $\Lambda_{\chi}$, as expected.

Figure \ref{fig:alpha}(f) displays the effect of changing the initialization scale 
$\Lambda_{\chi}$ and the reference scale for the bosonic masses in the MF integration.
This means that we now allow for higher cutoff scales than $\Lambda_{\chi} = 1.2\ \mathrm{GeV}$
as the starting point for the integration of momentum modes in the UV. Moreover, we also change 
the IR cutoff scale $k$ of the integration to values larger than zero ($k > 0$; this is called the 
``reference scale''), at which we preset the bosonic and fermionic masses as the outcome of the MF 
integration (in contrast to the physical scale of $k = 0$). While the results presented in Sec.\ 
\ref{sec:results} are produced by taking the IR limit $k = 0$ as the reference scale, i.e., with 
the IR bosonic masses $M_{\pi}$ and $M_{\sigma}$ determining the Taylor coefficients $\alpha_{1,\Lambda_{\chi}}$ 
and $\alpha_{2,\Lambda_{\chi}}$ (cf.\ the line named ``original''), we also investigate alternative 
choices for $\Lambda_{\chi}$ and the IR reference scale $k > 0$. Despite the cutoff independence 
of the renormalized effective potential (\ref{eq:Vren}), we expect certain inaccuracies in the Taylor 
coefficients themselves for the reason that the current squared minimum $\sigma_{0}^{2}$ of the 
``flowing'' potential $V(\sigma^{2})$ deviates more and more from the squared expansion point
$\varphi_{0}^{2} \gtrsim f_{\pi}^{2}$ as the scale increases, $\sigma_{0}^{2} \ll \varphi_{0}^{2}$.\
We therefore compare the ``original'' integration to cases where we determine $\alpha_{1}$ and 
$\alpha_{2}$ at scales larger than $\Lambda_{\chi} = 1.2\ \mathrm{GeV}$, namely, at $\Lambda_{\chi} 
= 1.25\ \mathrm{GeV}$ and $\Lambda_{\chi} = 1.3\ \mathrm{GeV}$, using also reference scales away 
from zero. Being more precise, the four respective curves in Fig.\ \ref{fig:alpha}(f) depict the 
MF integration for taking the scales $k = 1.2\ \mathrm{GeV}$ (lines without further specification) 
and $k = 1\ \mathrm{GeV}$ [lines with the specification ``$(\varphi_{0}')$''] as the reference scale 
for determining the Taylor coefficients (at $\Lambda_{\chi} = 1.25\ \mathrm{GeV}$ and $\Lambda_{\chi} 
= 1.3\ \mathrm{GeV}$), where we additionally shift the expansion point in the latter cases, 
$\varphi_{0} \rightarrow \varphi_{0}' = \sigma_{0}|_{k\, =\, 1\, \mathrm{GeV}}$. One observes that, 
for the reference scale lying at $k = 1.2\ \mathrm{GeV}$, the IR $\sigma$-mass at $k = 0$ substantially 
differs by about $250\ \mathrm{MeV}$ (cf.\ solid and dotted lines). This effect is again weakened 
by taking the reference scale of $k = 1\ \mathrm{GeV}$ instead (and shifting the expansion point
accordingly). Eventually, the importance of higher Taylor coefficients can be assessed 
by the deviations stemming from the different initialization scales of $1.25\ \mathrm{GeV}$ 
and $1.3\ \mathrm{GeV}$, compare the corresponding curves (even larger initialization 
scales of e.g.\ $1.4\ \mathrm{GeV}$ or $\mathrm{1.5}\ \mathrm{GeV}$ are not considered
as the coefficient $\alpha_{2}$ becomes negative there, signaling inaccuracies in the
employed approach). In total, we understand these observed differences in the IR $\sigma$-mass 
mainly as consequences of neglecting the higher Taylor coefficients at the initialization scale 
$\Lambda_{\chi} = 1.2\ \mathrm{GeV}$.\ Nevertheless, the MF integration apparently generates 
a large isoscalar mass in the IR ($M_{\sigma} > 1\ \mathrm{GeV}$) and this particular 
feature is independent of the choice of the initialization and reference scales as studied in 
Fig.\ \ref{fig:alpha}(f).

\begin{figure}[t]
	\centering
	\includegraphics[scale=1.0]{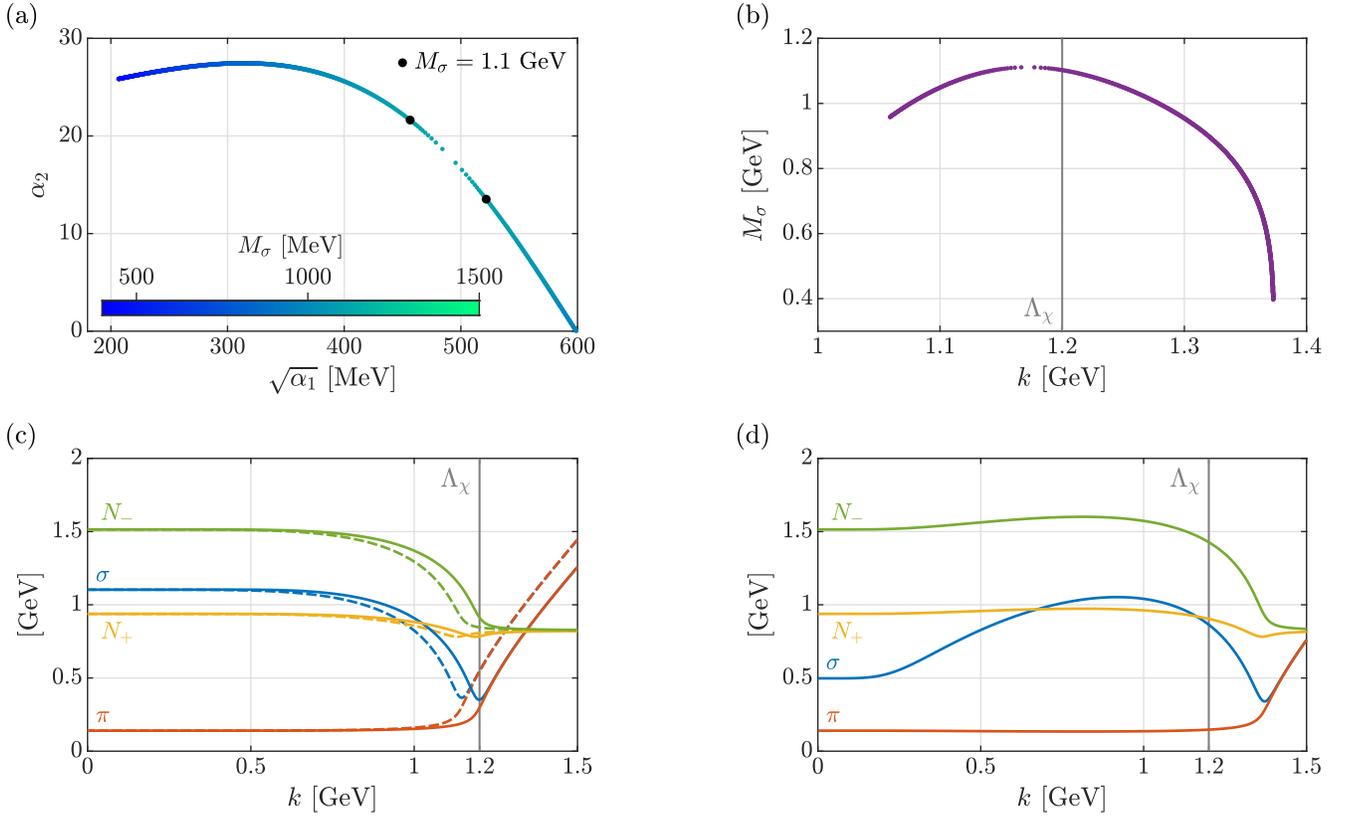}
	\caption{One-loop integration and determination of the initial Taylor 
	coefficients. (a) Numeric roots of Eqs.\ (\ref{eq:deta1}) and 
	(\ref{eq:deta2}) as functions of the IR isoscalar mass $M_{\sigma}$
	within the one-loop approximation (rescaled to $\sqrt{\alpha_{1}}$ and 
	$\alpha_{2}$); the color code covers the range of $400\ \mathrm{MeV} \le 
	M_{\sigma} \le 1500\ \mathrm{MeV}$ and the black dots indicate the two 
	solutions found for $M_{\sigma} = 1.1\ \mathrm{GeV}$. (b) Numeric determination 
	of the isoscalar mass $M_{\sigma}$ as a function of the $k$-value of the chiral-symmetry 
	breaking scale. (c) and (d) Cases of two/one solution(s) to Eqs.\ (\ref{eq:deta1}) 
	and (\ref{eq:deta2}) for a given value of the $\sigma$-mass [(c) 
	$M_{\sigma} = 1103\ \mathrm{MeV}$; (d) $M_{\sigma} = 498\ \mathrm{MeV}$].}
	\label{fig:Msigma_oneloop}
\end{figure}

In the one-loop approximation, the Taylor coefficients $\alpha_{1,\Lambda_{\chi}}$ and 
$\alpha_{2,\Lambda_{\chi}}$ enter the right sides of Eqs.\ (\ref{eq:deta1}) and (\ref{eq:deta2}) 
through the loop function $F_{\Lambda_{\chi}}$, which includes the bosonic loop contribution
[the boson masses in Eqs.\ (\ref{eq:pionmass}) and (\ref{eq:sigmamass}) depend on $\alpha_{1}$
and $\alpha_{2}$], i.e.,
\begin{IEEEeqnarray}{rCl}
	F_{\Lambda_{\chi}}\!\left(\sigma^{2}\right) & \equiv & 
	F_{\Lambda_{\chi}}^{\mathrm{MF}}\!\left(\sigma^{2}\right)
	+ \frac{1}{16 \pi^{2}} \left\lbrace 
	3 \left[\alpha_{1,\Lambda_{\chi}} + \alpha_{2,\Lambda_{\chi}}
	\left(\sigma^{2} - \varphi_{0}^{2}\right)\right]^{2} 
	\left[\frac{\Lambda_{\chi}^{2}}{2 \left[\alpha_{1,\Lambda_{\chi}}  
	+ \alpha_{2,\Lambda_{\chi}}\left(\sigma^{2} - \varphi_{0}^{2}\right)\right]}
	\right. \right. \nonumber\\[0.3cm] & & \hspace{6cm}
	+\left. \ln \frac{2 \left[\alpha_{1,\Lambda_{\chi}} + \alpha_{2,\Lambda_{\chi}}
	\left(\sigma^{2} - \varphi_{0}^{2}\right)\right]}{\Lambda_{\chi}^{2}
	+ 2 \left[\alpha_{1,\Lambda_{\chi}} + \alpha_{2,\Lambda_{\chi}}
	\left(\sigma^{2} - \varphi_{0}^{2}\right)\right]} \right]
	\nonumber\\[0.3cm] & & \hspace{3cm}
	+ \left[\alpha_{1,\Lambda_{\chi}} + \alpha_{2,\Lambda_{\chi}}
	\left(3 \sigma^{2} - \varphi_{0}^{2}\right)\right]^{2} 
	\left[\frac{\Lambda_{\chi}^{2}}{2 \left[\alpha_{1,\Lambda_{\chi}}  
	+ \alpha_{2,\Lambda_{\chi}}\left(3 \sigma^{2} - \varphi_{0}^{2}\right)\right]}
	\right. \nonumber\\[0.3cm] & & \hspace{6cm}
	+ \left. \left. \! \ln \frac{2 \left[\alpha_{1,\Lambda_{\chi}} + \alpha_{2,\Lambda_{\chi}}
	\left(3 \sigma^{2} - \varphi_{0}^{2}\right)\right]}{\Lambda_{\chi}^{2}
	+ 2 \left[\alpha_{1,\Lambda_{\chi}} + \alpha_{2,\Lambda_{\chi}}
	\left(3 \sigma^{2} - \varphi_{0}^{2}\right)\right]} \right] \right\rbrace .
	\label{eq:loopfunone}
\end{IEEEeqnarray}
In contrast, the fermion masses (\ref{eq:fermionmasses}) and the corresponding loop function
in the MF approximation are independent of $\alpha_{1}$ and $\alpha_{2}$. With the bosonic 
loop function in Eq.\ (\ref{eq:loopfunone}) containing the Taylor coefficients, it is more 
complicated to solve Eqs.\ (\ref{eq:deta1}) and (\ref{eq:deta2}) for $\alpha_{1}$ and $\alpha_{2}$.\ 
We hence formulate these equations as a root-finding problem and numerically search for solutions.\ 
The technical details of this implementation are listed in Table \ref{tab:root}. Figure 
\ref{fig:Msigma_oneloop}(a) collects the numeric roots as a function of $M_{\sigma}$.\ For a 
small value of $M_{\sigma} \sim 500\ \mathrm{MeV}$, the roots are located in the top-left corner 
and tend towards a larger $\alpha_{1}$ for growing $M_{\sigma}$ (roughly up to $M_{\sigma} 
\sim 950\ \mathrm{MeV}$), whereas the value of the coefficient $\alpha_{2}$ does not change 
drastically.\ Starting from $M_{\sigma} \sim 950\ \mathrm{MeV}$, we find two solutions to the 
system of equations, both of which are physically equivalent.\ This means that the two parameter
sets consisting of $\alpha_{1}$ and $\alpha_{2}$ give the same physical outcome in the IR; they only 
disagree in the higher Taylor coefficients ($\alpha_{n}$, $n > 2$), which do not affect the
first and second derivatives of the effective potential at $\varphi_{0}^{2} \simeq f_{\pi}^{2}$.
Going to even higher isoscalar masses, the two distinct solutions approach each other from the
top-left and the bottom-right. Above values of $M_{\sigma} \sim 1110\ \mathrm{MeV}$, we no longer 
find numeric solutions.
\begin{table*}
	\caption{\label{tab:root}Implementation of numeric root finding in the
	one-loop approximation.\ We employ the dimensionless rescaled variables 
	$\kappa_{1}$ and $\kappa_{2}$ (as defined in the first column). The
	variables $\kappa_{1}$ and $\kappa_{2}$ are sampled in equal spaces of 
	their fourth and square root, respectively (with $\kappa_{1,\mathrm{min}} 
	= \varepsilon$ and $\kappa_{1,\mathrm{max}} = 1$; $\kappa_{2,\mathrm{min}} 
	= \varepsilon$ and $\kappa_{2,\mathrm{max}} = 1$; $\epsilon = 10^{-6}$).
	The equations are numerically solved for isoscalar masses in the interval
	of $400\ \mathrm{MeV} \le M_{\sigma} \le 1500\ \mathrm{MeV}$ (in steps
	of $1\ \mathrm{MeV}$; cf.\ Fig.\ \ref{fig:Msigma_oneloop}).}
	\begin{ruledtabular}
		\begin{tabular}{llcc}
		& & & Number of sampling points \\
		Variables & \multicolumn{1}{c}{Equations} & Sampling intervals 
		& (equally spaced) \\
		\colrule\\[-0.25cm]
		$\kappa_{1} = \alpha_{1}/\Lambda_{\chi}^{2}$ &
		Eq.\ (\ref{eq:deta1}): 
		$\kappa_{1} - \left(M_{\pi}^{2}/2 - 
		F_{\Lambda_{\chi}}'\right)/\Lambda_{\chi}^{2} = 0$ & 
		$\kappa_{1}^{1/4} \in \left[\varepsilon^{1/4},1\right]$ & 
		$N_{\kappa_{1}} = 100$ \\[0.15cm]
		$\kappa_{2} = \alpha_{2}f_{\pi}^{2}/\Lambda_{\chi}^{2}$ &
		Eq.\ (\ref{eq:deta2}): 
		$\kappa_{2} - \left[\left(M_{\sigma}^{2} - M_{\pi}^{2}\right)/4 
		- F_{\Lambda_{\chi}}'' f_{\pi}^{2}\right]/\Lambda_{\chi}^{2} = 0$ & 
		$\kappa_{2}^{1/2} \in \left[\varepsilon^{1/2},1\right]$ & 
		$N_{\kappa_{2}} = 100$
		\end{tabular}
	\end{ruledtabular}
\end{table*}

Transferring these findings to the determination of the isoscalar mass $M_{\sigma}$ by adjusting 
the chiral-symmetry breaking scale, we see in Fig.\ \ref{fig:Msigma_oneloop}(b) that the local 
minimum in the $M_{\sigma}$-evolution (for $k > 1\ \mathrm{GeV}$) coincides with the scale 
$\Lambda_{\chi}$ at $M_{\sigma} \simeq 1.1\ \mathrm{GeV}$ [indicated as black dots in Fig.\ 
\ref{fig:Msigma_oneloop}(a)]. The second solutions above $M_{\sigma} \simeq 950\ \mathrm{MeV}$ add 
the data points in the top-left corner, exhibiting a breaking scale below $\Lambda_{\chi}$.\ Contrarily, 
for light isoscalar masses ($M_{\sigma} \sim 500\ \mathrm{MeV}$), the breaking scale is moved to
larger $k$-values, ending up at $k \simeq 1370\ \mathrm{MeV}$ for $M_{\sigma} \simeq 400\ 
\mathrm{MeV}$. Figures \ref{fig:Msigma_oneloop}(c) and \ref{fig:Msigma_oneloop}(d) represent the 
case with two equivalent solutions ($M_{\sigma} \sim 1.1\ \mathrm{GeV}$) and the one with a single 
solution ($M_{\sigma} \simeq 500\ \mathrm{MeV}$), respectively.\ Similar to the FRG flows in Fig.\ 
\ref{fig:masses}, the $M_{\sigma}$-evolution in Fig.\ \ref{fig:Msigma_oneloop}(d) shows a 
``down-bending,'' but the actual chiral breaking scale does not match the requested value of 
$\Lambda_{\chi}$.\ In summary, the large $\sigma$-masses in the MF and one-loop approximations 
result from insisting on chiral symmetry breaking occurring at $\Lambda_{\chi} = 1.2\ \mathrm{GeV}
\simeq 4\pi f_{\pi}$ and vice versa.

\section{Higher-derivative pion self-interactions}
\label{sec:LECs}

The elimination of the $\theta$-field by the EOM (\ref{eq:eom}) generates various higher-derivative
pion self-interactions. Additionally to Eq.\ (\ref{eq:termstructures}), one finds $13$ (possible) 
structures at $\mathcal{O}\!\left(p^{4}\right)$,
\begin{IEEEeqnarray}{rCl}
	\mathcal{O}\!\left(p^{4}\right)\!\colon & \quad & 
	\Pi^{2} \Pi\cdot \Box^{2}\Pi, \quad
	\Pi^{2} \left(\partial_{\mu}\Pi\right)\cdot \partial^{\mu}\Box \Pi, \quad
	\Pi^{2} \left(\Box\Pi\right)^{2} , \quad
	\Pi^{2} \left(\partial_{\mu}\partial_{\nu}\Pi\right)^{2}, \quad
	\Pi\cdot \left(\partial_{\mu}\Pi\right) 
	\Pi \cdot \partial^{\mu}\Box \Pi, \nonumber\\[0.1cm]
	& & \Pi \cdot \left(\partial_{\mu}\Pi\right) 
	\left(\partial^{\mu}\Pi\right) \cdot \Box \Pi, \quad
	\Pi \cdot \left(\partial_{\mu}\Pi\right) 
	\left(\partial_{\nu}\Pi\right) \cdot \partial_{\mu}\partial_{\nu}\Pi, \quad
	\left(\Pi \cdot \Box\Pi\right)^{2}, \quad
	\Pi \cdot \left(\Box\Pi\right) \left(\partial_{\mu}\Pi\right)^{2}, \nonumber\\[0.1cm]
	& & \left(\Pi\cdot \partial_{\mu}\partial_{\nu}\Pi\right)^{2}, \quad
	\Pi\cdot \left(\partial_{\mu}\partial_{\nu}\Pi\right)
	\left(\partial_{\mu}\Pi\right)\cdot\left(\partial_{\nu}\Pi\right), \quad
	\left[\left(\partial_{\mu}\Pi\right) \cdot \partial^{\mu}\Pi\right]^{2}, \quad
	\left[\left(\partial_{\mu}\Pi\right) \cdot \partial_{\nu}\Pi\right]^{2}, \quad
	\label{eq:termstructuresp4}
\end{IEEEeqnarray}
as well as $54$, $221$, and $883$ possibilities at $\smash{\mathcal{O}\!\left(p^{6}\right)}$, 
$\smash{\mathcal{O}\!\left(p^{8}\right)}$, and $\smash{\mathcal{O}\!\left(p^{10}\right)}$, 
respectively [due to the large numbers, we do not list the structures of
$\smash{\mathcal{O}\!\left(p^{6}\right)}$ and higher]. As already advertised in Sec.\ 
\ref{sec:scattering}, we obtain $3335$ possibilities at $\smash{\mathcal{O}\!
\left(p^{12}\right)}$, where we finally stopped the expansion in the pion momentum $p$.
In order to keep track of the large amount of terms, we employed again the powerful
algebra tool \texttt{FeynCalc} \cite{Mertig:1990an, *Shtabovenko:2016sxi, *Shtabovenko:2020gxv}.

The pion self-interactions of the effective action $\Gamma_{\mathrm{sol}}$, listed in Eqs.\ 
(\ref{eq:termstructures}) and (\ref{eq:termstructuresp4}), are parametrized by the (nonzero) 
low-energy couplings
\begin{IEEEeqnarray}{rCl}
	\mathcal{O}\!\left(p^{0}\right)\!\colon & \quad & 
	\mathfrak{C}_{1} = \frac{M_{\pi}^{2}(\epsilon + 1)}{8 f_{\pi}^{2}}, \\[0.2cm]
	\mathcal{O}\!\left(p^{2}\right)\!\colon & \quad & 
	\mathfrak{C}_{2} = - \frac{2\epsilon + 1}{4 f_{\pi}^{2}} , \qquad
	\mathfrak{C}_{3} = \frac{\epsilon^{2}}{2 f_{\pi}^{2}}, \\[0.2cm]
	\mathcal{O}\!\left(p^{4}\right)\!\colon & \quad & 
	\mathfrak{C}_{5} \equiv \mathfrak{C}_{6} \equiv \mathfrak{C}_{9} 
	= - \frac{\epsilon^{3}}{f_{\pi}^{2} M_{\pi}^{2}}, \qquad
	\mathfrak{C}_{7} = - \frac{2\epsilon^{2}(\epsilon + 1)}{f_{\pi}^{2} M_{\pi}^{2}}, \qquad
	\mathfrak{C}_{8} = - \frac{\epsilon^{3}}{2 f_{\pi}^{2} M_{\pi}^{2}}, \qquad
	\mathfrak{C}_{12} = \frac{\epsilon (1 - \epsilon^{2})}{2 f_{\pi}^{2} M_{\pi}^{2}}, \qquad \\
	& \vdots & \nonumber
\end{IEEEeqnarray}
where the indices of the couplings refer to the respective sortings of the structures in Eqs.\ 
(\ref{eq:termstructures}) and (\ref{eq:termstructuresp4}). Hence, only a specific coordinate-dependent
subset of all possibilities is generated.\footnote{The generated subset of term structures can considerably 
be reduced by means of partial-integration identities \cite{Divotgey:2019xea, Eser:2020phd}, which
however is irrelevant for the analytic and numeric results for the scattering lengths.} Again, we do 
not give the numerous couplings of $\smash{\mathcal{O}\!\left(p^{6}\right)}$ and higher---the overall 
strategy remains the same for all $p$-orders. The corresponding contributions to the scattering lengths 
read
\begin{equation}
	\begin{aligned}
	\mathcal{O}\!\left(p^{0}\right)\!\colon & \quad & 
	\Delta_{0}a_{0}^{0} & = \frac{5}{4\pi}\;\! \mathfrak{C}_{1}, \qquad &
	\Delta_{0}a_{0}^{2} & = \frac{1}{2\pi}\;\! \mathfrak{C}_{1}, \\[0.1cm]
	\mathcal{O}\!\left(p^{2}\right)\!\colon & \quad &
	\Delta_{2}a_{0}^{0} & = - \frac{M_{\pi}^{2}}{4\pi}\left(\mathfrak{C}_{2} 
	- 3 \mathfrak{C}_{3}\right), \qquad &
	\Delta_{2}a_{0}^{2} & = \frac{M_{\pi}^{2}}{2\pi}\;\! \mathfrak{C}_{2}, \\[0.1cm]
	\mathcal{O}\!\left(p^{4}\right)\!\colon & \quad &
	\Delta_{4}a_{0}^{0} & = \frac{M_{\pi}^{4}}{4\pi}\left[5 
	\left(\mathfrak{C}_{8} + \mathfrak{C}_{12}\right)
	+ \mathfrak{C}_{9} - 3 \left(\mathfrak{C}_{5} + \mathfrak{C}_{6} 
	+ \mathfrak{C}_{7}\right)\right], \qquad &
	\Delta_{4}a_{0}^{2} & = \frac{M_{\pi}^{4}}{2\pi}
	\left(\mathfrak{C}_{8} - \mathfrak{C}_{9} + \mathfrak{C}_{12}\right),
	\end{aligned}
\end{equation}
such that
\begin{IEEEeqnarray}{rCl}
	a_{0}^{0} & = & \sum_{i\,\mathrm{even}} \Delta_{i} a_{0}^{0}
	= \frac{M_{\pi}^{2}}{32 \pi f_{\pi}^{2}}\left(7 + 29\epsilon 
	+ 60\epsilon^{2} + 48\epsilon^{3} \right) + \cdots , \\[0.2cm]
	a_{0}^{2} & = & \sum_{i\,\mathrm{even}} \Delta_{i} a_{0}^{2}
	= - \frac{M_{\pi}^{2}}{16 \pi f_{\pi}^{2}}\left(1 - \epsilon\right).
\end{IEEEeqnarray}
These are the $\mathcal{O}(\epsilon)$-accurate scattering lengths, cf.\ Eqs.\ (\ref{eq:eomexpansionp40})
and (\ref{eq:eomexpansionp42}). In the limit $M_{\sigma} \rightarrow \infty$ ($\epsilon \rightarrow 0$ 
for nonzero and finite $M_{\pi}$), we have
\begin{equation}
	\mathcal{O}\!\left(p^{0}\right)\!\colon \quad \lim_{\epsilon\, \rightarrow\, 0} 
	\mathfrak{C}_{1} = \frac{M_{\pi}^{2}}{8 f_{\pi}^{2}}
	\equiv \frac{\tilde{h}}{8 f_{\pi}^{3}}, \qquad
	\mathcal{O}\!\left(p^{2}\right)\!\colon \quad \lim_{\epsilon\, \rightarrow\, 0} 
	\mathfrak{C}_{2} = - \frac{1}{4 f_{\pi}^{2}},
\end{equation}
which are the unique low-energy couplings of the NLSM (with explicit chiral symmetry breaking)
up to $\smash{\mathcal{O}\!\left(\Pi^{4},\partial^{2}\right)}$ in stereographic pions, see once 
more Eqs.\ (\ref{eq:nlsm}) to (\ref{eq:mpion}).

\bibliography{references}

\begin{thebibliography}{210}%
\makeatletter
\providecommand \@ifxundefined [1]{%
 \@ifx{#1\undefined}
}%
\providecommand \@ifnum [1]{%
 \ifnum #1\expandafter \@firstoftwo
 \else \expandafter \@secondoftwo
 \fi
}%
\providecommand \@ifx [1]{%
 \ifx #1\expandafter \@firstoftwo
 \else \expandafter \@secondoftwo
 \fi
}%
\providecommand \natexlab [1]{#1}%
\providecommand \enquote  [1]{``#1''}%
\providecommand \bibnamefont  [1]{#1}%
\providecommand \bibfnamefont [1]{#1}%
\providecommand \citenamefont [1]{#1}%
\providecommand \href@noop [0]{\@secondoftwo}%
\providecommand \href [0]{\begingroup \@sanitize@url \@href}%
\providecommand \@href[1]{\@@startlink{#1}\@@href}%
\providecommand \@@href[1]{\endgroup#1\@@endlink}%
\providecommand \@sanitize@url [0]{\catcode `\\12\catcode `\$12\catcode
  `\&12\catcode `\#12\catcode `\^12\catcode `\_12\catcode `\%12\relax}%
\providecommand \@@startlink[1]{}%
\providecommand \@@endlink[0]{}%
\providecommand \url  [0]{\begingroup\@sanitize@url \@url }%
\providecommand \@url [1]{\endgroup\@href {#1}{\urlprefix }}%
\providecommand \urlprefix  [0]{URL }%
\providecommand \Eprint [0]{\href }%
\providecommand \doibase [0]{http://dx.doi.org/}%
\providecommand \selectlanguage [0]{\@gobble}%
\providecommand \bibinfo  [0]{\@secondoftwo}%
\providecommand \bibfield  [0]{\@secondoftwo}%
\providecommand \translation [1]{[#1]}%
\providecommand \BibitemOpen [0]{}%
\providecommand \bibitemStop [0]{}%
\providecommand \bibitemNoStop [0]{.\EOS\space}%
\providecommand \EOS [0]{\spacefactor3000\relax}%
\providecommand \BibitemShut  [1]{\csname bibitem#1\endcsname}%
\let\auto@bib@innerbib\@empty
%</preamble>
\bibitem [{\citenamefont {Heisenberg}(1932)}]{Heisenberg:1932dw}%
  \BibitemOpen
  \bibfield  {author} {\bibinfo {author} {\bibfnamefont {W.}~\bibnamefont
  {Heisenberg}},\ }\href {\doibase 10.1007/BF01342433} {\bibfield  {journal}
  {\bibinfo  {journal} {Z. Phys.}\ }\textbf {\bibinfo {volume} {77}},\ \bibinfo
  {pages} {1} (\bibinfo {year} {1932})}\BibitemShut {NoStop}%
\bibitem [{\citenamefont {Wigner}(1937)}]{Wigner:1936dx}%
  \BibitemOpen
  \bibfield  {author} {\bibinfo {author} {\bibfnamefont {E.}~\bibnamefont
  {Wigner}},\ }\href {\doibase 10.1103/PhysRev.51.106} {\bibfield  {journal}
  {\bibinfo  {journal} {Phys. Rev.}\ }\textbf {\bibinfo {volume} {51}},\
  \bibinfo {pages} {106} (\bibinfo {year} {1937})}\BibitemShut {NoStop}%
\bibitem [{\citenamefont {Zyla}\ \emph {et~al.}(2020)\citenamefont {Zyla} \emph
  {et~al.}}]{ParticleDataGroup:2020ssz}%
  \BibitemOpen
  \bibfield  {author} {\bibinfo {author} {\bibfnamefont {P.~A.}\ \bibnamefont
  {Zyla}} \emph {et~al.} (\bibinfo {collaboration} {Particle Data Group}),\
  }\href {\doibase 10.1093/ptep/ptaa104} {\bibfield  {journal} {\bibinfo
  {journal} {PTEP}\ }\textbf {\bibinfo {volume} {2020}},\ \bibinfo {pages}
  {083C01} (\bibinfo {year} {2020})}\BibitemShut {NoStop}%
\bibitem [{\citenamefont {Schwinger}(1957)}]{Schwinger:1957em}%
  \BibitemOpen
  \bibfield  {author} {\bibinfo {author} {\bibfnamefont {J.~S.}\ \bibnamefont
  {Schwinger}},\ }\href {\doibase 10.1016/0003-4916(57)90015-5} {\bibfield
  {journal} {\bibinfo  {journal} {Annals Phys.}\ }\textbf {\bibinfo {volume}
  {2}},\ \bibinfo {pages} {407} (\bibinfo {year} {1957})}\BibitemShut {NoStop}%
\bibitem [{\citenamefont {Gell-Mann}\ and\ \citenamefont
  {Levy}(1960)}]{Gell-Mann:1960mvl}%
  \BibitemOpen
  \bibfield  {author} {\bibinfo {author} {\bibfnamefont {M.}~\bibnamefont
  {Gell-Mann}}\ and\ \bibinfo {author} {\bibfnamefont {M.}~\bibnamefont
  {Levy}},\ }\href {\doibase 10.1007/BF02859738} {\bibfield  {journal}
  {\bibinfo  {journal} {Nuovo Cim.}\ }\textbf {\bibinfo {volume} {16}},\
  \bibinfo {pages} {705} (\bibinfo {year} {1960})}\BibitemShut {NoStop}%
\bibitem [{\citenamefont {Weinberg}(1967)}]{Weinberg:1966fm}%
  \BibitemOpen
  \bibfield  {author} {\bibinfo {author} {\bibfnamefont {S.}~\bibnamefont
  {Weinberg}},\ }\href {\doibase 10.1103/PhysRevLett.18.188} {\bibfield
  {journal} {\bibinfo  {journal} {Phys. Rev. Lett.}\ }\textbf {\bibinfo
  {volume} {18}},\ \bibinfo {pages} {188} (\bibinfo {year} {1967})}\BibitemShut
  {NoStop}%
\bibitem [{\citenamefont {Gursey}(1960)}]{Gursey:1959yy}%
  \BibitemOpen
  \bibfield  {author} {\bibinfo {author} {\bibfnamefont {F.}~\bibnamefont
  {Gursey}},\ }\href {\doibase 10.1007/BF02860276} {\bibfield  {journal}
  {\bibinfo  {journal} {Nuovo Cim.}\ }\textbf {\bibinfo {volume} {16}},\
  \bibinfo {pages} {230} (\bibinfo {year} {1960})}\BibitemShut {NoStop}%
\bibitem [{\citenamefont {Schwinger}(1967)}]{Schwinger:1967tc}%
  \BibitemOpen
  \bibfield  {author} {\bibinfo {author} {\bibfnamefont {J.~S.}\ \bibnamefont
  {Schwinger}},\ }\href {\doibase 10.1016/0370-2693(67)90277-8} {\bibfield
  {journal} {\bibinfo  {journal} {Phys. Lett. B}\ }\textbf {\bibinfo {volume}
  {24}},\ \bibinfo {pages} {473} (\bibinfo {year} {1967})}\BibitemShut
  {NoStop}%
\bibitem [{\citenamefont {Weinberg}(1968)}]{Weinberg:1968de}%
  \BibitemOpen
  \bibfield  {author} {\bibinfo {author} {\bibfnamefont {S.}~\bibnamefont
  {Weinberg}},\ }\href {\doibase 10.1103/PhysRev.166.1568} {\bibfield
  {journal} {\bibinfo  {journal} {Phys. Rev.}\ }\textbf {\bibinfo {volume}
  {166}},\ \bibinfo {pages} {1568} (\bibinfo {year} {1968})}\BibitemShut
  {NoStop}%
\bibitem [{\citenamefont {Zinn-Justin}(1996)}]{Zinn-Justin:1996khx}%
  \BibitemOpen
  \bibfield  {author} {\bibinfo {author} {\bibfnamefont {J.}~\bibnamefont
  {Zinn-Justin}},\ }\href@noop {} {\bibfield  {journal} {\bibinfo  {journal}
  {Int. Ser. Monogr. Phys.}\ }\textbf {\bibinfo {volume} {92}},\ \bibinfo
  {pages} {1} (\bibinfo {year} {1996})}\BibitemShut {NoStop}%
\bibitem [{\citenamefont {Gasser}\ and\ \citenamefont
  {Leutwyler}(1984)}]{Gasser:1983yg}%
  \BibitemOpen
  \bibfield  {author} {\bibinfo {author} {\bibfnamefont {J.}~\bibnamefont
  {Gasser}}\ and\ \bibinfo {author} {\bibfnamefont {H.}~\bibnamefont
  {Leutwyler}},\ }\href {\doibase 10.1016/0003-4916(84)90242-2} {\bibfield
  {journal} {\bibinfo  {journal} {Annals Phys.}\ }\textbf {\bibinfo {volume}
  {158}},\ \bibinfo {pages} {142} (\bibinfo {year} {1984})}\BibitemShut
  {NoStop}%
\bibitem [{\citenamefont {Gasser}\ and\ \citenamefont
  {Leutwyler}(1985)}]{Gasser:1984gg}%
  \BibitemOpen
  \bibfield  {author} {\bibinfo {author} {\bibfnamefont {J.}~\bibnamefont
  {Gasser}}\ and\ \bibinfo {author} {\bibfnamefont {H.}~\bibnamefont
  {Leutwyler}},\ }\href {\doibase 10.1016/0550-3213(85)90492-4} {\bibfield
  {journal} {\bibinfo  {journal} {Nucl. Phys. B}\ }\textbf {\bibinfo {volume}
  {250}},\ \bibinfo {pages} {465} (\bibinfo {year} {1985})}\BibitemShut
  {NoStop}%
\bibitem [{\citenamefont {Leutwyler}(1994)}]{Leutwyler:1993iq}%
  \BibitemOpen
  \bibfield  {author} {\bibinfo {author} {\bibfnamefont {H.}~\bibnamefont
  {Leutwyler}},\ }\href {\doibase 10.1006/aphy.1994.1094} {\bibfield  {journal}
  {\bibinfo  {journal} {Annals Phys.}\ }\textbf {\bibinfo {volume} {235}},\
  \bibinfo {pages} {165} (\bibinfo {year} {1994})},\ \Eprint
  {http://arxiv.org/abs/hep-ph/9311274} {arXiv:hep-ph/9311274} \BibitemShut
  {NoStop}%
\bibitem [{\citenamefont {Scherer}(2003)}]{Scherer:2002tk}%
  \BibitemOpen
  \bibfield  {author} {\bibinfo {author} {\bibfnamefont {S.}~\bibnamefont
  {Scherer}},\ }\href@noop {} {\bibfield  {journal} {\bibinfo  {journal} {Adv.
  Nucl. Phys.}\ }\textbf {\bibinfo {volume} {27}},\ \bibinfo {pages} {277}
  (\bibinfo {year} {2003})},\ \Eprint {http://arxiv.org/abs/hep-ph/0210398}
  {arXiv:hep-ph/0210398} \BibitemShut {NoStop}%
\bibitem [{\citenamefont {Bijnens}(2007)}]{Bijnens:2006zp}%
  \BibitemOpen
  \bibfield  {author} {\bibinfo {author} {\bibfnamefont {J.}~\bibnamefont
  {Bijnens}},\ }\href {\doibase 10.1016/j.ppnp.2006.08.002} {\bibfield
  {journal} {\bibinfo  {journal} {Prog. Part. Nucl. Phys.}\ }\textbf {\bibinfo
  {volume} {58}},\ \bibinfo {pages} {521} (\bibinfo {year} {2007})},\ \Eprint
  {http://arxiv.org/abs/hep-ph/0604043} {arXiv:hep-ph/0604043} \BibitemShut
  {NoStop}%
\bibitem [{\citenamefont {Leutwyler}(2012)}]{Leutwyler:2012sco}%
  \BibitemOpen
  \bibfield  {author} {\bibinfo {author} {\bibfnamefont {H.}~\bibnamefont
  {Leutwyler}},\ }\href {\doibase doi:10.4249/scholarpedia.8708} {\bibfield
  {journal} {\bibinfo  {journal} {Scholarpedia}\ }\textbf {\bibinfo {volume}
  {7}},\ \bibinfo {pages} {(10):8708} (\bibinfo {year} {2012})}\BibitemShut
  {NoStop}%
\bibitem [{\citenamefont {Coleman}\ \emph {et~al.}(1969)\citenamefont
  {Coleman}, \citenamefont {Wess},\ and\ \citenamefont
  {Zumino}}]{Coleman:1969sm}%
  \BibitemOpen
  \bibfield  {author} {\bibinfo {author} {\bibfnamefont {S.~R.}\ \bibnamefont
  {Coleman}}, \bibinfo {author} {\bibfnamefont {J.}~\bibnamefont {Wess}}, \
  and\ \bibinfo {author} {\bibfnamefont {B.}~\bibnamefont {Zumino}},\ }\href
  {\doibase 10.1103/PhysRev.177.2239} {\bibfield  {journal} {\bibinfo
  {journal} {Phys. Rev.}\ }\textbf {\bibinfo {volume} {177}},\ \bibinfo {pages}
  {2239} (\bibinfo {year} {1969})}\BibitemShut {NoStop}%
\bibitem [{\citenamefont {Callan}\ \emph {et~al.}(1969)\citenamefont {Callan},
  \citenamefont {Coleman}, \citenamefont {Wess},\ and\ \citenamefont
  {Zumino}}]{Callan:1969sn}%
  \BibitemOpen
  \bibfield  {author} {\bibinfo {author} {\bibfnamefont {C.~G.}\ \bibnamefont
  {Callan}, \bibfnamefont {Jr.}}, \bibinfo {author} {\bibfnamefont {S.~R.}\
  \bibnamefont {Coleman}}, \bibinfo {author} {\bibfnamefont {J.}~\bibnamefont
  {Wess}}, \ and\ \bibinfo {author} {\bibfnamefont {B.}~\bibnamefont
  {Zumino}},\ }\href {\doibase 10.1103/PhysRev.177.2247} {\bibfield  {journal}
  {\bibinfo  {journal} {Phys. Rev.}\ }\textbf {\bibinfo {volume} {177}},\
  \bibinfo {pages} {2247} (\bibinfo {year} {1969})}\BibitemShut {NoStop}%
\bibitem [{\citenamefont {Meetz}(1969)}]{Meetz:1969as}%
  \BibitemOpen
  \bibfield  {author} {\bibinfo {author} {\bibfnamefont {K.}~\bibnamefont
  {Meetz}},\ }\href {\doibase 10.1063/1.1664881} {\bibfield  {journal}
  {\bibinfo  {journal} {J. Math. Phys.}\ }\textbf {\bibinfo {volume} {10}},\
  \bibinfo {pages} {589} (\bibinfo {year} {1969})}\BibitemShut {NoStop}%
\bibitem [{\citenamefont {Isham}(1969)}]{Isham:1969ci}%
  \BibitemOpen
  \bibfield  {author} {\bibinfo {author} {\bibfnamefont {C.~J.}\ \bibnamefont
  {Isham}},\ }\href {\doibase 10.1007/BF02755023} {\bibfield  {journal}
  {\bibinfo  {journal} {Nuovo Cim. A}\ }\textbf {\bibinfo {volume} {59}},\
  \bibinfo {pages} {356} (\bibinfo {year} {1969})}\BibitemShut {NoStop}%
\bibitem [{\citenamefont {Peczkis}(1977)}]{Peczkis:1977xw}%
  \BibitemOpen
  \bibfield  {author} {\bibinfo {author} {\bibfnamefont {M.}~\bibnamefont
  {Peczkis}},\ }\href {\doibase 10.1016/0034-4877(77)90062-3} {\bibfield
  {journal} {\bibinfo  {journal} {Rept. Math. Phys.}\ }\textbf {\bibinfo
  {volume} {11}},\ \bibinfo {pages} {211} (\bibinfo {year} {1977})}\BibitemShut
  {NoStop}%
\bibitem [{\citenamefont {Wetterich}(1993)}]{Wetterich:1992yh}%
  \BibitemOpen
  \bibfield  {author} {\bibinfo {author} {\bibfnamefont {C.}~\bibnamefont
  {Wetterich}},\ }\href {\doibase 10.1016/0370-2693(93)90726-X} {\bibfield
  {journal} {\bibinfo  {journal} {Phys. Lett. B}\ }\textbf {\bibinfo {volume}
  {301}},\ \bibinfo {pages} {90} (\bibinfo {year} {1993})},\ \Eprint
  {http://arxiv.org/abs/1710.05815} {arXiv:1710.05815 [hep-th]} \BibitemShut
  {NoStop}%
\bibitem [{\citenamefont {Ellwanger}(1994)}]{Ellwanger:1993mw}%
  \BibitemOpen
  \bibfield  {author} {\bibinfo {author} {\bibfnamefont {U.}~\bibnamefont
  {Ellwanger}},\ }\href {\doibase 10.1007/BF01555911} {\bibfield  {journal}
  {\bibinfo  {journal} {Z. Phys. C}\ }\textbf {\bibinfo {volume} {62}},\
  \bibinfo {pages} {503} (\bibinfo {year} {1994})},\ \Eprint
  {http://arxiv.org/abs/hep-ph/9308260} {arXiv:hep-ph/9308260} \BibitemShut
  {NoStop}%
\bibitem [{\citenamefont {Morris}(1994)}]{Morris:1993qb}%
  \BibitemOpen
  \bibfield  {author} {\bibinfo {author} {\bibfnamefont {T.~R.}\ \bibnamefont
  {Morris}},\ }\href {\doibase 10.1142/S0217751X94000972} {\bibfield  {journal}
  {\bibinfo  {journal} {Int. J. Mod. Phys. A}\ }\textbf {\bibinfo {volume}
  {9}},\ \bibinfo {pages} {2411} (\bibinfo {year} {1994})},\ \Eprint
  {http://arxiv.org/abs/hep-ph/9308265} {arXiv:hep-ph/9308265} \BibitemShut
  {NoStop}%
\bibitem [{\citenamefont {Pawlowski}(2007)}]{Pawlowski:2005xe}%
  \BibitemOpen
  \bibfield  {author} {\bibinfo {author} {\bibfnamefont {J.~M.}\ \bibnamefont
  {Pawlowski}},\ }\href {\doibase 10.1016/j.aop.2007.01.007} {\bibfield
  {journal} {\bibinfo  {journal} {Annals Phys.}\ }\textbf {\bibinfo {volume}
  {322}},\ \bibinfo {pages} {2831} (\bibinfo {year} {2007})},\ \Eprint
  {http://arxiv.org/abs/hep-th/0512261} {arXiv:hep-th/0512261} \BibitemShut
  {NoStop}%
\bibitem [{\citenamefont {Parganlija}\ \emph {et~al.}(2010)\citenamefont
  {Parganlija}, \citenamefont {Giacosa},\ and\ \citenamefont
  {Rischke}}]{Parganlija:2010fz}%
  \BibitemOpen
  \bibfield  {author} {\bibinfo {author} {\bibfnamefont {D.}~\bibnamefont
  {Parganlija}}, \bibinfo {author} {\bibfnamefont {F.}~\bibnamefont {Giacosa}},
  \ and\ \bibinfo {author} {\bibfnamefont {D.~H.}\ \bibnamefont {Rischke}},\
  }\href {\doibase 10.1103/PhysRevD.82.054024} {\bibfield  {journal} {\bibinfo
  {journal} {Phys. Rev. D}\ }\textbf {\bibinfo {volume} {82}},\ \bibinfo
  {pages} {054024} (\bibinfo {year} {2010})},\ \Eprint
  {http://arxiv.org/abs/1003.4934} {arXiv:1003.4934 [hep-ph]} \BibitemShut
  {NoStop}%
\bibitem [{\citenamefont {Parganlija}\ \emph {et~al.}(2013)\citenamefont
  {Parganlija}, \citenamefont {Kovacs}, \citenamefont {Wolf}, \citenamefont
  {Giacosa},\ and\ \citenamefont {Rischke}}]{Parganlija:2012fy}%
  \BibitemOpen
  \bibfield  {author} {\bibinfo {author} {\bibfnamefont {D.}~\bibnamefont
  {Parganlija}}, \bibinfo {author} {\bibfnamefont {P.}~\bibnamefont {Kovacs}},
  \bibinfo {author} {\bibfnamefont {G.}~\bibnamefont {Wolf}}, \bibinfo {author}
  {\bibfnamefont {F.}~\bibnamefont {Giacosa}}, \ and\ \bibinfo {author}
  {\bibfnamefont {D.~H.}\ \bibnamefont {Rischke}},\ }\href {\doibase
  10.1103/PhysRevD.87.014011} {\bibfield  {journal} {\bibinfo  {journal} {Phys.
  Rev. D}\ }\textbf {\bibinfo {volume} {87}},\ \bibinfo {pages} {014011}
  (\bibinfo {year} {2013})},\ \Eprint {http://arxiv.org/abs/1208.0585}
  {arXiv:1208.0585 [hep-ph]} \BibitemShut {NoStop}%
\bibitem [{\citenamefont {Divotgey}\ \emph {et~al.}(2018)\citenamefont
  {Divotgey}, \citenamefont {Kovacs}, \citenamefont {Giacosa},\ and\
  \citenamefont {Rischke}}]{Divotgey:2016pst}%
  \BibitemOpen
  \bibfield  {author} {\bibinfo {author} {\bibfnamefont {F.}~\bibnamefont
  {Divotgey}}, \bibinfo {author} {\bibfnamefont {P.}~\bibnamefont {Kovacs}},
  \bibinfo {author} {\bibfnamefont {F.}~\bibnamefont {Giacosa}}, \ and\
  \bibinfo {author} {\bibfnamefont {D.~H.}\ \bibnamefont {Rischke}},\ }\href
  {\doibase 10.1140/epja/i2018-12458-9} {\bibfield  {journal} {\bibinfo
  {journal} {Eur. Phys. J. A}\ }\textbf {\bibinfo {volume} {54}},\ \bibinfo
  {pages} {5} (\bibinfo {year} {2018})},\ \Eprint
  {http://arxiv.org/abs/1605.05154} {arXiv:1605.05154 [hep-ph]} \BibitemShut
  {NoStop}%
\bibitem [{\citenamefont {Lakaschus}\ \emph {et~al.}(2019)\citenamefont
  {Lakaschus}, \citenamefont {Mauldin}, \citenamefont {Giacosa},\ and\
  \citenamefont {Rischke}}]{Lakaschus:2018rki}%
  \BibitemOpen
  \bibfield  {author} {\bibinfo {author} {\bibfnamefont {P.}~\bibnamefont
  {Lakaschus}}, \bibinfo {author} {\bibfnamefont {J.~L.~P.}\ \bibnamefont
  {Mauldin}}, \bibinfo {author} {\bibfnamefont {F.}~\bibnamefont {Giacosa}}, \
  and\ \bibinfo {author} {\bibfnamefont {D.~H.}\ \bibnamefont {Rischke}},\
  }\href {\doibase 10.1103/PhysRevC.99.045203} {\bibfield  {journal} {\bibinfo
  {journal} {Phys. Rev. C}\ }\textbf {\bibinfo {volume} {99}},\ \bibinfo
  {pages} {045203} (\bibinfo {year} {2019})},\ \Eprint
  {http://arxiv.org/abs/1807.03735} {arXiv:1807.03735 [hep-ph]} \BibitemShut
  {NoStop}%
\bibitem [{\citenamefont {Bessis}\ and\ \citenamefont
  {Zinn-Justin}(1972)}]{Bessis:1972sn}%
  \BibitemOpen
  \bibfield  {author} {\bibinfo {author} {\bibfnamefont {D.}~\bibnamefont
  {Bessis}}\ and\ \bibinfo {author} {\bibfnamefont {J.}~\bibnamefont
  {Zinn-Justin}},\ }\href {\doibase 10.1103/PhysRevD.5.1313} {\bibfield
  {journal} {\bibinfo  {journal} {Phys. Rev. D}\ }\textbf {\bibinfo {volume}
  {5}},\ \bibinfo {pages} {1313} (\bibinfo {year} {1972})}\BibitemShut
  {NoStop}%
\bibitem [{\citenamefont {Jhung}\ and\ \citenamefont
  {Willey}(1974)}]{Jhung:1974fd}%
  \BibitemOpen
  \bibfield  {author} {\bibinfo {author} {\bibfnamefont {K.~S.}\ \bibnamefont
  {Jhung}}\ and\ \bibinfo {author} {\bibfnamefont {R.~S.}\ \bibnamefont
  {Willey}},\ }\href {\doibase 10.1103/PhysRevD.9.3132} {\bibfield  {journal}
  {\bibinfo  {journal} {Phys. Rev. D}\ }\textbf {\bibinfo {volume} {9}},\
  \bibinfo {pages} {3132} (\bibinfo {year} {1974})}\BibitemShut {NoStop}%
\bibitem [{\citenamefont {Appelquist}\ and\ \citenamefont
  {Bernard}(1981)}]{Appelquist:1980ae}%
  \BibitemOpen
  \bibfield  {author} {\bibinfo {author} {\bibfnamefont {T.}~\bibnamefont
  {Appelquist}}\ and\ \bibinfo {author} {\bibfnamefont {C.~W.}\ \bibnamefont
  {Bernard}},\ }\href {\doibase 10.1103/PhysRevD.23.425} {\bibfield  {journal}
  {\bibinfo  {journal} {Phys. Rev. D}\ }\textbf {\bibinfo {volume} {23}},\
  \bibinfo {pages} {425} (\bibinfo {year} {1981})}\BibitemShut {NoStop}%
\bibitem [{\citenamefont {Bentz}\ \emph {et~al.}(1997)\citenamefont {Bentz},
  \citenamefont {Matulla},\ and\ \citenamefont {Baier}}]{Bentz:1997ry}%
  \BibitemOpen
  \bibfield  {author} {\bibinfo {author} {\bibfnamefont {W.}~\bibnamefont
  {Bentz}}, \bibinfo {author} {\bibfnamefont {C.}~\bibnamefont {Matulla}}, \
  and\ \bibinfo {author} {\bibfnamefont {H.}~\bibnamefont {Baier}},\ }\href
  {\doibase 10.1103/PhysRevC.56.2280} {\bibfield  {journal} {\bibinfo
  {journal} {Phys. Rev. C}\ }\textbf {\bibinfo {volume} {56}},\ \bibinfo
  {pages} {2280} (\bibinfo {year} {1997})}\BibitemShut {NoStop}%
\bibitem [{\citenamefont {Brezin}\ and\ \citenamefont
  {Zinn-Justin}(1976{\natexlab{a}})}]{Brezin:1975sq}%
  \BibitemOpen
  \bibfield  {author} {\bibinfo {author} {\bibfnamefont {E.}~\bibnamefont
  {Brezin}}\ and\ \bibinfo {author} {\bibfnamefont {J.}~\bibnamefont
  {Zinn-Justin}},\ }\href {\doibase 10.1103/PhysRevLett.36.691} {\bibfield
  {journal} {\bibinfo  {journal} {Phys. Rev. Lett.}\ }\textbf {\bibinfo
  {volume} {36}},\ \bibinfo {pages} {691} (\bibinfo {year}
  {1976}{\natexlab{a}})}\BibitemShut {NoStop}%
\bibitem [{\citenamefont {Brezin}\ \emph {et~al.}(1976)\citenamefont {Brezin},
  \citenamefont {Zinn-Justin},\ and\ \citenamefont
  {Le~Guillou}}]{Brezin:1976ap}%
  \BibitemOpen
  \bibfield  {author} {\bibinfo {author} {\bibfnamefont {E.}~\bibnamefont
  {Brezin}}, \bibinfo {author} {\bibfnamefont {J.}~\bibnamefont {Zinn-Justin}},
  \ and\ \bibinfo {author} {\bibfnamefont {J.~C.}\ \bibnamefont {Le~Guillou}},\
  }\href {\doibase 10.1103/PhysRevD.14.2615} {\bibfield  {journal} {\bibinfo
  {journal} {Phys. Rev. D}\ }\textbf {\bibinfo {volume} {14}},\ \bibinfo
  {pages} {2615} (\bibinfo {year} {1976})}\BibitemShut {NoStop}%
\bibitem [{\citenamefont {Brezin}\ and\ \citenamefont
  {Zinn-Justin}(1976{\natexlab{b}})}]{Brezin:1976qa}%
  \BibitemOpen
  \bibfield  {author} {\bibinfo {author} {\bibfnamefont {E.}~\bibnamefont
  {Brezin}}\ and\ \bibinfo {author} {\bibfnamefont {J.}~\bibnamefont
  {Zinn-Justin}},\ }\href {\doibase 10.1103/PhysRevB.14.3110} {\bibfield
  {journal} {\bibinfo  {journal} {Phys. Rev. B}\ }\textbf {\bibinfo {volume}
  {14}},\ \bibinfo {pages} {3110} (\bibinfo {year}
  {1976}{\natexlab{b}})}\BibitemShut {NoStop}%
\bibitem [{\citenamefont {Eser}\ \emph {et~al.}(2018)\citenamefont {Eser},
  \citenamefont {Divotgey}, \citenamefont {Mitter},\ and\ \citenamefont
  {Rischke}}]{Eser:2018jqo}%
  \BibitemOpen
  \bibfield  {author} {\bibinfo {author} {\bibfnamefont {J.}~\bibnamefont
  {Eser}}, \bibinfo {author} {\bibfnamefont {F.}~\bibnamefont {Divotgey}},
  \bibinfo {author} {\bibfnamefont {M.}~\bibnamefont {Mitter}}, \ and\ \bibinfo
  {author} {\bibfnamefont {D.~H.}\ \bibnamefont {Rischke}},\ }\href {\doibase
  10.1103/PhysRevD.98.014024} {\bibfield  {journal} {\bibinfo  {journal} {Phys.
  Rev. D}\ }\textbf {\bibinfo {volume} {98}},\ \bibinfo {pages} {014024}
  (\bibinfo {year} {2018})},\ \Eprint {http://arxiv.org/abs/1804.01787}
  {arXiv:1804.01787 [hep-ph]} \BibitemShut {NoStop}%
\bibitem [{\citenamefont {Divotgey}\ \emph {et~al.}(2019)\citenamefont
  {Divotgey}, \citenamefont {Eser},\ and\ \citenamefont
  {Mitter}}]{Divotgey:2019xea}%
  \BibitemOpen
  \bibfield  {author} {\bibinfo {author} {\bibfnamefont {F.}~\bibnamefont
  {Divotgey}}, \bibinfo {author} {\bibfnamefont {J.}~\bibnamefont {Eser}}, \
  and\ \bibinfo {author} {\bibfnamefont {M.}~\bibnamefont {Mitter}},\ }\href
  {\doibase 10.1103/PhysRevD.99.054023} {\bibfield  {journal} {\bibinfo
  {journal} {Phys. Rev. D}\ }\textbf {\bibinfo {volume} {99}},\ \bibinfo
  {pages} {054023} (\bibinfo {year} {2019})},\ \Eprint
  {http://arxiv.org/abs/1901.02472} {arXiv:1901.02472 [hep-ph]} \BibitemShut
  {NoStop}%
\bibitem [{\citenamefont {Eser}\ \emph {et~al.}(2019)\citenamefont {Eser},
  \citenamefont {Divotgey},\ and\ \citenamefont {Mitter}}]{Eser:2019pvd}%
  \BibitemOpen
  \bibfield  {author} {\bibinfo {author} {\bibfnamefont {J.}~\bibnamefont
  {Eser}}, \bibinfo {author} {\bibfnamefont {F.}~\bibnamefont {Divotgey}}, \
  and\ \bibinfo {author} {\bibfnamefont {M.}~\bibnamefont {Mitter}},\ }\href
  {\doibase 10.22323/1.317.0060} {\bibfield  {journal} {\bibinfo  {journal}
  {PoS}\ }\textbf {\bibinfo {volume} {CD2018}},\ \bibinfo {pages} {060}
  (\bibinfo {year} {2019})},\ \Eprint {http://arxiv.org/abs/1902.04804}
  {arXiv:1902.04804 [hep-ph]} \BibitemShut {NoStop}%
\bibitem [{\citenamefont {Eser}(2020)}]{Eser:2020phd}%
  \BibitemOpen
  \bibfield  {author} {\bibinfo {author} {\bibfnamefont {J.}~\bibnamefont
  {Eser}},\ }\emph {\bibinfo {title} {Momentum-dependent pion self-interactions
  from quantum fluctuations}},\ \href@noop {} {\bibinfo {type}
  {Dissertation}},\ \bibinfo  {school} {Johann Wolfgang Goethe-Universit\"at
  Frankfurt am Main} (\bibinfo {year} {2020})\BibitemShut {NoStop}%
\bibitem [{\citenamefont {Divotgey}(2020)}]{Divotgey:2020phd}%
  \BibitemOpen
  \bibfield  {author} {\bibinfo {author} {\bibfnamefont {F.}~\bibnamefont
  {Divotgey}},\ }\emph {\bibinfo {title} {Niederenergiestudien effektiver
  Modelle stark wechselwirkender Systeme}},\ \href@noop {} {\bibinfo {type}
  {Dissertation}},\ \bibinfo  {school} {Johann Wolfgang Goethe-Universit\"at
  Frankfurt am Main} (\bibinfo {year} {2020})\BibitemShut {NoStop}%
\bibitem [{\citenamefont {Cichutek}\ \emph {et~al.}(2020)\citenamefont
  {Cichutek}, \citenamefont {Divotgey},\ and\ \citenamefont
  {Eser}}]{Cichutek:2020bli}%
  \BibitemOpen
  \bibfield  {author} {\bibinfo {author} {\bibfnamefont {N.}~\bibnamefont
  {Cichutek}}, \bibinfo {author} {\bibfnamefont {F.}~\bibnamefont {Divotgey}},
  \ and\ \bibinfo {author} {\bibfnamefont {J.}~\bibnamefont {Eser}},\ }\href
  {\doibase 10.1103/PhysRevD.102.034030} {\bibfield  {journal} {\bibinfo
  {journal} {Phys. Rev. D}\ }\textbf {\bibinfo {volume} {102}},\ \bibinfo
  {pages} {034030} (\bibinfo {year} {2020})},\ \Eprint
  {http://arxiv.org/abs/2006.12473} {arXiv:2006.12473 [hep-ph]} \BibitemShut
  {NoStop}%
\bibitem [{\citenamefont {Wilson}(1971{\natexlab{a}})}]{Wilson:1971bg}%
  \BibitemOpen
  \bibfield  {author} {\bibinfo {author} {\bibfnamefont {K.~G.}\ \bibnamefont
  {Wilson}},\ }\href {\doibase 10.1103/PhysRevB.4.3174} {\bibfield  {journal}
  {\bibinfo  {journal} {Phys. Rev. B}\ }\textbf {\bibinfo {volume} {4}},\
  \bibinfo {pages} {3174} (\bibinfo {year} {1971}{\natexlab{a}})}\BibitemShut
  {NoStop}%
\bibitem [{\citenamefont {Wilson}(1971{\natexlab{b}})}]{Wilson:1971dh}%
  \BibitemOpen
  \bibfield  {author} {\bibinfo {author} {\bibfnamefont {K.~G.}\ \bibnamefont
  {Wilson}},\ }\href {\doibase 10.1103/PhysRevB.4.3184} {\bibfield  {journal}
  {\bibinfo  {journal} {Phys. Rev. B}\ }\textbf {\bibinfo {volume} {4}},\
  \bibinfo {pages} {3184} (\bibinfo {year} {1971}{\natexlab{b}})}\BibitemShut
  {NoStop}%
\bibitem [{\citenamefont {Dupuis}\ \emph {et~al.}(2021)\citenamefont {Dupuis},
  \citenamefont {Canet}, \citenamefont {Eichhorn}, \citenamefont {Metzner},
  \citenamefont {Pawlowski}, \citenamefont {Tissier},\ and\ \citenamefont
  {Wschebor}}]{Dupuis:2020fhh}%
  \BibitemOpen
  \bibfield  {author} {\bibinfo {author} {\bibfnamefont {N.}~\bibnamefont
  {Dupuis}}, \bibinfo {author} {\bibfnamefont {L.}~\bibnamefont {Canet}},
  \bibinfo {author} {\bibfnamefont {A.}~\bibnamefont {Eichhorn}}, \bibinfo
  {author} {\bibfnamefont {W.}~\bibnamefont {Metzner}}, \bibinfo {author}
  {\bibfnamefont {J.~M.}\ \bibnamefont {Pawlowski}}, \bibinfo {author}
  {\bibfnamefont {M.}~\bibnamefont {Tissier}}, \ and\ \bibinfo {author}
  {\bibfnamefont {N.}~\bibnamefont {Wschebor}},\ }\href {\doibase
  10.1016/j.physrep.2021.01.001} {\bibfield  {journal} {\bibinfo  {journal}
  {Phys. Rept.}\ }\textbf {\bibinfo {volume} {910}},\ \bibinfo {pages} {1}
  (\bibinfo {year} {2021})},\ \Eprint {http://arxiv.org/abs/2006.04853}
  {arXiv:2006.04853 [cond-mat.stat-mech]} \BibitemShut {NoStop}%
\bibitem [{\citenamefont {Ellwanger}\ and\ \citenamefont
  {Wetterich}(1994)}]{Ellwanger:1994wy}%
  \BibitemOpen
  \bibfield  {author} {\bibinfo {author} {\bibfnamefont {U.}~\bibnamefont
  {Ellwanger}}\ and\ \bibinfo {author} {\bibfnamefont {C.}~\bibnamefont
  {Wetterich}},\ }\href {\doibase 10.1016/0550-3213(94)90568-1} {\bibfield
  {journal} {\bibinfo  {journal} {Nucl. Phys. B}\ }\textbf {\bibinfo {volume}
  {423}},\ \bibinfo {pages} {137} (\bibinfo {year} {1994})},\ \Eprint
  {http://arxiv.org/abs/hep-ph/9402221} {arXiv:hep-ph/9402221} \BibitemShut
  {NoStop}%
\bibitem [{\citenamefont {Gies}\ and\ \citenamefont
  {Wetterich}(2002)}]{Gies:2001nw}%
  \BibitemOpen
  \bibfield  {author} {\bibinfo {author} {\bibfnamefont {H.}~\bibnamefont
  {Gies}}\ and\ \bibinfo {author} {\bibfnamefont {C.}~\bibnamefont
  {Wetterich}},\ }\href {\doibase 10.1103/PhysRevD.65.065001} {\bibfield
  {journal} {\bibinfo  {journal} {Phys. Rev. D}\ }\textbf {\bibinfo {volume}
  {65}},\ \bibinfo {pages} {065001} (\bibinfo {year} {2002})},\ \Eprint
  {http://arxiv.org/abs/hep-th/0107221} {arXiv:hep-th/0107221} \BibitemShut
  {NoStop}%
\bibitem [{\citenamefont {Gies}\ and\ \citenamefont
  {Wetterich}(2004)}]{Gies:2002hq}%
  \BibitemOpen
  \bibfield  {author} {\bibinfo {author} {\bibfnamefont {H.}~\bibnamefont
  {Gies}}\ and\ \bibinfo {author} {\bibfnamefont {C.}~\bibnamefont
  {Wetterich}},\ }\href {\doibase 10.1103/PhysRevD.69.025001} {\bibfield
  {journal} {\bibinfo  {journal} {Phys. Rev. D}\ }\textbf {\bibinfo {volume}
  {69}},\ \bibinfo {pages} {025001} (\bibinfo {year} {2004})},\ \Eprint
  {http://arxiv.org/abs/hep-th/0209183} {arXiv:hep-th/0209183} \BibitemShut
  {NoStop}%
\bibitem [{\citenamefont {Floerchinger}\ and\ \citenamefont
  {Wetterich}(2009)}]{Floerchinger:2009uf}%
  \BibitemOpen
  \bibfield  {author} {\bibinfo {author} {\bibfnamefont {S.}~\bibnamefont
  {Floerchinger}}\ and\ \bibinfo {author} {\bibfnamefont {C.}~\bibnamefont
  {Wetterich}},\ }\href {\doibase 10.1016/j.physletb.2009.09.014} {\bibfield
  {journal} {\bibinfo  {journal} {Phys. Lett. B}\ }\textbf {\bibinfo {volume}
  {680}},\ \bibinfo {pages} {371} (\bibinfo {year} {2009})},\ \Eprint
  {http://arxiv.org/abs/0905.0915} {arXiv:0905.0915 [hep-th]} \BibitemShut
  {NoStop}%
\bibitem [{\citenamefont {Mitter}\ \emph {et~al.}(2015)\citenamefont {Mitter},
  \citenamefont {Pawlowski},\ and\ \citenamefont
  {Strodthoff}}]{Mitter:2014wpa}%
  \BibitemOpen
  \bibfield  {author} {\bibinfo {author} {\bibfnamefont {M.}~\bibnamefont
  {Mitter}}, \bibinfo {author} {\bibfnamefont {J.~M.}\ \bibnamefont
  {Pawlowski}}, \ and\ \bibinfo {author} {\bibfnamefont {N.}~\bibnamefont
  {Strodthoff}},\ }\href {\doibase 10.1103/PhysRevD.91.054035} {\bibfield
  {journal} {\bibinfo  {journal} {Phys. Rev. D}\ }\textbf {\bibinfo {volume}
  {91}},\ \bibinfo {pages} {054035} (\bibinfo {year} {2015})},\ \Eprint
  {http://arxiv.org/abs/1411.7978} {arXiv:1411.7978 [hep-ph]} \BibitemShut
  {NoStop}%
\bibitem [{\citenamefont {Braun}\ \emph {et~al.}(2016)\citenamefont {Braun},
  \citenamefont {Fister}, \citenamefont {Pawlowski},\ and\ \citenamefont
  {Rennecke}}]{Braun:2014ata}%
  \BibitemOpen
  \bibfield  {author} {\bibinfo {author} {\bibfnamefont {J.}~\bibnamefont
  {Braun}}, \bibinfo {author} {\bibfnamefont {L.}~\bibnamefont {Fister}},
  \bibinfo {author} {\bibfnamefont {J.~M.}\ \bibnamefont {Pawlowski}}, \ and\
  \bibinfo {author} {\bibfnamefont {F.}~\bibnamefont {Rennecke}},\ }\href
  {\doibase 10.1103/PhysRevD.94.034016} {\bibfield  {journal} {\bibinfo
  {journal} {Phys. Rev. D}\ }\textbf {\bibinfo {volume} {94}},\ \bibinfo
  {pages} {034016} (\bibinfo {year} {2016})},\ \Eprint
  {http://arxiv.org/abs/1412.1045} {arXiv:1412.1045 [hep-ph]} \BibitemShut
  {NoStop}%
\bibitem [{\citenamefont {Cyrol}\ \emph {et~al.}(2018)\citenamefont {Cyrol},
  \citenamefont {Mitter}, \citenamefont {Pawlowski},\ and\ \citenamefont
  {Strodthoff}}]{Cyrol:2017ewj}%
  \BibitemOpen
  \bibfield  {author} {\bibinfo {author} {\bibfnamefont {A.~K.}\ \bibnamefont
  {Cyrol}}, \bibinfo {author} {\bibfnamefont {M.}~\bibnamefont {Mitter}},
  \bibinfo {author} {\bibfnamefont {J.~M.}\ \bibnamefont {Pawlowski}}, \ and\
  \bibinfo {author} {\bibfnamefont {N.}~\bibnamefont {Strodthoff}},\ }\href
  {\doibase 10.1103/PhysRevD.97.054006} {\bibfield  {journal} {\bibinfo
  {journal} {Phys. Rev. D}\ }\textbf {\bibinfo {volume} {97}},\ \bibinfo
  {pages} {054006} (\bibinfo {year} {2018})},\ \Eprint
  {http://arxiv.org/abs/1706.06326} {arXiv:1706.06326 [hep-ph]} \BibitemShut
  {NoStop}%
\bibitem [{\citenamefont {Alkofer}\ \emph {et~al.}(2019)\citenamefont
  {Alkofer}, \citenamefont {Maas}, \citenamefont {Mian}, \citenamefont
  {Mitter}, \citenamefont {Par\'\i{}s-L\'opez}, \citenamefont {Pawlowski},\
  and\ \citenamefont {Wink}}]{Alkofer:2018guy}%
  \BibitemOpen
  \bibfield  {author} {\bibinfo {author} {\bibfnamefont {R.}~\bibnamefont
  {Alkofer}}, \bibinfo {author} {\bibfnamefont {A.}~\bibnamefont {Maas}},
  \bibinfo {author} {\bibfnamefont {W.~A.}\ \bibnamefont {Mian}}, \bibinfo
  {author} {\bibfnamefont {M.}~\bibnamefont {Mitter}}, \bibinfo {author}
  {\bibfnamefont {J.}~\bibnamefont {Par\'\i{}s-L\'opez}}, \bibinfo {author}
  {\bibfnamefont {J.~M.}\ \bibnamefont {Pawlowski}}, \ and\ \bibinfo {author}
  {\bibfnamefont {N.}~\bibnamefont {Wink}},\ }\href {\doibase
  10.1103/PhysRevD.99.054029} {\bibfield  {journal} {\bibinfo  {journal} {Phys.
  Rev. D}\ }\textbf {\bibinfo {volume} {99}},\ \bibinfo {pages} {054029}
  (\bibinfo {year} {2019})},\ \Eprint {http://arxiv.org/abs/1810.07955}
  {arXiv:1810.07955 [hep-ph]} \BibitemShut {NoStop}%
\bibitem [{\citenamefont {Fu}\ \emph {et~al.}(2020)\citenamefont {Fu},
  \citenamefont {Pawlowski},\ and\ \citenamefont {Rennecke}}]{Fu:2019hdw}%
  \BibitemOpen
  \bibfield  {author} {\bibinfo {author} {\bibfnamefont {W.-j.}\ \bibnamefont
  {Fu}}, \bibinfo {author} {\bibfnamefont {J.~M.}\ \bibnamefont {Pawlowski}}, \
  and\ \bibinfo {author} {\bibfnamefont {F.}~\bibnamefont {Rennecke}},\ }\href
  {\doibase 10.1103/PhysRevD.101.054032} {\bibfield  {journal} {\bibinfo
  {journal} {Phys. Rev. D}\ }\textbf {\bibinfo {volume} {101}},\ \bibinfo
  {pages} {054032} (\bibinfo {year} {2020})},\ \Eprint
  {http://arxiv.org/abs/1909.02991} {arXiv:1909.02991 [hep-ph]} \BibitemShut
  {NoStop}%
\bibitem [{\citenamefont {Jungnickel}\ and\ \citenamefont
  {Wetterich}(1998)}]{Jungnickel:1997yu}%
  \BibitemOpen
  \bibfield  {author} {\bibinfo {author} {\bibfnamefont {D.~U.}\ \bibnamefont
  {Jungnickel}}\ and\ \bibinfo {author} {\bibfnamefont {C.}~\bibnamefont
  {Wetterich}},\ }\href {\doibase 10.1007/s100520050161} {\bibfield  {journal}
  {\bibinfo  {journal} {Eur. Phys. J. C}\ }\textbf {\bibinfo {volume} {2}},\
  \bibinfo {pages} {557} (\bibinfo {year} {1998})},\ \Eprint
  {http://arxiv.org/abs/hep-ph/9704345} {arXiv:hep-ph/9704345} \BibitemShut
  {NoStop}%
\bibitem [{\citenamefont {Jendges}\ \emph {et~al.}(2006)\citenamefont
  {Jendges}, \citenamefont {Klein}, \citenamefont {Pirner},\ and\ \citenamefont
  {Schwenzer}}]{Jendges:2006yk}%
  \BibitemOpen
  \bibfield  {author} {\bibinfo {author} {\bibfnamefont {L.}~\bibnamefont
  {Jendges}}, \bibinfo {author} {\bibfnamefont {B.}~\bibnamefont {Klein}},
  \bibinfo {author} {\bibfnamefont {H.-J.}\ \bibnamefont {Pirner}}, \ and\
  \bibinfo {author} {\bibfnamefont {K.}~\bibnamefont {Schwenzer}},\ }\href@noop
  {} {\  (\bibinfo {year} {2006})},\ \Eprint
  {http://arxiv.org/abs/hep-ph/0608056} {arXiv:hep-ph/0608056} \BibitemShut
  {NoStop}%
\bibitem [{\citenamefont {Codello}\ \emph {et~al.}(2016)\citenamefont
  {Codello}, \citenamefont {Percacci}, \citenamefont {Rachwa\l{}},\ and\
  \citenamefont {Tonero}}]{Codello:2015oqa}%
  \BibitemOpen
  \bibfield  {author} {\bibinfo {author} {\bibfnamefont {A.}~\bibnamefont
  {Codello}}, \bibinfo {author} {\bibfnamefont {R.}~\bibnamefont {Percacci}},
  \bibinfo {author} {\bibfnamefont {L.}~\bibnamefont {Rachwa\l{}}}, \ and\
  \bibinfo {author} {\bibfnamefont {A.}~\bibnamefont {Tonero}},\ }\href
  {\doibase 10.1140/epjc/s10052-016-4063-3} {\bibfield  {journal} {\bibinfo
  {journal} {Eur. Phys. J. C}\ }\textbf {\bibinfo {volume} {76}},\ \bibinfo
  {pages} {226} (\bibinfo {year} {2016})},\ \Eprint
  {http://arxiv.org/abs/1505.03119} {arXiv:1505.03119 [hep-th]} \BibitemShut
  {NoStop}%
\bibitem [{\citenamefont {Codello}\ and\ \citenamefont
  {Percacci}(2009)}]{Codello:2008qq}%
  \BibitemOpen
  \bibfield  {author} {\bibinfo {author} {\bibfnamefont {A.}~\bibnamefont
  {Codello}}\ and\ \bibinfo {author} {\bibfnamefont {R.}~\bibnamefont
  {Percacci}},\ }\href {\doibase 10.1016/j.physletb.2009.01.032} {\bibfield
  {journal} {\bibinfo  {journal} {Phys. Lett. B}\ }\textbf {\bibinfo {volume}
  {672}},\ \bibinfo {pages} {280} (\bibinfo {year} {2009})},\ \Eprint
  {http://arxiv.org/abs/0810.0715} {arXiv:0810.0715 [hep-th]} \BibitemShut
  {NoStop}%
\bibitem [{\citenamefont {Fabbrichesi}\ \emph {et~al.}(2011)\citenamefont
  {Fabbrichesi}, \citenamefont {Percacci}, \citenamefont {Tonero},\ and\
  \citenamefont {Zanusso}}]{Fabbrichesi:2010xy}%
  \BibitemOpen
  \bibfield  {author} {\bibinfo {author} {\bibfnamefont {M.}~\bibnamefont
  {Fabbrichesi}}, \bibinfo {author} {\bibfnamefont {R.}~\bibnamefont
  {Percacci}}, \bibinfo {author} {\bibfnamefont {A.}~\bibnamefont {Tonero}}, \
  and\ \bibinfo {author} {\bibfnamefont {O.}~\bibnamefont {Zanusso}},\ }\href
  {\doibase 10.1103/PhysRevD.83.025016} {\bibfield  {journal} {\bibinfo
  {journal} {Phys. Rev. D}\ }\textbf {\bibinfo {volume} {83}},\ \bibinfo
  {pages} {025016} (\bibinfo {year} {2011})},\ \Eprint
  {http://arxiv.org/abs/1010.0912} {arXiv:1010.0912 [hep-ph]} \BibitemShut
  {NoStop}%
\bibitem [{\citenamefont {Bazzocchi}\ \emph {et~al.}(2011)\citenamefont
  {Bazzocchi}, \citenamefont {Fabbrichesi}, \citenamefont {Percacci},
  \citenamefont {Tonero},\ and\ \citenamefont {Vecchi}}]{Bazzocchi:2011vr}%
  \BibitemOpen
  \bibfield  {author} {\bibinfo {author} {\bibfnamefont {F.}~\bibnamefont
  {Bazzocchi}}, \bibinfo {author} {\bibfnamefont {M.}~\bibnamefont
  {Fabbrichesi}}, \bibinfo {author} {\bibfnamefont {R.}~\bibnamefont
  {Percacci}}, \bibinfo {author} {\bibfnamefont {A.}~\bibnamefont {Tonero}}, \
  and\ \bibinfo {author} {\bibfnamefont {L.}~\bibnamefont {Vecchi}},\ }\href
  {\doibase 10.1016/j.physletb.2011.10.029} {\bibfield  {journal} {\bibinfo
  {journal} {Phys. Lett. B}\ }\textbf {\bibinfo {volume} {705}},\ \bibinfo
  {pages} {388} (\bibinfo {year} {2011})},\ \Eprint
  {http://arxiv.org/abs/1105.1968} {arXiv:1105.1968 [hep-ph]} \BibitemShut
  {NoStop}%
\bibitem [{\citenamefont {Flore}\ \emph {et~al.}(2013)\citenamefont {Flore},
  \citenamefont {Wipf},\ and\ \citenamefont {Zanusso}}]{Flore:2012ma}%
  \BibitemOpen
  \bibfield  {author} {\bibinfo {author} {\bibfnamefont {R.}~\bibnamefont
  {Flore}}, \bibinfo {author} {\bibfnamefont {A.}~\bibnamefont {Wipf}}, \ and\
  \bibinfo {author} {\bibfnamefont {O.}~\bibnamefont {Zanusso}},\ }\href
  {\doibase 10.1103/PhysRevD.87.065019} {\bibfield  {journal} {\bibinfo
  {journal} {Phys. Rev. D}\ }\textbf {\bibinfo {volume} {87}},\ \bibinfo
  {pages} {065019} (\bibinfo {year} {2013})},\ \Eprint
  {http://arxiv.org/abs/1207.4499} {arXiv:1207.4499 [hep-th]} \BibitemShut
  {NoStop}%
\bibitem [{\citenamefont {Percacci}\ and\ \citenamefont
  {Safari}(2013)}]{Percacci:2013jpa}%
  \BibitemOpen
  \bibfield  {author} {\bibinfo {author} {\bibfnamefont {R.}~\bibnamefont
  {Percacci}}\ and\ \bibinfo {author} {\bibfnamefont {M.}~\bibnamefont
  {Safari}},\ }\href {\doibase 10.1103/PhysRevD.88.085007} {\bibfield
  {journal} {\bibinfo  {journal} {Phys. Rev. D}\ }\textbf {\bibinfo {volume}
  {88}},\ \bibinfo {pages} {085007} (\bibinfo {year} {2013})},\ \Eprint
  {http://arxiv.org/abs/1306.3918} {arXiv:1306.3918 [hep-th]} \BibitemShut
  {NoStop}%
\bibitem [{\citenamefont {Braun}(2010)}]{Braun:2009si}%
  \BibitemOpen
  \bibfield  {author} {\bibinfo {author} {\bibfnamefont {J.}~\bibnamefont
  {Braun}},\ }\href {\doibase 10.1103/PhysRevD.81.016008} {\bibfield  {journal}
  {\bibinfo  {journal} {Phys. Rev. D}\ }\textbf {\bibinfo {volume} {81}},\
  \bibinfo {pages} {016008} (\bibinfo {year} {2010})},\ \Eprint
  {http://arxiv.org/abs/0908.1543} {arXiv:0908.1543 [hep-ph]} \BibitemShut
  {NoStop}%
\bibitem [{\citenamefont {Braun}(2012)}]{Braun:2011pp}%
  \BibitemOpen
  \bibfield  {author} {\bibinfo {author} {\bibfnamefont {J.}~\bibnamefont
  {Braun}},\ }\href {\doibase 10.1088/0954-3899/39/3/033001} {\bibfield
  {journal} {\bibinfo  {journal} {J. Phys. G}\ }\textbf {\bibinfo {volume}
  {39}},\ \bibinfo {pages} {033001} (\bibinfo {year} {2012})},\ \Eprint
  {http://arxiv.org/abs/1108.4449} {arXiv:1108.4449 [hep-ph]} \BibitemShut
  {NoStop}%
\bibitem [{\citenamefont {Rennecke}(2015)}]{Rennecke:2015eba}%
  \BibitemOpen
  \bibfield  {author} {\bibinfo {author} {\bibfnamefont {F.}~\bibnamefont
  {Rennecke}},\ }\href {\doibase 10.1103/PhysRevD.92.076012} {\bibfield
  {journal} {\bibinfo  {journal} {Phys. Rev. D}\ }\textbf {\bibinfo {volume}
  {92}},\ \bibinfo {pages} {076012} (\bibinfo {year} {2015})},\ \Eprint
  {http://arxiv.org/abs/1504.03585} {arXiv:1504.03585 [hep-ph]} \BibitemShut
  {NoStop}%
\bibitem [{\citenamefont {Braun}\ \emph {et~al.}(2017)\citenamefont {Braun},
  \citenamefont {Leonhardt},\ and\ \citenamefont {Pospiech}}]{Braun:2017srn}%
  \BibitemOpen
  \bibfield  {author} {\bibinfo {author} {\bibfnamefont {J.}~\bibnamefont
  {Braun}}, \bibinfo {author} {\bibfnamefont {M.}~\bibnamefont {Leonhardt}}, \
  and\ \bibinfo {author} {\bibfnamefont {M.}~\bibnamefont {Pospiech}},\ }\href
  {\doibase 10.1103/PhysRevD.96.076003} {\bibfield  {journal} {\bibinfo
  {journal} {Phys. Rev. D}\ }\textbf {\bibinfo {volume} {96}},\ \bibinfo
  {pages} {076003} (\bibinfo {year} {2017})},\ \Eprint
  {http://arxiv.org/abs/1705.00074} {arXiv:1705.00074 [hep-ph]} \BibitemShut
  {NoStop}%
\bibitem [{\citenamefont {Braun}\ \emph {et~al.}(2018)\citenamefont {Braun},
  \citenamefont {Leonhardt},\ and\ \citenamefont {Pospiech}}]{Braun:2018bik}%
  \BibitemOpen
  \bibfield  {author} {\bibinfo {author} {\bibfnamefont {J.}~\bibnamefont
  {Braun}}, \bibinfo {author} {\bibfnamefont {M.}~\bibnamefont {Leonhardt}}, \
  and\ \bibinfo {author} {\bibfnamefont {M.}~\bibnamefont {Pospiech}},\ }\href
  {\doibase 10.1103/PhysRevD.97.076010} {\bibfield  {journal} {\bibinfo
  {journal} {Phys. Rev. D}\ }\textbf {\bibinfo {volume} {97}},\ \bibinfo
  {pages} {076010} (\bibinfo {year} {2018})},\ \Eprint
  {http://arxiv.org/abs/1801.08338} {arXiv:1801.08338 [hep-ph]} \BibitemShut
  {NoStop}%
\bibitem [{\citenamefont {Braun}\ \emph {et~al.}(2020)\citenamefont {Braun},
  \citenamefont {Leonhardt},\ and\ \citenamefont {Pospiech}}]{Braun:2019aow}%
  \BibitemOpen
  \bibfield  {author} {\bibinfo {author} {\bibfnamefont {J.}~\bibnamefont
  {Braun}}, \bibinfo {author} {\bibfnamefont {M.}~\bibnamefont {Leonhardt}}, \
  and\ \bibinfo {author} {\bibfnamefont {M.}~\bibnamefont {Pospiech}},\ }\href
  {\doibase 10.1103/PhysRevD.101.036004} {\bibfield  {journal} {\bibinfo
  {journal} {Phys. Rev. D}\ }\textbf {\bibinfo {volume} {101}},\ \bibinfo
  {pages} {036004} (\bibinfo {year} {2020})},\ \Eprint
  {http://arxiv.org/abs/1909.06298} {arXiv:1909.06298 [hep-ph]} \BibitemShut
  {NoStop}%
\bibitem [{\citenamefont {Braun}\ \emph {et~al.}(2019)\citenamefont {Braun},
  \citenamefont {Leonhardt},\ and\ \citenamefont {Pawlowski}}]{Braun:2018svj}%
  \BibitemOpen
  \bibfield  {author} {\bibinfo {author} {\bibfnamefont {J.}~\bibnamefont
  {Braun}}, \bibinfo {author} {\bibfnamefont {M.}~\bibnamefont {Leonhardt}}, \
  and\ \bibinfo {author} {\bibfnamefont {J.~M.}\ \bibnamefont {Pawlowski}},\
  }\href {\doibase 10.21468/SciPostPhys.6.5.056} {\bibfield  {journal}
  {\bibinfo  {journal} {SciPost Phys.}\ }\textbf {\bibinfo {volume} {6}},\
  \bibinfo {pages} {056} (\bibinfo {year} {2019})},\ \Eprint
  {http://arxiv.org/abs/1806.04432} {arXiv:1806.04432 [hep-ph]} \BibitemShut
  {NoStop}%
\bibitem [{\citenamefont {Jung}\ and\ \citenamefont {von
  Smekal}(2019)}]{Jung:2019nnr}%
  \BibitemOpen
  \bibfield  {author} {\bibinfo {author} {\bibfnamefont {C.}~\bibnamefont
  {Jung}}\ and\ \bibinfo {author} {\bibfnamefont {L.}~\bibnamefont {von
  Smekal}},\ }\href {\doibase 10.1103/PhysRevD.100.116009} {\bibfield
  {journal} {\bibinfo  {journal} {Phys. Rev. D}\ }\textbf {\bibinfo {volume}
  {100}},\ \bibinfo {pages} {116009} (\bibinfo {year} {2019})},\ \Eprint
  {http://arxiv.org/abs/1909.13712} {arXiv:1909.13712 [hep-ph]} \BibitemShut
  {NoStop}%
\bibitem [{\citenamefont {Jungnickel}\ and\ \citenamefont
  {Wetterich}(1996)}]{Jungnickel:1995fp}%
  \BibitemOpen
  \bibfield  {author} {\bibinfo {author} {\bibfnamefont {D.~U.}\ \bibnamefont
  {Jungnickel}}\ and\ \bibinfo {author} {\bibfnamefont {C.}~\bibnamefont
  {Wetterich}},\ }\href {\doibase 10.1103/PhysRevD.53.5142} {\bibfield
  {journal} {\bibinfo  {journal} {Phys. Rev. D}\ }\textbf {\bibinfo {volume}
  {53}},\ \bibinfo {pages} {5142} (\bibinfo {year} {1996})},\ \Eprint
  {http://arxiv.org/abs/hep-ph/9505267} {arXiv:hep-ph/9505267} \BibitemShut
  {NoStop}%
\bibitem [{\citenamefont {Berges}\ \emph {et~al.}(1999)\citenamefont {Berges},
  \citenamefont {Jungnickel},\ and\ \citenamefont {Wetterich}}]{Berges:1997eu}%
  \BibitemOpen
  \bibfield  {author} {\bibinfo {author} {\bibfnamefont {J.}~\bibnamefont
  {Berges}}, \bibinfo {author} {\bibfnamefont {D.~U.}\ \bibnamefont
  {Jungnickel}}, \ and\ \bibinfo {author} {\bibfnamefont {C.}~\bibnamefont
  {Wetterich}},\ }\href {\doibase 10.1103/PhysRevD.59.034010} {\bibfield
  {journal} {\bibinfo  {journal} {Phys. Rev. D}\ }\textbf {\bibinfo {volume}
  {59}},\ \bibinfo {pages} {034010} (\bibinfo {year} {1999})},\ \Eprint
  {http://arxiv.org/abs/hep-ph/9705474} {arXiv:hep-ph/9705474} \BibitemShut
  {NoStop}%
\bibitem [{\citenamefont {Berges}\ \emph {et~al.}(2000)\citenamefont {Berges},
  \citenamefont {Jungnickel},\ and\ \citenamefont {Wetterich}}]{Berges:1998sd}%
  \BibitemOpen
  \bibfield  {author} {\bibinfo {author} {\bibfnamefont {J.}~\bibnamefont
  {Berges}}, \bibinfo {author} {\bibfnamefont {D.-U.}\ \bibnamefont
  {Jungnickel}}, \ and\ \bibinfo {author} {\bibfnamefont {C.}~\bibnamefont
  {Wetterich}},\ }\href {\doibase 10.1007/s100520000275} {\bibfield  {journal}
  {\bibinfo  {journal} {Eur. Phys. J. C}\ }\textbf {\bibinfo {volume} {13}},\
  \bibinfo {pages} {323} (\bibinfo {year} {2000})},\ \Eprint
  {http://arxiv.org/abs/hep-ph/9811347} {arXiv:hep-ph/9811347} \BibitemShut
  {NoStop}%
\bibitem [{\citenamefont {Schaefer}\ and\ \citenamefont
  {Wambach}(2005)}]{Schaefer:2004en}%
  \BibitemOpen
  \bibfield  {author} {\bibinfo {author} {\bibfnamefont {B.-J.}\ \bibnamefont
  {Schaefer}}\ and\ \bibinfo {author} {\bibfnamefont {J.}~\bibnamefont
  {Wambach}},\ }\href {\doibase 10.1016/j.nuclphysa.2005.04.012} {\bibfield
  {journal} {\bibinfo  {journal} {Nucl. Phys. A}\ }\textbf {\bibinfo {volume}
  {757}},\ \bibinfo {pages} {479} (\bibinfo {year} {2005})},\ \Eprint
  {http://arxiv.org/abs/nucl-th/0403039} {arXiv:nucl-th/0403039} \BibitemShut
  {NoStop}%
\bibitem [{\citenamefont {Schaefer}\ and\ \citenamefont
  {Wambach}(2008)}]{Schaefer:2006sr}%
  \BibitemOpen
  \bibfield  {author} {\bibinfo {author} {\bibfnamefont {B.-J.}\ \bibnamefont
  {Schaefer}}\ and\ \bibinfo {author} {\bibfnamefont {J.}~\bibnamefont
  {Wambach}},\ }\href {\doibase 10.1134/S1063779608070083} {\bibfield
  {journal} {\bibinfo  {journal} {Phys. Part. Nucl.}\ }\textbf {\bibinfo
  {volume} {39}},\ \bibinfo {pages} {1025} (\bibinfo {year} {2008})},\ \Eprint
  {http://arxiv.org/abs/hep-ph/0611191} {arXiv:hep-ph/0611191} \BibitemShut
  {NoStop}%
\bibitem [{\citenamefont {Herbst}\ \emph {et~al.}(2011)\citenamefont {Herbst},
  \citenamefont {Pawlowski},\ and\ \citenamefont {Schaefer}}]{Herbst:2010rf}%
  \BibitemOpen
  \bibfield  {author} {\bibinfo {author} {\bibfnamefont {T.~K.}\ \bibnamefont
  {Herbst}}, \bibinfo {author} {\bibfnamefont {J.~M.}\ \bibnamefont
  {Pawlowski}}, \ and\ \bibinfo {author} {\bibfnamefont {B.-J.}\ \bibnamefont
  {Schaefer}},\ }\href {\doibase 10.1016/j.physletb.2010.12.003} {\bibfield
  {journal} {\bibinfo  {journal} {Phys. Lett. B}\ }\textbf {\bibinfo {volume}
  {696}},\ \bibinfo {pages} {58} (\bibinfo {year} {2011})},\ \Eprint
  {http://arxiv.org/abs/1008.0081} {arXiv:1008.0081 [hep-ph]} \BibitemShut
  {NoStop}%
\bibitem [{\citenamefont {Herbst}\ \emph {et~al.}(2014)\citenamefont {Herbst},
  \citenamefont {Mitter}, \citenamefont {Pawlowski}, \citenamefont {Schaefer},\
  and\ \citenamefont {Stiele}}]{Herbst:2013ufa}%
  \BibitemOpen
  \bibfield  {author} {\bibinfo {author} {\bibfnamefont {T.~K.}\ \bibnamefont
  {Herbst}}, \bibinfo {author} {\bibfnamefont {M.}~\bibnamefont {Mitter}},
  \bibinfo {author} {\bibfnamefont {J.~M.}\ \bibnamefont {Pawlowski}}, \bibinfo
  {author} {\bibfnamefont {B.-J.}\ \bibnamefont {Schaefer}}, \ and\ \bibinfo
  {author} {\bibfnamefont {R.}~\bibnamefont {Stiele}},\ }\href {\doibase
  10.1016/j.physletb.2014.02.045} {\bibfield  {journal} {\bibinfo  {journal}
  {Phys. Lett. B}\ }\textbf {\bibinfo {volume} {731}},\ \bibinfo {pages} {248}
  (\bibinfo {year} {2014})},\ \Eprint {http://arxiv.org/abs/1308.3621}
  {arXiv:1308.3621 [hep-ph]} \BibitemShut {NoStop}%
\bibitem [{\citenamefont {Mitter}\ and\ \citenamefont
  {Schaefer}(2014)}]{Mitter:2013fxa}%
  \BibitemOpen
  \bibfield  {author} {\bibinfo {author} {\bibfnamefont {M.}~\bibnamefont
  {Mitter}}\ and\ \bibinfo {author} {\bibfnamefont {B.-J.}\ \bibnamefont
  {Schaefer}},\ }\href {\doibase 10.1103/PhysRevD.89.054027} {\bibfield
  {journal} {\bibinfo  {journal} {Phys. Rev. D}\ }\textbf {\bibinfo {volume}
  {89}},\ \bibinfo {pages} {054027} (\bibinfo {year} {2014})},\ \Eprint
  {http://arxiv.org/abs/1308.3176} {arXiv:1308.3176 [hep-ph]} \BibitemShut
  {NoStop}%
\bibitem [{\citenamefont {Pawlowski}\ and\ \citenamefont
  {Rennecke}(2014)}]{Pawlowski:2014zaa}%
  \BibitemOpen
  \bibfield  {author} {\bibinfo {author} {\bibfnamefont {J.~M.}\ \bibnamefont
  {Pawlowski}}\ and\ \bibinfo {author} {\bibfnamefont {F.}~\bibnamefont
  {Rennecke}},\ }\href {\doibase 10.1103/PhysRevD.90.076002} {\bibfield
  {journal} {\bibinfo  {journal} {Phys. Rev. D}\ }\textbf {\bibinfo {volume}
  {90}},\ \bibinfo {pages} {076002} (\bibinfo {year} {2014})},\ \Eprint
  {http://arxiv.org/abs/1403.1179} {arXiv:1403.1179 [hep-ph]} \BibitemShut
  {NoStop}%
\bibitem [{\citenamefont {Eser}\ \emph {et~al.}(2015)\citenamefont {Eser},
  \citenamefont {Grahl},\ and\ \citenamefont {Rischke}}]{Eser:2015pka}%
  \BibitemOpen
  \bibfield  {author} {\bibinfo {author} {\bibfnamefont {J.}~\bibnamefont
  {Eser}}, \bibinfo {author} {\bibfnamefont {M.}~\bibnamefont {Grahl}}, \ and\
  \bibinfo {author} {\bibfnamefont {D.~H.}\ \bibnamefont {Rischke}},\ }\href
  {\doibase 10.1103/PhysRevD.92.096008} {\bibfield  {journal} {\bibinfo
  {journal} {Phys. Rev. D}\ }\textbf {\bibinfo {volume} {92}},\ \bibinfo
  {pages} {096008} (\bibinfo {year} {2015})},\ \Eprint
  {http://arxiv.org/abs/1508.06928} {arXiv:1508.06928 [hep-ph]} \BibitemShut
  {NoStop}%
\bibitem [{\citenamefont {Jung}\ \emph {et~al.}(2017)\citenamefont {Jung},
  \citenamefont {Rennecke}, \citenamefont {Tripolt}, \citenamefont {von
  Smekal},\ and\ \citenamefont {Wambach}}]{Jung:2016yxl}%
  \BibitemOpen
  \bibfield  {author} {\bibinfo {author} {\bibfnamefont {C.}~\bibnamefont
  {Jung}}, \bibinfo {author} {\bibfnamefont {F.}~\bibnamefont {Rennecke}},
  \bibinfo {author} {\bibfnamefont {R.-A.}\ \bibnamefont {Tripolt}}, \bibinfo
  {author} {\bibfnamefont {L.}~\bibnamefont {von Smekal}}, \ and\ \bibinfo
  {author} {\bibfnamefont {J.}~\bibnamefont {Wambach}},\ }\href {\doibase
  10.1103/PhysRevD.95.036020} {\bibfield  {journal} {\bibinfo  {journal} {Phys.
  Rev. D}\ }\textbf {\bibinfo {volume} {95}},\ \bibinfo {pages} {036020}
  (\bibinfo {year} {2017})},\ \Eprint {http://arxiv.org/abs/1610.08754}
  {arXiv:1610.08754 [hep-ph]} \BibitemShut {NoStop}%
\bibitem [{\citenamefont {Rennecke}\ and\ \citenamefont
  {Schaefer}(2017)}]{Rennecke:2016tkm}%
  \BibitemOpen
  \bibfield  {author} {\bibinfo {author} {\bibfnamefont {F.}~\bibnamefont
  {Rennecke}}\ and\ \bibinfo {author} {\bibfnamefont {B.-J.}\ \bibnamefont
  {Schaefer}},\ }\href {\doibase 10.1103/PhysRevD.96.016009} {\bibfield
  {journal} {\bibinfo  {journal} {Phys. Rev. D}\ }\textbf {\bibinfo {volume}
  {96}},\ \bibinfo {pages} {016009} (\bibinfo {year} {2017})},\ \Eprint
  {http://arxiv.org/abs/1610.08748} {arXiv:1610.08748 [hep-ph]} \BibitemShut
  {NoStop}%
\bibitem [{\citenamefont {Fu}\ \emph {et~al.}(2016)\citenamefont {Fu},
  \citenamefont {Pawlowski}, \citenamefont {Rennecke},\ and\ \citenamefont
  {Schaefer}}]{Fu:2016tey}%
  \BibitemOpen
  \bibfield  {author} {\bibinfo {author} {\bibfnamefont {W.-j.}\ \bibnamefont
  {Fu}}, \bibinfo {author} {\bibfnamefont {J.~M.}\ \bibnamefont {Pawlowski}},
  \bibinfo {author} {\bibfnamefont {F.}~\bibnamefont {Rennecke}}, \ and\
  \bibinfo {author} {\bibfnamefont {B.-J.}\ \bibnamefont {Schaefer}},\ }\href
  {\doibase 10.1103/PhysRevD.94.116020} {\bibfield  {journal} {\bibinfo
  {journal} {Phys. Rev. D}\ }\textbf {\bibinfo {volume} {94}},\ \bibinfo
  {pages} {116020} (\bibinfo {year} {2016})},\ \Eprint
  {http://arxiv.org/abs/1608.04302} {arXiv:1608.04302 [hep-ph]} \BibitemShut
  {NoStop}%
\bibitem [{\citenamefont {Tripolt}\ \emph {et~al.}(2018)\citenamefont
  {Tripolt}, \citenamefont {Schaefer}, \citenamefont {von Smekal},\ and\
  \citenamefont {Wambach}}]{Tripolt:2017zgc}%
  \BibitemOpen
  \bibfield  {author} {\bibinfo {author} {\bibfnamefont {R.-A.}\ \bibnamefont
  {Tripolt}}, \bibinfo {author} {\bibfnamefont {B.-J.}\ \bibnamefont
  {Schaefer}}, \bibinfo {author} {\bibfnamefont {L.}~\bibnamefont {von
  Smekal}}, \ and\ \bibinfo {author} {\bibfnamefont {J.}~\bibnamefont
  {Wambach}},\ }\href {\doibase 10.1103/PhysRevD.97.034022} {\bibfield
  {journal} {\bibinfo  {journal} {Phys. Rev. D}\ }\textbf {\bibinfo {volume}
  {97}},\ \bibinfo {pages} {034022} (\bibinfo {year} {2018})},\ \Eprint
  {http://arxiv.org/abs/1709.05991} {arXiv:1709.05991 [hep-ph]} \BibitemShut
  {NoStop}%
\bibitem [{\citenamefont {Resch}\ \emph {et~al.}(2019)\citenamefont {Resch},
  \citenamefont {Rennecke},\ and\ \citenamefont {Schaefer}}]{Resch:2017vjs}%
  \BibitemOpen
  \bibfield  {author} {\bibinfo {author} {\bibfnamefont {S.}~\bibnamefont
  {Resch}}, \bibinfo {author} {\bibfnamefont {F.}~\bibnamefont {Rennecke}}, \
  and\ \bibinfo {author} {\bibfnamefont {B.-J.}\ \bibnamefont {Schaefer}},\
  }\href {\doibase 10.1103/PhysRevD.99.076005} {\bibfield  {journal} {\bibinfo
  {journal} {Phys. Rev. D}\ }\textbf {\bibinfo {volume} {99}},\ \bibinfo
  {pages} {076005} (\bibinfo {year} {2019})},\ \Eprint
  {http://arxiv.org/abs/1712.07961} {arXiv:1712.07961 [hep-ph]} \BibitemShut
  {NoStop}%
\bibitem [{\citenamefont {Otto}\ \emph
  {et~al.}(2020{\natexlab{a}})\citenamefont {Otto}, \citenamefont {Oertel},\
  and\ \citenamefont {Schaefer}}]{Otto:2019zjy}%
  \BibitemOpen
  \bibfield  {author} {\bibinfo {author} {\bibfnamefont {K.}~\bibnamefont
  {Otto}}, \bibinfo {author} {\bibfnamefont {M.}~\bibnamefont {Oertel}}, \ and\
  \bibinfo {author} {\bibfnamefont {B.-J.}\ \bibnamefont {Schaefer}},\ }\href
  {\doibase 10.1103/PhysRevD.101.103021} {\bibfield  {journal} {\bibinfo
  {journal} {Phys. Rev. D}\ }\textbf {\bibinfo {volume} {101}},\ \bibinfo
  {pages} {103021} (\bibinfo {year} {2020}{\natexlab{a}})},\ \Eprint
  {http://arxiv.org/abs/1910.11929} {arXiv:1910.11929 [hep-ph]} \BibitemShut
  {NoStop}%
\bibitem [{\citenamefont {Otto}\ \emph
  {et~al.}(2020{\natexlab{b}})\citenamefont {Otto}, \citenamefont {Oertel},\
  and\ \citenamefont {Schaefer}}]{Otto:2020hoz}%
  \BibitemOpen
  \bibfield  {author} {\bibinfo {author} {\bibfnamefont {K.}~\bibnamefont
  {Otto}}, \bibinfo {author} {\bibfnamefont {M.}~\bibnamefont {Oertel}}, \ and\
  \bibinfo {author} {\bibfnamefont {B.-J.}\ \bibnamefont {Schaefer}},\ }\href
  {\doibase 10.1140/epjst/e2020-000155-y} {\bibfield  {journal} {\bibinfo
  {journal} {Eur. Phys. J. ST}\ }\textbf {\bibinfo {volume} {229}},\ \bibinfo
  {pages} {3629} (\bibinfo {year} {2020}{\natexlab{b}})},\ \Eprint
  {http://arxiv.org/abs/2007.07394} {arXiv:2007.07394 [hep-ph]} \BibitemShut
  {NoStop}%
\bibitem [{\citenamefont {Jung}\ \emph {et~al.}(2021)\citenamefont {Jung},
  \citenamefont {Otto}, \citenamefont {Tripolt},\ and\ \citenamefont {von
  Smekal}}]{Jung:2021ipc}%
  \BibitemOpen
  \bibfield  {author} {\bibinfo {author} {\bibfnamefont {C.}~\bibnamefont
  {Jung}}, \bibinfo {author} {\bibfnamefont {J.-H.}\ \bibnamefont {Otto}},
  \bibinfo {author} {\bibfnamefont {R.-A.}\ \bibnamefont {Tripolt}}, \ and\
  \bibinfo {author} {\bibfnamefont {L.}~\bibnamefont {von Smekal}},\
  }\href@noop {} {\  (\bibinfo {year} {2021})},\ \Eprint
  {http://arxiv.org/abs/2107.10748} {arXiv:2107.10748 [hep-ph]} \BibitemShut
  {NoStop}%
\bibitem [{\citenamefont {Detar}\ and\ \citenamefont
  {Kunihiro}(1989)}]{Detar:1988kn}%
  \BibitemOpen
  \bibfield  {author} {\bibinfo {author} {\bibfnamefont {C.~E.}\ \bibnamefont
  {Detar}}\ and\ \bibinfo {author} {\bibfnamefont {T.}~\bibnamefont
  {Kunihiro}},\ }\href {\doibase 10.1103/PhysRevD.39.2805} {\bibfield
  {journal} {\bibinfo  {journal} {Phys. Rev. D}\ }\textbf {\bibinfo {volume}
  {39}},\ \bibinfo {pages} {2805} (\bibinfo {year} {1989})}\BibitemShut
  {NoStop}%
\bibitem [{\citenamefont {Hatsuda}\ and\ \citenamefont
  {Prakash}(1989)}]{Hatsuda:1988mv}%
  \BibitemOpen
  \bibfield  {author} {\bibinfo {author} {\bibfnamefont {T.}~\bibnamefont
  {Hatsuda}}\ and\ \bibinfo {author} {\bibfnamefont {M.}~\bibnamefont
  {Prakash}},\ }\href {\doibase 10.1016/0370-2693(89)91040-X} {\bibfield
  {journal} {\bibinfo  {journal} {Phys. Lett. B}\ }\textbf {\bibinfo {volume}
  {224}},\ \bibinfo {pages} {11} (\bibinfo {year} {1989})}\BibitemShut
  {NoStop}%
\bibitem [{\citenamefont {Jido}\ \emph {et~al.}(2000)\citenamefont {Jido},
  \citenamefont {Nemoto}, \citenamefont {Oka},\ and\ \citenamefont
  {Hosaka}}]{Jido:1998av}%
  \BibitemOpen
  \bibfield  {author} {\bibinfo {author} {\bibfnamefont {D.}~\bibnamefont
  {Jido}}, \bibinfo {author} {\bibfnamefont {Y.}~\bibnamefont {Nemoto}},
  \bibinfo {author} {\bibfnamefont {M.}~\bibnamefont {Oka}}, \ and\ \bibinfo
  {author} {\bibfnamefont {A.}~\bibnamefont {Hosaka}},\ }\href {\doibase
  10.1016/S0375-9474(99)00844-1} {\bibfield  {journal} {\bibinfo  {journal}
  {Nucl. Phys. A}\ }\textbf {\bibinfo {volume} {671}},\ \bibinfo {pages} {471}
  (\bibinfo {year} {2000})},\ \Eprint {http://arxiv.org/abs/hep-ph/9805306}
  {arXiv:hep-ph/9805306} \BibitemShut {NoStop}%
\bibitem [{\citenamefont {Jido}\ \emph {et~al.}(2001)\citenamefont {Jido},
  \citenamefont {Oka},\ and\ \citenamefont {Hosaka}}]{Jido:2001nt}%
  \BibitemOpen
  \bibfield  {author} {\bibinfo {author} {\bibfnamefont {D.}~\bibnamefont
  {Jido}}, \bibinfo {author} {\bibfnamefont {M.}~\bibnamefont {Oka}}, \ and\
  \bibinfo {author} {\bibfnamefont {A.}~\bibnamefont {Hosaka}},\ }\href
  {\doibase 10.1143/PTP.106.873} {\bibfield  {journal} {\bibinfo  {journal}
  {Prog. Theor. Phys.}\ }\textbf {\bibinfo {volume} {106}},\ \bibinfo {pages}
  {873} (\bibinfo {year} {2001})},\ \Eprint
  {http://arxiv.org/abs/hep-ph/0110005} {arXiv:hep-ph/0110005} \BibitemShut
  {NoStop}%
\bibitem [{\citenamefont {Shifman}\ \emph {et~al.}(1978)\citenamefont
  {Shifman}, \citenamefont {Vainshtein},\ and\ \citenamefont
  {Zakharov}}]{Shifman:1978zn}%
  \BibitemOpen
  \bibfield  {author} {\bibinfo {author} {\bibfnamefont {M.~A.}\ \bibnamefont
  {Shifman}}, \bibinfo {author} {\bibfnamefont {A.~I.}\ \bibnamefont
  {Vainshtein}}, \ and\ \bibinfo {author} {\bibfnamefont {V.~I.}\ \bibnamefont
  {Zakharov}},\ }\href {\doibase 10.1016/0370-2693(78)90481-1} {\bibfield
  {journal} {\bibinfo  {journal} {Phys. Lett. B}\ }\textbf {\bibinfo {volume}
  {78}},\ \bibinfo {pages} {443} (\bibinfo {year} {1978})}\BibitemShut
  {NoStop}%
\bibitem [{\citenamefont {Ji}(1995)}]{Ji:1994av}%
  \BibitemOpen
  \bibfield  {author} {\bibinfo {author} {\bibfnamefont {X.-D.}\ \bibnamefont
  {Ji}},\ }\href {\doibase 10.1103/PhysRevLett.74.1071} {\bibfield  {journal}
  {\bibinfo  {journal} {Phys. Rev. Lett.}\ }\textbf {\bibinfo {volume} {74}},\
  \bibinfo {pages} {1071} (\bibinfo {year} {1995})},\ \Eprint
  {http://arxiv.org/abs/hep-ph/9410274} {arXiv:hep-ph/9410274} \BibitemShut
  {NoStop}%
\bibitem [{\citenamefont {Yang}\ \emph {et~al.}(2016)\citenamefont {Yang},
  \citenamefont {Alexandru}, \citenamefont {Draper}, \citenamefont {Liang},\
  and\ \citenamefont {Liu}}]{Yang:2015uis}%
  \BibitemOpen
  \bibfield  {author} {\bibinfo {author} {\bibfnamefont {Y.-B.}\ \bibnamefont
  {Yang}}, \bibinfo {author} {\bibfnamefont {A.}~\bibnamefont {Alexandru}},
  \bibinfo {author} {\bibfnamefont {T.}~\bibnamefont {Draper}}, \bibinfo
  {author} {\bibfnamefont {J.}~\bibnamefont {Liang}}, \ and\ \bibinfo {author}
  {\bibfnamefont {K.-F.}\ \bibnamefont {Liu}} (\bibinfo {collaboration}
  {xQCD}),\ }\href {\doibase 10.1103/PhysRevD.94.054503} {\bibfield  {journal}
  {\bibinfo  {journal} {Phys. Rev. D}\ }\textbf {\bibinfo {volume} {94}},\
  \bibinfo {pages} {054503} (\bibinfo {year} {2016})},\ \Eprint
  {http://arxiv.org/abs/1511.09089} {arXiv:1511.09089 [hep-lat]} \BibitemShut
  {NoStop}%
\bibitem [{\citenamefont {Yang}\ \emph {et~al.}(2018)\citenamefont {Yang},
  \citenamefont {Liang}, \citenamefont {Bi}, \citenamefont {Chen},
  \citenamefont {Draper}, \citenamefont {Liu},\ and\ \citenamefont
  {Liu}}]{Yang:2018nqn}%
  \BibitemOpen
  \bibfield  {author} {\bibinfo {author} {\bibfnamefont {Y.-B.}\ \bibnamefont
  {Yang}}, \bibinfo {author} {\bibfnamefont {J.}~\bibnamefont {Liang}},
  \bibinfo {author} {\bibfnamefont {Y.-J.}\ \bibnamefont {Bi}}, \bibinfo
  {author} {\bibfnamefont {Y.}~\bibnamefont {Chen}}, \bibinfo {author}
  {\bibfnamefont {T.}~\bibnamefont {Draper}}, \bibinfo {author} {\bibfnamefont
  {K.-F.}\ \bibnamefont {Liu}}, \ and\ \bibinfo {author} {\bibfnamefont
  {Z.}~\bibnamefont {Liu}},\ }\href {\doibase 10.1103/PhysRevLett.121.212001}
  {\bibfield  {journal} {\bibinfo  {journal} {Phys. Rev. Lett.}\ }\textbf
  {\bibinfo {volume} {121}},\ \bibinfo {pages} {212001} (\bibinfo {year}
  {2018})},\ \Eprint {http://arxiv.org/abs/1808.08677} {arXiv:1808.08677
  [hep-lat]} \BibitemShut {NoStop}%
\bibitem [{\citenamefont {Walecka}(1974)}]{Walecka:1974qa}%
  \BibitemOpen
  \bibfield  {author} {\bibinfo {author} {\bibfnamefont {J.~D.}\ \bibnamefont
  {Walecka}},\ }\href {\doibase 10.1016/0003-4916(74)90208-5} {\bibfield
  {journal} {\bibinfo  {journal} {Annals Phys.}\ }\textbf {\bibinfo {volume}
  {83}},\ \bibinfo {pages} {491} (\bibinfo {year} {1974})}\BibitemShut
  {NoStop}%
\bibitem [{\citenamefont {Lee}\ and\ \citenamefont {Wick}(1974)}]{Lee:1974ma}%
  \BibitemOpen
  \bibfield  {author} {\bibinfo {author} {\bibfnamefont {T.~D.}\ \bibnamefont
  {Lee}}\ and\ \bibinfo {author} {\bibfnamefont {G.~C.}\ \bibnamefont {Wick}},\
  }\href {\doibase 10.1103/PhysRevD.9.2291} {\bibfield  {journal} {\bibinfo
  {journal} {Phys. Rev. D}\ }\textbf {\bibinfo {volume} {9}},\ \bibinfo {pages}
  {2291} (\bibinfo {year} {1974})}\BibitemShut {NoStop}%
\bibitem [{\citenamefont {Weyrich}\ \emph {et~al.}(2015)\citenamefont
  {Weyrich}, \citenamefont {Strodthoff},\ and\ \citenamefont {von
  Smekal}}]{Weyrich:2015hha}%
  \BibitemOpen
  \bibfield  {author} {\bibinfo {author} {\bibfnamefont {J.}~\bibnamefont
  {Weyrich}}, \bibinfo {author} {\bibfnamefont {N.}~\bibnamefont {Strodthoff}},
  \ and\ \bibinfo {author} {\bibfnamefont {L.}~\bibnamefont {von Smekal}},\
  }\href {\doibase 10.1103/PhysRevC.92.015214} {\bibfield  {journal} {\bibinfo
  {journal} {Phys. Rev. C}\ }\textbf {\bibinfo {volume} {92}},\ \bibinfo
  {pages} {015214} (\bibinfo {year} {2015})},\ \Eprint
  {http://arxiv.org/abs/1504.02697} {arXiv:1504.02697 [nucl-th]} \BibitemShut
  {NoStop}%
\bibitem [{\citenamefont {Tripolt}\ \emph {et~al.}(2021)\citenamefont
  {Tripolt}, \citenamefont {Jung}, \citenamefont {von Smekal},\ and\
  \citenamefont {Wambach}}]{Tripolt:2021jtp}%
  \BibitemOpen
  \bibfield  {author} {\bibinfo {author} {\bibfnamefont {R.-A.}\ \bibnamefont
  {Tripolt}}, \bibinfo {author} {\bibfnamefont {C.}~\bibnamefont {Jung}},
  \bibinfo {author} {\bibfnamefont {L.}~\bibnamefont {von Smekal}}, \ and\
  \bibinfo {author} {\bibfnamefont {J.}~\bibnamefont {Wambach}},\ }\href@noop
  {} {\  (\bibinfo {year} {2021})},\ \Eprint {http://arxiv.org/abs/2105.00861}
  {arXiv:2105.00861 [hep-ph]} \BibitemShut {NoStop}%
\bibitem [{\citenamefont {Gallas}\ \emph {et~al.}(2010)\citenamefont {Gallas},
  \citenamefont {Giacosa},\ and\ \citenamefont {Rischke}}]{Gallas:2009qp}%
  \BibitemOpen
  \bibfield  {author} {\bibinfo {author} {\bibfnamefont {S.}~\bibnamefont
  {Gallas}}, \bibinfo {author} {\bibfnamefont {F.}~\bibnamefont {Giacosa}}, \
  and\ \bibinfo {author} {\bibfnamefont {D.~H.}\ \bibnamefont {Rischke}},\
  }\href {\doibase 10.1103/PhysRevD.82.014004} {\bibfield  {journal} {\bibinfo
  {journal} {Phys. Rev. D}\ }\textbf {\bibinfo {volume} {82}},\ \bibinfo
  {pages} {014004} (\bibinfo {year} {2010})},\ \Eprint
  {http://arxiv.org/abs/0907.5084} {arXiv:0907.5084 [hep-ph]} \BibitemShut
  {NoStop}%
\bibitem [{\citenamefont {Nemoto}\ \emph {et~al.}(1998)\citenamefont {Nemoto},
  \citenamefont {Jido}, \citenamefont {Oka},\ and\ \citenamefont
  {Hosaka}}]{Nemoto:1998um}%
  \BibitemOpen
  \bibfield  {author} {\bibinfo {author} {\bibfnamefont {Y.}~\bibnamefont
  {Nemoto}}, \bibinfo {author} {\bibfnamefont {D.}~\bibnamefont {Jido}},
  \bibinfo {author} {\bibfnamefont {M.}~\bibnamefont {Oka}}, \ and\ \bibinfo
  {author} {\bibfnamefont {A.}~\bibnamefont {Hosaka}},\ }\href {\doibase
  10.1103/PhysRevD.57.4124} {\bibfield  {journal} {\bibinfo  {journal} {Phys.
  Rev. D}\ }\textbf {\bibinfo {volume} {57}},\ \bibinfo {pages} {4124}
  (\bibinfo {year} {1998})},\ \Eprint {http://arxiv.org/abs/hep-ph/9710445}
  {arXiv:hep-ph/9710445} \BibitemShut {NoStop}%
\bibitem [{\citenamefont {Bramon}\ \emph {et~al.}(2004)\citenamefont {Bramon},
  \citenamefont {Escribano},\ and\ \citenamefont
  {Lucio~Martinez}}]{Bramon:2003xq}%
  \BibitemOpen
  \bibfield  {author} {\bibinfo {author} {\bibfnamefont {A.}~\bibnamefont
  {Bramon}}, \bibinfo {author} {\bibfnamefont {R.}~\bibnamefont {Escribano}}, \
  and\ \bibinfo {author} {\bibfnamefont {J.~L.}\ \bibnamefont
  {Lucio~Martinez}},\ }\href {\doibase 10.1103/PhysRevD.69.074008} {\bibfield
  {journal} {\bibinfo  {journal} {Phys. Rev. D}\ }\textbf {\bibinfo {volume}
  {69}},\ \bibinfo {pages} {074008} (\bibinfo {year} {2004})},\ \Eprint
  {http://arxiv.org/abs/hep-ph/0312338} {arXiv:hep-ph/0312338} \BibitemShut
  {NoStop}%
\bibitem [{\citenamefont {Zschiesche}\ \emph {et~al.}(2007)\citenamefont
  {Zschiesche}, \citenamefont {Tolos}, \citenamefont {Schaffner-Bielich},\ and\
  \citenamefont {Pisarski}}]{Zschiesche:2006zj}%
  \BibitemOpen
  \bibfield  {author} {\bibinfo {author} {\bibfnamefont {D.}~\bibnamefont
  {Zschiesche}}, \bibinfo {author} {\bibfnamefont {L.}~\bibnamefont {Tolos}},
  \bibinfo {author} {\bibfnamefont {J.}~\bibnamefont {Schaffner-Bielich}}, \
  and\ \bibinfo {author} {\bibfnamefont {R.~D.}\ \bibnamefont {Pisarski}},\
  }\href {\doibase 10.1103/PhysRevC.75.055202} {\bibfield  {journal} {\bibinfo
  {journal} {Phys. Rev. C}\ }\textbf {\bibinfo {volume} {75}},\ \bibinfo
  {pages} {055202} (\bibinfo {year} {2007})},\ \Eprint
  {http://arxiv.org/abs/nucl-th/0608044} {arXiv:nucl-th/0608044} \BibitemShut
  {NoStop}%
\bibitem [{\citenamefont {Wilms}\ \emph {et~al.}(2007)\citenamefont {Wilms},
  \citenamefont {Giacosa},\ and\ \citenamefont {Rischke}}]{Wilms:2007uc}%
  \BibitemOpen
  \bibfield  {author} {\bibinfo {author} {\bibfnamefont {S.}~\bibnamefont
  {Wilms}}, \bibinfo {author} {\bibfnamefont {F.}~\bibnamefont {Giacosa}}, \
  and\ \bibinfo {author} {\bibfnamefont {D.~H.}\ \bibnamefont {Rischke}},\
  }\href {\doibase 10.1142/S0218301307007982} {\bibfield  {journal} {\bibinfo
  {journal} {Int. J. Mod. Phys. E}\ }\textbf {\bibinfo {volume} {16}},\
  \bibinfo {pages} {2388} (\bibinfo {year} {2007})},\ \Eprint
  {http://arxiv.org/abs/nucl-th/0702076} {arXiv:nucl-th/0702076} \BibitemShut
  {NoStop}%
\bibitem [{\citenamefont {Dexheimer}\ \emph
  {et~al.}(2008{\natexlab{a}})\citenamefont {Dexheimer}, \citenamefont
  {Schramm},\ and\ \citenamefont {Zschiesche}}]{Dexheimer:2007tn}%
  \BibitemOpen
  \bibfield  {author} {\bibinfo {author} {\bibfnamefont {V.}~\bibnamefont
  {Dexheimer}}, \bibinfo {author} {\bibfnamefont {S.}~\bibnamefont {Schramm}},
  \ and\ \bibinfo {author} {\bibfnamefont {D.}~\bibnamefont {Zschiesche}},\
  }\href {\doibase 10.1103/PhysRevC.77.025803} {\bibfield  {journal} {\bibinfo
  {journal} {Phys. Rev. C}\ }\textbf {\bibinfo {volume} {77}},\ \bibinfo
  {pages} {025803} (\bibinfo {year} {2008}{\natexlab{a}})},\ \Eprint
  {http://arxiv.org/abs/0710.4192} {arXiv:0710.4192 [nucl-th]} \BibitemShut
  {NoStop}%
\bibitem [{\citenamefont {Dexheimer}\ \emph
  {et~al.}(2008{\natexlab{b}})\citenamefont {Dexheimer}, \citenamefont
  {Pagliara}, \citenamefont {Tolos}, \citenamefont {Schaffner-Bielich},\ and\
  \citenamefont {Schramm}}]{Dexheimer:2008cv}%
  \BibitemOpen
  \bibfield  {author} {\bibinfo {author} {\bibfnamefont {V.}~\bibnamefont
  {Dexheimer}}, \bibinfo {author} {\bibfnamefont {G.}~\bibnamefont {Pagliara}},
  \bibinfo {author} {\bibfnamefont {L.}~\bibnamefont {Tolos}}, \bibinfo
  {author} {\bibfnamefont {J.}~\bibnamefont {Schaffner-Bielich}}, \ and\
  \bibinfo {author} {\bibfnamefont {S.}~\bibnamefont {Schramm}},\ }\href
  {\doibase 10.1140/epja/i2008-10652-0} {\bibfield  {journal} {\bibinfo
  {journal} {Eur. Phys. J. A}\ }\textbf {\bibinfo {volume} {38}},\ \bibinfo
  {pages} {105} (\bibinfo {year} {2008}{\natexlab{b}})},\ \Eprint
  {http://arxiv.org/abs/0805.3301} {arXiv:0805.3301 [nucl-th]} \BibitemShut
  {NoStop}%
\bibitem [{\citenamefont {Hayano}\ and\ \citenamefont
  {Hatsuda}(2010)}]{Hayano:2008vn}%
  \BibitemOpen
  \bibfield  {author} {\bibinfo {author} {\bibfnamefont {R.~S.}\ \bibnamefont
  {Hayano}}\ and\ \bibinfo {author} {\bibfnamefont {T.}~\bibnamefont
  {Hatsuda}},\ }\href {\doibase 10.1103/RevModPhys.82.2949} {\bibfield
  {journal} {\bibinfo  {journal} {Rev. Mod. Phys.}\ }\textbf {\bibinfo {volume}
  {82}},\ \bibinfo {pages} {2949} (\bibinfo {year} {2010})},\ \Eprint
  {http://arxiv.org/abs/0812.1702} {arXiv:0812.1702 [nucl-ex]} \BibitemShut
  {NoStop}%
\bibitem [{\citenamefont {Sasaki}\ and\ \citenamefont
  {Mishustin}(2010)}]{Sasaki:2010bp}%
  \BibitemOpen
  \bibfield  {author} {\bibinfo {author} {\bibfnamefont {C.}~\bibnamefont
  {Sasaki}}\ and\ \bibinfo {author} {\bibfnamefont {I.}~\bibnamefont
  {Mishustin}},\ }\href {\doibase 10.1103/PhysRevC.82.035204} {\bibfield
  {journal} {\bibinfo  {journal} {Phys. Rev. C}\ }\textbf {\bibinfo {volume}
  {82}},\ \bibinfo {pages} {035204} (\bibinfo {year} {2010})},\ \Eprint
  {http://arxiv.org/abs/1005.4811} {arXiv:1005.4811 [hep-ph]} \BibitemShut
  {NoStop}%
\bibitem [{\citenamefont {Sasaki}\ \emph {et~al.}(2011)\citenamefont {Sasaki},
  \citenamefont {Lee}, \citenamefont {Paeng},\ and\ \citenamefont
  {Rho}}]{Sasaki:2011ff}%
  \BibitemOpen
  \bibfield  {author} {\bibinfo {author} {\bibfnamefont {C.}~\bibnamefont
  {Sasaki}}, \bibinfo {author} {\bibfnamefont {H.~K.}\ \bibnamefont {Lee}},
  \bibinfo {author} {\bibfnamefont {W.-G.}\ \bibnamefont {Paeng}}, \ and\
  \bibinfo {author} {\bibfnamefont {M.}~\bibnamefont {Rho}},\ }\href {\doibase
  10.1103/PhysRevD.84.034011} {\bibfield  {journal} {\bibinfo  {journal} {Phys.
  Rev. D}\ }\textbf {\bibinfo {volume} {84}},\ \bibinfo {pages} {034011}
  (\bibinfo {year} {2011})},\ \Eprint {http://arxiv.org/abs/1103.0184}
  {arXiv:1103.0184 [hep-ph]} \BibitemShut {NoStop}%
\bibitem [{\citenamefont {Giacosa}(2012)}]{Giacosa:2011qd}%
  \BibitemOpen
  \bibfield  {author} {\bibinfo {author} {\bibfnamefont {F.}~\bibnamefont
  {Giacosa}},\ }\href {\doibase 10.1016/j.ppnp.2011.12.039} {\bibfield
  {journal} {\bibinfo  {journal} {Prog. Part. Nucl. Phys.}\ }\textbf {\bibinfo
  {volume} {67}},\ \bibinfo {pages} {332} (\bibinfo {year} {2012})},\ \Eprint
  {http://arxiv.org/abs/1111.4944} {arXiv:1111.4944 [hep-ph]} \BibitemShut
  {NoStop}%
\bibitem [{\citenamefont {Gallas}\ \emph {et~al.}(2011)\citenamefont {Gallas},
  \citenamefont {Giacosa},\ and\ \citenamefont {Pagliara}}]{Gallas:2011qp}%
  \BibitemOpen
  \bibfield  {author} {\bibinfo {author} {\bibfnamefont {S.}~\bibnamefont
  {Gallas}}, \bibinfo {author} {\bibfnamefont {F.}~\bibnamefont {Giacosa}}, \
  and\ \bibinfo {author} {\bibfnamefont {G.}~\bibnamefont {Pagliara}},\ }\href
  {\doibase 10.1016/j.nuclphysa.2011.09.008} {\bibfield  {journal} {\bibinfo
  {journal} {Nucl. Phys. A}\ }\textbf {\bibinfo {volume} {872}},\ \bibinfo
  {pages} {13} (\bibinfo {year} {2011})},\ \Eprint
  {http://arxiv.org/abs/1105.5003} {arXiv:1105.5003 [hep-ph]} \BibitemShut
  {NoStop}%
\bibitem [{\citenamefont {Steinheimer}\ \emph {et~al.}(2011)\citenamefont
  {Steinheimer}, \citenamefont {Schramm},\ and\ \citenamefont
  {Stocker}}]{Steinheimer:2011ea}%
  \BibitemOpen
  \bibfield  {author} {\bibinfo {author} {\bibfnamefont {J.}~\bibnamefont
  {Steinheimer}}, \bibinfo {author} {\bibfnamefont {S.}~\bibnamefont
  {Schramm}}, \ and\ \bibinfo {author} {\bibfnamefont {H.}~\bibnamefont
  {Stocker}},\ }\href {\doibase 10.1103/PhysRevC.84.045208} {\bibfield
  {journal} {\bibinfo  {journal} {Phys. Rev. C}\ }\textbf {\bibinfo {volume}
  {84}},\ \bibinfo {pages} {045208} (\bibinfo {year} {2011})},\ \Eprint
  {http://arxiv.org/abs/1108.2596} {arXiv:1108.2596 [hep-ph]} \BibitemShut
  {NoStop}%
\bibitem [{\citenamefont {Paeng}\ \emph {et~al.}(2012)\citenamefont {Paeng},
  \citenamefont {Lee}, \citenamefont {Rho},\ and\ \citenamefont
  {Sasaki}}]{Paeng:2011hy}%
  \BibitemOpen
  \bibfield  {author} {\bibinfo {author} {\bibfnamefont {W.-G.}\ \bibnamefont
  {Paeng}}, \bibinfo {author} {\bibfnamefont {H.~K.}\ \bibnamefont {Lee}},
  \bibinfo {author} {\bibfnamefont {M.}~\bibnamefont {Rho}}, \ and\ \bibinfo
  {author} {\bibfnamefont {C.}~\bibnamefont {Sasaki}},\ }\href {\doibase
  10.1103/PhysRevD.85.054022} {\bibfield  {journal} {\bibinfo  {journal} {Phys.
  Rev. D}\ }\textbf {\bibinfo {volume} {85}},\ \bibinfo {pages} {054022}
  (\bibinfo {year} {2012})},\ \Eprint {http://arxiv.org/abs/1109.5431}
  {arXiv:1109.5431 [hep-ph]} \BibitemShut {NoStop}%
\bibitem [{\citenamefont {Dexheimer}\ \emph {et~al.}(2013)\citenamefont
  {Dexheimer}, \citenamefont {Steinheimer}, \citenamefont {Negreiros},\ and\
  \citenamefont {Schramm}}]{Dexheimer:2012eu}%
  \BibitemOpen
  \bibfield  {author} {\bibinfo {author} {\bibfnamefont {V.}~\bibnamefont
  {Dexheimer}}, \bibinfo {author} {\bibfnamefont {J.}~\bibnamefont
  {Steinheimer}}, \bibinfo {author} {\bibfnamefont {R.}~\bibnamefont
  {Negreiros}}, \ and\ \bibinfo {author} {\bibfnamefont {S.}~\bibnamefont
  {Schramm}},\ }\href {\doibase 10.1103/PhysRevC.87.015804} {\bibfield
  {journal} {\bibinfo  {journal} {Phys. Rev. C}\ }\textbf {\bibinfo {volume}
  {87}},\ \bibinfo {pages} {015804} (\bibinfo {year} {2013})},\ \Eprint
  {http://arxiv.org/abs/1206.3086} {arXiv:1206.3086 [astro-ph.HE]} \BibitemShut
  {NoStop}%
\bibitem [{\citenamefont {Gallas}\ and\ \citenamefont
  {Giacosa}(2014)}]{Gallas:2013ipa}%
  \BibitemOpen
  \bibfield  {author} {\bibinfo {author} {\bibfnamefont {S.}~\bibnamefont
  {Gallas}}\ and\ \bibinfo {author} {\bibfnamefont {F.}~\bibnamefont
  {Giacosa}},\ }\href {\doibase 10.1142/S0217751X14500985} {\bibfield
  {journal} {\bibinfo  {journal} {Int. J. Mod. Phys. A}\ }\textbf {\bibinfo
  {volume} {29}},\ \bibinfo {pages} {1450098} (\bibinfo {year} {2014})},\
  \Eprint {http://arxiv.org/abs/1308.4817} {arXiv:1308.4817 [hep-ph]}
  \BibitemShut {NoStop}%
\bibitem [{\citenamefont {Heinz}\ \emph {et~al.}(2015)\citenamefont {Heinz},
  \citenamefont {Giacosa},\ and\ \citenamefont {Rischke}}]{Heinz:2013hza}%
  \BibitemOpen
  \bibfield  {author} {\bibinfo {author} {\bibfnamefont {A.}~\bibnamefont
  {Heinz}}, \bibinfo {author} {\bibfnamefont {F.}~\bibnamefont {Giacosa}}, \
  and\ \bibinfo {author} {\bibfnamefont {D.~H.}\ \bibnamefont {Rischke}},\
  }\href {\doibase 10.1016/j.nuclphysa.2014.09.027} {\bibfield  {journal}
  {\bibinfo  {journal} {Nucl. Phys. A}\ }\textbf {\bibinfo {volume} {933}},\
  \bibinfo {pages} {34} (\bibinfo {year} {2015})},\ \Eprint
  {http://arxiv.org/abs/1312.3244} {arXiv:1312.3244 [nucl-th]} \BibitemShut
  {NoStop}%
\bibitem [{\citenamefont {Paeng}\ \emph {et~al.}(2013)\citenamefont {Paeng},
  \citenamefont {Lee}, \citenamefont {Rho},\ and\ \citenamefont
  {Sasaki}}]{Paeng:2013xya}%
  \BibitemOpen
  \bibfield  {author} {\bibinfo {author} {\bibfnamefont {W.-G.}\ \bibnamefont
  {Paeng}}, \bibinfo {author} {\bibfnamefont {H.~K.}\ \bibnamefont {Lee}},
  \bibinfo {author} {\bibfnamefont {M.}~\bibnamefont {Rho}}, \ and\ \bibinfo
  {author} {\bibfnamefont {C.}~\bibnamefont {Sasaki}},\ }\href {\doibase
  10.1103/PhysRevD.88.105019} {\bibfield  {journal} {\bibinfo  {journal} {Phys.
  Rev. D}\ }\textbf {\bibinfo {volume} {88}},\ \bibinfo {pages} {105019}
  (\bibinfo {year} {2013})},\ \Eprint {http://arxiv.org/abs/1303.2898}
  {arXiv:1303.2898 [nucl-th]} \BibitemShut {NoStop}%
\bibitem [{\citenamefont {Benic}\ \emph {et~al.}(2015)\citenamefont {Benic},
  \citenamefont {Mishustin},\ and\ \citenamefont {Sasaki}}]{Benic:2015pia}%
  \BibitemOpen
  \bibfield  {author} {\bibinfo {author} {\bibfnamefont {S.}~\bibnamefont
  {Benic}}, \bibinfo {author} {\bibfnamefont {I.}~\bibnamefont {Mishustin}}, \
  and\ \bibinfo {author} {\bibfnamefont {C.}~\bibnamefont {Sasaki}},\ }\href
  {\doibase 10.1103/PhysRevD.91.125034} {\bibfield  {journal} {\bibinfo
  {journal} {Phys. Rev. D}\ }\textbf {\bibinfo {volume} {91}},\ \bibinfo
  {pages} {125034} (\bibinfo {year} {2015})},\ \Eprint
  {http://arxiv.org/abs/1502.05969} {arXiv:1502.05969 [hep-ph]} \BibitemShut
  {NoStop}%
\bibitem [{\citenamefont {Motohiro}\ \emph {et~al.}(2015)\citenamefont
  {Motohiro}, \citenamefont {Kim},\ and\ \citenamefont
  {Harada}}]{Motohiro:2015taa}%
  \BibitemOpen
  \bibfield  {author} {\bibinfo {author} {\bibfnamefont {Y.}~\bibnamefont
  {Motohiro}}, \bibinfo {author} {\bibfnamefont {Y.}~\bibnamefont {Kim}}, \
  and\ \bibinfo {author} {\bibfnamefont {M.}~\bibnamefont {Harada}},\ }\href
  {\doibase 10.1103/PhysRevC.92.025201} {\bibfield  {journal} {\bibinfo
  {journal} {Phys. Rev. C}\ }\textbf {\bibinfo {volume} {92}},\ \bibinfo
  {pages} {025201} (\bibinfo {year} {2015})},\ \bibinfo {note} {[Erratum:
  Phys.Rev.C 95, 059903 (2017)]},\ \Eprint {http://arxiv.org/abs/1505.00988}
  {arXiv:1505.00988 [nucl-th]} \BibitemShut {NoStop}%
\bibitem [{\citenamefont {Olbrich}\ \emph {et~al.}(2016)\citenamefont
  {Olbrich}, \citenamefont {Z\'et\'enyi}, \citenamefont {Giacosa},\ and\
  \citenamefont {Rischke}}]{Olbrich:2015gln}%
  \BibitemOpen
  \bibfield  {author} {\bibinfo {author} {\bibfnamefont {L.}~\bibnamefont
  {Olbrich}}, \bibinfo {author} {\bibfnamefont {M.}~\bibnamefont
  {Z\'et\'enyi}}, \bibinfo {author} {\bibfnamefont {F.}~\bibnamefont
  {Giacosa}}, \ and\ \bibinfo {author} {\bibfnamefont {D.~H.}\ \bibnamefont
  {Rischke}},\ }\href {\doibase 10.1103/PhysRevD.93.034021} {\bibfield
  {journal} {\bibinfo  {journal} {Phys. Rev. D}\ }\textbf {\bibinfo {volume}
  {93}},\ \bibinfo {pages} {034021} (\bibinfo {year} {2016})},\ \Eprint
  {http://arxiv.org/abs/1511.05035} {arXiv:1511.05035 [hep-ph]} \BibitemShut
  {NoStop}%
\bibitem [{\citenamefont {Mukherjee}\ \emph
  {et~al.}(2017{\natexlab{a}})\citenamefont {Mukherjee}, \citenamefont
  {Steinheimer},\ and\ \citenamefont {Schramm}}]{Mukherjee:2016nhb}%
  \BibitemOpen
  \bibfield  {author} {\bibinfo {author} {\bibfnamefont {A.}~\bibnamefont
  {Mukherjee}}, \bibinfo {author} {\bibfnamefont {J.}~\bibnamefont
  {Steinheimer}}, \ and\ \bibinfo {author} {\bibfnamefont {S.}~\bibnamefont
  {Schramm}},\ }\href {\doibase 10.1103/PhysRevC.96.025205} {\bibfield
  {journal} {\bibinfo  {journal} {Phys. Rev. C}\ }\textbf {\bibinfo {volume}
  {96}},\ \bibinfo {pages} {025205} (\bibinfo {year} {2017}{\natexlab{a}})},\
  \Eprint {http://arxiv.org/abs/1611.10144} {arXiv:1611.10144 [nucl-th]}
  \BibitemShut {NoStop}%
\bibitem [{\citenamefont {Mukherjee}\ \emph
  {et~al.}(2017{\natexlab{b}})\citenamefont {Mukherjee}, \citenamefont
  {Schramm}, \citenamefont {Steinheimer},\ and\ \citenamefont
  {Dexheimer}}]{Mukherjee:2017jzi}%
  \BibitemOpen
  \bibfield  {author} {\bibinfo {author} {\bibfnamefont {A.}~\bibnamefont
  {Mukherjee}}, \bibinfo {author} {\bibfnamefont {S.}~\bibnamefont {Schramm}},
  \bibinfo {author} {\bibfnamefont {J.}~\bibnamefont {Steinheimer}}, \ and\
  \bibinfo {author} {\bibfnamefont {V.}~\bibnamefont {Dexheimer}},\ }\href
  {\doibase 10.1051/0004-6361/201731505} {\bibfield  {journal} {\bibinfo
  {journal} {Astron. Astrophys.}\ }\textbf {\bibinfo {volume} {608}},\ \bibinfo
  {pages} {A110} (\bibinfo {year} {2017}{\natexlab{b}})},\ \Eprint
  {http://arxiv.org/abs/1706.09191} {arXiv:1706.09191 [nucl-th]} \BibitemShut
  {NoStop}%
\bibitem [{\citenamefont {Suenaga}(2018)}]{Suenaga:2017wbb}%
  \BibitemOpen
  \bibfield  {author} {\bibinfo {author} {\bibfnamefont {D.}~\bibnamefont
  {Suenaga}},\ }\href {\doibase 10.1103/PhysRevC.97.045203} {\bibfield
  {journal} {\bibinfo  {journal} {Phys. Rev. C}\ }\textbf {\bibinfo {volume}
  {97}},\ \bibinfo {pages} {045203} (\bibinfo {year} {2018})},\ \Eprint
  {http://arxiv.org/abs/1704.03630} {arXiv:1704.03630 [nucl-th]} \BibitemShut
  {NoStop}%
\bibitem [{\citenamefont {Takeda}\ \emph
  {et~al.}(2018{\natexlab{a}})\citenamefont {Takeda}, \citenamefont {Kim},\
  and\ \citenamefont {Harada}}]{Takeda:2017mrm}%
  \BibitemOpen
  \bibfield  {author} {\bibinfo {author} {\bibfnamefont {Y.}~\bibnamefont
  {Takeda}}, \bibinfo {author} {\bibfnamefont {Y.}~\bibnamefont {Kim}}, \ and\
  \bibinfo {author} {\bibfnamefont {M.}~\bibnamefont {Harada}},\ }\href
  {\doibase 10.1103/PhysRevC.97.065202} {\bibfield  {journal} {\bibinfo
  {journal} {Phys. Rev. C}\ }\textbf {\bibinfo {volume} {97}},\ \bibinfo
  {pages} {065202} (\bibinfo {year} {2018}{\natexlab{a}})},\ \Eprint
  {http://arxiv.org/abs/1704.04357} {arXiv:1704.04357 [nucl-th]} \BibitemShut
  {NoStop}%
\bibitem [{\citenamefont {Paeng}\ \emph {et~al.}(2017)\citenamefont {Paeng},
  \citenamefont {Kuo}, \citenamefont {Lee}, \citenamefont {Ma},\ and\
  \citenamefont {Rho}}]{Paeng:2017qvp}%
  \BibitemOpen
  \bibfield  {author} {\bibinfo {author} {\bibfnamefont {W.-G.}\ \bibnamefont
  {Paeng}}, \bibinfo {author} {\bibfnamefont {T.~T.~S.}\ \bibnamefont {Kuo}},
  \bibinfo {author} {\bibfnamefont {H.~K.}\ \bibnamefont {Lee}}, \bibinfo
  {author} {\bibfnamefont {Y.-L.}\ \bibnamefont {Ma}}, \ and\ \bibinfo {author}
  {\bibfnamefont {M.}~\bibnamefont {Rho}},\ }\href {\doibase
  10.1103/PhysRevD.96.014031} {\bibfield  {journal} {\bibinfo  {journal} {Phys.
  Rev. D}\ }\textbf {\bibinfo {volume} {96}},\ \bibinfo {pages} {014031}
  (\bibinfo {year} {2017})},\ \Eprint {http://arxiv.org/abs/1704.02775}
  {arXiv:1704.02775 [nucl-th]} \BibitemShut {NoStop}%
\bibitem [{\citenamefont {Marczenko}\ and\ \citenamefont
  {Sasaki}(2018)}]{Marczenko:2017huu}%
  \BibitemOpen
  \bibfield  {author} {\bibinfo {author} {\bibfnamefont {M.}~\bibnamefont
  {Marczenko}}\ and\ \bibinfo {author} {\bibfnamefont {C.}~\bibnamefont
  {Sasaki}},\ }\href {\doibase 10.1103/PhysRevD.97.036011} {\bibfield
  {journal} {\bibinfo  {journal} {Phys. Rev. D}\ }\textbf {\bibinfo {volume}
  {97}},\ \bibinfo {pages} {036011} (\bibinfo {year} {2018})},\ \Eprint
  {http://arxiv.org/abs/1711.05521} {arXiv:1711.05521 [hep-ph]} \BibitemShut
  {NoStop}%
\bibitem [{\citenamefont {Sasaki}(2018)}]{Sasaki:2017glk}%
  \BibitemOpen
  \bibfield  {author} {\bibinfo {author} {\bibfnamefont {C.}~\bibnamefont
  {Sasaki}},\ }\href {\doibase 10.1016/j.nuclphysa.2018.01.004} {\bibfield
  {journal} {\bibinfo  {journal} {Nucl. Phys. A}\ }\textbf {\bibinfo {volume}
  {970}},\ \bibinfo {pages} {388} (\bibinfo {year} {2018})},\ \Eprint
  {http://arxiv.org/abs/1707.05081} {arXiv:1707.05081 [hep-ph]} \BibitemShut
  {NoStop}%
\bibitem [{\citenamefont {Marczenko}\ \emph {et~al.}(2018)\citenamefont
  {Marczenko}, \citenamefont {Blaschke}, \citenamefont {Redlich},\ and\
  \citenamefont {Sasaki}}]{Marczenko:2018jui}%
  \BibitemOpen
  \bibfield  {author} {\bibinfo {author} {\bibfnamefont {M.}~\bibnamefont
  {Marczenko}}, \bibinfo {author} {\bibfnamefont {D.}~\bibnamefont {Blaschke}},
  \bibinfo {author} {\bibfnamefont {K.}~\bibnamefont {Redlich}}, \ and\
  \bibinfo {author} {\bibfnamefont {C.}~\bibnamefont {Sasaki}},\ }\href
  {\doibase 10.1103/PhysRevD.98.103021} {\bibfield  {journal} {\bibinfo
  {journal} {Phys. Rev. D}\ }\textbf {\bibinfo {volume} {98}},\ \bibinfo
  {pages} {103021} (\bibinfo {year} {2018})},\ \Eprint
  {http://arxiv.org/abs/1805.06886} {arXiv:1805.06886 [nucl-th]} \BibitemShut
  {NoStop}%
\bibitem [{\citenamefont {Takeda}\ \emph
  {et~al.}(2018{\natexlab{b}})\citenamefont {Takeda}, \citenamefont {Abuki},\
  and\ \citenamefont {Harada}}]{Takeda:2018ldi}%
  \BibitemOpen
  \bibfield  {author} {\bibinfo {author} {\bibfnamefont {Y.}~\bibnamefont
  {Takeda}}, \bibinfo {author} {\bibfnamefont {H.}~\bibnamefont {Abuki}}, \
  and\ \bibinfo {author} {\bibfnamefont {M.}~\bibnamefont {Harada}},\ }\href
  {\doibase 10.1103/PhysRevD.97.094032} {\bibfield  {journal} {\bibinfo
  {journal} {Phys. Rev. D}\ }\textbf {\bibinfo {volume} {97}},\ \bibinfo
  {pages} {094032} (\bibinfo {year} {2018}{\natexlab{b}})},\ \Eprint
  {http://arxiv.org/abs/1803.06779} {arXiv:1803.06779 [hep-ph]} \BibitemShut
  {NoStop}%
\bibitem [{\citenamefont {Yamazaki}\ and\ \citenamefont
  {Harada}(2019{\natexlab{a}})}]{Yamazaki:2018stk}%
  \BibitemOpen
  \bibfield  {author} {\bibinfo {author} {\bibfnamefont {T.}~\bibnamefont
  {Yamazaki}}\ and\ \bibinfo {author} {\bibfnamefont {M.}~\bibnamefont
  {Harada}},\ }\href {\doibase 10.1103/PhysRevD.99.034012} {\bibfield
  {journal} {\bibinfo  {journal} {Phys. Rev. D}\ }\textbf {\bibinfo {volume}
  {99}},\ \bibinfo {pages} {034012} (\bibinfo {year} {2019}{\natexlab{a}})},\
  \Eprint {http://arxiv.org/abs/1809.02359} {arXiv:1809.02359 [hep-ph]}
  \BibitemShut {NoStop}%
\bibitem [{\citenamefont {Yamazaki}\ and\ \citenamefont
  {Harada}(2019{\natexlab{b}})}]{Yamazaki:2019tuo}%
  \BibitemOpen
  \bibfield  {author} {\bibinfo {author} {\bibfnamefont {T.}~\bibnamefont
  {Yamazaki}}\ and\ \bibinfo {author} {\bibfnamefont {M.}~\bibnamefont
  {Harada}},\ }\href {\doibase 10.1103/PhysRevC.100.025205} {\bibfield
  {journal} {\bibinfo  {journal} {Phys. Rev. C}\ }\textbf {\bibinfo {volume}
  {100}},\ \bibinfo {pages} {025205} (\bibinfo {year} {2019}{\natexlab{b}})},\
  \Eprint {http://arxiv.org/abs/1901.02167} {arXiv:1901.02167 [nucl-th]}
  \BibitemShut {NoStop}%
\bibitem [{\citenamefont {Marczenko}\ \emph {et~al.}(2019)\citenamefont
  {Marczenko}, \citenamefont {Blaschke}, \citenamefont {Redlich},\ and\
  \citenamefont {Sasaki}}]{Marczenko:2019trv}%
  \BibitemOpen
  \bibfield  {author} {\bibinfo {author} {\bibfnamefont {M.}~\bibnamefont
  {Marczenko}}, \bibinfo {author} {\bibfnamefont {D.}~\bibnamefont {Blaschke}},
  \bibinfo {author} {\bibfnamefont {K.}~\bibnamefont {Redlich}}, \ and\
  \bibinfo {author} {\bibfnamefont {C.}~\bibnamefont {Sasaki}},\ }\href
  {\doibase 10.3390/universe5080180} {\bibfield  {journal} {\bibinfo  {journal}
  {Universe}\ }\textbf {\bibinfo {volume} {5}},\ \bibinfo {pages} {180}
  (\bibinfo {year} {2019})},\ \Eprint {http://arxiv.org/abs/1905.04974}
  {arXiv:1905.04974 [nucl-th]} \BibitemShut {NoStop}%
\bibitem [{\citenamefont {Suenaga}\ and\ \citenamefont
  {Lakaschus}(2020)}]{Suenaga:2019urn}%
  \BibitemOpen
  \bibfield  {author} {\bibinfo {author} {\bibfnamefont {D.}~\bibnamefont
  {Suenaga}}\ and\ \bibinfo {author} {\bibfnamefont {P.}~\bibnamefont
  {Lakaschus}},\ }\href {\doibase 10.1103/PhysRevC.101.035209} {\bibfield
  {journal} {\bibinfo  {journal} {Phys. Rev. C}\ }\textbf {\bibinfo {volume}
  {101}},\ \bibinfo {pages} {035209} (\bibinfo {year} {2020})},\ \Eprint
  {http://arxiv.org/abs/1908.10509} {arXiv:1908.10509 [nucl-th]} \BibitemShut
  {NoStop}%
\bibitem [{\citenamefont {Marczenko}\ \emph {et~al.}(2020)\citenamefont
  {Marczenko}, \citenamefont {Blaschke}, \citenamefont {Redlich},\ and\
  \citenamefont {Sasaki}}]{Marczenko:2020jma}%
  \BibitemOpen
  \bibfield  {author} {\bibinfo {author} {\bibfnamefont {M.}~\bibnamefont
  {Marczenko}}, \bibinfo {author} {\bibfnamefont {D.}~\bibnamefont {Blaschke}},
  \bibinfo {author} {\bibfnamefont {K.}~\bibnamefont {Redlich}}, \ and\
  \bibinfo {author} {\bibfnamefont {C.}~\bibnamefont {Sasaki}},\ }\href
  {\doibase 10.1051/0004-6361/202038211} {\bibfield  {journal} {\bibinfo
  {journal} {Astron. Astrophys.}\ }\textbf {\bibinfo {volume} {643}},\ \bibinfo
  {pages} {A82} (\bibinfo {year} {2020})},\ \Eprint
  {http://arxiv.org/abs/2004.09566} {arXiv:2004.09566 [astro-ph.HE]}
  \BibitemShut {NoStop}%
\bibitem [{\citenamefont {Minamikawa}\ \emph
  {et~al.}(2021{\natexlab{a}})\citenamefont {Minamikawa}, \citenamefont
  {Kojo},\ and\ \citenamefont {Harada}}]{Minamikawa:2020jfj}%
  \BibitemOpen
  \bibfield  {author} {\bibinfo {author} {\bibfnamefont {T.}~\bibnamefont
  {Minamikawa}}, \bibinfo {author} {\bibfnamefont {T.}~\bibnamefont {Kojo}}, \
  and\ \bibinfo {author} {\bibfnamefont {M.}~\bibnamefont {Harada}},\ }\href
  {\doibase 10.1103/PhysRevC.103.045205} {\bibfield  {journal} {\bibinfo
  {journal} {Phys. Rev. C}\ }\textbf {\bibinfo {volume} {103}},\ \bibinfo
  {pages} {045205} (\bibinfo {year} {2021}{\natexlab{a}})},\ \Eprint
  {http://arxiv.org/abs/2011.13684} {arXiv:2011.13684 [nucl-th]} \BibitemShut
  {NoStop}%
\bibitem [{\citenamefont {Marczenko}\ \emph {et~al.}(2021)\citenamefont
  {Marczenko}, \citenamefont {Redlich},\ and\ \citenamefont
  {Sasaki}}]{Marczenko:2020omo}%
  \BibitemOpen
  \bibfield  {author} {\bibinfo {author} {\bibfnamefont {M.}~\bibnamefont
  {Marczenko}}, \bibinfo {author} {\bibfnamefont {K.}~\bibnamefont {Redlich}},
  \ and\ \bibinfo {author} {\bibfnamefont {C.}~\bibnamefont {Sasaki}},\ }\href
  {\doibase 10.1103/PhysRevD.103.054035} {\bibfield  {journal} {\bibinfo
  {journal} {Phys. Rev. D}\ }\textbf {\bibinfo {volume} {103}},\ \bibinfo
  {pages} {054035} (\bibinfo {year} {2021})},\ \Eprint
  {http://arxiv.org/abs/2012.15535} {arXiv:2012.15535 [hep-ph]} \BibitemShut
  {NoStop}%
\bibitem [{\citenamefont {Minamikawa}\ \emph
  {et~al.}(2021{\natexlab{b}})\citenamefont {Minamikawa}, \citenamefont
  {Kojo},\ and\ \citenamefont {Harada}}]{Minamikawa:2021fln}%
  \BibitemOpen
  \bibfield  {author} {\bibinfo {author} {\bibfnamefont {T.}~\bibnamefont
  {Minamikawa}}, \bibinfo {author} {\bibfnamefont {T.}~\bibnamefont {Kojo}}, \
  and\ \bibinfo {author} {\bibfnamefont {M.}~\bibnamefont {Harada}},\
  }\href@noop {} {\  (\bibinfo {year} {2021}{\natexlab{b}})},\ \Eprint
  {http://arxiv.org/abs/2107.14545} {arXiv:2107.14545 [nucl-th]} \BibitemShut
  {NoStop}%
\bibitem [{\citenamefont {Glozman}\ \emph {et~al.}(2012)\citenamefont
  {Glozman}, \citenamefont {Lang},\ and\ \citenamefont
  {Schrock}}]{Glozman:2012fj}%
  \BibitemOpen
  \bibfield  {author} {\bibinfo {author} {\bibfnamefont {L.~Y.}\ \bibnamefont
  {Glozman}}, \bibinfo {author} {\bibfnamefont {C.~B.}\ \bibnamefont {Lang}}, \
  and\ \bibinfo {author} {\bibfnamefont {M.}~\bibnamefont {Schrock}},\ }\href
  {\doibase 10.1103/PhysRevD.86.014507} {\bibfield  {journal} {\bibinfo
  {journal} {Phys. Rev. D}\ }\textbf {\bibinfo {volume} {86}},\ \bibinfo
  {pages} {014507} (\bibinfo {year} {2012})},\ \Eprint
  {http://arxiv.org/abs/1205.4887} {arXiv:1205.4887 [hep-lat]} \BibitemShut
  {NoStop}%
\bibitem [{\citenamefont {Aarts}\ \emph {et~al.}(2015)\citenamefont {Aarts},
  \citenamefont {Allton}, \citenamefont {Hands}, \citenamefont {J\"ager},
  \citenamefont {Praki},\ and\ \citenamefont {Skullerud}}]{Aarts:2015mma}%
  \BibitemOpen
  \bibfield  {author} {\bibinfo {author} {\bibfnamefont {G.}~\bibnamefont
  {Aarts}}, \bibinfo {author} {\bibfnamefont {C.}~\bibnamefont {Allton}},
  \bibinfo {author} {\bibfnamefont {S.}~\bibnamefont {Hands}}, \bibinfo
  {author} {\bibfnamefont {B.}~\bibnamefont {J\"ager}}, \bibinfo {author}
  {\bibfnamefont {C.}~\bibnamefont {Praki}}, \ and\ \bibinfo {author}
  {\bibfnamefont {J.-I.}\ \bibnamefont {Skullerud}},\ }\href {\doibase
  10.1103/PhysRevD.92.014503} {\bibfield  {journal} {\bibinfo  {journal} {Phys.
  Rev. D}\ }\textbf {\bibinfo {volume} {92}},\ \bibinfo {pages} {014503}
  (\bibinfo {year} {2015})},\ \Eprint {http://arxiv.org/abs/1502.03603}
  {arXiv:1502.03603 [hep-lat]} \BibitemShut {NoStop}%
\bibitem [{\citenamefont {Aarts}\ \emph {et~al.}(2017)\citenamefont {Aarts},
  \citenamefont {Allton}, \citenamefont {De~Boni}, \citenamefont {Hands},
  \citenamefont {J\"ager}, \citenamefont {Praki},\ and\ \citenamefont
  {Skullerud}}]{Aarts:2017rrl}%
  \BibitemOpen
  \bibfield  {author} {\bibinfo {author} {\bibfnamefont {G.}~\bibnamefont
  {Aarts}}, \bibinfo {author} {\bibfnamefont {C.}~\bibnamefont {Allton}},
  \bibinfo {author} {\bibfnamefont {D.}~\bibnamefont {De~Boni}}, \bibinfo
  {author} {\bibfnamefont {S.}~\bibnamefont {Hands}}, \bibinfo {author}
  {\bibfnamefont {B.}~\bibnamefont {J\"ager}}, \bibinfo {author} {\bibfnamefont
  {C.}~\bibnamefont {Praki}}, \ and\ \bibinfo {author} {\bibfnamefont {J.-I.}\
  \bibnamefont {Skullerud}},\ }\href {\doibase 10.1007/JHEP06(2017)034}
  {\bibfield  {journal} {\bibinfo  {journal} {JHEP}\ }\textbf {\bibinfo
  {volume} {06}},\ \bibinfo {pages} {034} (\bibinfo {year} {2017})},\ \Eprint
  {http://arxiv.org/abs/1703.09246} {arXiv:1703.09246 [hep-lat]} \BibitemShut
  {NoStop}%
\bibitem [{\citenamefont {Berges}\ \emph {et~al.}(2003)\citenamefont {Berges},
  \citenamefont {Jungnickel},\ and\ \citenamefont {Wetterich}}]{Berges:1998ha}%
  \BibitemOpen
  \bibfield  {author} {\bibinfo {author} {\bibfnamefont {J.}~\bibnamefont
  {Berges}}, \bibinfo {author} {\bibfnamefont {D.~U.}\ \bibnamefont
  {Jungnickel}}, \ and\ \bibinfo {author} {\bibfnamefont {C.}~\bibnamefont
  {Wetterich}},\ }\href {\doibase 10.1142/S0217751X03014034} {\bibfield
  {journal} {\bibinfo  {journal} {Int. J. Mod. Phys. A}\ }\textbf {\bibinfo
  {volume} {18}},\ \bibinfo {pages} {3189} (\bibinfo {year} {2003})},\ \Eprint
  {http://arxiv.org/abs/hep-ph/9811387} {arXiv:hep-ph/9811387} \BibitemShut
  {NoStop}%
\bibitem [{\citenamefont {Floerchinger}\ and\ \citenamefont
  {Wetterich}(2012)}]{Floerchinger:2012xd}%
  \BibitemOpen
  \bibfield  {author} {\bibinfo {author} {\bibfnamefont {S.}~\bibnamefont
  {Floerchinger}}\ and\ \bibinfo {author} {\bibfnamefont {C.}~\bibnamefont
  {Wetterich}},\ }\href {\doibase 10.1016/j.nuclphysa.2012.07.009} {\bibfield
  {journal} {\bibinfo  {journal} {Nucl. Phys. A}\ }\textbf {\bibinfo {volume}
  {890-891}},\ \bibinfo {pages} {11} (\bibinfo {year} {2012})},\ \Eprint
  {http://arxiv.org/abs/1202.1671} {arXiv:1202.1671 [nucl-th]} \BibitemShut
  {NoStop}%
\bibitem [{\citenamefont {Drews}\ \emph {et~al.}(2013)\citenamefont {Drews},
  \citenamefont {Hell}, \citenamefont {Klein},\ and\ \citenamefont
  {Weise}}]{Drews:2013hha}%
  \BibitemOpen
  \bibfield  {author} {\bibinfo {author} {\bibfnamefont {M.}~\bibnamefont
  {Drews}}, \bibinfo {author} {\bibfnamefont {T.}~\bibnamefont {Hell}},
  \bibinfo {author} {\bibfnamefont {B.}~\bibnamefont {Klein}}, \ and\ \bibinfo
  {author} {\bibfnamefont {W.}~\bibnamefont {Weise}},\ }\href {\doibase
  10.1103/PhysRevD.88.096011} {\bibfield  {journal} {\bibinfo  {journal} {Phys.
  Rev. D}\ }\textbf {\bibinfo {volume} {88}},\ \bibinfo {pages} {096011}
  (\bibinfo {year} {2013})},\ \Eprint {http://arxiv.org/abs/1308.5596}
  {arXiv:1308.5596 [hep-ph]} \BibitemShut {NoStop}%
\bibitem [{\citenamefont {Drews}\ and\ \citenamefont
  {Weise}(2014)}]{Drews:2014wba}%
  \BibitemOpen
  \bibfield  {author} {\bibinfo {author} {\bibfnamefont {M.}~\bibnamefont
  {Drews}}\ and\ \bibinfo {author} {\bibfnamefont {W.}~\bibnamefont {Weise}},\
  }\href {\doibase 10.1016/j.physletb.2014.09.051} {\bibfield  {journal}
  {\bibinfo  {journal} {Phys. Lett. B}\ }\textbf {\bibinfo {volume} {738}},\
  \bibinfo {pages} {187} (\bibinfo {year} {2014})},\ \Eprint
  {http://arxiv.org/abs/1404.0882} {arXiv:1404.0882 [nucl-th]} \BibitemShut
  {NoStop}%
\bibitem [{\citenamefont {Drews}\ and\ \citenamefont
  {Weise}(2015)}]{Drews:2014spa}%
  \BibitemOpen
  \bibfield  {author} {\bibinfo {author} {\bibfnamefont {M.}~\bibnamefont
  {Drews}}\ and\ \bibinfo {author} {\bibfnamefont {W.}~\bibnamefont {Weise}},\
  }\href {\doibase 10.1103/PhysRevC.91.035802} {\bibfield  {journal} {\bibinfo
  {journal} {Phys. Rev. C}\ }\textbf {\bibinfo {volume} {91}},\ \bibinfo
  {pages} {035802} (\bibinfo {year} {2015})},\ \Eprint
  {http://arxiv.org/abs/1412.7655} {arXiv:1412.7655 [nucl-th]} \BibitemShut
  {NoStop}%
\bibitem [{\citenamefont {Drews}\ and\ \citenamefont
  {Weise}(2017)}]{Drews:2016wpi}%
  \BibitemOpen
  \bibfield  {author} {\bibinfo {author} {\bibfnamefont {M.}~\bibnamefont
  {Drews}}\ and\ \bibinfo {author} {\bibfnamefont {W.}~\bibnamefont {Weise}},\
  }\href {\doibase 10.1016/j.ppnp.2016.10.002} {\bibfield  {journal} {\bibinfo
  {journal} {Prog. Part. Nucl. Phys.}\ }\textbf {\bibinfo {volume} {93}},\
  \bibinfo {pages} {69} (\bibinfo {year} {2017})},\ \Eprint
  {http://arxiv.org/abs/1610.07568} {arXiv:1610.07568 [nucl-th]} \BibitemShut
  {NoStop}%
\bibitem [{\citenamefont {Fej\H{o}s}\ and\ \citenamefont
  {Hosaka}(2017)}]{Fejos:2017kpq}%
  \BibitemOpen
  \bibfield  {author} {\bibinfo {author} {\bibfnamefont {G.}~\bibnamefont
  {Fej\H{o}s}}\ and\ \bibinfo {author} {\bibfnamefont {A.}~\bibnamefont
  {Hosaka}},\ }\href {\doibase 10.1103/PhysRevD.95.116011} {\bibfield
  {journal} {\bibinfo  {journal} {Phys. Rev. D}\ }\textbf {\bibinfo {volume}
  {95}},\ \bibinfo {pages} {116011} (\bibinfo {year} {2017})},\ \Eprint
  {http://arxiv.org/abs/1701.03717} {arXiv:1701.03717 [hep-ph]} \BibitemShut
  {NoStop}%
\bibitem [{\citenamefont {Fejos}\ and\ \citenamefont
  {Hosaka}(2018)}]{Fejos:2018dyy}%
  \BibitemOpen
  \bibfield  {author} {\bibinfo {author} {\bibfnamefont {G.}~\bibnamefont
  {Fejos}}\ and\ \bibinfo {author} {\bibfnamefont {A.}~\bibnamefont {Hosaka}},\
  }\href {\doibase 10.1103/PhysRevD.98.036009} {\bibfield  {journal} {\bibinfo
  {journal} {Phys. Rev. D}\ }\textbf {\bibinfo {volume} {98}},\ \bibinfo
  {pages} {036009} (\bibinfo {year} {2018})},\ \Eprint
  {http://arxiv.org/abs/1805.08713} {arXiv:1805.08713 [nucl-th]} \BibitemShut
  {NoStop}%
\bibitem [{\citenamefont {Weise}(2018)}]{Weise:2018ukn}%
  \BibitemOpen
  \bibfield  {author} {\bibinfo {author} {\bibfnamefont {W.}~\bibnamefont
  {Weise}},\ }\href {\doibase 10.1142/S0218301318400049} {\bibfield  {journal}
  {\bibinfo  {journal} {Int. J. Mod. Phys. E}\ }\textbf {\bibinfo {volume}
  {27}},\ \bibinfo {pages} {1840004} (\bibinfo {year} {2018})},\ \Eprint
  {http://arxiv.org/abs/1811.09682} {arXiv:1811.09682 [nucl-th]} \BibitemShut
  {NoStop}%
\bibitem [{\citenamefont {Leonhardt}\ \emph {et~al.}(2020)\citenamefont
  {Leonhardt}, \citenamefont {Pospiech}, \citenamefont {Schallmo},
  \citenamefont {Braun}, \citenamefont {Drischler}, \citenamefont {Hebeler},\
  and\ \citenamefont {Schwenk}}]{Leonhardt:2019fua}%
  \BibitemOpen
  \bibfield  {author} {\bibinfo {author} {\bibfnamefont {M.}~\bibnamefont
  {Leonhardt}}, \bibinfo {author} {\bibfnamefont {M.}~\bibnamefont {Pospiech}},
  \bibinfo {author} {\bibfnamefont {B.}~\bibnamefont {Schallmo}}, \bibinfo
  {author} {\bibfnamefont {J.}~\bibnamefont {Braun}}, \bibinfo {author}
  {\bibfnamefont {C.}~\bibnamefont {Drischler}}, \bibinfo {author}
  {\bibfnamefont {K.}~\bibnamefont {Hebeler}}, \ and\ \bibinfo {author}
  {\bibfnamefont {A.}~\bibnamefont {Schwenk}},\ }\href {\doibase
  10.1103/PhysRevLett.125.142502} {\bibfield  {journal} {\bibinfo  {journal}
  {Phys. Rev. Lett.}\ }\textbf {\bibinfo {volume} {125}},\ \bibinfo {pages}
  {142502} (\bibinfo {year} {2020})},\ \Eprint
  {http://arxiv.org/abs/1907.05814} {arXiv:1907.05814 [nucl-th]} \BibitemShut
  {NoStop}%
\bibitem [{\citenamefont {Friman}\ and\ \citenamefont
  {Weise}(2019)}]{Friman:2019ncm}%
  \BibitemOpen
  \bibfield  {author} {\bibinfo {author} {\bibfnamefont {B.}~\bibnamefont
  {Friman}}\ and\ \bibinfo {author} {\bibfnamefont {W.}~\bibnamefont {Weise}},\
  }\href {\doibase 10.1103/PhysRevC.100.065807} {\bibfield  {journal} {\bibinfo
   {journal} {Phys. Rev. C}\ }\textbf {\bibinfo {volume} {100}},\ \bibinfo
  {pages} {065807} (\bibinfo {year} {2019})},\ \Eprint
  {http://arxiv.org/abs/1908.09722} {arXiv:1908.09722 [nucl-th]} \BibitemShut
  {NoStop}%
\bibitem [{\citenamefont {Brandes}\ \emph {et~al.}(2021)\citenamefont
  {Brandes}, \citenamefont {Kaiser},\ and\ \citenamefont
  {Weise}}]{Brandes:2021pti}%
  \BibitemOpen
  \bibfield  {author} {\bibinfo {author} {\bibfnamefont {L.}~\bibnamefont
  {Brandes}}, \bibinfo {author} {\bibfnamefont {N.}~\bibnamefont {Kaiser}}, \
  and\ \bibinfo {author} {\bibfnamefont {W.}~\bibnamefont {Weise}},\ }\href
  {\doibase 10.1140/epja/s10050-021-00528-2} {\bibfield  {journal} {\bibinfo
  {journal} {Eur. Phys. J. A}\ }\textbf {\bibinfo {volume} {57}},\ \bibinfo
  {pages} {243} (\bibinfo {year} {2021})},\ \Eprint
  {http://arxiv.org/abs/2103.06096} {arXiv:2103.06096 [nucl-th]} \BibitemShut
  {NoStop}%
\bibitem [{\citenamefont {Nambu}(1960)}]{Nambu:1960xd}%
  \BibitemOpen
  \bibfield  {author} {\bibinfo {author} {\bibfnamefont {Y.}~\bibnamefont
  {Nambu}},\ }\href {\doibase 10.1103/PhysRevLett.4.380} {\bibfield  {journal}
  {\bibinfo  {journal} {Phys. Rev. Lett.}\ }\textbf {\bibinfo {volume} {4}},\
  \bibinfo {pages} {380} (\bibinfo {year} {1960})}\BibitemShut {NoStop}%
\bibitem [{\citenamefont {Goldstone}(1961)}]{Goldstone:1961eq}%
  \BibitemOpen
  \bibfield  {author} {\bibinfo {author} {\bibfnamefont {J.}~\bibnamefont
  {Goldstone}},\ }\href {\doibase 10.1007/BF02812722} {\bibfield  {journal}
  {\bibinfo  {journal} {Nuovo Cim.}\ }\textbf {\bibinfo {volume} {19}},\
  \bibinfo {pages} {154} (\bibinfo {year} {1961})}\BibitemShut {NoStop}%
\bibitem [{\citenamefont {Goldstone}\ \emph {et~al.}(1962)\citenamefont
  {Goldstone}, \citenamefont {Salam},\ and\ \citenamefont
  {Weinberg}}]{Goldstone:1962es}%
  \BibitemOpen
  \bibfield  {author} {\bibinfo {author} {\bibfnamefont {J.}~\bibnamefont
  {Goldstone}}, \bibinfo {author} {\bibfnamefont {A.}~\bibnamefont {Salam}}, \
  and\ \bibinfo {author} {\bibfnamefont {S.}~\bibnamefont {Weinberg}},\ }\href
  {\doibase 10.1103/PhysRev.127.965} {\bibfield  {journal} {\bibinfo  {journal}
  {Phys. Rev.}\ }\textbf {\bibinfo {volume} {127}},\ \bibinfo {pages} {965}
  (\bibinfo {year} {1962})}\BibitemShut {NoStop}%
\bibitem [{\citenamefont {Bijnens}\ and\ \citenamefont
  {Ecker}(2014)}]{Bijnens:2014lea}%
  \BibitemOpen
  \bibfield  {author} {\bibinfo {author} {\bibfnamefont {J.}~\bibnamefont
  {Bijnens}}\ and\ \bibinfo {author} {\bibfnamefont {G.}~\bibnamefont
  {Ecker}},\ }\href {\doibase 10.1146/annurev-nucl-102313-025528} {\bibfield
  {journal} {\bibinfo  {journal} {Ann. Rev. Nucl. Part. Sci.}\ }\textbf
  {\bibinfo {volume} {64}},\ \bibinfo {pages} {149} (\bibinfo {year} {2014})},\
  \Eprint {http://arxiv.org/abs/1405.6488} {arXiv:1405.6488 [hep-ph]}
  \BibitemShut {NoStop}%
\bibitem [{\citenamefont {Manohar}\ and\ \citenamefont
  {Georgi}(1984)}]{Manohar:1983md}%
  \BibitemOpen
  \bibfield  {author} {\bibinfo {author} {\bibfnamefont {A.}~\bibnamefont
  {Manohar}}\ and\ \bibinfo {author} {\bibfnamefont {H.}~\bibnamefont
  {Georgi}},\ }\href {\doibase 10.1016/0550-3213(84)90231-1} {\bibfield
  {journal} {\bibinfo  {journal} {Nucl. Phys. B}\ }\textbf {\bibinfo {volume}
  {234}},\ \bibinfo {pages} {189} (\bibinfo {year} {1984})}\BibitemShut
  {NoStop}%
\bibitem [{\citenamefont {Weinberg}(1966)}]{Weinberg:1966kf}%
  \BibitemOpen
  \bibfield  {author} {\bibinfo {author} {\bibfnamefont {S.}~\bibnamefont
  {Weinberg}},\ }\href {\doibase 10.1103/PhysRevLett.17.616} {\bibfield
  {journal} {\bibinfo  {journal} {Phys. Rev. Lett.}\ }\textbf {\bibinfo
  {volume} {17}},\ \bibinfo {pages} {616} (\bibinfo {year} {1966})}\BibitemShut
  {NoStop}%
\bibitem [{\citenamefont {Bijnens}\ \emph {et~al.}(1996)\citenamefont
  {Bijnens}, \citenamefont {Colangelo}, \citenamefont {Ecker}, \citenamefont
  {Gasser},\ and\ \citenamefont {Sainio}}]{Bijnens:1995yn}%
  \BibitemOpen
  \bibfield  {author} {\bibinfo {author} {\bibfnamefont {J.}~\bibnamefont
  {Bijnens}}, \bibinfo {author} {\bibfnamefont {G.}~\bibnamefont {Colangelo}},
  \bibinfo {author} {\bibfnamefont {G.}~\bibnamefont {Ecker}}, \bibinfo
  {author} {\bibfnamefont {J.}~\bibnamefont {Gasser}}, \ and\ \bibinfo {author}
  {\bibfnamefont {M.~E.}\ \bibnamefont {Sainio}},\ }\href {\doibase
  10.1016/0370-2693(96)00165-7} {\bibfield  {journal} {\bibinfo  {journal}
  {Phys. Lett. B}\ }\textbf {\bibinfo {volume} {374}},\ \bibinfo {pages} {210}
  (\bibinfo {year} {1996})},\ \Eprint {http://arxiv.org/abs/hep-ph/9511397}
  {arXiv:hep-ph/9511397} \BibitemShut {NoStop}%
\bibitem [{\citenamefont {Bijnens}\ \emph {et~al.}(1997)\citenamefont
  {Bijnens}, \citenamefont {Colangelo}, \citenamefont {Ecker}, \citenamefont
  {Gasser},\ and\ \citenamefont {Sainio}}]{Bijnens:1997vq}%
  \BibitemOpen
  \bibfield  {author} {\bibinfo {author} {\bibfnamefont {J.}~\bibnamefont
  {Bijnens}}, \bibinfo {author} {\bibfnamefont {G.}~\bibnamefont {Colangelo}},
  \bibinfo {author} {\bibfnamefont {G.}~\bibnamefont {Ecker}}, \bibinfo
  {author} {\bibfnamefont {J.}~\bibnamefont {Gasser}}, \ and\ \bibinfo {author}
  {\bibfnamefont {M.~E.}\ \bibnamefont {Sainio}},\ }\href {\doibase
  10.1016/S0550-3213(97)00621-4} {\bibfield  {journal} {\bibinfo  {journal}
  {Nucl. Phys. B}\ }\textbf {\bibinfo {volume} {508}},\ \bibinfo {pages} {263}
  (\bibinfo {year} {1997})},\ \bibinfo {note} {[Erratum: Nucl.Phys.B 517,
  639--639 (1998)]},\ \Eprint {http://arxiv.org/abs/hep-ph/9707291}
  {arXiv:hep-ph/9707291} \BibitemShut {NoStop}%
\bibitem [{\citenamefont {Ananthanarayan}\ \emph {et~al.}(2001)\citenamefont
  {Ananthanarayan}, \citenamefont {Colangelo}, \citenamefont {Gasser},\ and\
  \citenamefont {Leutwyler}}]{Ananthanarayan:2000ht}%
  \BibitemOpen
  \bibfield  {author} {\bibinfo {author} {\bibfnamefont {B.}~\bibnamefont
  {Ananthanarayan}}, \bibinfo {author} {\bibfnamefont {G.}~\bibnamefont
  {Colangelo}}, \bibinfo {author} {\bibfnamefont {J.}~\bibnamefont {Gasser}}, \
  and\ \bibinfo {author} {\bibfnamefont {H.}~\bibnamefont {Leutwyler}},\ }\href
  {\doibase 10.1016/S0370-1573(01)00009-6} {\bibfield  {journal} {\bibinfo
  {journal} {Phys. Rept.}\ }\textbf {\bibinfo {volume} {353}},\ \bibinfo
  {pages} {207} (\bibinfo {year} {2001})},\ \Eprint
  {http://arxiv.org/abs/hep-ph/0005297} {arXiv:hep-ph/0005297} \BibitemShut
  {NoStop}%
\bibitem [{\citenamefont {Colangelo}\ \emph {et~al.}(2000)\citenamefont
  {Colangelo}, \citenamefont {Gasser},\ and\ \citenamefont
  {Leutwyler}}]{Colangelo:2000jc}%
  \BibitemOpen
  \bibfield  {author} {\bibinfo {author} {\bibfnamefont {G.}~\bibnamefont
  {Colangelo}}, \bibinfo {author} {\bibfnamefont {J.}~\bibnamefont {Gasser}}, \
  and\ \bibinfo {author} {\bibfnamefont {H.}~\bibnamefont {Leutwyler}},\ }\href
  {\doibase 10.1016/S0370-2693(00)00898-4} {\bibfield  {journal} {\bibinfo
  {journal} {Phys. Lett. B}\ }\textbf {\bibinfo {volume} {488}},\ \bibinfo
  {pages} {261} (\bibinfo {year} {2000})},\ \Eprint
  {http://arxiv.org/abs/hep-ph/0007112} {arXiv:hep-ph/0007112} \BibitemShut
  {NoStop}%
\bibitem [{\citenamefont {Colangelo}\ \emph {et~al.}(2001)\citenamefont
  {Colangelo}, \citenamefont {Gasser},\ and\ \citenamefont
  {Leutwyler}}]{Colangelo:2001df}%
  \BibitemOpen
  \bibfield  {author} {\bibinfo {author} {\bibfnamefont {G.}~\bibnamefont
  {Colangelo}}, \bibinfo {author} {\bibfnamefont {J.}~\bibnamefont {Gasser}}, \
  and\ \bibinfo {author} {\bibfnamefont {H.}~\bibnamefont {Leutwyler}},\ }\href
  {\doibase 10.1016/S0550-3213(01)00147-X} {\bibfield  {journal} {\bibinfo
  {journal} {Nucl. Phys. B}\ }\textbf {\bibinfo {volume} {603}},\ \bibinfo
  {pages} {125} (\bibinfo {year} {2001})},\ \Eprint
  {http://arxiv.org/abs/hep-ph/0103088} {arXiv:hep-ph/0103088} \BibitemShut
  {NoStop}%
\bibitem [{\citenamefont {Greiner}\ and\ \citenamefont
  {Muller}(1989)}]{Greiner:1989eu}%
  \BibitemOpen
  \bibfield  {author} {\bibinfo {author} {\bibfnamefont {W.}~\bibnamefont
  {Greiner}}\ and\ \bibinfo {author} {\bibfnamefont {B.}~\bibnamefont
  {Muller}},\ }\href@noop {} {\emph {\bibinfo {title} {{Theoretical physics.
  Vol. 2: Quantum mechanics. Symmetries}}}}\ (\bibinfo {year}
  {1989})\BibitemShut {NoStop}%
\bibitem [{\citenamefont {Koch}(1997)}]{Koch:1997ei}%
  \BibitemOpen
  \bibfield  {author} {\bibinfo {author} {\bibfnamefont {V.}~\bibnamefont
  {Koch}},\ }\href {\doibase 10.1142/S0218301397000147} {\bibfield  {journal}
  {\bibinfo  {journal} {Int. J. Mod. Phys. E}\ }\textbf {\bibinfo {volume}
  {6}},\ \bibinfo {pages} {203} (\bibinfo {year} {1997})},\ \Eprint
  {http://arxiv.org/abs/nucl-th/9706075} {arXiv:nucl-th/9706075} \BibitemShut
  {NoStop}%
\bibitem [{\citenamefont {Sannino}\ and\ \citenamefont
  {Schechter}(1995)}]{Sannino:1995ik}%
  \BibitemOpen
  \bibfield  {author} {\bibinfo {author} {\bibfnamefont {F.}~\bibnamefont
  {Sannino}}\ and\ \bibinfo {author} {\bibfnamefont {J.}~\bibnamefont
  {Schechter}},\ }\href {\doibase 10.1103/PhysRevD.52.96} {\bibfield  {journal}
  {\bibinfo  {journal} {Phys. Rev. D}\ }\textbf {\bibinfo {volume} {52}},\
  \bibinfo {pages} {96} (\bibinfo {year} {1995})},\ \Eprint
  {http://arxiv.org/abs/hep-ph/9501417} {arXiv:hep-ph/9501417} \BibitemShut
  {NoStop}%
\bibitem [{\citenamefont {Harada}\ \emph {et~al.}(1996)\citenamefont {Harada},
  \citenamefont {Sannino},\ and\ \citenamefont {Schechter}}]{Harada:1995dc}%
  \BibitemOpen
  \bibfield  {author} {\bibinfo {author} {\bibfnamefont {M.}~\bibnamefont
  {Harada}}, \bibinfo {author} {\bibfnamefont {F.}~\bibnamefont {Sannino}}, \
  and\ \bibinfo {author} {\bibfnamefont {J.}~\bibnamefont {Schechter}},\ }\href
  {\doibase 10.1103/PhysRevD.54.1991} {\bibfield  {journal} {\bibinfo
  {journal} {Phys. Rev. D}\ }\textbf {\bibinfo {volume} {54}},\ \bibinfo
  {pages} {1991} (\bibinfo {year} {1996})},\ \Eprint
  {http://arxiv.org/abs/hep-ph/9511335} {arXiv:hep-ph/9511335} \BibitemShut
  {NoStop}%
\bibitem [{\citenamefont {Scadron}(1999)}]{Scadron:1999qd}%
  \BibitemOpen
  \bibfield  {author} {\bibinfo {author} {\bibfnamefont {M.~D.}\ \bibnamefont
  {Scadron}},\ }\href {\doibase 10.1142/S0217732399001449} {\bibfield
  {journal} {\bibinfo  {journal} {Mod. Phys. Lett. A}\ }\textbf {\bibinfo
  {volume} {14}},\ \bibinfo {pages} {1349} (\bibinfo {year} {1999})},\ \Eprint
  {http://arxiv.org/abs/hep-ph/9910243} {arXiv:hep-ph/9910243} \BibitemShut
  {NoStop}%
\bibitem [{\citenamefont {Lucio}\ \emph {et~al.}(1999)\citenamefont {Lucio},
  \citenamefont {Napsuciale},\ and\ \citenamefont
  {Ruiz-Altaba}}]{Lucio:1999ha}%
  \BibitemOpen
  \bibfield  {author} {\bibinfo {author} {\bibfnamefont {J.~L.}\ \bibnamefont
  {Lucio}}, \bibinfo {author} {\bibfnamefont {M.}~\bibnamefont {Napsuciale}}, \
  and\ \bibinfo {author} {\bibfnamefont {M.}~\bibnamefont {Ruiz-Altaba}},\
  }\href@noop {} {\  (\bibinfo {year} {1999})},\ \Eprint
  {http://arxiv.org/abs/hep-ph/9903420} {arXiv:hep-ph/9903420} \BibitemShut
  {NoStop}%
\bibitem [{\citenamefont {Black}\ \emph {et~al.}(2001)\citenamefont {Black},
  \citenamefont {Fariborz}, \citenamefont {Moussa}, \citenamefont {Nasri},\
  and\ \citenamefont {Schechter}}]{Black:2000qq}%
  \BibitemOpen
  \bibfield  {author} {\bibinfo {author} {\bibfnamefont {D.}~\bibnamefont
  {Black}}, \bibinfo {author} {\bibfnamefont {A.~H.}\ \bibnamefont {Fariborz}},
  \bibinfo {author} {\bibfnamefont {S.}~\bibnamefont {Moussa}}, \bibinfo
  {author} {\bibfnamefont {S.}~\bibnamefont {Nasri}}, \ and\ \bibinfo {author}
  {\bibfnamefont {J.}~\bibnamefont {Schechter}},\ }\href {\doibase
  10.1103/PhysRevD.64.014031} {\bibfield  {journal} {\bibinfo  {journal} {Phys.
  Rev. D}\ }\textbf {\bibinfo {volume} {64}},\ \bibinfo {pages} {014031}
  (\bibinfo {year} {2001})},\ \Eprint {http://arxiv.org/abs/hep-ph/0012278}
  {arXiv:hep-ph/0012278} \BibitemShut {NoStop}%
\bibitem [{\citenamefont {Scadron}\ \emph {et~al.}(2006)\citenamefont
  {Scadron}, \citenamefont {Kleefeld},\ and\ \citenamefont
  {Rupp}}]{Scadron:2006mq}%
  \BibitemOpen
  \bibfield  {author} {\bibinfo {author} {\bibfnamefont {M.~D.}\ \bibnamefont
  {Scadron}}, \bibinfo {author} {\bibfnamefont {F.}~\bibnamefont {Kleefeld}}, \
  and\ \bibinfo {author} {\bibfnamefont {G.}~\bibnamefont {Rupp}},\ }\href@noop
  {} {\  (\bibinfo {year} {2006})},\ \Eprint
  {http://arxiv.org/abs/hep-ph/0601196} {arXiv:hep-ph/0601196} \BibitemShut
  {NoStop}%
\bibitem [{\citenamefont {Fariborz}\ \emph {et~al.}(2007)\citenamefont
  {Fariborz}, \citenamefont {Jora},\ and\ \citenamefont
  {Schechter}}]{Fariborz:2007km}%
  \BibitemOpen
  \bibfield  {author} {\bibinfo {author} {\bibfnamefont {A.~H.}\ \bibnamefont
  {Fariborz}}, \bibinfo {author} {\bibfnamefont {R.}~\bibnamefont {Jora}}, \
  and\ \bibinfo {author} {\bibfnamefont {J.}~\bibnamefont {Schechter}},\ }\href
  {\doibase 10.1103/PhysRevD.76.114001} {\bibfield  {journal} {\bibinfo
  {journal} {Phys. Rev. D}\ }\textbf {\bibinfo {volume} {76}},\ \bibinfo
  {pages} {114001} (\bibinfo {year} {2007})},\ \Eprint
  {http://arxiv.org/abs/0708.3402} {arXiv:0708.3402 [hep-ph]} \BibitemShut
  {NoStop}%
\bibitem [{\citenamefont {Kramer}(1969)}]{Kramer:1969gw}%
  \BibitemOpen
  \bibfield  {author} {\bibinfo {author} {\bibfnamefont {G.}~\bibnamefont
  {Kramer}},\ }\href {\doibase 10.1103/PhysRev.177.2515} {\bibfield  {journal}
  {\bibinfo  {journal} {Phys. Rev.}\ }\textbf {\bibinfo {volume} {177}},\
  \bibinfo {pages} {2515} (\bibinfo {year} {1969})}\BibitemShut {NoStop}%
\bibitem [{\citenamefont {Basdevant}\ and\ \citenamefont
  {Lee}(1970)}]{Basdevant:1970nu}%
  \BibitemOpen
  \bibfield  {author} {\bibinfo {author} {\bibfnamefont {J.~I.}\ \bibnamefont
  {Basdevant}}\ and\ \bibinfo {author} {\bibfnamefont {B.~W.}\ \bibnamefont
  {Lee}},\ }\href {\doibase 10.1103/PhysRevD.2.1680} {\bibfield  {journal}
  {\bibinfo  {journal} {Phys. Rev. D}\ }\textbf {\bibinfo {volume} {2}},\
  \bibinfo {pages} {1680} (\bibinfo {year} {1970})}\BibitemShut {NoStop}%
\bibitem [{\citenamefont {Geddes}\ and\ \citenamefont
  {Graham}(1975)}]{Geddes:1975cf}%
  \BibitemOpen
  \bibfield  {author} {\bibinfo {author} {\bibfnamefont {H.~B.}\ \bibnamefont
  {Geddes}}\ and\ \bibinfo {author} {\bibfnamefont {R.~H.}\ \bibnamefont
  {Graham}},\ }\href {\doibase 10.1103/PhysRevD.12.855} {\bibfield  {journal}
  {\bibinfo  {journal} {Phys. Rev. D}\ }\textbf {\bibinfo {volume} {12}},\
  \bibinfo {pages} {855} (\bibinfo {year} {1975})}\BibitemShut {NoStop}%
\bibitem [{\citenamefont {Geddes}\ and\ \citenamefont
  {Graham}(1976)}]{Geddes:1976qf}%
  \BibitemOpen
  \bibfield  {author} {\bibinfo {author} {\bibfnamefont {H.~B.}\ \bibnamefont
  {Geddes}}\ and\ \bibinfo {author} {\bibfnamefont {R.~H.}\ \bibnamefont
  {Graham}},\ }\href {\doibase 10.1103/PhysRevD.13.56} {\bibfield  {journal}
  {\bibinfo  {journal} {Phys. Rev. D}\ }\textbf {\bibinfo {volume} {13}},\
  \bibinfo {pages} {56} (\bibinfo {year} {1976})}\BibitemShut {NoStop}%
\bibitem [{\citenamefont {Aouissat}\ \emph {et~al.}(1995)\citenamefont
  {Aouissat}, \citenamefont {Rapp}, \citenamefont {Chanfray}, \citenamefont
  {Schuck},\ and\ \citenamefont {Wambach}}]{Aouissat:1994sx}%
  \BibitemOpen
  \bibfield  {author} {\bibinfo {author} {\bibfnamefont {Z.}~\bibnamefont
  {Aouissat}}, \bibinfo {author} {\bibfnamefont {R.}~\bibnamefont {Rapp}},
  \bibinfo {author} {\bibfnamefont {G.}~\bibnamefont {Chanfray}}, \bibinfo
  {author} {\bibfnamefont {P.}~\bibnamefont {Schuck}}, \ and\ \bibinfo {author}
  {\bibfnamefont {J.}~\bibnamefont {Wambach}},\ }\href {\doibase
  10.1016/0375-9474(94)00429-Q} {\bibfield  {journal} {\bibinfo  {journal}
  {Nucl. Phys. A}\ }\textbf {\bibinfo {volume} {581}},\ \bibinfo {pages} {471}
  (\bibinfo {year} {1995})},\ \Eprint {http://arxiv.org/abs/nucl-th/9406010}
  {arXiv:nucl-th/9406010} \BibitemShut {NoStop}%
\bibitem [{\citenamefont {Soto}\ \emph {et~al.}(2013)\citenamefont {Soto},
  \citenamefont {Talavera},\ and\ \citenamefont {Tarrus}}]{Soto:2011ap}%
  \BibitemOpen
  \bibfield  {author} {\bibinfo {author} {\bibfnamefont {J.}~\bibnamefont
  {Soto}}, \bibinfo {author} {\bibfnamefont {P.}~\bibnamefont {Talavera}}, \
  and\ \bibinfo {author} {\bibfnamefont {J.}~\bibnamefont {Tarrus}},\ }\href
  {\doibase 10.1016/j.nuclphysb.2012.09.005} {\bibfield  {journal} {\bibinfo
  {journal} {Nucl. Phys. B}\ }\textbf {\bibinfo {volume} {866}},\ \bibinfo
  {pages} {270} (\bibinfo {year} {2013})},\ \Eprint
  {http://arxiv.org/abs/1110.6156} {arXiv:1110.6156 [hep-ph]} \BibitemShut
  {NoStop}%
\bibitem [{\citenamefont {Fariborz}\ \emph {et~al.}(2009)\citenamefont
  {Fariborz}, \citenamefont {Park}, \citenamefont {Schechter},\ and\
  \citenamefont {Naeem~Shahid}}]{Fariborz:2009wf}%
  \BibitemOpen
  \bibfield  {author} {\bibinfo {author} {\bibfnamefont {A.~H.}\ \bibnamefont
  {Fariborz}}, \bibinfo {author} {\bibfnamefont {N.~W.}\ \bibnamefont {Park}},
  \bibinfo {author} {\bibfnamefont {J.}~\bibnamefont {Schechter}}, \ and\
  \bibinfo {author} {\bibfnamefont {M.}~\bibnamefont {Naeem~Shahid}},\ }\href
  {\doibase 10.1103/PhysRevD.80.113001} {\bibfield  {journal} {\bibinfo
  {journal} {Phys. Rev. D}\ }\textbf {\bibinfo {volume} {80}},\ \bibinfo
  {pages} {113001} (\bibinfo {year} {2009})},\ \Eprint
  {http://arxiv.org/abs/0907.0482} {arXiv:0907.0482 [hep-ph]} \BibitemShut
  {NoStop}%
\bibitem [{\citenamefont {van Beveren}\ \emph {et~al.}(2002)\citenamefont {van
  Beveren}, \citenamefont {Kleefeld}, \citenamefont {Rupp},\ and\ \citenamefont
  {Scadron}}]{vanBeveren:2002mc}%
  \BibitemOpen
  \bibfield  {author} {\bibinfo {author} {\bibfnamefont {E.}~\bibnamefont {van
  Beveren}}, \bibinfo {author} {\bibfnamefont {F.}~\bibnamefont {Kleefeld}},
  \bibinfo {author} {\bibfnamefont {G.}~\bibnamefont {Rupp}}, \ and\ \bibinfo
  {author} {\bibfnamefont {M.~D.}\ \bibnamefont {Scadron}},\ }\href {\doibase
  10.1142/S0217732302007326} {\bibfield  {journal} {\bibinfo  {journal} {Mod.
  Phys. Lett. A}\ }\textbf {\bibinfo {volume} {17}},\ \bibinfo {pages} {1673}
  (\bibinfo {year} {2002})},\ \Eprint {http://arxiv.org/abs/hep-ph/0204139}
  {arXiv:hep-ph/0204139} \BibitemShut {NoStop}%
\bibitem [{\citenamefont {Black}\ \emph {et~al.}(2009)\citenamefont {Black},
  \citenamefont {Fariborz}, \citenamefont {Jora}, \citenamefont {Park},
  \citenamefont {Schechter},\ and\ \citenamefont
  {Naeem~Shahid}}]{Black:2009bi}%
  \BibitemOpen
  \bibfield  {author} {\bibinfo {author} {\bibfnamefont {D.}~\bibnamefont
  {Black}}, \bibinfo {author} {\bibfnamefont {A.~H.}\ \bibnamefont {Fariborz}},
  \bibinfo {author} {\bibfnamefont {R.}~\bibnamefont {Jora}}, \bibinfo {author}
  {\bibfnamefont {N.~W.}\ \bibnamefont {Park}}, \bibinfo {author}
  {\bibfnamefont {J.}~\bibnamefont {Schechter}}, \ and\ \bibinfo {author}
  {\bibfnamefont {M.}~\bibnamefont {Naeem~Shahid}},\ }\href {\doibase
  10.1142/S0217732309031533} {\bibfield  {journal} {\bibinfo  {journal} {Mod.
  Phys. Lett. A}\ }\textbf {\bibinfo {volume} {24}},\ \bibinfo {pages} {2285}
  (\bibinfo {year} {2009})},\ \Eprint {http://arxiv.org/abs/0904.2161}
  {arXiv:0904.2161 [hep-ph]} \BibitemShut {NoStop}%
\bibitem [{\citenamefont {Weinberg}(2013)}]{Weinberg:1996kr}%
  \BibitemOpen
  \bibfield  {author} {\bibinfo {author} {\bibfnamefont {S.}~\bibnamefont
  {Weinberg}},\ }\href@noop {} {\emph {\bibinfo {title} {{The quantum theory of
  fields. Vol. 2: Modern applications}}}}\ (\bibinfo  {publisher} {Cambridge
  University Press},\ \bibinfo {year} {2013})\BibitemShut {NoStop}%
\bibitem [{\citenamefont {Litim}\ and\ \citenamefont
  {Pawlowski}(2002{\natexlab{a}})}]{Litim:2001ky}%
  \BibitemOpen
  \bibfield  {author} {\bibinfo {author} {\bibfnamefont {D.~F.}\ \bibnamefont
  {Litim}}\ and\ \bibinfo {author} {\bibfnamefont {J.~M.}\ \bibnamefont
  {Pawlowski}},\ }\href {\doibase 10.1103/PhysRevD.65.081701} {\bibfield
  {journal} {\bibinfo  {journal} {Phys. Rev. D}\ }\textbf {\bibinfo {volume}
  {65}},\ \bibinfo {pages} {081701} (\bibinfo {year} {2002}{\natexlab{a}})},\
  \Eprint {http://arxiv.org/abs/hep-th/0111191} {arXiv:hep-th/0111191}
  \BibitemShut {NoStop}%
\bibitem [{\citenamefont {Litim}\ and\ \citenamefont
  {Pawlowski}(2002{\natexlab{b}})}]{Litim:2002xm}%
  \BibitemOpen
  \bibfield  {author} {\bibinfo {author} {\bibfnamefont {D.~F.}\ \bibnamefont
  {Litim}}\ and\ \bibinfo {author} {\bibfnamefont {J.~M.}\ \bibnamefont
  {Pawlowski}},\ }\href {\doibase 10.1103/PhysRevD.66.025030} {\bibfield
  {journal} {\bibinfo  {journal} {Phys. Rev. D}\ }\textbf {\bibinfo {volume}
  {66}},\ \bibinfo {pages} {025030} (\bibinfo {year} {2002}{\natexlab{b}})},\
  \Eprint {http://arxiv.org/abs/hep-th/0202188} {arXiv:hep-th/0202188}
  \BibitemShut {NoStop}%
\bibitem [{\citenamefont {Strodthoff}\ \emph {et~al.}(2012)\citenamefont
  {Strodthoff}, \citenamefont {Schaefer},\ and\ \citenamefont {von
  Smekal}}]{Strodthoff:2011tz}%
  \BibitemOpen
  \bibfield  {author} {\bibinfo {author} {\bibfnamefont {N.}~\bibnamefont
  {Strodthoff}}, \bibinfo {author} {\bibfnamefont {B.-J.}\ \bibnamefont
  {Schaefer}}, \ and\ \bibinfo {author} {\bibfnamefont {L.}~\bibnamefont {von
  Smekal}},\ }\href {\doibase 10.1103/PhysRevD.85.074007} {\bibfield  {journal}
  {\bibinfo  {journal} {Phys. Rev. D}\ }\textbf {\bibinfo {volume} {85}},\
  \bibinfo {pages} {074007} (\bibinfo {year} {2012})},\ \Eprint
  {http://arxiv.org/abs/1112.5401} {arXiv:1112.5401 [hep-ph]} \BibitemShut
  {NoStop}%
\bibitem [{\citenamefont {Stoll}\ \emph {et~al.}(2021)\citenamefont {Stoll},
  \citenamefont {Zorbach}, \citenamefont {Koenigstein}, \citenamefont {Steil},\
  and\ \citenamefont {Rechenberger}}]{Stoll:2021ori}%
  \BibitemOpen
  \bibfield  {author} {\bibinfo {author} {\bibfnamefont {J.}~\bibnamefont
  {Stoll}}, \bibinfo {author} {\bibfnamefont {N.}~\bibnamefont {Zorbach}},
  \bibinfo {author} {\bibfnamefont {A.}~\bibnamefont {Koenigstein}}, \bibinfo
  {author} {\bibfnamefont {M.~J.}\ \bibnamefont {Steil}}, \ and\ \bibinfo
  {author} {\bibfnamefont {S.}~\bibnamefont {Rechenberger}},\ }\href@noop {} {\
   (\bibinfo {year} {2021})},\ \Eprint {http://arxiv.org/abs/2108.10616}
  {arXiv:2108.10616 [hep-ph]} \BibitemShut {NoStop}%
\bibitem [{\citenamefont {Gasser}(2009)}]{Gasser:2009zz}%
  \BibitemOpen
  \bibfield  {author} {\bibinfo {author} {\bibfnamefont {J.}~\bibnamefont
  {Gasser}},\ }\href {\doibase 10.22323/1.069.0029} {\bibfield  {journal}
  {\bibinfo  {journal} {PoS}\ }\textbf {\bibinfo {volume} {EFT09}},\ \bibinfo
  {pages} {029} (\bibinfo {year} {2009})}\BibitemShut {NoStop}%
\bibitem [{\citenamefont {Gell-Mann}\ \emph {et~al.}(1968)\citenamefont
  {Gell-Mann}, \citenamefont {Oakes},\ and\ \citenamefont
  {Renner}}]{Gell-Mann:1968hlm}%
  \BibitemOpen
  \bibfield  {author} {\bibinfo {author} {\bibfnamefont {M.}~\bibnamefont
  {Gell-Mann}}, \bibinfo {author} {\bibfnamefont {R.~J.}\ \bibnamefont
  {Oakes}}, \ and\ \bibinfo {author} {\bibfnamefont {B.}~\bibnamefont
  {Renner}},\ }\href {\doibase 10.1103/PhysRev.175.2195} {\bibfield  {journal}
  {\bibinfo  {journal} {Phys. Rev.}\ }\textbf {\bibinfo {volume} {175}},\
  \bibinfo {pages} {2195} (\bibinfo {year} {1968})}\BibitemShut {NoStop}%
\bibitem [{\citenamefont {Schaefer}\ and\ \citenamefont
  {Pirner}(1997)}]{Schaefer:1997nd}%
  \BibitemOpen
  \bibfield  {author} {\bibinfo {author} {\bibfnamefont {B.-J.}\ \bibnamefont
  {Schaefer}}\ and\ \bibinfo {author} {\bibfnamefont {H.-J.}\ \bibnamefont
  {Pirner}},\ }\href {\doibase 10.1016/S0375-9474(97)00601-5} {\bibfield
  {journal} {\bibinfo  {journal} {Nucl. Phys. A}\ }\textbf {\bibinfo {volume}
  {627}},\ \bibinfo {pages} {481} (\bibinfo {year} {1997})},\ \Eprint
  {http://arxiv.org/abs/hep-ph/9706258} {arXiv:hep-ph/9706258} \BibitemShut
  {NoStop}%
\bibitem [{\citenamefont {Loewe}\ and\ \citenamefont
  {Martinez}(2008)}]{Loewe:2008ui}%
  \BibitemOpen
  \bibfield  {author} {\bibinfo {author} {\bibfnamefont {M.}~\bibnamefont
  {Loewe}}\ and\ \bibinfo {author} {\bibfnamefont {C.~V.}\ \bibnamefont
  {Martinez}},\ }\href {\doibase 10.1103/PhysRevD.78.069902} {\bibfield
  {journal} {\bibinfo  {journal} {Phys. Rev. D}\ }\textbf {\bibinfo {volume}
  {77}},\ \bibinfo {pages} {105006} (\bibinfo {year} {2008})},\ \bibinfo {note}
  {[Erratum: Phys.Rev.D 78, 069902 (2008)]},\ \Eprint
  {http://arxiv.org/abs/0801.2176} {arXiv:0801.2176 [hep-ph]} \BibitemShut
  {NoStop}%
\bibitem [{\citenamefont {Loewe}\ \emph {et~al.}(2018)\citenamefont {Loewe},
  \citenamefont {Monje},\ and\ \citenamefont {Zamora}}]{Loewe:2017kiw}%
  \BibitemOpen
  \bibfield  {author} {\bibinfo {author} {\bibfnamefont {M.}~\bibnamefont
  {Loewe}}, \bibinfo {author} {\bibfnamefont {L.}~\bibnamefont {Monje}}, \ and\
  \bibinfo {author} {\bibfnamefont {R.}~\bibnamefont {Zamora}},\ }\href
  {\doibase 10.1103/PhysRevD.97.056023} {\bibfield  {journal} {\bibinfo
  {journal} {Phys. Rev. D}\ }\textbf {\bibinfo {volume} {97}},\ \bibinfo
  {pages} {056023} (\bibinfo {year} {2018})},\ \Eprint
  {http://arxiv.org/abs/1712.10047} {arXiv:1712.10047 [hep-ph]} \BibitemShut
  {NoStop}%
\bibitem [{\citenamefont {Loewe}\ \emph
  {et~al.}(2019{\natexlab{a}})\citenamefont {Loewe}, \citenamefont {Monje},
  \citenamefont {Mu\~noz}, \citenamefont {Raya},\ and\ \citenamefont
  {Zamora}}]{Loewe:2019zwq}%
  \BibitemOpen
  \bibfield  {author} {\bibinfo {author} {\bibfnamefont {M.}~\bibnamefont
  {Loewe}}, \bibinfo {author} {\bibfnamefont {L.}~\bibnamefont {Monje}},
  \bibinfo {author} {\bibfnamefont {E.}~\bibnamefont {Mu\~noz}}, \bibinfo
  {author} {\bibfnamefont {A.}~\bibnamefont {Raya}}, \ and\ \bibinfo {author}
  {\bibfnamefont {R.}~\bibnamefont {Zamora}},\ }\href {\doibase
  10.1103/PhysRevD.99.056002} {\bibfield  {journal} {\bibinfo  {journal} {Phys.
  Rev. D}\ }\textbf {\bibinfo {volume} {99}},\ \bibinfo {pages} {056002}
  (\bibinfo {year} {2019}{\natexlab{a}})},\ \Eprint
  {http://arxiv.org/abs/1901.03256} {arXiv:1901.03256 [hep-ph]} \BibitemShut
  {NoStop}%
\bibitem [{\citenamefont {Loewe}\ \emph
  {et~al.}(2019{\natexlab{b}})\citenamefont {Loewe}, \citenamefont {Mu\~noz},\
  and\ \citenamefont {Zamora}}]{Loewe:2019xtn}%
  \BibitemOpen
  \bibfield  {author} {\bibinfo {author} {\bibfnamefont {M.}~\bibnamefont
  {Loewe}}, \bibinfo {author} {\bibfnamefont {E.}~\bibnamefont {Mu\~noz}}, \
  and\ \bibinfo {author} {\bibfnamefont {R.}~\bibnamefont {Zamora}},\ }\href
  {\doibase 10.1103/PhysRevD.100.116006} {\bibfield  {journal} {\bibinfo
  {journal} {Phys. Rev. D}\ }\textbf {\bibinfo {volume} {100}},\ \bibinfo
  {pages} {116006} (\bibinfo {year} {2019}{\natexlab{b}})},\ \Eprint
  {http://arxiv.org/abs/1905.03783} {arXiv:1905.03783 [hep-ph]} \BibitemShut
  {NoStop}%
\bibitem [{\citenamefont {Grossi}\ and\ \citenamefont
  {Wink}(2019)}]{Grossi:2019urj}%
  \BibitemOpen
  \bibfield  {author} {\bibinfo {author} {\bibfnamefont {E.}~\bibnamefont
  {Grossi}}\ and\ \bibinfo {author} {\bibfnamefont {N.}~\bibnamefont {Wink}},\
  }\href@noop {} {\  (\bibinfo {year} {2019})},\ \Eprint
  {http://arxiv.org/abs/1903.09503} {arXiv:1903.09503 [hep-th]} \BibitemShut
  {NoStop}%
\bibitem [{\citenamefont {Grossi}\ \emph {et~al.}(2021)\citenamefont {Grossi},
  \citenamefont {Ihssen}, \citenamefont {Pawlowski},\ and\ \citenamefont
  {Wink}}]{Grossi:2021ksl}%
  \BibitemOpen
  \bibfield  {author} {\bibinfo {author} {\bibfnamefont {E.}~\bibnamefont
  {Grossi}}, \bibinfo {author} {\bibfnamefont {F.~J.}\ \bibnamefont {Ihssen}},
  \bibinfo {author} {\bibfnamefont {J.~M.}\ \bibnamefont {Pawlowski}}, \ and\
  \bibinfo {author} {\bibfnamefont {N.}~\bibnamefont {Wink}},\ }\href {\doibase
  10.1103/PhysRevD.104.016028} {\bibfield  {journal} {\bibinfo  {journal}
  {Phys. Rev. D}\ }\textbf {\bibinfo {volume} {104}},\ \bibinfo {pages}
  {016028} (\bibinfo {year} {2021})},\ \Eprint
  {http://arxiv.org/abs/2102.01602} {arXiv:2102.01602 [hep-ph]} \BibitemShut
  {NoStop}%
\bibitem [{\citenamefont {Koenigstein}\ \emph
  {et~al.}(2021{\natexlab{a}})\citenamefont {Koenigstein}, \citenamefont
  {Steil}, \citenamefont {Wink}, \citenamefont {Grossi}, \citenamefont {Braun},
  \citenamefont {Buballa},\ and\ \citenamefont
  {Rischke}}]{Koenigstein:2021syz}%
  \BibitemOpen
  \bibfield  {author} {\bibinfo {author} {\bibfnamefont {A.}~\bibnamefont
  {Koenigstein}}, \bibinfo {author} {\bibfnamefont {M.~J.}\ \bibnamefont
  {Steil}}, \bibinfo {author} {\bibfnamefont {N.}~\bibnamefont {Wink}},
  \bibinfo {author} {\bibfnamefont {E.}~\bibnamefont {Grossi}}, \bibinfo
  {author} {\bibfnamefont {J.}~\bibnamefont {Braun}}, \bibinfo {author}
  {\bibfnamefont {M.}~\bibnamefont {Buballa}}, \ and\ \bibinfo {author}
  {\bibfnamefont {D.~H.}\ \bibnamefont {Rischke}},\ }\href@noop {} {\
  (\bibinfo {year} {2021}{\natexlab{a}})},\ \Eprint
  {http://arxiv.org/abs/2108.02504} {arXiv:2108.02504 [cond-mat.stat-mech]}
  \BibitemShut {NoStop}%
\bibitem [{\citenamefont {Koenigstein}\ \emph
  {et~al.}(2021{\natexlab{b}})\citenamefont {Koenigstein}, \citenamefont
  {Steil}, \citenamefont {Wink}, \citenamefont {Grossi},\ and\ \citenamefont
  {Braun}}]{Koenigstein:2021rxj}%
  \BibitemOpen
  \bibfield  {author} {\bibinfo {author} {\bibfnamefont {A.}~\bibnamefont
  {Koenigstein}}, \bibinfo {author} {\bibfnamefont {M.~J.}\ \bibnamefont
  {Steil}}, \bibinfo {author} {\bibfnamefont {N.}~\bibnamefont {Wink}},
  \bibinfo {author} {\bibfnamefont {E.}~\bibnamefont {Grossi}}, \ and\ \bibinfo
  {author} {\bibfnamefont {J.}~\bibnamefont {Braun}},\ }\href@noop {} {\
  (\bibinfo {year} {2021}{\natexlab{b}})},\ \Eprint
  {http://arxiv.org/abs/2108.10085} {arXiv:2108.10085 [cond-mat.stat-mech]}
  \BibitemShut {NoStop}%
\bibitem [{\citenamefont {Steil}\ and\ \citenamefont
  {Koenigstein}(2021)}]{Steil:2021cbu}%
  \BibitemOpen
  \bibfield  {author} {\bibinfo {author} {\bibfnamefont {M.~J.}\ \bibnamefont
  {Steil}}\ and\ \bibinfo {author} {\bibfnamefont {A.}~\bibnamefont
  {Koenigstein}},\ }\href@noop {} {\  (\bibinfo {year} {2021})},\ \Eprint
  {http://arxiv.org/abs/2108.04037} {arXiv:2108.04037 [cond-mat.stat-mech]}
  \BibitemShut {NoStop}%
\bibitem [{\citenamefont {Ecker}\ \emph {et~al.}(1989)\citenamefont {Ecker},
  \citenamefont {Gasser}, \citenamefont {Pich},\ and\ \citenamefont
  {de~Rafael}}]{Ecker:1988te}%
  \BibitemOpen
  \bibfield  {author} {\bibinfo {author} {\bibfnamefont {G.}~\bibnamefont
  {Ecker}}, \bibinfo {author} {\bibfnamefont {J.}~\bibnamefont {Gasser}},
  \bibinfo {author} {\bibfnamefont {A.}~\bibnamefont {Pich}}, \ and\ \bibinfo
  {author} {\bibfnamefont {E.}~\bibnamefont {de~Rafael}},\ }\href {\doibase
  10.1016/0550-3213(89)90346-5} {\bibfield  {journal} {\bibinfo  {journal}
  {Nucl. Phys. B}\ }\textbf {\bibinfo {volume} {321}},\ \bibinfo {pages} {311}
  (\bibinfo {year} {1989})}\BibitemShut {NoStop}%
\bibitem [{\citenamefont {Giacosa}\ \emph {et~al.}(2018)\citenamefont
  {Giacosa}, \citenamefont {Koenigstein},\ and\ \citenamefont
  {Pisarski}}]{Giacosa:2017pos}%
  \BibitemOpen
  \bibfield  {author} {\bibinfo {author} {\bibfnamefont {F.}~\bibnamefont
  {Giacosa}}, \bibinfo {author} {\bibfnamefont {A.}~\bibnamefont
  {Koenigstein}}, \ and\ \bibinfo {author} {\bibfnamefont {R.~D.}\ \bibnamefont
  {Pisarski}},\ }\href {\doibase 10.1103/PhysRevD.97.091901} {\bibfield
  {journal} {\bibinfo  {journal} {Phys. Rev. D}\ }\textbf {\bibinfo {volume}
  {97}},\ \bibinfo {pages} {091901} (\bibinfo {year} {2018})},\ \Eprint
  {http://arxiv.org/abs/1709.07454} {arXiv:1709.07454 [hep-ph]} \BibitemShut
  {NoStop}%
\bibitem [{\citenamefont {Adler}(1965{\natexlab{a}})}]{Adler:1964um}%
  \BibitemOpen
  \bibfield  {author} {\bibinfo {author} {\bibfnamefont {S.~L.}\ \bibnamefont
  {Adler}},\ }\href {\doibase 10.1103/PhysRev.137.B1022} {\bibfield  {journal}
  {\bibinfo  {journal} {Phys. Rev.}\ }\textbf {\bibinfo {volume} {137}},\
  \bibinfo {pages} {B1022} (\bibinfo {year} {1965}{\natexlab{a}})}\BibitemShut
  {NoStop}%
\bibitem [{\citenamefont {Adler}(1965{\natexlab{b}})}]{Adler:1965ga}%
  \BibitemOpen
  \bibfield  {author} {\bibinfo {author} {\bibfnamefont {S.~L.}\ \bibnamefont
  {Adler}},\ }\href {\doibase 10.1103/PhysRev.139.B1638} {\bibfield  {journal}
  {\bibinfo  {journal} {Phys. Rev.}\ }\textbf {\bibinfo {volume} {139}},\
  \bibinfo {pages} {B1638} (\bibinfo {year} {1965}{\natexlab{b}})}\BibitemShut
  {NoStop}%
\bibitem [{\citenamefont {Litim}(2001)}]{Litim:2001up}%
  \BibitemOpen
  \bibfield  {author} {\bibinfo {author} {\bibfnamefont {D.~F.}\ \bibnamefont
  {Litim}},\ }\href {\doibase 10.1103/PhysRevD.64.105007} {\bibfield  {journal}
  {\bibinfo  {journal} {Phys. Rev. D}\ }\textbf {\bibinfo {volume} {64}},\
  \bibinfo {pages} {105007} (\bibinfo {year} {2001})},\ \Eprint
  {http://arxiv.org/abs/hep-th/0103195} {arXiv:hep-th/0103195} \BibitemShut
  {NoStop}%
\bibitem [{\citenamefont {Huber}\ and\ \citenamefont
  {Braun}(2012)}]{Huber:2011qr}%
  \BibitemOpen
  \bibfield  {author} {\bibinfo {author} {\bibfnamefont {M.~Q.}\ \bibnamefont
  {Huber}}\ and\ \bibinfo {author} {\bibfnamefont {J.}~\bibnamefont {Braun}},\
  }\href {\doibase 10.1016/j.cpc.2012.01.014} {\bibfield  {journal} {\bibinfo
  {journal} {Comput. Phys. Commun.}\ }\textbf {\bibinfo {volume} {183}},\
  \bibinfo {pages} {1290} (\bibinfo {year} {2012})},\ \Eprint
  {http://arxiv.org/abs/1102.5307} {arXiv:1102.5307 [hep-th]} \BibitemShut
  {NoStop}%
\bibitem [{\citenamefont {Huber}\ \emph {et~al.}(2020)\citenamefont {Huber},
  \citenamefont {Cyrol},\ and\ \citenamefont {Pawlowski}}]{Huber:2019dkb}%
  \BibitemOpen
  \bibfield  {author} {\bibinfo {author} {\bibfnamefont {M.~Q.}\ \bibnamefont
  {Huber}}, \bibinfo {author} {\bibfnamefont {A.~K.}\ \bibnamefont {Cyrol}}, \
  and\ \bibinfo {author} {\bibfnamefont {J.~M.}\ \bibnamefont {Pawlowski}},\
  }\href {\doibase 10.1016/j.cpc.2019.107058} {\bibfield  {journal} {\bibinfo
  {journal} {Comput. Phys. Commun.}\ }\textbf {\bibinfo {volume} {248}},\
  \bibinfo {pages} {107058} (\bibinfo {year} {2020})},\ \Eprint
  {http://arxiv.org/abs/1908.02760} {arXiv:1908.02760 [hep-ph]} \BibitemShut
  {NoStop}%
\bibitem [{\citenamefont {Cyrol}\ \emph {et~al.}(2017)\citenamefont {Cyrol},
  \citenamefont {Mitter},\ and\ \citenamefont {Strodthoff}}]{Cyrol:2016zqb}%
  \BibitemOpen
  \bibfield  {author} {\bibinfo {author} {\bibfnamefont {A.~K.}\ \bibnamefont
  {Cyrol}}, \bibinfo {author} {\bibfnamefont {M.}~\bibnamefont {Mitter}}, \
  and\ \bibinfo {author} {\bibfnamefont {N.}~\bibnamefont {Strodthoff}},\
  }\href {\doibase 10.1016/j.cpc.2017.05.024} {\bibfield  {journal} {\bibinfo
  {journal} {Comput. Phys. Commun.}\ }\textbf {\bibinfo {volume} {219}},\
  \bibinfo {pages} {346} (\bibinfo {year} {2017})},\ \Eprint
  {http://arxiv.org/abs/1610.09331} {arXiv:1610.09331 [hep-ph]} \BibitemShut
  {NoStop}%
\bibitem [{\citenamefont {Mertig}\ \emph {et~al.}(1991)\citenamefont {Mertig},
  \citenamefont {Bohm},\ and\ \citenamefont {Denner}}]{Mertig:1990an}%
  \BibitemOpen
  \bibfield  {author} {\bibinfo {author} {\bibfnamefont {R.}~\bibnamefont
  {Mertig}}, \bibinfo {author} {\bibfnamefont {M.}~\bibnamefont {Bohm}}, \ and\
  \bibinfo {author} {\bibfnamefont {A.}~\bibnamefont {Denner}},\ }\href
  {\doibase 10.1016/0010-4655(91)90130-D} {\bibfield  {journal} {\bibinfo
  {journal} {Comput. Phys. Commun.}\ }\textbf {\bibinfo {volume} {64}},\
  \bibinfo {pages} {345} (\bibinfo {year} {1991})}\BibitemShut {NoStop}%
\bibitem [{\citenamefont {Shtabovenko}\ \emph {et~al.}(2016)\citenamefont
  {Shtabovenko}, \citenamefont {Mertig},\ and\ \citenamefont
  {Orellana}}]{Shtabovenko:2016sxi}%
  \BibitemOpen
  \bibfield  {author} {\bibinfo {author} {\bibfnamefont {V.}~\bibnamefont
  {Shtabovenko}}, \bibinfo {author} {\bibfnamefont {R.}~\bibnamefont {Mertig}},
  \ and\ \bibinfo {author} {\bibfnamefont {F.}~\bibnamefont {Orellana}},\
  }\href {\doibase 10.1016/j.cpc.2016.06.008} {\bibfield  {journal} {\bibinfo
  {journal} {Comput. Phys. Commun.}\ }\textbf {\bibinfo {volume} {207}},\
  \bibinfo {pages} {432} (\bibinfo {year} {2016})},\ \Eprint
  {http://arxiv.org/abs/1601.01167} {arXiv:1601.01167 [hep-ph]} \BibitemShut
  {NoStop}%
\bibitem [{\citenamefont {Shtabovenko}\ \emph {et~al.}(2020)\citenamefont
  {Shtabovenko}, \citenamefont {Mertig},\ and\ \citenamefont
  {Orellana}}]{Shtabovenko:2020gxv}%
  \BibitemOpen
  \bibfield  {author} {\bibinfo {author} {\bibfnamefont {V.}~\bibnamefont
  {Shtabovenko}}, \bibinfo {author} {\bibfnamefont {R.}~\bibnamefont {Mertig}},
  \ and\ \bibinfo {author} {\bibfnamefont {F.}~\bibnamefont {Orellana}},\
  }\href {\doibase 10.1016/j.cpc.2020.107478} {\bibfield  {journal} {\bibinfo
  {journal} {Comput. Phys. Commun.}\ }\textbf {\bibinfo {volume} {256}},\
  \bibinfo {pages} {107478} (\bibinfo {year} {2020})},\ \Eprint
  {http://arxiv.org/abs/2001.04407} {arXiv:2001.04407 [hep-ph]} \BibitemShut
  {NoStop}%
\end{thebibliography}%

\end{document}